\documentclass[11pt,a4paper,fancyhdr]{article}
\usepackage{etex}
\pdfoutput=1
\usepackage[bookmarks=false]{hyperref} %,pagebackref,hidelinks
\usepackage{graphicx}
\date{Version 2, November 2016}
\graphicspath{{pics/}}
\usepackage{mathptmx} 
\usepackage[scaled=.90]{helvet}
\usepackage{courier}
\usepackage[round]{natbib}
\newcommand{\picscale}{1.0}

\usepackage{ifthen}

\usepackage{amsmath}
\usepackage{amssymb}
\usepackage{amsthm}
\usepackage{amsxtra}
\usepackage{thmtools}

\usepackage{listings}

\usepackage{wrapfig}
\usepackage{tcolorbox}
\tcbuselibrary{theorems}
\tcbuselibrary{breakable}

\usepackage{xcolor}
\makeatletter
\def\leftharpoonupfill{$\m@th \mathord\leftharpoonup \mkern-9mu
  \cleaders\hbox{$\mkern-5mu \smash- \mkern-5mu$}\hfill
  \mkern-9mu \smash-$}  % p.357, \leftarrowfill

\def\vglue{\afterassignment\vgl@\skip@=}

\def\overleftharpoonup#1{{%
  \vbox{\ialign{##\crcr % p.359, \overleftarrow
    \leftharpoonupfill\crcr\noalign{\kern-\p@\nointerlineskip\vglue 0.5pt}
    $\hfil\displaystyle{#1}\hfil$\crcr}}}}

\let\hook=\overleftharpoonup  % c'est simple comme bonjour
\makeatother

\usepackage{cancel}
\newcommand\hcancel[2][red]{\setbox0=\hbox{$#2$}%
\rlap{\raisebox{.45\ht0}{\textcolor{#1}{\rule{\wd0}{1pt}}}}#2}
\usepackage{color}
\definecolor{diff.del}{rgb}{0.0, 0.6, 0.13} % for (hacked) latexdiff, the colour of deleted text (dark green) 
\definecolor{diff.add}{rgb}{0.0, 0, 1.0} % for (hacked) latexdiff, the colour of added text (blue)  
\usepackage{accents}
\newlength{\dhatheight}
\newcommand{\dhat}[2][0.35]{%
    \settoheight{\dhatheight}{\ensuremath{\widehat{#2}}}%
    \addtolength{\dhatheight}{-#1ex}%
    \widehat{\vphantom{\rule{1pt}{\dhatheight}}%
    \smash{\widehat{#2}}}}
\newcommand{\dtilde}[2][0.35]{%
    \settoheight{\dhatheight}{\ensuremath{\widetilde{#2}}}%
    \addtolength{\dhatheight}{-#1ex}%
    \widetilde{\vphantom{\rule{1pt}{\dhatheight}}%
    \smash{\widetilde{#2}}}}

\newcommand{\dhatv}[1]{\dhat{#1\,}\!}
\newcommand{\dtildev}[1]{\dtilde{#1\,}\!}

\usepackage[ligature,reserved,shorthand]{semantic}
\newcommand{\varstyle}[1]{\ensuremath{\mathit{#1}}}
\reservestyle{\var}{\varstyle}
\newcommand{\conststyle}[1]{\ensuremath{\mathrm{#1}}}
\reservestyle{\const}{\conststyle}
\newcommand{\bconststyle}[1]{\ensuremath{\mathbf{#1}}}
\reservestyle{\bconst}{\bconststyle}
\newcommand{\wordstyle}[1]{\ensuremath{\operatorname{#1}}}
\reservestyle{\word}{\wordstyle}
\newcommand{\bracketstyle}[1]{\ensuremath{\operatorname{\mathsf{#1}}}}
\reservestyle{\bracket}{\bracketstyle}
\newcommand{\blackboardstyle}[1]{\ensuremath{\mathbb{#1}}}
\reservestyle{\blackboard}{\blackboardstyle}
\newcommand{\sansstyle}[1]{\ensuremath{\mathsf{#1}}}
\reservestyle{\sans}{\sansstyle}

\reservestyle{\labelname}{\conststyle}

\newcommand{\orderstyle}[1]{\ensuremath{\operatorname{\mathsf{#1}}}}
\reservestyle{\order}{\orderstyle}
\newcommand{\Modalitystyle}[1]{\ensuremath{\blackboardstyle{#1}}}
\reservestyle{\Modality}{\Modalitystyle}

\newcommand{\verythinspace}{\hspace{0.5pt}}
\newcommand{\myrel}[1]{\verythinspace{#1}\verythinspace}
\newcommand{\myarith}[1]{\verythinspace{#1}\verythinspace}

\mathlig{--*}{\mathbin{-\!\circledast}}
\mathlig{-**}{\mathbin{-\!\!\circledast}}
\mathlig{-*}{\mathbin{-\!\!\star}}
\mathlig{-/}{\rightharpoondown}
\mathlig{|->}{\mapsto}
\mathlig{||-}{\Vdash}
\mathlig{|-}{\vdash}
\mathlig{|=}{\vDash}
\mathlig{|*}{\exists}
\mathlig{|>}{\overset{\rightarrow}{\lor}}
\mathlig{|/}{\downarrow}
\mathlig{||/}{\Downarrow}
\mathlig{|||}{|}
\mathlig{||}{\|}
\mathlig{|}{\lor}
\mathlig{@*}{\forall}
\mathlig{@+}{\oplus}
\mathlig{@@@@}{@}
\mathlig{@}{\land}
\mathlig{/->}{\hookrightarrow}
\mathlig{->}{\rightarrow}
\mathlig{=>}{\Rightarrow}
\mathlig{==>}{\Rrightarrow}
\mathlig{==}{=}
\mathlig{=@=}{\triangleq}
\mathlig{!/->}{\not\hookrightarrow}
\mathlig{!=>}{\not\Rightarrow}
\mathlig{!=}{\myrel{\neq}}
\mathlig{!|-}{\not\vdash}
\mathlig{!||-}{\not\Vdash}
\mathlig{!|=}{\not\,\vDash}
\mathlig{!|*}{\neg\exists}%necessary before !|
\mathlig{!|}{\not\!\lor}
\mathlig{!@*}{\neg\forall}%necessary before !@
\mathlig{!@}{\not\!\!\land}
\mathlig{!}{\neg}
\mathlig{>=}{\myrel{\geq}}
\mathlig{>->}{\rightarrowtail}
\mathlig{>}{\myrel{>}}
\mathlig{~>}{\rightsquigarrow}
\mathlig{~<>}{\leftrightsquigarrow}
\mathlig{~<}{\leftsquigarrow}
\mathlig{<->}{\leftrightarrow}
\mathlig{<<=}{\Leftarrow}
\mathlig{<=>}{\Leftrightarrow}
\mathlig{<=}{\myrel{\leq}}
\mathlig{<}{\myrel{<}}
\mathlig{:::}{\!\colon\!\!}
\mathlig{::}{\colon\!}
\mathlig{:=}{:=}
\mathlig{=}{\myrel{=}}

\mathlig{*}{\myarith{\times}}
\mathlig{+}{\myarith{+}}
\mathlig{-}{\myarith{-}}
\newcommand{\hstrut}[1]{\rule{#1}{0pt}}
\newcommand{\vstrut}[1]{\rule{0pt}{#1}}

\newcommand {\cols}[1][*{50}{l}]{\begin{array}{#1}}
\newcommand {\sloc}{\end{array}}
\newcommand {\squeezecols}[1]{\setlength{\arraycolsep}{#1}}

\newcommand {\BRA}[1][*{50}{l}]{\left(\!\!\cols[#1]}
\newcommand {\KET}{\sloc\!\!\right)}

\newcommand {\SQBRA}[1][*{50}{l}]{\left[\cols[#1]}
\newcommand {\SQKET}{\sloc\right]}
\newcommand {\BRACE}[1][*{50}{l}]{\left\{\!\!\cols[#1]}
\newcommand {\ECARB}{\sloc\!\!\right\}}

\newcommand {\FRAME}[1][l]{\text{Framed:}\left\langle\begin{array}{#1}}
\newcommand {\EMARF}{\end{array}\right\rangle}

\newcommand{\pindent}{\hstrut{10pt}}
\newenvironment{progindent}[1][*{50}{l}]{\pindent\cols[#1]}{\sloc}

\newcommand{\eqnlabel}[1]{\label{eqn:#1}}
\newcommand{\eqnref}[1]{(\ref{eqn:#1})}

\newcommand{\figlabel}[1]{\label{fig:#1}}
\newcommand{\figref}[1]{fig.\;\ref{fig:#1}}
\newcommand{\Figref}[1]{Fig.\;\ref{fig:#1}}

\newcommand{\seclabel}[1]{\label{sec:#1}}
\newcommand{\secref}[1]{\ifcsname r@sec:#1\endcsname section \ref{sec:#1}\else appendix \ref{appx:#1}\fi}

\newcommand{\appxlabel}[1]{\label{appx:#1}}
\newcommand{\appxref}[1]{\secref{#1}}

\newcommand{\partlabel}[1]{\label{part:#1}}
\newcommand{\partref}[1]{part \ref{part:#1}}

\newcommand{\tablabel}[1]{\label{tab:#1}}
\newcommand{\tabref}[1]{table \ref{tab:#1}}
\newcommand{\Tabref}[1]{Table \ref{tab:#1}}

\newcommand{\defref}[1]{definition \ref{def:#1}}
\newcommand{\Defref}[1]{Definition \ref{def:#1}}

\newcommand{\ruleref}[1]{rule \ref{rule:#1}}
\newcommand{\Ruleref}[1]{Rule \ref{rule:#1}}

\newcommand{\asdlabel}[1]{\label{asd:#1}}
\newcommand{\asdref}[1]{aside \ref{asd:#1}}

\colorlet{assertcolor}{blue!70!white}

\newcommand{\assert}[1]{{\color{assertcolor}\left\{ #1 \right\}}}
\newcommand{\assertd}[1]{{\color{assertcolor}\left\{ #1 \right\}}}

\newcommand{\Assertd}[1]{{\color{assertcolor}\BRACE #1 \ECARB\vspace{3pt}}}
\newcommand{\ipre}[1]{{\color{assertcolor}\left[ #1 \right]}}

\newcommand{\pre}[1]{\assert{#1}\,}

\newcommand{\post}[1]{\,\assert{#1}}

\newcommand{\guarantee}[1]{\setword{guar}\SQBRA[r@{\hspace{2.5pt}}|@{\hspace{2.5pt}}l] #1 \SQKET  \vspace{3pt}}
\newcommand{\emptyguarantee}{\setword{guar}\SQBRA\, \SQKET  \vspace{3pt}}
\newcommand{\rely}[1]{\setword{rely}\SQBRA[r@{\hspace{2.5pt}}|@{\hspace{2.5pt}}l] #1 \SQKET}

\newcommand{\intfspace}[1][2.5pt]{\hspace{#1}}
\newcommand{\interferenced}[3][]{\ensuremath{{\iarg {#1}} #2 \mid #3 }}

\newcommand{\interferenceg}[3][]{\ensuremath{{\iarg {#1}} #2 & \;#3}}
\newcommand{\iarg}[1]{\ifthenelse{\equal{#1}{}}{}{[#1].\,}}

\newcommand{\sometimesincestart}{\lozenge\llap{{\raise.17ex\hbox{$\scriptstyle\sim$}}}}

\newcommand{\alwayssincestart}{\square\llap{{\raise.19ex\hbox{$\scriptstyle\sim$}}}}

\newcommand{\thr}[1]{\begin{array}[t]{c} \text{#1} \vspace{3pt}\\ \cols}
\newcommand{\rht}{\sloc\sloc}
\newcommand{\adcols}{\cols[r@{\intfspace}l]}
\newcommand{\lthr}[1]{\begin{array}[t]{c} \text{#1} \vspace{3pt}\\ \adcols}
\newcommand{\rhtl}{\sloc\sloc}

\newcommand{\macro}[3][]{#2#1 =@= #3}

\usepackage{changepage}

\newcommand{\hbra}{%
\hbox to .9995      \columnwidth{\vrule width0.3mm height 1.8mm depth-0.3mm
                    \leaders\hrule height1.8mm depth-1.5mm\hfill
                    \vrule width0.3mm height 1.8mm depth-0.3mm}}
\newcommand{\hket}{%
\vspace{-.5em}\hbox to .9995 
                   \columnwidth{\vrule width0.3mm height1.5mm
                   \leaders\hrule height0.3mm\hfill
                   \vrule width0.3mm height1.5mm}}

\newsavebox{\hbralab}
\newcommand{\hbral}[1]{%
\hbox to .9995      \columnwidth{\vrule width0.3mm height 1.8mm depth-0.3mm
                    \vrule height1.8mm depth-1.5mm width2mm
			        \raisebox{.6mm}{#1\!}
                    \leaders\hrule height1.8mm depth-1.5mm\hfill
                    \vrule width0.3mm height 1.8mm depth-0.3mm}}

\newenvironment{displayed}[1]{%

\smallskip%
{\noindent\hbral{\textit{\textbf{#1}}}}\nopagebreak%
\begin{adjustwidth}{1.5mm}{1.5mm}
}{%
\nopagebreak \end{adjustwidth}\nopagebreak
\vspace{-.1em}\hket
\smallskip
}

\newcounter{asidectr}
\newenvironment{aside}[1][]{\refstepcounter{asidectr}%
							\begin{displayed}{Aside \arabic{asidectr}%
											 {\ifthenelse{\equal{#1}{}}{}{: #1}}}
						 }%
						 {\end{displayed}}

\usepackage{tikz}
\usetikzlibrary{arrows,shapes}
\usetikzlibrary{matrix,positioning,calc}
\usetikzlibrary{shapes.geometric,decorations.pathreplacing,shapes.misc}
\mathchardef\ordinarycolon\mathcode`\:

\usepackage{dot2texi}

\usepackage[maxfloats=36]{morefloats}

\newcommand{\nonterminal}[1]{\ensuremath{\mathit{#1}}}

\newcommand{\csource}{source}
\newcommand{\ctarget}{target}

\usepackage{proof}
\newcommand{\reasonscale}{0.8}
\newcommand{\inferR}[3][]{\infer[\scalebox{\reasonscale}{$#1$}]{#2}{#3}}

\newcommand{\compilingmark}{(C)}
\newcommand{\problemmark}{(P)}
\newcommand{\semanticsmark}{(S)}

\newtcbtheorem[list inside=defn]{defn}{Definition}%
{top=3pt,bottom=3pt,lowerbox=ignored,colframe=blue!50!black,colback=blue!10!white,fonttitle=\bfseries}{def}
\newcommand{\definition}[3]{\begin{defn}{#1}{#2} #3 \end{defn}}
\newtcbtheorem[list inside=rulee]{rulee}{Rule}%
{top=3pt,bottom=3pt,lowerbox=ignored,colframe=red!50!black,colback=red!10!white,fonttitle=\bfseries}{rule}
\newcommand{\ruledef}[3]{\begin{rulee}{#1}{#2} #3 \end{rulee}}
\newenvironment{colsdefn}
  {\begin{tcolorbox}[before skip=3pt,breakable=true,top=3pt,bottom=0pt,lowerbox=ignored,colframe=blue!50!black,colback=blue!10!white, colbacktitle=blue!20!white]$$\cols[rcll]}
  {\sloc$$\end{tcolorbox}}

\newenvironment{keeptogether}{%
\begin{tcolorbox}[left=0pt,right=0pt,top=0pt,bottom=0pt,leftrule=0pt,rightrule=0pt,toprule=0pt,bottomrule=0pt,colback=white,colframe=white]}{%
\end{tcolorbox}}

\var{vs,rs}
\var{aux,raux}
\var{r1,r2,r3,r4,r5,r6,r7,r8,r9,r10}
\var{C1,C1s,C2,C2s,C3,C3s,P0,P1,P2,Q0,Q1,Q2,R0,R1,R2,S1,S2}
\var{r,r'[\hook{r}],x,x'[\hook{x}],y,y'[\hook{y}],z,z'[\hook{z}],n,n'[\hook{n}],v,v'[\hook{v}]}
\var{vhat[\widehat{v}],xhat[\widehat{x}],yhat[\widehat{y}],zhat[\widehat{z}]}
\var{vtilde[\widetilde{v}],xtilde[\widetilde{x}],ytilde[\widetilde{y}],ztilde[\widetilde{z}]}
\var{ytilde[\widetilde{y}]}
\var{reg}

\var{init[\alpha],final[\omega],beta[\beta],gamma[\gamma],delta[\delta]} % labels for assertions

\var{msg[m],flag[f],msg'[\hook{m}],flag'[\hook{f}],msghat[\widehat{m}],flaghat[\widehat{f}],msgtilde[\widetilde{m}],flagtilde[\widetilde{f}]}
\var{flag1[g],flag1'[\hook{g}],flag1hat[\widehat{g}]} 
\var{latch[lh],latchhat[\widehat{lh}],latch'[\hook{lh}],token[tk],tokenhat[\widehat{tk}],token'[\hook{tk}]}
\var{latch0[lh0],latch1[lh1],latch2[lh2],latch3[lh3],token0[tk0],token1[tk1],token2[tk2],token3[tk3],auxt,raux}

\var{aux0}

\const{nil,true,false}

\bracket[\mathopen]{if,while,While,assert}
\bracket[\mathrel]{then,else,do,until,since}
\bracket[\mathclose]{fi,od}
\bracket{skip}
\bracket{lwsync,sync,isync,eieio,ctrlisync,dep,ddep,adep,lwarx,stwcx[stwcx.],ldrex,strex,CAS}
\bracket{dsb,isb}
\bracket{MFENCE}

\order{lo,Lo,bo,Bo,uo,Uo,so,So,go,Go,rf,Rf,co,Co,po,Po,data,RRdep}

\Modality{U,Univ[U],Bfr[B],sofar[S\mathsf{ofar}],Sofar[S\mathsf{ofar}],ouat[O\mathsf{uat}],Ouat[O\mathsf{uat}],Evw[F\mathsf{andw}],fandw[F\mathsf{andw}],Fandw[F\mathsf{andw}],modality[M\mathsf{odality}],Modality[M\mathsf{odality}]}

\word{sp,Sp,hat,tilde,sat,valid,cv}

\labelname{laba[a],labb[b],labc[c],labd[d],labe[e],labf[f],labg[g],labh[h],labi[i],labj[j]}
\labelname{labk[k],labl[l],labm[m],labn[n],labo[o],labp[p],labq[q],labr[r],labs[s],labt[t]}
\labelname{labu[u],labv[v],labw[w],labx[x],laby[y],labz[z]}
\labelname{labA[A],labB[B],labC[C],labD[D],labE[E],labF[F],labG[G],labH[H],labI[I],labJ[J]}
\labelname{labK[K],labL[L],labM[M],labN[N],labO[O],labP[P],labQ[Q],labR[R],labS[S],labT[T]}
\labelname{intf}

\newcommand{\nvec}[1]{\ensuremath{\overline{\setvar{#1}}}}

\newcounter{rowcount}
\usepackage{mdwlist}
\usepackage{setspace}
\usepackage{schemata}

\usepackage{fancyhdr}
\makeatletter
\renewcommand{\@makefntext}[1]{\setlength{\parindent}{0pt}%
\begin{list}{}{\setlength{\labelwidth}{1em}%
  \setlength{\leftmargin}{\labelwidth}%
  \setlength{\labelsep}{3pt}\setlength{\itemsep}{0pt}%
  \setlength{\parsep}{0pt}\setlength{\topsep}{0pt}%
  \footnotesize}\item[\hfill\@makefnmark]#1%
\end{list}}
\makeatother

\title{New Lace and Arsenic: adventures in weak memory with a program logic (v2)}
\author{Richard Bornat, Middlesex University\ (R.Bornat@mdx.ac.uk) \\
Jade Alglave, Microsoft Research \\
Matthew Parkinson, Microsoft Research}

\renewcommand{\baselinestretch}{1.0}

\begin{document}
%!TEX root = ./Paper.tex
\maketitle
\thispagestyle{empty}
%\begin{center} Version 1.0 \end{center}
\begin{abstract}
\noindent
We describe a program logic for weak memory (also known as relaxed memory). The logic is based on Hoare logic within a thread, and rely/guarantee between threads. It is presented via examples, giving proofs of many weak-memory litmus tests. It extends to coherence but not yet to synchronised assignment (compare-and-swap, load-logical/store-conditional). It deals with conditionals and loops but not yet arrays or heap.

\noindent
The logic uses a version of Hoare logic within threads, and a version of rely/guarantee between threads, with five stability rules to handle various kinds of parallelism (external, internal, propagation-free and two kinds of in-flight parallelism). There are $\mathbb{B}$ and $\mathbb{U}$ modalities to handle propagation, and temporal modalities $\mathsf{since}$, $\mathbb{S}\mathsf{ofar}$ and $\mathbb{O}\mathsf{uat}$ to deal with global coherence (SC per location).

\noindent
The logic is presented by example. Proofs and unproofs of about thirty weak-memory examples, including many litmus tests in various guises, are dealt with in detail. There is a proof of a version of the token ring.
\end{abstract}

\paragraph{A note on authorship}

This paper reports the joint work of its authors. But the words in the paper were written by Richard Bornat. Any opprobrium, bug reports, complaints, and observations about sins of commission or omission should be directed at him.

\paragraph{A note on version 2}

The correspondence with Herding Cats has been clarified. The stability rules have been simplified: in particular the \<sat> and $x=\<xhat>$ tests have been eliminated from external stability checks. The embedding is simplified and has a more transparent relation to the mechanisms of the logic. Definitions of \<U>, \<sofar> and \<ouat> have been considerably altered. The description of modalities and the treatment of termination has been reworked. Many proofs are reconstructed. A comprehensive summary of the logic is an appendix.

\clearpage
\pagestyle{plain}
\pagenumbering{roman}
\renewcommand{\baselinestretch}{0.7}
\tableofcontents
\clearpage
\phantomsection
\renewcommand{\listtheoremname}{List of Definitions}
\addcontentsline{toc}{section}{\listtheoremname}
\tcblistof[\section*]{defn}{\listtheoremname}
\renewcommand{\listtheoremname}{List of Rules}
\addcontentsline{toc}{section}{\listtheoremname}
\tcblistof[\section*]{rulee}{\listtheoremname}
\renewcommand{\baselinestretch}{1.0}
\clearpage
\pagestyle{fancy}
\pagenumbering{arabic}
\part{Background}
\thispagestyle{empty}
\clearpage 
\hstrut{50pt} \vspace{-38pt} \\
%!TEX root = ./Paper.tex

\section{Introduction}

\paragraph{But you see $\dots$ Well, insanity runs in my family. It practically gallops.}

Concurrent programming is in a weird place. We write our programs in sequential languages (e.g. C~\citep{C2011}, C++~\citep{C++2014} or Java~\citep{Java2015}) but modern hardware (e.g. ARM~\citep{ARM2010}, IBM Power~\citep{Power2009}, Intel x86~\citep{x862009} and any GPU) is mostly parallel: commands may not be carried out in the order they are written, and writes to memory may be substantially reordered. The consequences of parallelism are hidden from the naive user: ARM, Power and x86, for example, each ensure that a single isolated thread\footnote{As is common in the weak-memory research community, we use `thread' to mean `hardware thread', the smallest unit of processing supported by hardware. A single processor may have many cores, each of which may support several threads, each thread with its own registers.} -- one which doesn't share memory with other threads -- is the same initial state to final state mechanism that it would be on sequential hardware. But threads which share memory can observe the non-sequential behaviour of each other. %In concurrent execution parallelism really matters. 

The effects of instruction-execution reordering and reordering of writes to memory are called \emph{weak memory} or \emph{relaxed execution}. We say weak memory.

\paragraph{Oh darling, just because Teddy's a little strange, that doesn't mean $\dots$}

To write a shared-memory concurent program that is to run on weak-memory hardware, a programmer must understand where reordering needs to be constrained, and know how to constrain it with special instructions called \emph{fences} or \emph{barriers}, and/or with special assembly-code instruction-ordering tricks. There's a persuasive argument, put forward in~\citep{BoehmAdveYouDontKnowJack2012}, that shared-memory concurrency is so complicated that users should remain ignorant and leave the difficult work to the experts, but those who write concurrency libraries, define languages or write compilers need formal tool support to ensure that they are getting the constraints right.

%%%But there are so far no corresponding program logics so far, let alone a logic
%%%that encompasses relaxed architectures in general. There is an attempt to
%%%define C so that programs can run with comparable effects on any relaxed
%%%architecture~(C11 ref) but like earlier work on the Java memory model~{(refs
%%%to JMM and Sevcik at least)} it appears that it may be flawed~{(Alexey et
%%%al?)}, and in any case it has no program logic.\richard{But oh dear Victor's
%%%C11 separation logic?}
%%
In the absence of formal models for concurrent systems, concurrent programs are often described as if they were to execute on a sequentially-consistent machine~\citep{LamportHowtomakeamultiprocessorcomputerthatcorrectlyexecutesmultiprocessprograms1979}, i.e. as if their executions were interleavings of the executions of separate threads. To execute those programs on weak-memory hardware we need to know whether and how to constrain the hardware, and how to do so on different architectures and on different implementations of those architectures.

Our focus is on programming, and on verification of programs. We worked for three years on a program logic for weak memory, which we called Lace logic. We targeted IBM Power, because it is one of the weakest and one of the most studied of modern architectures. But our logic ended up too complicated to be palatable, and with too many corner-cases for us to believe that a proof of soundness was possible. 

Nevertheless we learnt a great deal about how to reason about programs under weak memory. Inspired by~\citet{CrarySullivanACalculusforRelaxedMemory2015} we realised belatedly that we might make a simpler logic which deals with a wide variety of weak-memory architectures. This paper presents our first steps in that direction; this version retraces and revises those initial steps. 

To back up, and somewhat validate, the work presented in this paper, we have a proof checker called Arsenic~\citep{ArsenicOnGitHub} and tens of worked examples. All our example proofs have been mechanically transliterated from mechanically-checked proofs (sometimes \LaTeX-tweaked a little to fit on the page). 

Another milestone would be a proof that our New Lace logic is sound with respect to existing models of concurrent systems, such as those put forward by~\citep{AlglaveetalHerdingcats2014}. We are not there yet; the concrete boots of soundness remain unfilled. One of our difficulties is that our logic is \emph{thread-local} or \emph{compositional} -- a program proof is made up of separate proofs of its threads -- whereas those modals deal with \emph{global} properties of program executions. A soundness proof of our logic would likely have two stages: a proof of soundness against a compositional model of thread execution, and a proof of soundness of that model against global models of hardware. We have not yet attempted either part of that exercise.

%%\paragraph{Outline} 
%%TODO

%!TEX root = ./Paper.tex

\section{Choose Your Own Adventure}

Weak memory has surprising properties, and to deal with it our logic must be innovative. We expect that most of our readership will have a reasonable familiarity with Hoare logic, but not so many will have experience of concurrency proofs and rely/guarantee, and very few will understand weak memory. We have decided to present through examples, explaining as we go, rather than to present logic first and examples later. Some topics are dealt with incrementally, with forward references to the complete description later in the paper; it's not necessary to follow those references on a first reading.

We cover a great deal of ground over very many pages, and we expect that our readers may wish, at first, to skip some of our exposition. Our main purpose is to say what our logic is, via worked examples. We also have to justify our design, to explain the semantic background, show how it relates to existing hardware, and explain how we deal with some of the difficult problems which are well-known to researchers in the area. To help you navigate, we have labelled our sections and subsections with particular marks.
\begin{itemize*}
\item Unmarked sections are mainstream;
\item \compilingmark{}: how to relate laced programs to machine-code;
\item \problemmark{}: how we deal with well-known problematic examples;
\item \semanticsmark{}: questions of semantics.
\end{itemize*}
For the brave and knowledgeable, there is a summary of our definitions and rules in \appxref{summary}. There is also \appxref{embedding} which explains the embedding of our logic into the Z3 SMT engine to produce the Arsenic proof-checker.

%!TEX root = ./Paper.tex

\var{reg,var,expr,command,seq,label,knot}
\begin{table}
\caption{programming language}
\centering
\begin{tabular}{lcl}
\multicolumn{3}{c}{\hstrut{50pt}} \\
\nonterminal{program} &{:}{:}{=}& \nonterminal{passert}\ (\ \nonterminal{threads}\ )\ \nonterminal{passert} \vspace{3pt} \\

\nonterminal{passert} &{:}{:}{=}& `\{'\ \nonterminal{label}\ ${:}$\ \nonterminal{assertion}\ `\}' \vspace{3pt} \\

\nonterminal{threads} 
					&{:}{:}{=}& \nonterminal{thread}\\
				  	&$\mid$& \nonterminal{thread} `$\mid\mid$' \nonterminal{threads}\vspace{3pt} \\
					
\nonterminal{thread}	
					&{:}{:}{=}	& \nonterminal{seq}\vspace{3pt} \\
					
\nonterminal{seq} 	&{:}{:}{=}& \nonterminal{command} \\
				  	&$\mid$& \nonterminal{command}\text{ ; } \nonterminal{seq}  \\
					&$\mid$		&  \vspace{3pt} \\

\nonterminal{command} 
					&{:}{:}{=}& \nonterminal{label}\ {:}\ \<skip> \\
				  	&$\mid$& \nonterminal{label}\ {:}\ \nonterminal{assign} \\
				  	&$\mid$& \nonterminal{label}\ {:}\ \<assert>\ \nonterminal{assertion}  \\
				  	&$\mid$& \<if>\ \nonterminal{label}\ {:}\ \nonterminal{expr}\ \<then>\ \nonterminal{seq}\ \<fi> \\
				  	&$\mid$& \<if>\ \nonterminal{label}\ {:}\ \nonterminal{expr}\ \<then>\ \nonterminal{seq}\ \<else>\ \nonterminal{seq}\ \<fi> \\
				  	&$\mid$& \<while>\ \nonterminal{label}\ {:}\ \nonterminal{expr}\ \<do>\ \nonterminal{seq}\ \<od> \\
				  	&$\mid$& \<do>\ \nonterminal{seq}\ \<until>\ \nonterminal{label}\ {:}\ \nonterminal{expr} \vspace{3pt} \\

\nonterminal{assign} 
					&{:}{:}{=}& \nonterminal{write}\\
				  	&$\mid$& \nonterminal{read}\\
				  	&$\mid$& \nonterminal{calculation}\vspace{3pt} \\

\nonterminal{write} 
					&{:}{:}{=}& \nonterminal{var}\ {:}{=}\ \nonterminal{expr} \\
				  	&$\mid$& \nonterminal{var}\ {:}{=}\ \nonterminal{expr}\ [\ ,\ \nonterminal{auxexpr}\ ]$^{\text{+}}$\\
				  	&$\mid$& \nonterminal{var}\ [\ ,\ \nonterminal{auxvar}\ ]$^{n}$
							 \ {:}{=}\ \nonterminal{expr}\ [\ ,\ \nonterminal{auxexpr}\ ]$^{n}$ \\
				  	&$\mid$& \nonterminal{auxvar}\ {:}{=}\ \nonterminal{auxexpr} \vspace{3pt} \\

\nonterminal{read} 
					&{:}{:}{=}&  \nonterminal{reg}\ {:}{=}\ \nonterminal{var} \\
					&$\mid$&     \nonterminal{reg}\ [\ ,\ \nonterminal{auxreg}\ ]$^{\text{+}}$
							 \ {:}{=}\ \nonterminal{var}\vspace{3pt} \\

\nonterminal{calculation} 
					&{:}{:}{=}&  \nonterminal{reg}\ {:}{=}\ \nonterminal{expr}  \\
					&$\mid$&     \nonterminal{auxreg}\ {:}{=}\ \nonterminal{auxexpr} \vspace{3pt}\\

\nonterminal{reg} 	&{:}{:}{=}& 
\begin{minipage}{300pt}
any name starting with `r', but not starting with ``raux''
\end{minipage} \vspace{3pt} \\

\nonterminal{auxreg} 	&{:}{:}{=}& 
\begin{minipage}{300pt}
any name starting with ``raux''
\end{minipage} \vspace{3pt} \\

\nonterminal{label} &{:}{:}{=}& 
\begin{minipage}{300pt}
any name
\end{minipage} \vspace{3pt} \\

\nonterminal{var} &{:}{:}{=}& 
\begin{minipage}{300pt}
\schema{}{any name starting with a lower-case letter other than `r', but not starting with ``aux''}
\end{minipage} \vspace{3pt} \\

\nonterminal{auxvar} &{:}{:}{=}& 
\begin{minipage}{300pt}
any name starting with ``aux''
\end{minipage} \vspace{3pt} \\

\nonterminal{logicalvar} &{:}{:}{=}& 
\begin{minipage}{300pt}
any name starting with an upper-case letter
\end{minipage} \vspace{3pt} \\

\nonterminal{expr} &{:}{:}{=}& 
\begin{minipage}{300pt}
\schema{}{any expression whose operands are regs, constants (integers, \<true>, \<false>) and logvars, and whose operators are arithmetic or Boolean-arithmetic ($!, @, |, =>, <=>$)}
\end{minipage} \vspace{3pt} \\

\nonterminal{expr} &{:}{:}{=}& 
\begin{minipage}{300pt}
as expr, but allowing auxregs
\end{minipage} \vspace{3pt} \\

\nonterminal{assertion} &{:}{:}{=}& 
%$\left\{
\begin{minipage}{300pt}
\schema{}{any predicate-calculus expression; may include modalities \<Bfr>, \<U>, \<sofar>, \<ouat>, \<since> and coherence assertions $\setvar{var}_{c}(\nonterminal{expr},\nonterminal{expr})$%; inequalities using registers and constants.
}\end{minipage}
\end{tabular}
\tablabel{proglang}
\end{table}

\begin{table}
\caption{Laced threads}
\centering
\begin{tabular}{lcl}
\multicolumn{3}{c}{\hstrut{50pt}} \\
\nonterminal{thread}	
					&{:}{:}{=}	& \nonterminal{guarantee}\ \nonterminal{seq}\ \nonterminal{threadpost}\ \nonterminal{rely} \vspace{3pt} \\

\nonterminal{guarantee}
					&{:}{:}{=}	& \setword{guar}\ `['\ \nonterminal{interferences}\ `]' \vspace{3pt} \\
					
\nonterminal{rely}	&{:}{:}{=}	& \setword{rely}\ `['\ \nonterminal{interferences}\ `]' \\
					&$\mid$		&  \vspace{3pt} \\

\nonterminal{interferences}	
					&{:}{:}{=}	& \nonterminal{interference} \\
					&$\mid$		& \nonterminal{interference}\ ;\ \nonterminal{interferences} \\
					&$\mid$		&  \vspace{3pt} \\

\nonterminal{interference}
					&{:}{:}{=}	& \nonterminal{assertion}\ `$\mid$'\ \<var>\ ${:=}$\ \<expr> \\
					&$\mid$		& `['\ \nonterminal{names}\ `]'\ .\ \nonterminal{assertion}\ `$\mid$'\ \<var>\ ${:=}$\ \<expr> \vspace{3pt} \\

\nonterminal{names}
					&{:}{:}{=}	& \nonterminal{name} \\
					&$\mid$		& \nonterminal{name}\ , \nonterminal{names} \vspace{3pt} \\
\nonterminal{threadpost}
					&{:}{:}{=}	& \nonterminal{simpleknot} \\
					&$\mid$		&  \vspace{3pt} \\
						  
\nonterminal{simpleknot}	
					&{:}{:}{=}	& `\{'\ \nonterminal{stitches}\ `\}' \\
					&$\mid$		& \nonterminal{simpleknot}\ $|$\ \nonterminal{simpleknot} \\
					&$\mid$		& \nonterminal{simpleknot}\ $|>$\ \nonterminal{simpleknot} \vspace{3pt} \\

\nonterminal{knot}	&{:}{:}{=}	& \nonterminal{simpleknot} \nonterminal{intfpre} \\
					&$\mid$		&  \vspace{3pt} \\

\nonterminal{intfpre} 
					&{:}{:}{=}	& {}`['\ \nonterminal{assertion}\ `]' \\
					&$\mid$		&  \vspace{3pt} \\

\nonterminal{stitches} 
					&{:}{:}{=}	& \nonterminal{stitch} \\
					&$\mid$		& \nonterminal{stitch}\ ;\ \nonterminal{stitches} \\
					&$\mid$		&  \vspace{3pt} \\

\nonterminal{stitch} 
					&{:}{:}{=}	& \nonterminal{labref}\ \nonterminal{ordering}\ \nonterminal{sourcepost} \ 
									${:}$\ \nonterminal{assertion} \vspace{3pt} \\

\nonterminal{labref}
					&{:}{:}{=}	& \nonterminal{label} \\
					&$\mid$		& \nonterminal{label}$_{t}$ \\
					&$\mid$		& \nonterminal{label}$_{f}$\vspace{3pt} \\
										
\nonterminal{ordering}
					&{:}{:}{=}	& \<lo> \\
					&$\mid$		& \<bo> \\
					&$\mid$		& \<uo> \\
					&$\mid$		& \<go> \vspace{3pt} \\
					
\nonterminal{sourcepost}
					&{:}{:}{=}	& `\{'\ \nonterminal{assertion}\ `\}' \\
					&$\mid$		&  \vspace{3pt} \\

\nonterminal{command} &{:}{:}{=}& \nonterminal{knot}\ \nonterminal{label}\ {:}\ \<skip> \\
				  &$\mid$& \nonterminal{knot}\ \nonterminal{label}\ {:}\ \nonterminal{assign} \\
				  &$\mid$& \nonterminal{knot}\ \nonterminal{label}\ {:}\ \<assert>\ \nonterminal{assertion}  \\
				  &$\mid$& \<if>\ \nonterminal{knot}\ \nonterminal{label}\ {:}\ \nonterminal{expr}\ \<then>\ \nonterminal{seq}\ \<fi> \\
				  &$\mid$& \<if>\ \nonterminal{knot}\ \nonterminal{label}\ {:}\ \nonterminal{expr}\ \<then>\ \nonterminal{seq}\ 
				           \<else>\ \nonterminal{seq}\ \<fi> \\
				  &$\mid$& \<while>\ \nonterminal{knot}\ \nonterminal{label}\ {:}\ \nonterminal{expr}\ \<do>\ \nonterminal{seq}\ \<od> \\
				  &$\mid$& \<do>\ \nonterminal{seq}\ \<until>\ \nonterminal{knot}\ \nonterminal{label}\ {:}\ \nonterminal{expr} \vspace{3pt} \\

\end{tabular}
\tablabel{lacedprograms}
\end{table}

\section{\semanticsmark{} \compilingmark{} Notation}

Our programs are written in a notation given in \tabref{proglang}. Our notation is inspired by that of separation logic~\citep{ReynoldsSeparationlogic2002}, with local registers and shared variables, and assuming no aliasing.\footnote{Though inspired by separation logic to concentrate on single writes, we do not yet approach the problem of the heap.} Programs manipulate integers and Booleans. Access to memory is only in assignment commands, and each assignment makes at most one memory access. There are various forms of auxiliary assignments.

Each \<expr>, in an assignment or a control expression, is entirely local, referring only to non-auxiliary registers, constants and logical variables. Auxiliary \nonterminal{auxexpr}s may include auxiliary variables, but may only be assigned to auxiliary registers and variables. Tuples consisting of an \<expr> and one or more \nonterminal{auxexpr}s may be assigned to a regular variable; the value assigned may be read simultaneously into a regular register and one or more auxiliary registers. A regular variable and one or more auxiliary variables may be assigned simultaneously.

Primitive commands -- \<skip>s, assignments, \<assert>s and control expressions -- are labelled; labels are unique within a thread. There are no declarations. Programs are usually preceded by a labelled initial assertion and followed by a labelled final assertion.

Ours is not a high-level language, and in this paper we do not consider optimisations that might be attempted by a compiler. We note that optimisation in weak memory is problematic (see \secref{LibraryAbstractionandOptimisation}). What we have is close to the machine, a sort of architecture-independent assembly code with structured commands.

In introducing our examples we deal with the annotations that are necessary for programming, introducing them incrementally and explaining their meanings. For reference we give in \tabref{lacedprograms} a definition of laced threads -- threads adorned with constraints for weak memory, with an interference guarantee and an optional interference rely. It is not necessary on first reading to study this table, nor at first to understand the concepts of constraint, ordering, interference, guarantee and rely. Some of the non-terminals in \tabref{lacedprograms} take their definitions from \tabref{proglang}.

\section{\semanticsmark{} Related work}

A very great deal of work has been expended on models of weak memory, compilation for weak memory and verification of programs in languages designed to work on weak memory. We do not attempt a complete bibliography, but we list work which is immediately relevant to ours.

\begin{itemize}
\item On models of hardware: 
\citet{AlglaveetalTheSemanticsofPowerandARMMultiprocessorMachineCode2009},
\citet{SarkaretalThesemanticsofx86CCmultiprocessormachinecode2009},
\citet{OwensSarkarSewellABetterx86MemoryModelx86TSO2009},
\citet{Sewelletalx86TSOARigorousandUsableProgrammersModelforx86Multiprocessors2010}.
\citet{alglave2011litmus},
\citet{SarkaretalThesemanticsofx86CCmultiprocessormachinecode2009,SarkaretalUnderstandingPOWERmultiprocessors2011},
\citet{AlglaveMarangetStabilityinWeakMemoryModels2011},
\citet{AlglaveetalFencesinWeakMemoryModelsExtendedVersion2012},
\citet{MadorHaimetalAnAxiomaticMemoryModelforPOWERMultiprocessors2012},
\citet{AlglaveetalHerdingcats2014,AlglaveetalGPUConcurrencyWeakBehavioursandProgrammingAssumptions2015}.

\item On the semantics of programming languages for weak memory: 
\citet{BoehmAdveFoundationsoftheC++concurrencymemorymodel2008}, 
\citet{BattyetalMathematizingC++concurrency2011}, 
\citet{BoehmAdveYouDontKnowJack2012}, 
\citet{BoehmDemskyOutlawingghosts2014}, 
\citet{SarkaretalSynchronisingCstrokeCplusplusandPOWER2012}, 
\citet{VafeiadisetalCommonCompilerOptimisationsareInvalidintheC11MemoryModel2015},
\citep{VafeiadisFormalreasoningabouttheC11weakmemorymodel2015}.

\item On verification of programs in weak-memory-savvy languages:
\citet{battyetalLibraryabstractionC++},
\citet{LahavVafeiadisOwicki-GriesReasoningforWeakMemoryModels2015},
\citet{TassarottiDreyerVafeiadisVerifyingread-copy-updateinalogicforweakmemory2015},
\citet{TuronVafeiadisDreyerGPSnavigatingweakmemorywithghostsprotocolsandseparation2014},
\citet{VafeiadisNarayanRelaxedseparationlogicaprogramlogicforC11concurrency2013},
\citet{NorrisDemskyCDSchecker:checkingconcurrentdatastructureswrittenwithC/C++atomics2013},
\citet{AlglaveetalSoftwareVerificationforWeakMemoryviaProgramTransformation2013},
\citet{BurckhardtMusuvathiEffectiveProgramVerificationforRelaxedMemoryModels2008},
\citet{AtigBouajjaniBurckhardtMusuvathiOntheverificationproblemforweakmemorymodels2010},
\citet{AtigBouajjaniBurckhardtMusuvathiWhatsDecidableaboutWeakMemoryModels2012}.

\item On compilation for weak memory:
\citet{KupersteinVechevYahavAutomaticinferenceofmemoryfences2010},
\citet{LindenP.WolperAverificationbasedapproachtomemoryfenceinsertion2011},
\citet{SevciketalRelaxed-memoryconcurrencyandverifiedcompilation2011},
\citet{VafeiadisZappaNardelliVerifyingFenceEliminationOptimisations2011},
\citet{AtigBouajjaniParlatoGettingRidofStoreBuffersinTSOAnalysis2011},
\citet{AbdullaetalCounterExampleGuidedFenceInsertion2012},
\citet{AbdullaetalMemoraxaPreciseandSoundToolforAutomaticFenceInsertion2013}.

\end{itemize}

Our work relates particularly to the work on models of hardware. We take logical reasoning techniques for SC concurrency developed by \citet{Owickithesis1975}, \citet{OwickiGriesAxiomaticprooftechniqueforparallelprograms1976,OwickiGriesVerifyingpropertiesofparallelprograms1976} and \citet{JonesRelyguarantee1983}, and, using intuitions derived from those models and the experiments which underpin them, port those techniques to weak memory. 

We hope that we'll be able in the future to add various features to our logic, as was done for SC concurrency with Separation Logic~\citep{ReynoldsSeparationlogic2002,CalcagnoOHearnYangConcAbsSepLogic2006},
Concurrent Abstract Predicates~\citep{DinsdaleYoungetalConcurrentAbstractPredicates2010}, and RGSep~\citep{VafeiadisParkinsonMarriageofrelyguaranteeandseparation2007,Vafeiadisthesis2007}.

\subsection{Related work in program logics}

We note several attempts to reason about programs running on TSO hardware (x86, notably, is TSO) and to reason about C11 programs.

In reasoning about TSO, \citet{RidgeArely-guaranteeproofsystemforx86-TSO2010} made a program logic in which the write buffers described by~\citet{OwensSarkarSewellABetterx86MemoryModelx86TSO2009} were dealt with explicitly. \citet{WehrmanBerdineAproposalforweak-memorylocalreasoning2011} abstracted the frame buffers in a substructural logic which used a frame rule to handle the effect of delayed writes to memory. Sieczkowskie \citep{sieczkowski2013towards,SieczkowskietalAseparationlogicforfictionalsequentialconsistency2015} abstracted the write buffers entirely in a temporal logic which employed a `fiction of sequential consistency', building on the concurrent abstract predicates notion~\citep{DinsdaleYoungetalConcurrentAbstractPredicates2010}.

In reasoning about C11, \citet{battyetalLibraryabstractionC++} showed that the possibility of `satisfaction cycles' made abstraction of libraries which use relaxed atomics impossible (see also our discussion in \secref{localspec}). Vafeiadis et al.~\citep{%
VafeiadisZappaNardelliVerifyingFenceEliminationOptimisations2011,%
VafeiadisNarayanRelaxedseparationlogicaprogramlogicforC11concurrency2013,%
TuronVafeiadisDreyerGPSnavigatingweakmemorywithghostsprotocolsandseparation2014,%
VafeiadisetalCommonCompilerOptimisationsareInvalidintheC11MemoryModel2015,%
VafeiadisFormalreasoningabouttheC11weakmemorymodel2015,LahavVafeiadisOwicki-GriesReasoningforWeakMemoryModels2015,%
TassarottiDreyerVafeiadisVerifyingread-copy-updateinalogicforweakmemory2015}
have reasonably therefore focussed on the release-acquire fragment of C11. Their logic effectively covers the same area as the \<bo>, \<lo> fragment which we deal with in sections \ref{sec:bolobegins} to \ref{sec:boloends} of \partref{LacingEmbroideryModality} of this paper. But their reach is far beyond ours: they employ the `ownership transfer' notion of separation to deal with release-acquire and are able then to make proofs of important algorithms, especially including RCU~\citep{McKenneyExploitingDeferredDestruction2004}. We envy their work.

%%\begin{table}
%%\caption{Relation to other logics}
%%\centering
%%\begin{tabular}{c|c|c}
%%\multicolumn{3}{c}{\hstrut{50pt}} \\
%%Tom Ridge's TSO logic & & \\
%%Josh and Ian's TSO logic & & \\
%%iCAP-TSO & & \\
%%RSL+GPS & & \\
%%\end{tabular}
%%\tablabel{otherlogics}
%%\end{table}
%%More precisely \jade{this is where Matt knows =)}

\section{\compilingmark{} A logic for all seasonings}

%%On a sequentially consistent (SC) machine there are assertions at each
%%semicolon; we employ explicit \emph{ordering requirements} between commands,
%%which we call \stitches. Each \stitch{} bears an assertion. Our \stitches, take
%%inspiration from the \cats{} work~\cite{amt14}, as well as the \rmc{}
%%work~\cite{}. $\<lo>$, $\<bo>$, and $\<uo>$.

\begin{table}
\caption{The orderings of lace logic and their equivalents in the cats model and in hardware}
\centering
\begin{tabular}{r|c|c|c|c|}
\multicolumn{5}{c}{\hstrut{50pt}} \\
\multicolumn{1}{c}{}   		
		&  \multicolumn{1}{c}{cats} 	
					& \multicolumn{1}{c}{x86} 			
									& \multicolumn{1}{c}{Power} 		& \multicolumn{1}{c}{ARM} \\
\cline{2-5}
\<lo>	& ii/ci		& implicit	 	& \begin{minipage}{150pt}\raggedright
									    \hstrut{50pt}\\[-2pt]
										implicit (register assign$->$use);
										implicit (same location, read$->$assign / assign$->$read / assign$->$assign);
										otherwise data (calc$->$calc),
										addr (read$->$read, read$->$assign) 
										or rfi-addr (assign$->$assign, assign$->$read)	
										or ctrl-\<isync> (control-expr$->$read)\\[-6pt]
										\hstrut{50pt}
									  \end{minipage}		& \begin{minipage}{150pt}\raggedright
																			implicit (register assign$->$use);
																			implicit (same location, assign$->$read);
																			otherwise otherwise data (calc$->$calc),
																			addr (read$->$read, read$->$assign) 
																			or rfi-addr (assign$->$assign, assign$->$read)	
																			or ctrl-\<isb> (read$->$read)
									  									  \end{minipage}\\
\cline{2-5}
\<bo>	& lwfence 	& implicit	 	& \vstrut{8pt}\<lwsync>				& dsb \\
\cline{2-5}
\<uo>	& ffence 	& MFENCE 		& \<sync>							& dsb \\
\cline{2-5}
\<go>	& ic 		& implicit		& ctrl								& \begin{minipage}{150pt}\raggedright
									    								    \hstrut{50pt}\\[-2pt]
																			implicit (same location, read$->$assign or assign$->$assign);
																			otherwise rfi-ctrl or ctrl\\[-6pt]
																			\hstrut{50pt}
									  									  \end{minipage}\\

\cline{2-5}
\end{tabular}
\tablabel{orderingequivalences}
\end{table}
Our logic is based on programs explicitly laced with constraints, each imposing an ordering between commands or between control expressions and commands. Those orderings are introduced and explained in \partref{LacingEmbroideryModality}. \Tabref{orderingequivalences} shows our understanding (not yet backed up by a soundness proof) of the correspondence of our orderings to the treatment in~\citep{AlglaveetalHerdingcats2014} and to programs running on the three major processor architectures. We expect that there will be correspondences to GPU architectures as well, but we don't know them yet.

In various places we have referred to compilation of laced programs into machine code. We expect that the correspondences in \tabref{orderingequivalences} would be the basis of any such compiler, but we don't expect it to be easy to make a compiler. In particular fence/barrier placement is a hard problem~\citep{BurckhardtAlurMartinCheckFence2007,HuynhRoychoudhuryMemorymodelsensitivebytecodeverification2007,AtigBouajjaniParlatoGettingRidofStoreBuffersinTSOAnalysis2011}, but dealing with multiple constraints in precondition knots is also by no means straightforward.

\clearpage
\part{Lacing, embroidery, modality}
\partlabel{LacingEmbroideryModality}
\thispagestyle{empty}
\clearpage
%!TEX root = ./Paper.tex

\section{Sequentially-consistent concurrency}

Before we consider our treatment of weak memory, we explain the basics of our reasoning, using an example program which will be the basis of many later examples.

\emph{Sequentially-consistent} -- henceforth \emph{SC} -- execution~\citep{LamportHowtomakeamultiprocessorcomputerthatcorrectlyexecutesmultiprocessprograms1979}, is what machines did before weak memory upset the applecart. Within a thread, SC execution follows the \emph{sequential order} -- henceforth \<so> -- of commands. Within in a program, thread executions are \emph{interleaved}. Because there is a single shared memory, each read $r:=x$ in the interleaving takes its value from the latest preceding write $x:=E$, no matter which thread made that write, or from the initial state, if there is no preceding write. If we prefix the interleaving with writes which establish the initial state, we can say uniformly that each read in the interleaving takes its value from the latest preceding write to the same variable.

\begin{figure}
\centering
$$\cols[c]
\assert{\<init>::\<flag>=0} \vspace{3pt} \\
\BRA[l||l]
\thr{Sender (0)}
  \<laba>:: \<msg>:=1; \\
  \<labb>:: \<flag>:=1
\rht
&
\thr{Receiver (1)}
  \<labc>:: \<r1>:=\<flag>; \\
  \<labd>:: \<r2>:=\<msg>
\rht
\KET \vspace{3pt} \\
\assert{\<final>::(1:::\<r1>)=1=>(1:::\<r2>)=1} \\
\sloc$$
\caption{MP: shared-variable message-passing}
\figlabel{MPunlaced}
\end{figure}

Consider SC execution of the two-thread program in \figref{MPunlaced}. The initial-state assertion \<init>\footnote{In our logic the program precondition must be labelled, in order that we can refer to in when placing constraints on execution. By convention we use the label \<init>. The program postcondition doesn't really need a label, but for symmetry we require the label \<final>.} requires that execution begins with 0 in variable \<flag>. The sender (thread 0) writes 1 into message variable \<msg> and then writes 1 into flag variable \<flag>; the receiver (thread 1) reads from \<flag> and then from \<msg>. If the execution interleaving is `\<init>,\<laba>,\<labb>,\<labc>,\<labd>' then command \<labc> must read 1 from the write of command \<labb>, and \<labd> must read 1 from the write of \<laba>. In all other interleavings \<labc> precedes \<labb> and so will read 0 from \<init>, and command \<labd> may or may not read 1, depending on whether or not it follows a. So we can only be certain that \<labd> reads 1 when \<labc> has already read 1. The final-state assertion \<final> states this formally, using the notation $(i:::\!\setvar{reg})$ for `register $\setvar{reg}$ in thread $i$'. 

Informal reasoning about execution of this example is easy, but in more complex cases it is very difficult to consider all possible interleavings. We'd like to prove formally that every execution of the program which starts in a state satisfying precondition \<init> will, \emph{if} it terminates, do so in a state satisfying postcondition \<final>. We shan't be able to prove that every execution starting from \<init> \emph{must} terminate: even in such a simple example, our logic doesn't run to it. 

\begin{figure}
\squeezecols{3.5pt}
\centering
$$\cols[c]
\assert{\<init>::\<flag>=0} \vspace{3pt}\\
\BRA[l||l]
\thr{Sender (0)} 
	\guarantee{\interferenceg{\<true>}{\<msg>:=1} \\
			   \interferenceg{\<msg>=1}{\<flag>:=1} 
			  } \\
	\quad \assert{\<true>} \\ \<laba>::\<msg>:=1; \\ 
	\quad \assert{\<msg>=1} \\ \<labb>::\<flag>:=1 \\
	\quad \assert{\<true>}
\rht
&
\thr{Receiver (1)} 
	\emptyguarantee \\
	\quad \assert{\<flag>=1=>\<msg>=1} \\ \<labc>::\<r1>:=\<flag>; \\ 
	\quad \assert{\<r1>=1=>\<msg>=1} \\ \<labd>::\<r2>:=\<msg> \\
	\quad \assert{\<r1>=1=>\<r2>=1}
\rht
\KET \vspace{3pt}\\ 
\assert{\<final>:: (1:::\<r1>)=1=>(1:::\<r2>)=1} 
\sloc$$
\caption{SC proof of MP}
\figlabel{MPSCproof}
\end{figure}

\subsection{Interference and stability in SC}

Our SC-concurrency proof of MP is shown in \figref{MPSCproof}. There are state assertions enclosed in braces before, between and after the commands of each thread; in each triple $\pre{P}C\post{Q}$ precondition $P$, command $C$ and postcondition $Q$ are formally related by the logic's command-execution rules. In multi-thread programs the assertions are further constrained, because one thread may change the values of shared variables used by another. 

Interleaving, from the point of view of a thread, means that other threads' commands execute in the gaps before, between and after its own command executions, just where we place the proof assertions. We say that those interpolated executions are \emph{parallel} with those assertions, and \emph{interfere} with them by assigning values to variables on which they may depend. We require that assertions are \emph{stable} against all parallel command executions, which means that when the assertion is \<true>, none of the interference of any possibly-interpolated command can make it \<false>. Only the initial and final assertions (\<init> and \<final> in \figref{MPSCproof}, for example) are automatically stable, because they hold when no thread is executing.

If we can invent stable assertions which also fit the command-execution rules then we have a proof of concurrent execution. That's tricky, as we shall see, but it's what we need for a shared-variable concurrency proof. 

Those are the bare bones of the longest-established method for verifying concurrent programs, a modification of Hoare logic~\citep{HoareAxiomaticbasisforcomputerprogramming1969} devised by Owicki and Gries~\citep{Owickithesis1975,OwickiGriesAxiomaticprooftechniqueforparallelprograms1976,OwickiGriesVerifyingpropertiesofparallelprograms1976}. Jones's rely/guarantee~\citep{JonesRelyguarantee1983} is a later development, abstracting interference from the commands and assertions of a thread, designed as a refinement method for concurrent-program development. The word `stability' is more modern still~\citep{XudeRoeverHeTherely-guaranteemethodforverifyingsharedvariableconcurrentprograms1997}, supplanting the earlier `interference-free'. 

We take from Jones the ideas of a \emph{guarantee} which summarises the interference made by a thread, and a \emph{rely} which summarises the interference of other threads which it must withstand. Like Owicki and Gries we concentrate on verifying proofs and we focus on the interference of individual assignments. 

\subsection{A checklist for proofs}

It's not enough simply to present a proof like that in \figref{MPSCproof}. We have to check that its assertions fit the command-execution rules and are stable against interference from other threads. Our checklist is in \tabref{checklist}: check \ref{impcheck} works on single commands within a thread; check \ref{stabcheck} is about interference and stability; check \ref{guarcheck} is about making a summary of a thread's interference, which simplifies stability reasoning in other threads.

\begin{table}
\caption{Per-thread concurrency-proof checklist\vspace{10pt}}
\centering
\vstrut{3pt}\vspace{-3pt}\\
\setcounter{rowcount}{0}
\begin{tcolorbox}
\begin{tabular}{l}
$\!\!\!\!\!\!$\refstepcounter{rowcount}\therowcount.
\label{impcheck}Check that each assertion is implied by the postcondition of the immediately-preceding component. \\
$\!\!\!\!\!\!$\refstepcounter{rowcount}\therowcount.
\label{stabcheck}Check that each assertion is stable under the interference of parallel assignments. \\
$\!\!\!\!\!\!$\refstepcounter{rowcount}\therowcount.
\label{guarcheck}Check that the stable interference of each variable assignment is included in the thread's guarantee. \\ 
\end{tabular}
\end{tcolorbox}
\tablabel{checklist}
\end{table}

In each thread of each of our proofs there is a summary of its interference in a \emph{guarantee} clause, prefixed `\setword{guar}'. In an SC guarantee we write $\interferenced{Q}{x:=E}$ to describe the interference of an assignment $x:=E$ with precondition $Q$: that description tells us that when $Q$ holds the guaranteeing thread may execute $x:=E$. The guarantee of the sender in \figref{MPSCproof} claims that it makes two kinds of interference; the empty guarantee of the receiver claims that it doesn't make any interference at all. Each of those claims must be checked.

For check \ref{impcheck} we need to calculate postconditions of components from their preconditions. In this example the only components are assignments and the initial assertion. (In later examples the control expressions of conditionals and loops are also components and have pre and postconditions: see \secref{structcom}.) 
\definition{Postcondition of a component}{postconditions}{The postcondition of the initial-state assertion is the assertion itself. The postcondition of \<skip> is its precondition. The postcondition of  $\<assert> P$ is the conjunction of its precondition and $P$. For assignments we use \emph{strongest postconditions}~\citep{DijkstraDisciplineofprogramming1976}.}
Strongest postconditions are given by Floyd's assignment rule~\citep{FloydAssigningmeaningstoprograms1967}. We adapt it to the various forms of our assignments: a write $x:=E$; a calculation $r:=E$; and a read $r:=x$.
\definition{Strongest postcondition \setword{sp}}{sp}{$$\cols[lcl]
\<sp>(P,\;x:=E) &=@=& P[x\backslash \<x'>]@x=E \\
\<sp>(P,\;r:=E) &=@=& P[r\backslash \<r'>]@r=E[r\backslash \<r'>] \\
\<sp>(P,\;r:=x) &=@=& P[r\backslash \<r'>]@r=x
\sloc$$ } 
-- $\<x'>$ is the pre-assignment value of $x$ and $\<r'>$ the pre-assignment value of $r$. Floyd used existential quantification to quotient references to pre-assignment values, but since \<x'> and \<r'> can't appear in our proof assertions we don't need to quotient. Note that according to the syntax of our language in \tabref{proglang} there can be no occurrences of $x$ in $E$ or $r$ in $x$, which means that we can avoid substituting in either in the first and third lines of the definition. %We shall see in \secref{coherence} that \defref{sp} needs some modification to handle temporal assertions, but for the time being it is perfectly adequate.

Interference in SC execution happens when an assertion already holds and a variable assignment from another thread muscles in, changing the values of variables which underpin it.\footnote{In this paper we deal only with registers and shared variables. In the future we hope to deal also with the heap, building perhaps on the techniques of RGSep~\citep{VafeiadisParkinsonMarriageofrelyguaranteeandseparation2007,Vafeiadisthesis2007}.} Registers are private to a thread, so register assignments don't interfere with other threads' assertions. %The SC stability rule is
\ruledef{SC stability}{SCstability}
{$$\cols
	\text{In SC execution, } P \text{ is stable against } \interferenced{Q}{x:=E} \text{ if} \vspace{2pt}\\
	\quad \<sp>(P@Q,\;x:=E)=>P
\sloc$$}
-- suppose $P$ holds; suppose $Q$ holds as well, so $x:=E$ may be executed; we require that if it is executed, then $P$ still holds afterwards. 

There's a circularity in the notion of stability: thread A's assertions must be stable against the interference of thread B, using B's precondition-assertions in the stability check, and B's assertions must be stable against the interference of A, using A's precondition-assertions. This is quite ok semantically, and we shan't worry about its subtleties. Circularity can simplify proofs, so we make use of it.

\subsection{Checking the proof of SC message-passing}

We begin our check in the sender of \figref{MPSCproof}. Before command \<laba> executes, \<true> certainly holds, so we can use it as a precondition. Following our checklist we note: the initial state \<init> implies \<true> (check \ref{impcheck}); there isn't any interference, because the receiver's guarantee is empty (check \ref{stabcheck}); there's an entry in the sender's guarantee with \<laba>'s precondition \<true> and \<laba>'s command $\<msg>:=1$ (check \ref{guarcheck}). 

To deal with the precondition of command \<labb>, we must calculate the postcondition of \<laba>. \Defref{sp} gives the strongest postcondition $\<true>@\<msg>=1$, which implies $\<msg>=1$, the precondition we've used for \<labb> (check \ref{impcheck}); stability is immediate, as before (check \ref{stabcheck}); there's an entry in the guarantee with \<labb>'s command and precondition (check \ref{guarcheck}). And that's it for the sender: we've checked all its assertions with checks \ref{impcheck} and \ref{stabcheck}; we've checked all its interference with check \ref{guarcheck}.

The sender writes once to \<msg> and once to \<flag>, but its guarantee doesn't say that. From the guarantee it appears that it might interfere repeatedly and at any time with $\<msg>:=1$, and it might interfere repeatedly with $\<flag>:=1$ whenever $\<msg>=1$. We shall see that the receiver's assertions are stable despite this overstatement of interference. Sometimes it's useful to give more constraining interference preconditions, and to make sure that interference occurs only once, but not in this example. 

The clever bit of our proof is the choice of $\<flag>=1=>\<msg>=1$ as precondition for command \<labc> in the receiver. Informally, considering the execution of the sender, when \<flag> is 1, \<msg> must already be 1. Formally, we must show (check \ref{impcheck}) that the precondition of \<labc> is implied by the initial state: that's easy because the initial state gives $\<flag>=0$, so the precondition is trivially satisfied. Then we have to show (check \ref{stabcheck}) that the assertion is stable against the guaranteee of the sender. \Ruleref{SCstability} generates two checks, one for each of the sender's interferences. The first check is passed because the interference produces $\<msg>=1$:
\begin{equation}
\cols
	& \<sp>((\<flag>=1=>\<msg>=1)@\<true>,\;\<msg>:=1) \vspace{2pt}\\
=	& (\<flag>=1=>\<msg'>=1)@\<true>@\<msg>=1 \\
=> 	& \<msg>=1 \\
=> 	& \<flag>=1=>\<msg>=1
\sloc
\eqnlabel{MPSCstab1}
\end{equation}
and the second because the interference can only happen when $\<msg>=1$ holds already:
\begin{equation}
\cols
	& \<sp>((\<flag>=1=>\<msg>=1)@\<msg>=1,\;\<flag>:=1) \vspace{2pt}\\
=	& (\<flag'>=1=>\<msg>=1)@\<msg>=1@\<flag>=1 \\
=> 	& \<msg>=1 \\
=> 	& \<flag>=1=>\<msg>=1
\sloc
\eqnlabel{MPSCstab2}
\end{equation}
We don't have to put anything in the receiver's guarantee because \<labc>  assigns to a register (check \ref{guarcheck}).

The strongest postcondition of command \<labc>, by \defref{sp}, is 
%\begin{equation}
$(\<flag>=1=>\<msg>=1)@\<r1>=\<flag>$,
%\eqnlabel{MPSCthread1post1}
%\end{equation}
which implies $\<r1>=1=>\<msg>=1$, the precondition of \<labd> (check \ref{impcheck}); that in turn is stable against the first line of the sender's guarantee because $\<msg>=1$ is stable, and against the second line because it doesn't mention \<flag> (check \ref{stabcheck}); there's nothing to put in the guarantee for a register assignment (check \ref{guarcheck}).

We need a postcondition for the receiver, to justify the final assertion \<final>. The postcondition of \<labd> is
%\begin{equation}
$(\<r1>=1=>\<msg>=1)@\<r2>=\<msg>$,
%\eqnlabel{MPSCthread1post2}
%\end{equation}
which implies $\<r1>=1=>\<r2>=1$, which is stable because it doesn't mention any variables at all. And that's it for the receiver: we've checked the assertions with checks \ref{impcheck} and \ref{stabcheck}; there weren't any variable assignments for check \ref{guarcheck}.

The final assertion \<final> is derived from the conjunction of the postconditions of the two threads, annotated where necessary to make it clear which thread's registers we're talking about. Informally, we use just the receiver's postcondition, because the sender's postcondition \<true> contributes nothing.

We have proved that under SC execution, and provided that it terminates, MP will do just what we think it ought to. We've been through the proof in great detail to show how the method works. Weak-memory execution introduces interesting new kinds of ordering and parallelism, but we shall use the same method, and the same checklist, to check our weak-memory proofs.

\section{The so tree and program order po}

In straight-line threads like those of \figref{MPunlaced}, sequential order \<so> is just textual order. In more complicated cases conditionals must be expanded and loops unrolled to generate a potentially infinite \<so> tree of command instances (see sections \ref{sec:conditionals} and \ref{sec:loops}). %The operation is entirely syntactic, each control-expression instance generating two arms in the tree. If there are loops the tree will necessarily be infinite. 

Every SC execution of a thread must follow a path in the \<so> tree, but there may be many executions that follow the same path, differing according to the values in their registers and variables. \emph{Program order} \<po> orders command-instance \emph{executions} within a thread execution -- i.e. \<po> is \<so> with values. 

\section{Weak-memory executions}
\seclabel{weakmemoryexecutions}

Weak-memory execution of a single thread is constrained to follow a \emph{reordering} of a path in the thread's \<so> tree. In our logic there are no other implicit constraints on the execution of a thread. That means that there can be significant parallelism between command executions even within a single thread, and the job of the programmer is to describe explicitly what constraints are needed to define the (usually partial) order they require. 

Weak memory doesn't rely on interleaving of executions to give values to reads. Instead writes are \emph{propagated} between threads -- but not necessarily in any particular order and not necessarily to all threads in the same order or at the same time. There is no longer a global interleaving or a single global shared-memory state, and at any instant a thread may not yet have been affected by all the writes made by other threads. In our logic we suppose that propagation isn't entirely unconstrained, but the issue is a little subtle, and we defer a full discussion until section \ref{sec:SCloc}. Most of the time it's reasonable to suppose that there are no implicit constraints on the reordering of propagation. 

\begin{figure}
\centering
\includegraphics[scale=\picscale]{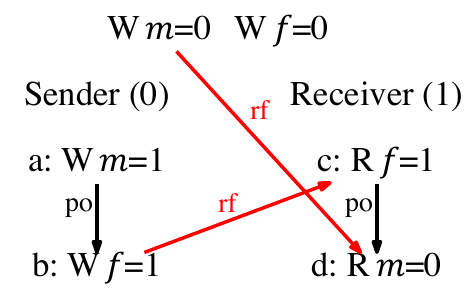}
\caption{MP litmus execution}
\figlabel{MPlitmus}
\end{figure}

So weak memory is not SC, and we illustrate that by presenting three classic \emph{litmus-test} examples~\citep{alglave2011litmus}. In each case a program which works under SC is observed to execute anomalously in weak memory. The first example is MP: on Power, on ARM and on Nvidia GPUs, experiment shows that the receiver can read 0 from \<msg> even though it reads 1 from \<flag>. The offending execution is pictured in \figref{MPlitmus}. The nodes are writes (W) and reads (R); the first row shows the writes that create the initial state (for simplicity, litmus tests initialise all variables); the two columns show the write and read actions of the threads; and the diagram explicitly shows \<rf> arrows from write to read. Note that the diagram shows \<po> ordering in each thread, but remember that the hardware takes little account of \<po>.

The diagram shows one \emph{possible} weak-memory execution of MP, which happens sometimes but not always (and that makes it, in John Reynolds' words, the worst kind of bug). It doesn't happen under SC, as we proved it couldn't and as generations of programmers who have used the algorithm know very well. But in weak memory the command executions of the sender can be reordered so that \<flag> is written before \<msg>; the command executions of the receiver can be reordered to read \<msg> before \<flag>; the sender's \<flag> write can be propagated to the receiver before its \<msg> write; or there can be some combination of some or all or none of those things. To make the program work reliably in weak memory we have to impose constraints on execution and propagation reordering, described in the next section.

\begin{figure}
\centering
$$\cols[c]
\assert{\<init>::x=y=0} \vspace{3pt} \\
\BRA[l||l]
\thr{Thread 0}
  \<laba>:: \<r1>:=y; \\
  \<labb>:: x:=1
\rht
&
\thr{Thread 1}
  \<labc>:: \<r1>:=x; \\
  \<labd>:: y:=1
\rht
\KET \vspace{3pt} \\
\assert{\<final>::!((0:::\<r1>)=1@(1:::\<r1>)=1)} 
\sloc$$
\caption{LB: a race with read before write}
\figlabel{LB}
\end{figure}

\begin{figure}
\centering
\includegraphics[scale=\picscale]{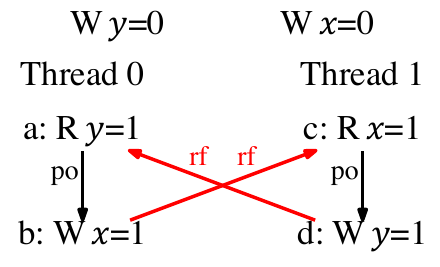}
\caption{LB litmus execution}
\figlabel{LBlitmus}
\end{figure}

\begin{figure}
\centering
$$\cols[c]
\assert{\<init>::x=0@y=0} \vspace{3pt} \\
\BRA[l||l]
\thr{Thread 0}
  \<laba>:: x:=1; \\
  \<labb>:: \<r1>:=y
\rht
&
\thr{Thread 1}
  \<labc>:: y:=1; \\
  \<labd>:: \<r1>:=x
\rht
\KET \vspace{3pt} \\
\assert{\<final>::!((0:::\<r1>)=0@(1:::\<r1>)=0)} 
\sloc$$
\caption{SB: a race with write before read}
\figlabel{SBunlaced}
\end{figure}

\begin{figure}
\centering
\includegraphics[scale=\picscale]{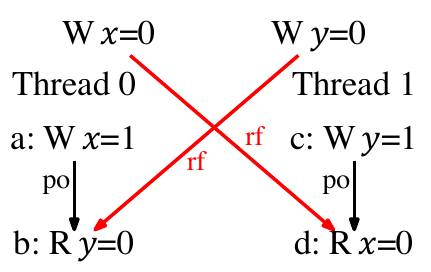}
\caption{SB litmus exécution}
\figlabel{SBlitmus}
\end{figure}

Our second example is LB, in \figref{LB}. Each thread reads from a variable -- $\<x>$ in thread 0, $\<y>$ in thread 1 -- \<so>-before it writes to the other variable. In SC it would be impossible that they could each read from the other's write: command \<labc> can only read 1 in an `\<init>,\<laba>,\<labb>,\<labc>,\<labd>' interleaving, in which \<laba> must read 0; vice-versa \<laba> can only read 1 in an `\<init>,\<labc>,\<labd>,\<laba>,\<labb>' interleaving, in which \<labc> must read 0. But in weak memory it can and does happen, just because command executions can be reordered within the threads. \Figref{LBlitmus} pictures the offending execution, which is observed on Power, ARM and Nvidia GPUs (at least). We deal with the problem in \secref{LB}.

Our third example is SB, in \figref{SBunlaced}. This time each thread writes to one variable -- $\<x>$ in thread 0, $\<y>$ in thread 1 -- \<so>-before reading from the other variable. In any SC interleaving one thread's write has to come first, and then the other thread's read must see 1. This code is part of Dekker's mutual-exclusion algorithm~\citep{EWD35-1}. In weak memory both threads can read 0, because of execution and/or propagation delays (even if each thread executes its commands in order, each may execute its read before the other's write has managed to reach it). The execution in \figref{SBlitmus} is observed on several architectures, including x86, Power and ARM. We show how we deal with the problem in \secref{SBetc}.

\section{Lacing to constrain weak memory, embroidery to make a proof}
\seclabel{bolobegins}

Weak memory, as we've described it, has very few constraints on reordering of execution and propagation. Any particular architecture may impose constraints of its own, but a programmer often needs to impose still more constraints by using special architecture-dependent instructions (barriers, fences) and/or coding devices (address and data dependencies). Programming is tricky because the effects of reordering are subtle, the architectural constraints are  subtle, and the effects of special instructions and coding devices are subtle. 

Following \citet{CrarySullivanACalculusforRelaxedMemory2015} we abstract from machine-dependent constraints and programming devices and \emph{lace} our programs with explicit architecture-independent constraints. Lace constraints can then be translated into programming devices for a particular architecture, or ignored if the architecture already imposes that constraint. 

Just like the architecture-dependent programming devices, lacing applies only within a thread, and only between components that are already \<so> ordered. Lacing strengthens the partial order of execution and/or propagation of the thread.

Lace logic's constraints apply between labelled components of a thread, and they are ordered -- \<laba>$->$\<labb>, \<labc>$->$\<labd>, etc. -- each with a \emph{\csource} and a \emph{\ctarget}. In straight-line threads like the MP sender and receiver, labelled components are the initial assertion and primitive commands, but we shall see that we also label control expressions in conditionals and loops.

There are four kinds of constraint -- \<lo>, \<bo>, \<uo> and \<go> -- each imposing a particular kind of ordering. In laced programs each target component has a \emph{knot} of constraints, enclosed in braces like a Hoare-logic precondition, each constraint giving an ordering and a source. In principle a knot may include several constraints, but in our first examples each will have only one.

\begin{figure}
\squeezecols{3pt}
\centering
$$\cols[c]
\assert{\<init>::\<flag>=0} \vspace{3pt} \\
\BRA[l||l]
\lthr{Sender (0)}
  \assertd{\<lo> \<init>} & \<laba>:: \<msg>:=1; \\
  \assertd{\<bo> \<laba>} & \<labb>:: \<flag>:=1
\rhtl
&
\lthr{Receiver (1)}
  \assertd{\<lo> \<init>} & \<labc>:: \<r1>:=\<flag>; \\
  \assertd{\<lo> \<labc>} & \<labd>:: \<r2>:=\<msg>
\rhtl
\KET \vspace{3pt} \\
\assert{\<final>::(1:::\<r1>)=1=>(1:::\<r2>)=1} 
\sloc$$
\caption{MP laced for weak memory}
\figlabel{MPlaced}
\end{figure}

Weak-memory execution is a tricky concept because commands don't `execute' all at once. Instead we talk of \emph{elaboration} of components and \emph{propagation} of writes. Elaboration is whatever is required to establish a component's postcondition in a thread's local state; we suppose that elaboration completes all at once. Variable assignments create writes, and for a variable assignment propagation follows elaboration; we don't say that it follows immediately, and we don't suppose that propagation completes all at once in every destination thread.

An \emph{elaboration order} \<lo> constraint ensures that its {\csource} elaborates before its {\ctarget}; as a consequence the source's postcondition is available to the precondition of the target. \<Lo> is the least we can ask for in order to be able to reason about states within a thread. In \figref{MPlaced}, command \<laba> is \<lo>-after the initial state \<init>, and \<init>, \<labc> and \<labd> are \<lo>-ordered. 

\<Lo> doesn't control propagation: it can ensure that assignments are elaborated in \<so> order, but their writes might still be propagated in another order, and they might be propagated in different orders to different destinations. A \emph{before order} \<bo> constraint controls both propagation and elaboration. Like \<lo> it imposes elaboration order between source and target, and in addition it ensures that the writes which underpin the source's postcondition will be propagated to any thread before the writes made by the target.\footnote{This wording hides a rather complicated operational mechanism. Not all the writes will be propagated to all threads, but no source-postcondition write will be propagated to any particular thread after the target write is propagated to that same thread. In our discussion the distinction between `in order' and `not out of order' will not matter until \secref{PMS}. Program logic, in this and in many other cases, abstracts and simplifies.} 

\<Bo> is the least we can ask for in order to control propagation. In \figref{MPlaced}, commands \<laba> and \<labb> are \<bo> ordered, so they will be elaborated in \<so> order, echoing SC execution, and the $\<msg>=1$ write will be propagated to the receiver before the $\<flag>=1$ write, echoing the effects of SC interleaving.

\emph{Universal order} \<uo> gives still more control over propagation, ensuring immediate propagation of the source-postcondition writes. It is discussed in \secref{SBetc}. 

\emph{Propagation order} \<go> ensures that the source is elaborated before the target's writes are propagated. It is discussed in \secref{localspec}.

Constraint orderings compose: $\langle\<lo>;\<lo>\rangle$ is \<lo>; both $\langle\<lo>;\<bo>\rangle$ and $\langle\<bo>;\<lo>\rangle$ are \<bo>; each of $\langle\<lo>;\<uo>\rangle$, $\langle\<bo>;\<uo>\rangle$, $\langle\<uo>;\<lo>\rangle$; $\langle\<uo>;\<bo>\rangle$ is \<uo>%; $\langle\<go>;\<uo>\rangle$ and $\langle\<uo>;\<go>\rangle$ are \<uo>
. Orderings nest: \<lo> includes \<bo> and \<uo>; \<bo> includes \<uo>; \<go> is on its own.

\begin{figure}
\squeezecols{3pt}
\centering
$$\cols[c]
\assert{\<init>::\<flag>=0} \vspace{3pt} \\
\BRA[l||l]
\lthr{Sender (0)}
  \multicolumn{2}{l}{\guarantee{\interferenceg{\<true>}{\<msg>:=1} \\ \interferenceg{\<Bfr>(\<msg>=1)}{\<flag>:=1}}} \\
  \assertd{\<lo> \<init>:: \<true>} & \<laba>:: \<msg>:=1;\\
  \assertd{\<bo> \<laba>:: \<Bfr>(\<msg>=1)} & \<labb>:: \<flag>:=1
\rhtl
&
\lthr{Receiver (1)}
  \multicolumn{2}{l}{\emptyguarantee} \\
  \assertd{\<lo> \<init>:: \<flag>=1=>\<msg>=1} & \<labc>:: \<r1>:=\<flag>;\\
  \assertd{\<lo> \<labc>:: \<r1>=1=>\<msg>=1} & \<labd>:: \<r2>:=\<msg> \\
  \assertd{\<lo> \<labd>:: \<r1>=1=>\<r2>=1}
\rhtl
\KET \vspace{3pt} \\
\assert{\<final>::(1:::\<r1>)=1=>(1:::\<r2>)=1} \\
\sloc$$
\caption{A weak memory proof of MP}
\figlabel{MPproof}
\end{figure}

\subsection{Embroidered constraints and the \setModality{B} modality}
\seclabel{embroideryandB}

In Hoare-logic proofs for SC we write assertions before, between and after commands. In Lace logic weak-memory proofs we \emph{embroider} constraints with assertions. \Figref{MPproof} is an embroidered version of \figref{MPlaced}; note the embroidered thread-postcondition knot after command \<labd> in the receiver. A novelty in this proof is the use of the \<Bfr> modality on the \<bo> constraint \<laba>$->$\<labb>, and also in the guarantee entry generated by \<labb>. The \<Bfr> modality is explained below, but note that if we ignore the \<Bfr>s then the assertions and the interference descriptions are just the same as those in the SC proof of \figref{MPSCproof}.

There is an important difference between SC assertions and weak-memory assertions. SC assertions are about the state of the global shared store, whereas weak-memory assertions are about a \emph{thread-local view} of the store, more or less about the values in the asserting thread's cache. It's entirely possible, for example, that at some instant in a weak-memory execution of MP, even when constrained as in \figref{MPlaced}, $\<msg>=1$ holds in the sender and $\<msg>=0$ in the receiver. That couldn't be true in SC.

The precondition of a command, in simple cases, is  the conjunction of its knot's embroideries (for a fuller discussion, dealing with disjunction of knots and \<go> constraints, see sections \ref{sec:conditionals}, \ref{sec:loops} and \ref{sec:localspec}). The postcondition of a component is derived, as in SC proofs, according to \defref{postconditions}.

In an SC proof each assertion is implied by the preceding component's postcondition, and is in turn the precondition of the next component. In a weak-memory proof there's a lacier connection. The embroidery of a constraint depends on the postcondition of the constraint's source, but the relationship depends on the kind of constraint. 

\ruledef{Inheritance of embroidery}{inheritance}{%!TEX root = ./Paper.tex
If a constraint's source postcondition is $P$ and the constraint's embroidery is $Q$ then we require \vspace{-5pt}
\begin{itemize*}
\item on an \<lo> or \<go> constraint, $P=>Q$;
\item on a \<bo> constraint, for some $R$, $P=>R$ and $\<Bfr>(R)=>Q$;
\item on a \<uo> constraint, for some $R$, $P=>R$ and $\<U>(R)=>Q$.
\end{itemize*}}
In many cases a \<bo> constraint will be embroidered $\<Bfr>(R)$ where $P=>R$. A similar remark applies to \<uo> constraints, but we defer further discussion of \<uo> until \secref{SBetc}.

The embroidery on a constraint must be stable against external and (we shall see) internal interference. That's a weak-memory novelty: embroidery stability is per constraint, not per pre or postcondition. That's not the only kind of stability in weak memory: interference preconditions in the guarantee must also be checked for stability -- but we'll come to that.

The \<Bfr> modality polices the propagation-ordering requirement of the \<bo> constraint. 
\begin{itemize}
\item $\<Bfr>(P)$ in a precondition asserts that $P$ has held continuously since $\<Bfr>(P)$ was established in a \<bo> constraint; 
\item \interferenced{\<Bfr>(Q)}{x:=E} in a guarantee asserts that the writes which underpin $Q$ will be propagated to an interfered-with thread before the interfering write $x=E$.
\end{itemize}

\begin{aside}[(S) (C) Whence the \setModality{B} modality?]
Weak-memory hardware constrains propagation by special \emph{barrier} instructions. A \<bo> constraint between two commands will require, on sufficiently-weak hardware such as Power and ARM, a barrier instruction to be executed between those commands. The hardware will then ensure that the writes made or witnessed before the barrier will be propagated to any particular thread before those made after the barrier (more precisely, writes made after the barrier will not be propagated before those made or witnessed before).

Hoare logic puns between program variables and similarly-named logical variables. Lace logic puns also between assertions and writes. If the (stable) assertion $P$ holds just before a barrier, then the thread has made or witnessed\footnote{Or might as well have witnessed \dots Once again program logic simplifies and abstracts} the writes which gave variables the values implied by that precondition. So long as $P$ continues to hold after the barrier we can use it to stand for those writes. $\<Bfr>(P)$ on a constraint asserts that $P$ has held \emph{continuously} since a barrier was executed and therefore the writes which underpin it will be propagated before any write made by the target of the constraint. Note that \<Bfr> it is a particular instance of the \<since> modality (\secref{since}).
\asdlabel{whenceB}
\end{aside}

Because the $P$ assertion must hold continuously since $\<Bfr>(P)$ was established, $\<Bfr>(P)$ has special treatment in substitution -- i.e. in computing strongest postconditions according to \defref{sp}.
\definition{Substitution and the \setModality{B} modality}{Bsubst}
{$$\cols[rcll] \<Bfr>(P)[x\backslash \<x'>] 	&==& \hook{\<Bfr>(P)}@P[x\backslash \<x'>] \vspace{2pt}\\
\hook{\<Bfr>(P)}@P 				&=>& \<Bfr>(P) \vspace{2pt}  \\
\<Bfr>(P)[r\backslash \<r'>] 	&==& \<Bfr>(P[r\backslash \<r'>])
 \sloc$$}
As before a hook means `in the previous state'. Then the first line of \defref{Bsubst} tells us how to substitute for a variable, and the second line says that if $P$ is unchanged then $\<Bfr>(P)$ continues to hold. The \<Bfr> modality is about writes to shared variables, so register substitution sees through it. %These definitions fit with the fact that $\<Bfr>(P)$ means `$P$ has held since a barrier event'.

%We note some important properties of \<Bfr>-modal assertions 
\definition{Properties of the \setModality{B} modality}{Bproperties}
{$$\cols[rcll]
\<Bfr>(P)			&=>& P  \\
\<Bfr>(P@Q)			&==& \<Bfr>(P)@\<Bfr>(Q)  \\
\<Bfr>(\<Bfr>(P))	&==& \<Bfr>(P) \\
\<Bfr>(P|Q) 		&<<=& \<Bfr>(P)|\<Bfr>(Q)	\\
\<Bfr>(P) 			&<<=& P						& \text{(if } P \text{ doesn't mention variables)} \\
\<Bfr>(P|Q) 		&=>& \<Bfr>(P)|\<Bfr>(Q)	& \text{(if } P \text{ or } Q \text{ don't mention variables)} \\
(P=>Q)				&=>& \<Bfr>(P)=>\<Bfr>(Q)	& \text{(if } P=>Q \text{ is a tautology)}
\sloc$$}
The first line of this definition allows a \<Bfr>-modal assertion in an interference precondition to assert that $P$ holds when the interference is received. These properties also apply, as we shall see, to the \<U> and \<Sofar> modalities, because all three are, behind the scenes, versions of the `since' modality of temporal logic and based on a $@*$ quantification.

When all the occurrences of variables in an interference precondition are inside \<Bfr> modalities, as they are in \figref{MPproof} and will be in all our examples until \secref{interferenceNstability}, we can use a simple treatment of stability, equivalent to the SC stability \ruleref{SCstability}. 
\ruledef{External stability against \setModality{B}-modal interference}{EXTstabilityBmodal}
{$$\cols
	\text{Constraint assertion } P \text{ is stable against external interference \interferenced{Q}{x:=E},} \\
	\text{ where every variable occurrence in } Q \text{ is enclosed in a \<Bfr> modality, if }\vspace{2pt}\\
	\qquad \<sp>(P@Q,\;x:=E)=>P
\sloc$$}
We shall see in \secref{interferenceNstability} that this is a special case of the general external stability \ruleref{EXTstability}.

\subsection{The same game on a new pitch}

The checklist of \tabref{checklist} is carefully worded to suit both SC and weak-memory concurrency proofs, in order to emphasise the continuity between earlier work and our own, but the switch from between-component assertions to constraint embroidery requires some adjustment. In weak memory check \ref{impcheck} applies to {\csource} postcondition and constraint embroidery, not entire preconditions as with SC, shifting the stability requirement to constraint embroidery. We have to allow for \ruleref{inheritance} wrapping \<bo> constraint embroidery in a \<Bfr> modality and \<uo> embroidery with \<U>. But despite all that, in principle it's the same checklist as before: check that preconditions derive from preceding postconditions, check that they are stable, check that the action of variable assignments is reflected in the guarantee. 

We are playing the same rely/guarantee Owicki-Gries game that we played with SC, in a world where there are new kinds of relationship between components and, as we shall see, new forms of parallelism and corresponding new instabilities. 

\section{Checking the proof of weak-memory message passing}

The lacing of \figref{MPlaced} is very tight. Commands must elaborate after the initial state is established, and in \<so> order in each thread. Propagation may not be reordered. We seem to have reimposed much of the ordering that SC imposed. It looks plausible that the example will work in weak memory. Plausible isn't proof, but the proof is very easy, given the work we already did for the SC case and the fact that \ruleref{EXTstabilityBmodal} lets us use the SC stability calculation.

In the sender of \figref{MPproof}, the check \ref{impcheck} tests are exactly as in the SC proof, even given the different treatment of the \<bo> constraint and its \<Bfr> modality: 
\begin{itemize*}
\item on the \<init>{$->$}\<laba> \<lo> constraint $\<flag>=0=>\<true>$; 
\item on the \<laba>{$->$}\<labb> \<bo> constraint $\<true>@\<msg>=1=>\<msg>=1$ and $\<Bfr>(\<msg>=1)=>\<Bfr>(\<msg>=1)$.
\end{itemize*}
Stability in check \ref{stabcheck} is immediate in each case because the receiver's guarantee is empty. Check \ref{guarcheck} is passed because each variable assignment, with its precondition (the conjunction of its knot's embroideries) is transcribed to the guarantee.

In the receiver the implications for check \ref{impcheck} are just as in the SC proof, and there are no variable assignments so check \ref{guarcheck} is immediate. Stability of the precondition of command \<labc> against \interferenced{\<true>}{\<msg>:=1} is exactly as \eqnref{MPSCstab1} (same assertion, same interference, same calculation). Stability against \interferenced{\<Bfr>(\<msg>=1)}{\<flag>:=1} involves the \<Bfr> modality but, as before, we find that $\<msg>=1$ is stable:
\begin{equation}
\cols
	& \<sp>((\<flag>=1=>\<msg>=1)@\<Bfr>(\<msg>=1),\;\<flag>:=1) \vspace{2pt}\\
=	& (\<flag'>=1=>\<msg>=1)@\hook{\<Bfr>(\<msg>=1)}@\<msg>=1@\<flag>=1 \\
=> 	& \<msg>=1 \\
=> 	& \<flag>=1=>\<msg>=1
\sloc
\eqnlabel{MPstab2}
\end{equation}
The remaining assertions are stable against the sender's interference because they don't mention \<flag> and because $\<msg>=1$ is stable. The checklist is completed; we have verified the proof. Note that in \eqnref{MPstab2} the \<Bfr> modality has allowed us to conclude, just as in the SC proof, that when the receiver sees $\<flag>=1$ it will already have seen $\<msg>=1$.

\subsection{\compilingmark{} Compiling laced programs}

The aim of our work is to provide rules that programmers can use to reliably fit parts of their program together. Providing proofs justifies confidence that the imposed constraints are effective. But we can use unembroidered lacing to describe an algorithm, we expect that programmers will do that, and we're sure that compilers would ignore embroidery even if we were to show it to them. So, given a laced program, how do we produce an implementation for a particular architecture?

In the examples in this paper we assume a wrapper which deals with initiation and termination of threads. We suppose that the wrapper sets up the initial state and propagates it to all threads before execution starts. We shall see in \secref{SBetc} that this requires a \<uo> constraint between the initial state and the commands of each thread: \tabref{orderingequivalences} claims that \<MFENCE> on Intel x86, \<sync> on Power and \<dsb> on ARM would be sufficient to provide it. Since \<uo> is included in \<bo> and \<lo>, the initial constraints are then in place for each thread of our examples. 

On x86 we could transcribe the commands of \figref{MPlaced} directly, since its constraints are implicit in the architecture. On Power, \<lwsync> between the commands of the sender would provide a \<bo> constraint, and an addr dependency would suffice to provide an \<lo> constraint in the receiver. ARM is similar, using \<dsb> rather than \<lwsync> to provide \<uo> order (which implies \<bo>). The Power and ARM compilations would then provide exactly what \citep{PowerARMLitmusTests} says is required to prohibit the execution pictured in \figref{MPlitmus}.

It is possible to imagine a compiler which could translate a laced program into a implementation program for a weak-memory target machine, though minimal barrier placement is a hard problem~\citep{BurckhardtAlurMartinCheckFence2007,HuynhRoychoudhuryMemorymodelsensitivebytecodeverification2007,AtigBouajjaniParlatoGettingRidofStoreBuffersinTSOAnalysis2011}. It's also possible to imagine a program which could check, given an implementation program and a laced program, that the implementation provides the constraints of the lacing. 

\section{New instabilities}

In the proof of weak-memory MP we considered the interference of one thread with another: that is, interference caused by \emph{external} parallelism. We needed the \<Bfr> modality to constrain propagation, but otherwise the weak-memory proof was very similar to the SC proof. But weak memory introduces other kinds of parallelism: 
\begin{description*}
\item[Internal (lo) parallelism] An assignment can be elaborated after the elaboration of the source and before the elaboration of the target of a constraint, interfering with the constraint's embroidery.
\item[In-flight (bo) parallelism] The writes made by variable assignments which are not \<bo> ordered may be propagated out of \<so> order, and then an \<so>-later assignment will interfere with the interference precondition of an \<so>-earlier.
\item[Universal (uo) parallelism] A variable assignment may undermine a thread-universal assertion in another thread. Dealt with in \secref{SBetc}.
\end{description*}
%%There is no weak-memory parallelism associated with \<go>. 

Note that \<lo> parallelism can apply to the embroidery of any constraint: it has to do with the \<lo> ordering of the interfering assignment and the constraint's source and target. Also, \<bo> parallelism is about the absence of \<bo> constraints%, and \<uo> parallelism, when we come to it, is about the \<U> modality, not about \<uo> constraints
.

\begin{figure}
% proofs/MP_loparallelB.unproof
$$\cols[c]
\assert{\<init>::\<msg>=\<flag>=0} \vspace{3pt} \\
\BRA[l||l]
\thr{Lo-parallel sender (0)}
  \guarantee{\interferenceg{\<Bfr>(\<msg>=\<flag>=0)}{\<msg>:=1}; \\ \interferenceg{\<Bfr>(\<msg>=\<flag>=0)}{\<flag>:=1}} \vspace{3pt} \vspace{3pt} \\
  \!\!\cols[r@{\intfspace}l]
  \assertd{\<bo> \<init>:: \hcancel{\<Bfr>(\<msg>=\<flag>=0)}} & \<laba>:: \<msg>:=1; \\
  \assertd{\<bo> \<init>:: \hcancel{\<Bfr>(\<msg>=\<flag>=0)}} & \<labb>:: \<flag>:=1
  \sloc
\rht
&
\thr{Blocked receiver (1)}
  \emptyguarantee \vspace{3pt} \vspace{3pt} \\
  \!\!\cols[r@{\intfspace}l]
  \assertd{\<lo> \<init>:: \<flag>=1=>\<msg>=0} & \<labc>:: \<r1>:=\<flag>; \\
  \assertd{\<lo> \<labc>:: \<r1>=1=>\<flag>=1@\<msg>=0} & \<labd>:: \<r2>:=\<msg>
  \sloc \vspace{3pt} \\
  \assertd{\<lo> \<labd>:: \<r1>=1=>\<r2>=0}
\rht
\KET \vspace{3pt} \\
\assert{\<final>::(1:::\<r1>)=1=>(1:::\<r2>)=0}
\sloc$$
\caption{An unproof which ignores \setorder{lo} parallelism}
\figlabel{loParallelismAbused}
\end{figure}

\subsection{Internal (lo) parallelism}
\seclabel{loparallelism}

Consider thread 0 of \figref{loParallelismAbused}. Each of its assignments is laced directly to the initial assertion \<init> using a \<bo> constraint; we can use \<bo> because the initial state is broadcast to all threads before any of them start executing. It appears that the embroidery is stable, just because there's no interference from the receiver, and the sender's guarantee is constructed as usual from the assignments and their preconditions. But note that this is an unproof, and we have struck through the embroidery in thread 0 because, as we shall see, it is unstable and therefore invalid.

Supposing for the moment that the embroidery were valid, we can imagine pairing this sender with the MP receiver of \figref{MPproof}. Then 
the precondition of the receiver's command \<labc> would be unstable against the second line of this sender's guarantee:
\begin{equation}
\cols
	& \<sp>((\<flag>=1=>\<msg>=1)@\<Bfr>(\<msg>=\<flag>=0),\;\<flag>:=1) \\
=	& ((\<flag'>=1=>\<msg>=1)@\hook{\<Bfr>(\<msg>=\<flag>=0)}@\<msg>=\<flag'>=0@\<flag>=1 \\
=>	& \<flag>=1@\<msg>=0 \\
\not=>	& \<flag>!=1|\<msg>=1 \\
==	& \<flag>=1=>\<msg>=1
\sloc
\end{equation}
So we can't prove successful message passing, which is as we should expect: experiment shows that the sender needs a \<bo> constraint between \<laba> and \<labb> in order to succeed, and we don't have that in \figref{loParallelismAbused}.

But message passing is not the point of this example. The sender's guarantee is nonsense: it claims that each of the sender's writes $\<msg>=1$ and $\<flag>=1$ has to arrive at other threads in the initial state $\<msg>=\<flag>=0$, so that if either arrives first and changes the state, the other is blocked. Even weak memory isn't \emph{that} silly. 

Thread 1 of \figref{loParallelismAbused} illustrates the silliness. The precondition of \<labc> is not the same as in the MP receiver of \figref{MPproof}: there we had $\<msg>=1$; here we have $\<msg>=0$. This precondition is stable under the stated interference of the \<lo>-parallel sender:
\begin{align}
&{\cols
	& \<sp>((\<flag>=1=>\<msg>=0)@\<Bfr>(\<msg>=\<flag>=0),\;\<msg>:=1) \\
=	& (\<flag>=1=>\<msg'>=0)@\hook{\<Bfr>(\<msg>=\<flag>=0)}@\<msg'>=\<flag>=0@\<msg>=1 \\
=>	& \<flag>=0 \\
=>	& \<flag>=1=>\<msg>=0
\sloc} \\
&{\cols
	& \<sp>((\<flag>=1=>\<msg>=0)@\<Bfr>(\<msg>=\<flag>=0),\;\<flag>:=1) \\
=	& (\<flag'>=1=>\<msg>=0)@\hook{\<Bfr>(\<msg>=\<flag>=0)}@\<msg>=\<flag'>=0@\<flag>=1 \\
=>	& \<msg>=0 \\
=>	& \<flag>=1=>\<msg>=0
\sloc}
\end{align}
The postcondition of \<labc> is $(\<flag>=1=>\<msg>=0)@\<r1>=\<flag>$, which implies $\<r1>=1=>\<flag>=1@\<msg>=0$, the precondition of \<labd>. That precondition is also stable against the sender's stated guarantee, each line of which requires $\<flag>=\<msg>=0$:
\begin{align}
&{\cols
	& \<sp>((\<r1>=1=>\<flag>=1@\<msg>=0)@\<Bfr>(\<msg>=\<flag>=0),\;\<msg>:=1) \\
=	& (\<r1>=1=>\<flag>=1@\<msg'>=0)@\hook{\<Bfr>(\<msg>=\<flag>=0)}@\<msg'>=\<flag>=0@\<msg>=1 \\
=>	& (\<r1>=1=>\<flag>=1)@\<flag>=0 \\
=>	& \<r1>!=1 \\
=>	& \<r1>=1=>\<flag>=1@\<msg>=0
\sloc} \eqnlabel{Loparallelismabusedex3}\\
&{\cols
	& \<sp>((\<r1>=1=>\<flag>=1@\<msg>=0)@\<Bfr>(\<msg>=\<flag>=0),\;\<flag>:=1) \\
=	& (\<r1>=1=>\<flag'>=1@\<msg>=0)@\hook{\<Bfr>(\<msg>=\<flag>=0)}@\<msg>=\<flag'>=0@\<flag>=1 \\
=>	& \<flag>=1@\<msg>=0 \\
=>	& \<r1>=1=>\<flag>=1@\<msg>=0
\sloc} \eqnlabel{Loparallelismabusedex4}
\end{align}
The postcondition of \<labd> is $(\<r1>=1=>\<flag>=1@\<msg>=0)@\<r2>=\<msg>$, which implies $\<r1>=1=>\<r2>=0$. So if command \<labc> reads $\<flag>=1$, command \<labd> \emph{cannot} read $\<msg>=1$. This must be nonsense: the sender writes $\<msg>=1$, and the receiver can't be blocked from seeing it. Put it another way: \eqnref{Loparallelismabusedex3} says that if the $m:=1$ interference arrives, then the $f:=1$ interference cannot have, and vice-versa with \eqnref{Loparallelismabusedex4}.

Our unproof overlooks weak-memory parallelism \emph{within} a thread, which allows an assignment to interfere with the embroidery of a constraint on another command. 
\definition{\setorder{Lo} parallelism}{LOparallelism}{An assignment \<laba> is \emph{lo parallel} with a constraint \<labb>$->$\<labc> if tbere is an \<so> path connecting \<laba>, \<labb> and \<labc> in which \<laba> is not constrained to come before \<labb> or after \<labc>.} 
%In a straight-line thread it's easy to see: any assignment which is \<so>-between the ends of a constraint, or \<so>- but not \<lo>-before the {\csource}, or \<so>- but not \<lo>-after the {\ctarget} is \<lo> parallel with it. 
We can use an SC stability calculation to check for \<lo> stability, because there are no propagation effects inside a thread.
\ruledef{LO stability}{LOstability}
{%!TEX root = ./Paper.tex
Constraint embroidery $P$ is LO stable against \<lo>-parallel assignment $A$ with precondition $Q$ if \vspace{2pt}\\
\hstrut{50pt} $\<sp>(P@Q,\;A)=>P$}
The rule applies to all kinds of assignments: variable writes $x:=E$, reads $r:=x$ and calculations $r:=E$. (But on some architectures \defref{LOparallelism} may not apply to reads and calculations: see the discussion of SCreg in \secref{SCreg}.)

Each of the assignments in the sender of \figref{loParallelismAbused} is \<lo>-parallel with the constraint of the other: \<laba> is \<so>-between the ends of the \<init>$->$\<labb> constraint, and \<labb> is \<so>- but not \<lo>-after the \<init>$->$\<laba> constraint. In neither case do we have stability. We test the \<init>$->$\<laba> embroidery against the interference of \<labb>:
\begin{equation}
\cols
	& \<sp>(\<Bfr>(\<msg>=0@\<flag>=0)@\<Bfr>(\<msg>=0@\<flag>=0),\;\<flag>:=1) \\
=	& \<sp>(\<Bfr>(\<msg>=0@\<flag>=0),\;\<flag>:=1) \vspace{2pt}\\
=	& \hook{\<Bfr>(\<msg>=0@\<flag>=0)}@\<msg>=0@\<flag'>=0@\<flag>=1 \\
\not=>	& \<Bfr>(\<msg>=0@\<flag>=0)
\sloc
\eqnlabel{loParallelismAbusedinstability1}
\end{equation}
The case of the \<init>$->$\<labb> embroidery and the interference of \<laba> is similar. 

\begin{figure}
\centering
% proofs/MP_loparallelC.unproof
$$\cols[c]
\assert{\<init>::\<msg>=\<flag>=0} \vspace{3pt} \\
\BRA[l||l]
\thr{Lo-parallel sender (0)}
  \guarantee{\interferenceg{\<Bfr>(\<msg>=0)}{\<msg>:=1}; \\ \interferenceg{\<Bfr>(\<flag>=0)}{\<flag>:=1}} \vspace{3pt} \vspace{3pt} \\
  \!\!\cols[r@{\intfspace}l]
  \assertd{\<bo> \<init>:: \<Bfr>(\<msg>=0)} & \<laba>:: \<msg>:=1; \\
  \assertd{\<bo> \<init>:: \<Bfr>(\<flag>=0)} & \<labb>:: \<flag>:=1
  \sloc
\rht
&
\thr{Unblocked receiver (1)}
  \emptyguarantee \vspace{3pt} \vspace{3pt} \\
  \!\!\cols[r@{\intfspace}l]
  \assertd{\<lo> \<init>:: \hcancel{\<flag>=1=>\<msg>=0}} & \<labc>:: \<r1>:=\<flag>; \\
  \assertd{\<lo> \<labc>:: \hcancel{\<r1>=1=>\<flag>=1@\<msg>=0}} & \<labd>:: \<r2>:=\<msg>
  \sloc \vspace{3pt} \\
  \assertd{\<lo> \<labd>:: \<r1>=1=>\<r2>=0}
\rht
\KET \vspace{3pt} \\
\assert{\<final>::(1:::\<r1>)=1=>(1:::\<r2>)=0}
\sloc$$
\caption{\setorder{Lo} parallelism and blocking removed, but still an unproof}
\figlabel{loparallelismdisabused}
\end{figure}

If we weaken the \<init>$->$\<laba> embroidery to $\<Bfr>(\<msg>=0)$ and the \<init>$->$\<labb> embroidery to $\<Bfr>(\<flag>=0)$, as in \figref{loparallelismdisabused}, we have \<lo> stability in the sender, since neither of those embroideries mentions the other assignment's variable. But now have external instability in the receiver. The \<init>$->$\<labc> embroidery is unstable against both lines of the sender's guarantee. Here's the \interferenced{\<Bfr>(\<flag>=0)}{\<flag>:=1} case:
\begin{equation}
\cols
	& \<sp>((\<flag>=1=>\<msg>=0)@\<Bfr>(\<flag>=0),\;\<flag>:=1) \\
=	& (\<flag'>=1=>\<msg>=0)@\hook{\<Bfr>(\<flag>=0)}@\<flag'>=0@\<flag>=1 \\
\not=>	& \<flag>!=1|\<msg>=0 \\
\sloc
%\eqnlabel{label}
\end{equation}
The \<labc>$->$\<labd> embroidery, which prevented the receiver from reading $\<msg>=1$ when $\<r1>=1$, is unstable against \interferenced{\<Bfr>(\<msg>=0)}{\<msg>:=1}:
\begin{equation}
\cols
	& \<sp>((\<r1>=1=>\<flag>=1@\<msg>=0)@\<Bfr>(\<msg>=0),\;\<msg>:=1) \\
=	& (\<r1>=1=>\<flag>=1@\<msg'>=0)@\hook{\<Bfr>(\<msg>=0)}@\<msg'>=0@\<msg>=1 \\
\not=>	& \<r1>!=1|\<flag>=1@\<msg>=0
\sloc
%\eqnlabel{label}
\end{equation}

Even though the silly blocking has gone, the sender's weakened guarantee still doesn't support reliable message-sending: the precondition of the MP receiver's \<labc> (see \figref{MPproof}) is unstable against the second line of the new sender's guarantee:
\begin{equation}
\cols
	& \<sp>((\<flag>=1=>\<msg>=1)@\<Bfr>(\<flag>=0),\;\<flag>:=1) \\
=	& ((\<flag'>=1=>\<msg>=1)@\hook{\<Bfr>(\<flag>=0)}@\<flag'>=0@\<flag>=1 \\
\not=>	& \<flag>!=1|\<msg>=1 \\
==	& \<flag>=1=>\<msg>=1
\sloc
\eqnlabel{Bfr.flag0.isntenough}
\end{equation}

Note that in the MP sender of \figref{MPproof} the assigments are ordered by \<bo> and therefore by \<lo>, so there is no \<lo> parallelism. In the MP receiver of \figref{MPproof} and the receiver of \figref{loParallelismAbused} and \figref{loparallelismdisabused} the assignments are ordered by \<lo>, so no \<lo> parallelism there either.

\begin{figure}
\centering
$$\cols[c]
\assert{\<init>::\<msg>=\<flag>=0} \vspace{3pt} \\
\BRA[l||l]
\lthr{Bo-parallel sender (0)}
  \multicolumn{2}{l}{\guarantee{\interferenceg{\hcancel{\<Bfr>(\<msg>=\<flag>=0)}}{\<msg>:=1} \\ \interferenceg{\<Bfr>(\<flag>=0)}{\<flag>:=1}}} \\
  \assertd{\<bo> \<init>:: \<Bfr>(\<msg>=\<flag>=0)} & \<laba>:: \<msg>:=1; \\
  \assertd{\<lo> \<laba>:: \<Bfr>(\<flag>=0)} & \<labb>:: \<flag>:=1
\rhtl
&
\lthr{Blocked receiver (1)}
  \multicolumn{2}{l}{\emptyguarantee} \\
  \assertd{\<lo> \<init>:: \<true>} & \<labc>:: \<r1>:=\<flag>; \\
  \assertd{\<lo> \<labc>:: \<r1>=1=>\<flag>=1} & \<labd>:: \<r2>:=\<msg>; \\
  \assertd{\<lo> \<labd>:: \<r1>=1@\<r2>=0=>\<flag>=1@\<msg>=0} & \<labe>:: \<r3>:=\<msg> \\
  \assertd{\<lo> \<labe>:: \<r1>=1@\<r2>=0=>\<r3>=0}
\rhtl
\KET \vspace{3pt} \\
\assert{\<final>::(1:::\<r1>)=1@(1:::\<r2>)=0=>(1:::\<r3>)=0} 
\sloc$$
\caption{An unproof which ignores \setorder{bo} parallelism}
\figlabel{boParallelismAbused}
\end{figure}

\subsection{In-flight (bo) parallelism}
\seclabel{boparallelism}

Consider thread 0 of \figref{boParallelismAbused}. We've eliminated \<lo> parallelism by including an \<lo> constraint between \<laba> and \<labb>. This time we do have stable embroidery, but we shall see that the interference precondition of the first line of the sender's guarantee, though derived transparently from the elaboration precondition of \<laba>, is invalid. If we were to trust that interference precondition, we %still don't have reliable message-passing: if we put the MP receiver of \figref{MPproof} against this sender, then the receiver's command \<labc> precondition is unstable against the second line of the sender's guarantee, just as in  \eqnref{Bfr.flag0.isntenough}. And once again we 
would have another nonsense guarantee: the assertions in the receiver are stable against it, but if thread 2's \<labc> reads $\<flag>=1$ and \<labd>  reads $\<msg>=0$ then, according to its precondition, \<labe> \emph{cannot} read \<msg>=1.

The problem is propagation reordering. In thread 0 there is no \<bo> constraint \<laba>$->$\<labb> to ensure that $\<msg>=1$ is propagated before $\<flag>=1$, and if they are propagated to thread 1 in the reverse order -- $\<flag>=1$ before $\<msg>=1$ -- then the precondition of the first line of the sender's guarantee, which insists that $\<flag>=0$, won't hold when $\<msg>=1$ arrives. In effect the interference generated by \<labb> is altering the interference generated by \<laba>. It's weak-memory \emph{in-flight} or \emph{bo-parallel} interference.
\definition{\setorder{Bo} parallelism}{BOparallel}{%!TEX root = ./Paper.tex
If variable assignments to distinct variables are \<so>- but not \<bo>-ordered then the \<so>-later is \emph{bo parallel} with the \<so>-earlier and will interfere with its interference precondition.}
Note the asymmetry of this definition: the second assignment is \<bo>-parallel with the first, but not the first with the second. For justification, see below. The restriction to distinct variables is because writes to the same variable can't overtake each other: for explanation, see \secref{interferenceNstability}.

As with all our examples until \secref{interferenceNstability}, we once again consider only the special case of \<Bfr>-modal interference preconditions. For the general rule, see \secref{interferenceNstability}.
\ruledef{BO stability with \setModality{B}-modal interference}{BOstabilityBmodal}
{$$\cols
	\text{Interference } \interferenced{P}{x:=E} \text{ is BO stable against \<bo>-parallel \interferenced{Q}{y:=F},} \\
	\text{ where both }P \text{ and } Q \text{ have no occurrences of variables outside a \<Bfr> modality, if }\vspace{2pt}\\
	\quad \<sp>(P@Q,\;x:=E)=>P
\sloc$$}

Using this rule, we check stability of \<laba>'s interference in \figref{boParallelismAbused} against the interference of \<labb>. It's unstable:
\begin{equation}
\setlength{\arraycolsep}{3.5pt}
\cols
	& \<sp>(\<Bfr>(\<msg>=0@\<flag>=0)@\<Bfr>(\<flag>=0),\;\<flag>:=1) \\
=	& \<sp>(\<Bfr>(\<msg>=0)@\<Bfr>(\<flag>=0),\;\<flag>:=1) \\
=	& \hook{\<Bfr>(\<msg>=0)}@\<msg>=0@\hook{\<Bfr>(\<flag>=0)}@\<flag'>=0@\<flag>=1 \\
=> 	& \<Bfr>(\<msg>=0)@\hook{\<Bfr>(\<flag>=0)}@\<flag'>=0@\<flag>=1 \\
\not=>	& \<Bfr>(\<msg>=0)@\<Bfr>(\<flag>=0) \\
=	& \<Bfr>(\<msg>=0@\<flag>=0)
\sloc
\eqnlabel{boParallelismAbusedinstability1}
\end{equation}
We shouldn't weaken \<laba>'s elaboration precondition, which is validly derived from the stable embroidery of its constraints, and gives us a postcondition which validates the elaboration precondition of \<labb>. But we must weaken \<laba>'s interference precondition to regain in-flight stability. %We've done that in \figref{boParallelismDisabused}.
\begin{figure}
\centering
% proofs/MP_boparallelC.unproof
$$\cols[c]
\assert{\<init>::\<msg>=\<flag>=0} \vspace{3pt} \\
\BRA[l||l]
\thr{Bo-parallel sender (0)}
  \guarantee{\interferenceg{\<Bfr>(\<msg>=0)}{\<msg>:=1}; \\ \interferenceg{\<Bfr>(\<flag>=0)}{\<flag>:=1}} \vspace{3pt} \\
  \!\!\cols[r@{\intfspace}l]
  \assertd{\<bo> \<init>:: \<Bfr>(\<msg>=\<flag>=0)} & \\ \ipre{\<Bfr>(\<msg>=0)} & \<laba>:: \<msg>:=1; \\
  \assertd{\<lo> \<laba>:: \<Bfr>(\<flag>=0)} & \<labb>:: \<flag>:=1
  \sloc
\rht
&
\thr{Unblocked receiver (1)}
  \emptyguarantee \vspace{3pt} \\
  \!\!\cols[r@{\intfspace}l]
  \assertd{\<lo> \<init>:: \<true>} & \<labc>:: \<r1>:=\<flag>; \\
  \assertd{\<lo> \<labc>:: \<r1>=1=>\<flag>=1} & \<labd>:: \<r2>:=\<msg>; \\
  \assertd{\<lo> \<labd>:: \hcancel{\<r1>=1@\<r2>=0=>\<flag>=1@\<msg>=0}} & \<labe>:: \<r3>:=\<msg>
  \sloc \\
  \assertd{\<lo> \<labe>:: \<r1>=1@\<r2>=0=>\<r3>=0}
\rht
\KET \vspace{3pt} \\
\assert{\<final>::(1:::\<r1>)=1@(1:::\<r2>)=0=>(1:::\<r3>)=0}
\sloc$$
\caption{\setorder{Bo} instability removed makes an unproof of silly blocking}
\figlabel{boParallelismDisabused}
\end{figure}

In order that we can check for \<bo>-parallel stability in our checklist, the laced program must declare what the interference generated by each assignment actually is. In this example there is a single line in the guarantee for each assignment, so it's quite easy to see what's going on, but things aren't always so simple. Our solution is to append, after the knot which describes the elaboration precondition, a square-bracketed interference precondition. When it's absent, the elaboration precondition is the interference precondition. So we've declared, in \figref{boParallelismDisabused}, that the interference precondition of \<laba> is $\<Bfr>(\<msg>=0)$, and transcribed that to the sender's guarantee.

\Figref{boParallelismDisabused} removes the \<bo> instability in the sender -- \<laba>'s precondition no longer mentions \<flag>, so is stable against the \<bo>-parallel interference of \<labb> -- and removes the silliness in the receiver -- the precondition of \<labe> is now unstable against the first line of the sender's guarantee.
\begin{equation}
\cols
	& \<sp>((\<r1>=1@\<r2>=0=>\<flag>=1@\<msg>=0)@\<Bfr>(\<msg>=0),\;\<msg>:=1) \\
=	& (\<r1>=1@\<r2>=0=>\<flag>=1@\<msg'>=0)@\hook{\<Bfr>(\<msg>=0)}@\<msg'>=0@\<msg>=1 \\
\not=>	& \<r1>=1@\<r2>=0=>\<flag>=1@\<msg>=0
\sloc
\end{equation}
We still don't get message passing between this sender and the MP receiver because of the instability demonstrated in \eqnref{Bfr.flag0.isntenough}.

\Defref{BOparallel} and \ruleref{BOstabilityBmodal} deal with the interference of one variable assignment with the interference precondition of an earlier assignment. We don't have to check in the opposite direction (stability of later against earlier) because the elaboration precondition of the later assignment, and therefore its interference precondition, already take account of the earlier assignment. In \figref{boParallelismDisabused}, for example, the elaboration precondition of command \<labb> cannot imply $\<Bfr>(m=0)$ and therefore neither can its interference precondition.

\subsection{Bo parallelism in the rely}

\<Lo> parallelism is entirely within a single thread, and \<bo> parallelism can occur within a single thread, as we have shown. But when the interferences of more than one thread converge on another, we have to guard against in-flight interference between their separate guarantees. So when forming a rely for one thread from the guarantees of other threads, we have to form the \<bo>-stable merge of those other threads' guarantees, weakening where necessary. Two examples of this will be seen in \secref{tokenring}; the rule we use will be revealed in \secref{interferenceNstability}.

\subsection{Still the same game}

Check \ref{stabcheck} of our checklist requires assertions to be stable against `parallel assignments', which now includes both external and \<lo> parallelism. Check \ref{guarcheck} requires `stable interference' in the guarantee; we have seen that we must check \<bo>-parallel assignments for interference stability. 

We are still playing the Owicki-Gries-Jones game: check that assertions are stable, check that the external effects of variable assignments are reflected in the guarantee.

\begin{figure}
\squeezecols{3.5pt}
\centering
% proofs/almostWRC.proof
$$\cols[c]
\assert{\<init>::\<flag>=0} \vspace{3pt} \\
\BRA[l||l||l]
\thr{Sender (0)}
  \assertd{\<lo> \<init>}\ \<laba>:: \<msg>:=1
\rht
&
\thr{Proxy (1)}
  \!\!\cols[r@{\intfspace}l]
  \assertd{\<lo> \<init>} & \<labb>:: \<r1>:=\<msg>; \\
  \assertd{\<bo> \<labb>} & \<labc>:: \<flag>:=\<r1>
  \sloc
\rht
&
\thr{Receiver (2)}
  \!\!\cols[r@{\intfspace}l]
  \assertd{\<lo> \<init>} & \<labd>:: \<r1>:=\<flag>; \\
  \assertd{\<lo> \<labd>} & \<labe>:: \<r2>:=\<msg>
  \sloc
\rht
\KET \vspace{3pt} \\
\assert{\<final>::(2:::\<r1>)=1=>(2:::\<r2>)=1}
\sloc$$
\caption{AlmostWRC: sending through a proxy}
\figlabel{almostWRC}
\end{figure}

\begin{figure}
\centering
% proofs/almostISA2.proof
$$\cols[c]
\assert{\<init>::\<flag>=\<flag1>=0} \vspace{3pt} \\
\BRA[l||l||l]
\thr{Sender (0)}
  \!\!\cols[r@{\intfspace}l]
  \assertd{\<lo> \<init>} & \<laba>:: \<msg>:=1; \\
  \assertd{\<bo> \<laba>} & \<labb>:: \<flag>:=1
  \sloc
\rht
&
\thr{Proxy (1)}
  \!\!\cols[r@{\intfspace}l]
  \assertd{\<lo> \<init>} & \<labc>:: \<r1>:=\<flag>; \\
  \assertd{\<lo> \<labc>} & \<labd>:: \<flag1>:=\<r1>
  \sloc
\rht
&
\thr{Receiver (2)}
  \!\!\cols[r@{\intfspace}l]
  \assertd{\<lo> \<init>} & \<labe>:: \<r1>:=\<flag1>; \\
  \assertd{\<lo> \<labe>} & \<labf>:: \<r2>:=\<msg>
  \sloc
\rht
\KET \vspace{3pt} \\
\assert{\<final>::(2:::\<r1>)=1=>(2:::\<r2>)=1}
\sloc$$
\caption{AlmostISA2: flagging through a proxy}
\figlabel{almostISA2}
\end{figure}

%%\begin{figure}
%%\centering
%%\includegraphics[scale=\picscale]{almostISA2.pdf}
%%\caption{AlmostISA2 litmus test}
%%\figlabel{litmus_almostISA2}
%%\end{figure}

\section{Reads and the B modality}

In the examples so far we have applied \<bo> constraints between variable assignments. But because \<bo> controls propagation of the writes underpinning the {\csource} postcondition, no matter what kind of component the source is, we can usefully use it in other contexts. \Figref{almostWRC} shows a variation of message-passing in which a proxy thread can receive the sender's message and pass it on to the receiver. Observe that the sender plus the proxy look rather like the sender of MP, extended across two threads -- but they're not identical, because the proxy in \figref{almostWRC} might not read $\<msg>=1$. 

We have seen in the MP example, in both SC and weak-memory treatments, that an interference precondition carries information about earlier writes in the same thread, so that we can assert an implication which is stable against that interference. The implication $\<flag>=1=>\<msg>=1$ in the MP proofs works off the interference description \interferenced{m=1}{f=1} in the SC case and \interferenced{\<Bfr>(m=1)}{f=1} in the weak memory case (using the fact that $\<Bfr>(m=1)=>m=1$ when the interference hits). It's possible in the same way to use \<Bfr>-modal assertions, effectively extending \<bo> between threads. \Figref{almostISA2} shows a variation of an example which is considered rather subtle in discussions of the Power architecture. The sender is as in MPlaced; the proxy thread reads the sender's flag \<flag> and notifies the receiver via a proxy flag \<flag1>. The subtlety is that there is no need for a \<bo> constraint between commands \<labc> and \<labd> in the proxy: \<lo> everywhere but in the sender is enough. This time it is the MP receiver that seems to be extended across two threads.

\subsection{Bo from read to write}
\seclabel{almostWRC}

\begin{figure}
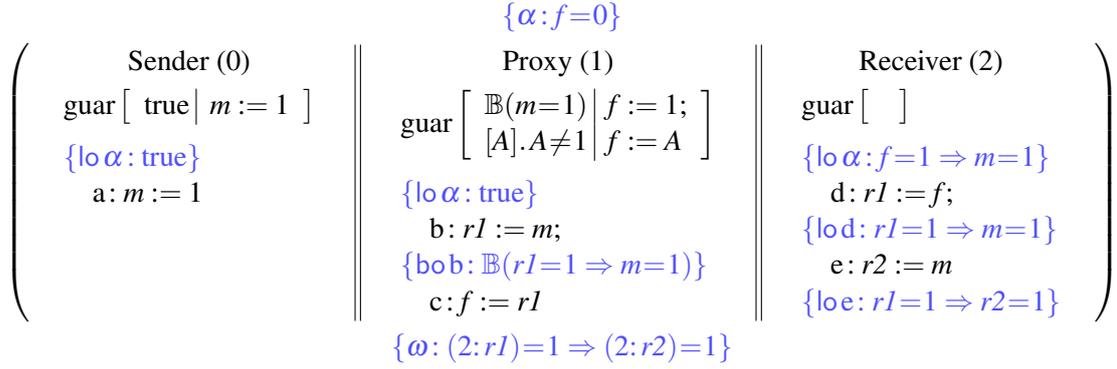

\centering
% proofs/almostWRC.proof
$$\cols[c]
\assert{\<init>::\<flag>=0} \vspace{3pt} \\
\BRA[l||l||l]
\thr{Sender (0)}
  \guarantee{\interferenceg{\<true>}{\<msg>:=1}} \vspace{3pt} \\
  \assertd{\<lo> \<init>:: \<true>} \\ \quad \<laba>:: \<msg>:=1
\rht
&
\thr{Proxy (1)}
  \guarantee{\interferenceg{\<Bfr>(\<msg>=1)}{\<flag>:=1}; \\ \interferenceg[A]{A!=1}{\<flag>:=A}} \vspace{3pt} \\
  \assertd{\<lo> \<init>:: \<true>} \\ \quad \<labb>:: \<r1>:=\<msg>; \\
  \assertd{\<bo> \<labb>:: \<Bfr>(\<r1>=1=>\<msg>=1)} \\ \quad \<labc>:: \<flag>:=\<r1>
\rht
&
\thr{Receiver (2)}
  \emptyguarantee \vspace{3pt} \\
  \assertd{\<lo> \<init>:: \<flag>=1=>\<msg>=1} \\ \quad \<labd>:: \<r1>:=\<flag>; \\
  \assertd{\<lo> \<labd>:: \<r1>=1=>\<msg>=1} \\ \quad \<labe>:: \<r2>:=\<msg> \\
  \assertd{\<lo> \<labe>:: \<r1>=1=>\<r2>=1}
\rht
\KET \vspace{3pt} \\
\assert{\<final>::(2:::\<r1>)=1=>(2:::\<r2>)=1}
\sloc$$
\caption{A proof of almostWRC}
\figlabel{almostWRCproof}
\end{figure}

\Figref{almostWRCproof} is a proof of almostWRC. The rely of the sender  -- the interference it can experience -- is the \<bo>-stable union of the guarantees of the proxy and the receiver: because the receiver's guarantee is empty, that means just the guarantee of the proxy. The precondition of command \<laba> is implied by the initial assertion \<init> (check \ref{impcheck}); it's stable because it's a constant (check \ref{stabcheck}); \<laba>'s interference is included in the guarantee (check \ref{guarcheck}).  

The rely of the proxy is in effect the guarantee of the sender. The precondition of \<labb> is implied by the initial assertion \<init> (check \ref{impcheck}), it's stable because it's a constant (check \ref{stabcheck}); \<labb> isn't a variable assignment, so there's nothing for the guarantee (check \ref{guarcheck}).

The postcondition of \<labb> is $\<true>@\<r1>=\<msg>$, which implies $\<r1>=1=>\<msg>=1$, which we wrap in \<Bfr> to embroider the \<bo> \<labb>$->$\<labc> constraint (check \ref{impcheck}). That embroidery is stable against the interference of the sender, just because $\<msg>=1$ is stable (check \ref{stabcheck}):
\begin{equation}
\cols
	& \<sp>(\<Bfr>(\<r1>=1=>\<msg>=1),\;\<msg>:=1) \\
=	& \hook{\<Bfr>(\<r1>=1=>\<msg>=1)}@(\<r1>=1=>\<msg'>=1)@\<msg>=1 \\
=>	& \hook{\<Bfr>(\<r1>=1=>\<msg>=1)}@\<msg>=1 \\
=>	& \hook{\<Bfr>(\<r1>=1=>\<msg>=1)}@(\<r1>=1=>\<msg>=1) \\
=>	& \<Bfr>(\<r1>=1=>\<msg>=1)
\sloc
\eqnlabel{almostWRCproxystab2}
\end{equation}

A novelty is the treatment of the interference generated by command \<labc>. Because precondition and assignment mention register \<r1>, which by definition is local to the thread, we must quotient the register, giving the interference description
\begin{equation}
\interferenced[\setvar{r1}]{\<Bfr>(\<r1>=1=>\<msg>=1)}{\<flag>:=\<r1>}
\eqnlabel{almostWRCproxyintf2}
\end{equation}
To check stability against quotiented interference, we must consider all the possible values of the quotiented registers:
\ruledef{Stability against quotiented interference}{quantifiedstability}{%!TEX root = ./Paper.tex
$P$ is stable against \interferenced[\nvec{ns}]{Q}{x:=E} if it is stable against \interferenced{Q[\nvec{ns}\backslash\nvec{fs}]}{x:=E[\nvec{ns}\backslash\nvec{fs}]} where $\nvec{fs}$ are fresh names. Similarly for internal interference \interferenced[\nvec{ns}]{Q}{r:=E} and \interferenced[\nvec{ns}]{Q}{r:=x}.
}

We can split \eqnref{almostWRCproxyintf2} for the two cases $\<r1>=1$ \eqnref{WRCRWdepintf1} and $\<r1>!=1$ \eqnref{WRCRWdepintf2}:
\begin{align}
\interferenced{\<Bfr>(\<msg>=1)}{&\;\<flag>:=1} \eqnlabel{WRCRWdepintf1} \\
\interferenced[\setvar{r1}]{\<r1>!=1}{&\;\<flag>:=\<r1>} \eqnlabel{WRCRWdepintf2}  
\end{align}
\eqnref{WRCRWdepintf1} is the first line of the proxy's guarantee. \eqnref{WRCRWdepintf2} is the second line, renaming \<r1> to $A$. That completes check \ref{guarcheck}, since there is no other assignment to be \<bo>-parallel with.

The rely of the receiver is the \<bo>-stable union of the guarantees of sender and proxy. In this example we can simply merge them:  the interferences of the proxy are immune to the interferences of the sender (in line 1, $\<msg>=1$ is stable, and in line 2, the precondition doesn't mention \<msg> or \<flag>); the interference preconditions of the sender don't mention \<flag>, which is the only variable the proxy writes. We have the same receiver with the same guarantee and embroidery as in the MP proof (\figref{MPproof}), but with a new rely we'll have to make some new checks.

We already showed, when checking \figref{MPproof}, that the receiver's embroidery is stable against the sender's guarantee and the first line of the proxy's guarantee. The second line of the proxy's guarantee is new, but the precondition of \<labd> contains the only assertion that mentions \<flag>. Stability holds because the interference establishes $\<flag>!=1$:
\begin{equation}
\cols
	& \<sp>((\<flag>=1=>\<msg>=1)@A!=1,\;\<flag>:=A) \\
=	& (\<flag'>=1=>\<msg>=1)@A!=1@\<flag>=A \\
=>	& \<flag>!=1 \\
=>	& \<flag>=1=>\<msg>=1
\sloc
\eqnlabel{almostWRCreceiverstab1}
\end{equation}
 
With that we've completed verification of the proof of almostWRC. Causally, the assignment $\<r1>:=\<msg>$ is \<bo>-before $\<flag>:=\<r1>$; if the proxy reads from $\<msg>:=1$ so that $\<r1>=1$, then $\<msg>:=1$ is also \<bo>-before $\<flag>:=1$, just as in \figref{MPproof}. Our proof agrees.

\subsection{The B modality transmitted by interference}
\seclabel{almostISA2}

\Figref{almostISA2proof} shows a proof of almostISA2. The sender is exactly as in the MP proof, and the proxy's guarantee doesn't interfere with any variables mentioned in the sender's assertions, so the earlier verification stands.

The proof-subtlety is in the precondition of command \<labc>: $\<flag>=1=>\<Bfr>(\<msg>=1)$ echoes $\<flag>=1=>\<msg>=1$ in the proofs of MP. It's trivially true in initial state \<init> because $\<flag>=0$, and is stable under the interference of the sender: 
\begin{equation}
\cols
	& \<sp>((\<flag>=1=>\<Bfr>(\<msg>=1))@\<true>,\<msg>:=1) \vspace{2pt}\\
=	& (\<flag>=1=>\hook{\<Bfr>(\<msg>=1)}@\<msg'>=1)@\<msg>=1 \\
=	& (\<flag>=1@\hook{\<Bfr>(\<msg>=1)}@\<msg'>=1@\<msg>=1)|\<flag>!=1@\<msg>=1 \\
=>	& \<Bfr>(\<msg>=1)|\<flag>!=1 \\
=	& \<flag>=1=>\<Bfr>(\<msg>=1)
\sloc
\eqnlabel{almostISA2stab1} \\
\end{equation}
\begin{equation}
\cols
	& \<sp>((\<flag>=1=>\<Bfr>(\<msg>=1))@\<Bfr>(\<msg>=1),\;\<flag>:=1) \vspace{2pt}\\
=	& (\<flag'>=1=>\<Bfr>(\<msg>=1))@\<Bfr>(\<msg>=1)@\<flag>=1 \\
=>	& \<Bfr>(\<msg>=1) \\
=> 	& \<flag>=1=>\<Bfr>(\<msg>=1)
\sloc
\eqnlabel{almostISA2stab2}
\end{equation}
The precondition of \<labd> is stable, because $\<msg>=1$ is stable. The interference is 
\begin{equation}
\interferenced[\setvar{r1}]{\<r1>=1=>\<Bfr>(\<msg>=1)}{\<flag1>:=\<r1>}
\eqnlabel{almostISA2intfd}
\end{equation}
which we can analyse just as we did \eqnref{almostWRCproxyintf2}, to give two entries in the proxy's guarantee.

The proof of the receiver is as in almostWRC, reading \<flag1> for \<flag>. Its rely is effectively the same as in almostWRC, with the same reading, since the second line of the sender's guarantee writes to \<flag>, which isn't used by the receiver. So our earlier verification stands.

Causally, if $\<r1>:=\<flag>$ reads from $\<flag>:=1$, which is \<bo>-after \<msg>:=1, then it, too, is \<bo>-after $\<msg>:=1$, and then so is $\<flag1>:=\<r1>$. \<Bo> extends from sender to proxy, and the $\<Bfr>(\<msg>=1)$ modality in the proof tracks the ordering.

\begin{figure}
\centering
% proofs/almostISA2.proof
$$\cols[c]
\assert{\<init>::\<flag>=\<flag1>=0} \vspace{3pt} \\
\BRA[l||l||l]
\thr{Sender (0)}
  \guarantee{\interferenceg{\<true>}{\<msg>:=1}; \\ \interferenceg{\<Bfr>(\<msg>=1)}{\<flag>:=1}} \vspace{3pt} \\
  \assertd{\<lo> \<init>:: \<true>} \\ \quad \<laba>:: \<msg>:=1; \\
  \assertd{\<bo> \<laba>:: \<Bfr>(\<msg>=1)} \\ \quad \<labb>:: \<flag>:=1
\rht
&
\thr{Proxy (1)}
  \guarantee{\interferenceg{\<Bfr>(\<msg>=1)}{\<flag1>:=1}; \\ \interferenceg[A]{A!=1}{\<flag1>:=A}} \vspace{3pt} \\
  \assertd{\<lo> \<init>:: \<flag>=1=>\<Bfr>(\<msg>=1)} \\ \quad \<labc>:: \<r1>:=\<flag>; \\
  \assertd{\<lo> \<labc>:: \<r1>=1=>\<Bfr>(\<msg>=1)} \\ \quad \<labd>:: \<flag1>:=\<r1>
\rht
&
\thr{Receiver (2)}
  \emptyguarantee \vspace{3pt} \\
  \assertd{\<lo> \<init>:: \<flag1>=1=>\<msg>=1} \\ \quad \<labe>:: \<r1>:=\<flag1>; \\
  \assertd{\<lo> \<labe>:: \<r1>=1=>\<msg>=1} \\ \quad \<labf>:: \<r2>:=\<msg> \\
  \assertd{\<lo> \<labf>:: \<r1>=1=>\<r2>=1}
\rht
\KET \vspace{3pt} \\
\assert{\<final>::(2:::\<r1>)=1=>(2:::\<r2>)=1}
\sloc$$
\caption{A proof of almostISA2}
\figlabel{almostISA2proof}
\end{figure}

\section{Auxiliary information in composite writes}
\seclabel{WRC}

Neither of the examples almostWRC and almostISA2 is quite the litmus test on which it is based. In each case the proxy reads a value and then writes it out again: if the value it reads is 1, it must write 1. In the WRC and ISA2 litmus tests of~\citep{PowerARMLitmusTests} the proxy always writes 1, and the proof challenge is to show that everything still works. \Figref{WRC}, then, is the real WRC litmus test. It's laced exactly like almostWRC, but it has a slightly more complicated postcondition because we're not interested in the case in which the proxy reads 0 from \<msg> and the receiver reads 1 from \<flag>. 

To make a proof we need somehow to distinguish the write-1-after-read-0 and write-1-after-read-1 interferences of the proxy. As usual in the Owick-Gries tradition, we resort to auxiliary assignments which record extra information to help us make a proof. It's necessary that the execution of the program, so far as regular variables and registers are concerned, is unaffected when auxiliary assignments are deleted. In SC proofs we mustn't assign the value of an auxiliary variable to a regular register, and we can't use auxiliary registers in conditional expressions. In weak memory proofs we must also control use of constraints between auxiliary assignments and regular components, which disappear when auxiliary assignments are deleted. The details of our treatment of constraints are given in \secref{auxiliaries}, but we needn't consider them in this example because it doesn't include any purely-auxiliary assignments.

In \figref{WRCaux} we've added auxiliary information to variable \<flag> -- so-called `auxiliary extension'. Command \<labc> writes 1, as in \figref{WRC}, but also the value that was read by command \<labb>. That is, it writes either $(1,0)$ or $(1,1)$: the first part is the regular part of the write, the second part an auxiliary value. We imagine that the extended value is propagated as a unit. Command \<labd> in the receiver reads the regular part into $\<r1>$ and ignores the auxiliary part. 

Note that in the initial state assertion we've had to change $\<flag>=0$ into $\<flag>=(0,0)$. But note also that the auxiliary extension to \<flag> is completely irrelevant to execution, so much so that the receiver doesn't even look at it. This is as it should be: auxiliary information is purely to facilitate proof. 

\begin{figure}
\centering
% proofs/WRC.proof
$$\cols[c]
\assert{\<init>::\<flag>=0} \vspace{3pt} \\
\BRA[l||l||l]
\thr{Sender (0)}
  \assertd{\<lo> \<init>}\ \<laba>:: \<msg>:=1
\rht
&
\thr{Proxy (1)}
  \!\!\cols[r@{\intfspace}l]
  \assertd{\<lo> \<init>} & \<labb>:: \<r1>:=\<msg>; \\
  \assertd{\<bo> \<labb>} & \<labc>:: \<flag>:=1
  \sloc
\rht
&
\thr{Receiver (2)}
  \!\!\cols[r@{\intfspace}l]
  \assertd{\<lo> \<init>} & \<labd>:: \<r1>:=\<flag>; \\
  \assertd{\<lo> \<labd>} & \<labe>:: \<r2>:=\<msg>
  \sloc
\rht
\KET \vspace{3pt} \\
\assert{\<final>::(1:::\<r1>)=1@(2:::\<r1>)=1=>(2:::\<r2>)=1}
\sloc$$
\caption{WRC: sending through an inattentive proxy}
\figlabel{WRC}
\end{figure}

\begin{figure}
\centering
% proofs/WRC.proof
$$\cols[c]
\assert{\<init>::\<flag>=(0,0)} \vspace{3pt} \\
\BRA[l||l||l]
\thr{Sender (0)}
  \assertd{\<lo> \<init>}\ \<laba>:: \<msg>:=1
\rht
&
\thr{Proxy (1)}
  \!\!\cols[r@{\intfspace}l]
  \assertd{\<lo> \<init>} & \<labb>:: \<r1>:=\<msg>; \\
  \assertd{\<bo> \<labb>} & \<labc>:: \<flag>:=1,\<r1>
  \sloc
\rht
&
\thr{Receiver (2)}
  \!\!\cols[r@{\intfspace}l]
  \assertd{\<lo> \<init>} & \<labd>:: \<r1>,\_:=\<flag>; \\
  \assertd{\<lo> \<labd>} & \<labe>:: \<r2>:=\<msg>
  \sloc
\rht
\KET \vspace{3pt} \\
\assert{\<final>::(1:::\<r1>)=1@(2:::\<r1>)=1=>(2:::\<r2>)=1}
\sloc$$
\caption{WRC with auxiliary extension to variable \setvar{f}} % have to say `f' rather than `flag'. I can see why ...
\figlabel{WRCaux}
\end{figure}

\begin{figure}
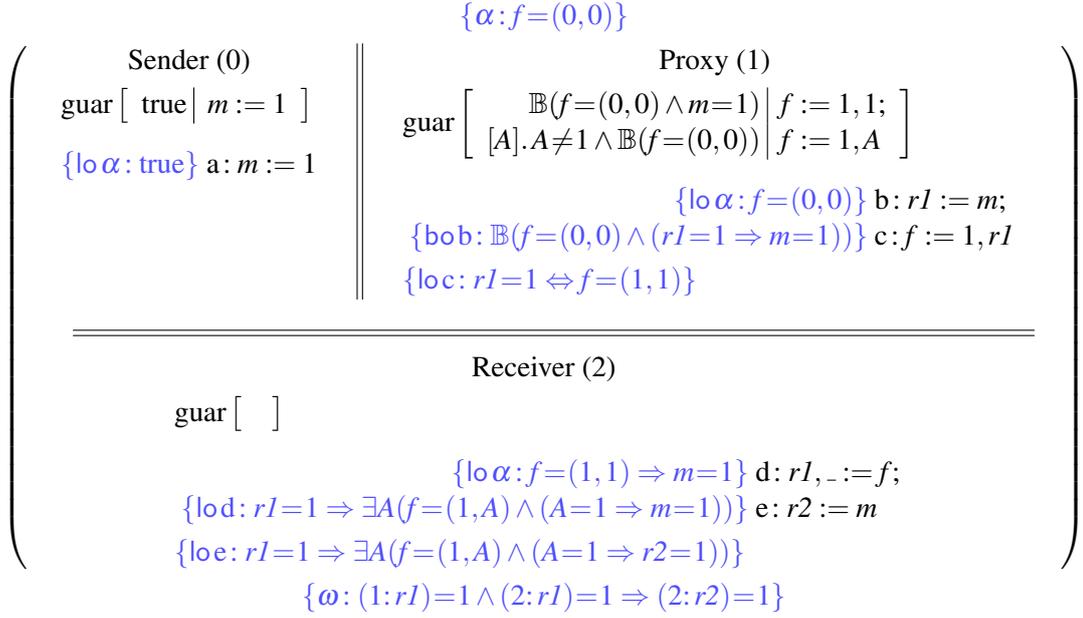

\centering
% proofs/WRC.proof
$$\cols[c]
\assert{\<init>::\<flag>=(0,0)} \vspace{3pt} \\
\BRA[l||l]
\thr{Sender (0)}
  \guarantee{\interferenceg{\<true>}{\<msg>:=1}} \vspace{3pt} \vspace{3pt} \\
  \assertd{\<lo> \<init>:: \<true>}\ \<laba>:: \<msg>:=1
\rht
&
\thr{Proxy (1)}
  \guarantee{\interferenceg{\<Bfr>(\<flag>=(0,0)@\<msg>=1)}{\<flag>:=1,1}; \\ \interferenceg[A]{A!=1@\<Bfr>(\<flag>=(0,0))}{\<flag>:=1,A}} \vspace{3pt} \vspace{3pt} \\
  \!\!\cols[r@{\intfspace}l]
  \assertd{\<lo> \<init>:: \<flag>=(0,0)} & \<labb>:: \<r1>:=\<msg>; \\
  \assertd{\<bo> \<labb>:: \<Bfr>(\<flag>=(0,0)@(\<r1>=1=>\<msg>=1))} & \<labc>:: \<flag>:=1,\<r1>
  \sloc \vspace{3pt} \\
  \assertd{\<lo> \<labc>:: \<r1>=1<=>\<flag>=(1,1)}
\rht \vspace{3pt} \\
\multicolumn{2}{c}{\begin{minipage}{350pt} $\cols \hstrut{350pt} \\[-8pt] \hline \hline \sloc$ \end{minipage}} \vspace{3pt} \\ 
\multicolumn{2}{c}{\thr{Receiver (2)}
  \emptyguarantee \vspace{3pt} \vspace{3pt} \\
  \!\!\cols[r@{\intfspace}l]
  \assertd{\<lo> \<init>:: \<flag>=(1,1)=>\<msg>=1} & \<labd>:: \<r1>,\_:=\<flag>; \\
  \assertd{\<lo> \<labd>:: \<r1>=1=>|* A (\<flag>=(1,A)@(A=1=>\<msg>=1))} & \<labe>:: \<r2>:=\<msg>
  \sloc \vspace{3pt} \\
  \assertd{\<lo> \<labe>:: \<r1>=1=>|* A (\<flag>=(1,A)@(A=1=>\<r2>=1))}
\rht}
\KET \vspace{3pt} \\
\assert{\<final>::(1:::\<r1>)=1@(2:::\<r1>)=1=>(2:::\<r2>)=1}
\sloc$$
\caption{A proof of WRC with auxiliary extension}
\figlabel{WRCproof}
\end{figure}

Figure \ref{fig:WRCproof} shows our proof. We stacked the threads to fit on the page. 

The sender's rely is the \<bo>-stable union of the guarantees of the proxy and the receiver, which is easy to make since the receiver's guarantee is empty. The rely doesn't matter anyway, because the sender's only assertion is the constant \<true>. The checks are just those we made for the almostWRC sender.

The proxy's rely is the union of the guarantee of the sender and the empty guarantee of the receiver -- again, no inter-thread \<bo> parallelism to detain us. The postcondition of \<labb> is $\<flag>=(0,0)@\<r1>=\<msg>$, which implies $\<flag>=(0,0)@(\<r1>=1=>\<msg>=1)$, which we wrap in a \<Bfr> modality to embroider the \<bo> \<labb>$->$\<labc> constraint. That embroidery is stable, since $\<flag>=(0,0)$ and $\<msg>=1$ are each stable under the sender's interference. The interference of \<labc> is therefore
\begin{equation}
\interferenced[\setvar{r1}]{\<Bfr>(\<flag>=(0,0)@(\<r1>=1=>\<msg>=1))}{\<flag>:=1,\<r1>}
\eqnlabel{WRCproxyintfdesc}
\end{equation}
The first line of the guarantee is the case when $\<r1>=1$, the second when $\<r1>!=1$. The strongest postcondition of \<labc> 
%%is 
%%\begin{equation}
%%\hook{\<Bfr>(\<flag>=(0,0)@(\<r1>=1=>\<msg>=1))}@\<flag'>=(0,0)@(\<r1>=1=>\<msg>=1)@\<flag>:=1,\<r1>
%%\eqnlabel{WRCproxyfinalsp}
%%\end{equation}
%%which 
implies $\<flag>=(1,\<r1>)$ and thus the thread postcondition, which is stable because neither sender nor receiver writes to \<flag>.

The rely of the receiver is the \<bo>-stable union of the guarantees of the sender and the proxy. There is no effect of inter-thread \<bo> parallelism: the sender's interference doesn't depend on \<flag>, and the proxy's interference is stable against $\<msg>:=1$. The precondition of \<labd> is very familiar, reading $\<flag>=(1,1)$ for $\<flag>=1$, and is stable for familiar reasons. The strongest postcondition of \<labd> is 
\begin{equation}
(\<flag>=(1,1)=>\<msg>=1)@\<r1>,\_=\<flag>
\eqnlabel{WRCreceiverpostconditiond}
\end{equation}
which implies the embroidery of \<labd>$->$\<labe>. That embroidery is stable against each component of the rely: 
\begin{equation}
\cols
	& \<sp>((\<r1>=1=>|* A (\<flag>=(1,A)@(A=1=>\<msg>=1)))@\<true>,\; \<msg>:=1) \\
=	& (\<r1>=1=>|* A (\<flag>=(1,A)@(A=1=>\<msg'>=1)))@\<msg>=1 \\
=>	& (\<r1>=1=>|* A (\<flag>=(1,A)))@\<msg>=1 \\
=>	& \<r1>=1=>|* A (\<flag>=(1,A)@\<msg>=1)\\
=> 	& \<r1>=1=>|* A (\<flag>=(1,A)@(A=1=>\<msg>=1))
\sloc
\end{equation}
\begin{equation}
\cols
	& \<sp>((\<r1>=1=>|* A (\<flag>=(1,A)@(A=1=>\<msg>=1)))@\<Bfr>(\<flag>=(0,0)@\<msg>=1),\; \<flag>:=1,1) \vspace{2pt}\\
=	& \BRA (\<r1>=1=>|* A (\<flag'>=(1,A)@(A=1=>\<msg>=1))) \\
	  @\; \hook{\<Bfr>(\<flag>=(0,0)@\<msg>=1)}@\<flag'>=(0,0)@\<msg>=1@\<flag>=(1,1)\KET \vspace{2pt}\\
=>	& \<flag>=(1,1)@\<msg>=1 \\
=>	& |* A (\<flag>=(1,A)@(A=1=>\<msg>=1)) \\
=> 	& \<r1>=1=>|* A (\<flag>=(1,A)@(A=1=>\<msg>=1))
\sloc
\end{equation}
\begin{equation}
\cols
	& \<sp>((\<r1>=1=>|* A (\<flag>=(1,A)@(A=1=>\<msg>=1)))@\<Bfr>(\<flag>=(0,0)@A!=1),\; \<flag>:=1,A) \vspace{2pt}\\
=	& \BRA (\<r1>=1=>|* A (\<flag'>=(1,A)@(A=1=>\<msg>=1))) \\
		   @\; \hook{\<Bfr>(\<flag>=(0,0)@A!=1)}@\<flag'>=(0,0)@A!=1@\<flag>=(1,A)\KET \vspace{2pt}\\
=>	& A!=1@\<flag>=(1,A) \\
=>	& |* A (\<flag>=(1,A)@A!=1) \\
=>	& |* A (\<flag>=(1,A)@(A=1=>\<msg>=1)) \\
=> 	& \<r1>=1=>|* A (\<flag>=(1,A)@(A=1=>\<msg>=1))
\sloc
\end{equation}
The strongest postcondition of \<labe> is $\<r1>=1=>|* A (\<flag>=(1,A)@(A=1=>\<msg>=1))@\<r2>=\<msg>$, which implies the $\<labe>{->}$postcondition embroidery; that embroidery is stable because $|* A (\<flag>=(1,A))$ and $\<r2>=1$ are stable. The program postcondition \<final> is implied by the conjunction of the final assertions of the proxy and the receiver and, as usual, \<true> from the sender. 

It's all rather neat. The implication $\<flag>=(1,1)=>\<msg>=1$ is implied in the proxy's interference, and the information $\<r1>=1<=>\<flag>=(1,1)$ in its postcondition. The stability of assertions in the receiver comes, more or less, from the stability of $\<msg>=1$ and $\<flag>=(1,1)$.

\section{The lo, bo fragment}

Although we've concentrated on simple, almost trivial examples, we've done enough to show how elaboration-order \<lo> and before-order \<bo> can be used to impose orderings on a weak-memory program, even in what are thought to be subtle litmus tests. \<Lo> prevents elaboration reordering; \<bo> prevents propagation reordering. Embroidered with assertions they can be used, according to well-established and well-understood reasoning principles, to show that a program has enough ordering to work. 

Well-established and well-understood principles they may be, but weak-memory concurrency proofs aren't simple or trivial. Even in SC it's hard to invent algorithms, let alone prove that they work. Weak memory is worse, as you will surely have observed.

Most of our examples have involved a great deal of \<lo> ordering, almost restoring SC in most cases. We haven't added as much as perhaps you might think, because some of that ordering will be implicit in particular weak-memory architectures: see \secref{interferenceNstability}. %In \secref{interferenceNstability} we shall see that on many weak-memory architectures there is an implicit \<lo> constraint between a register assignment and uses of the same register, and between a read from and a write to, or a write to and a read from, a single variable. 
And our examples so far are small, often with nothing going on but some weak memory effect. Counteracting that effect allows little room for internal parallelism. We shall, however, see examples of internal parallelism in \secref{tokenring}.

%!TEX root = ./Paper.tex

\section{Structured commands and disjunction of knots}
\seclabel{structcom}

Reordering of elaboration and/or propagation introduces uncertainty about program execution. Constraints remove unhelpful uncertainty. Structured commands -- conditionals and loops -- allow alternative \<so>-tree paths and introduce helpful uncertainty. But they also introduce complexity: conditionals allow instances of the same component on different paths, and loops allow multiple instances of the same component on the same path.

Hoare logic for SC gives structured commands pre and postconditions. We rely entirely on constraints that relate primitive components -- assignments, \<assert>s and control expressions. As usual we introduce our treatment by considering examples.

\subsection{Conditionals}
\seclabel{conditionals}

A conditional `$\<if> E \<then> \<C1> \<else> \<C2> \<fi>$' introduces alternative \<so>-tree paths: one path includes \<C1>, the other \<C2>. An SC rule can use disjunction of postconditions to deal with the problem 
\begin{equation}
\inferR[\text{SC conditional}]{\pre{P}\<if> E \<then> \<C1> \<else> \<C2> \<fi>\post{\<Q1>|\<Q2>}}
							  {\pre{P@E}\<C1>\post{\<Q1>} & \pre{P@!E}\<C2>\post{\<Q2>}}
\eqnlabel{SCconditionalrule}
\end{equation}
Our logic doesn't give pre and postconditions to conditionals so we can't use this technique. Instead we use disjunction of knots. 

\labelname{post}
\begin{figure}
\centering
\begin{minipage}{200pt}
$$\cols[c]
  \assert{\<init>::\<flag>=0} \vspace{5pt}\\
  \cols
    \assertd{\<lo> \<init>:: \<flag>=1=>\<msg>=1}\ \<labc>:: \<r1>:=\<flag>; \vspace{2pt} \\
    \<if> \assertd{\<lo> \<labc>:: \<r1>=1=>\<msg>=1} \<beta>:: \<r1>=1 \<then>  \vspace{2pt}\\
    \pindent\assertd{\<lo> \<beta>_{t}:: \<msg>=1}\ \<labd>:: \<r2>:=\<msg>  \vspace{2pt}\\
    \<fi> \vspace{3pt} \\
    \assertd{\<lo> \<labd>:: \<r2>=1} | \assertd{\<lo> \<beta>_{f}:: \<r1>!=1}
  \sloc \vspace{5pt} \\
  \assert{\<final>::(1:::\<r1>)=1=>(1:::\<r2>)=1}
\sloc$$
\end{minipage}
\hstrut{20pt}
\begin{minipage}{100pt}
\begin{tikzpicture}[>=latex',scale=0.35]
	\input{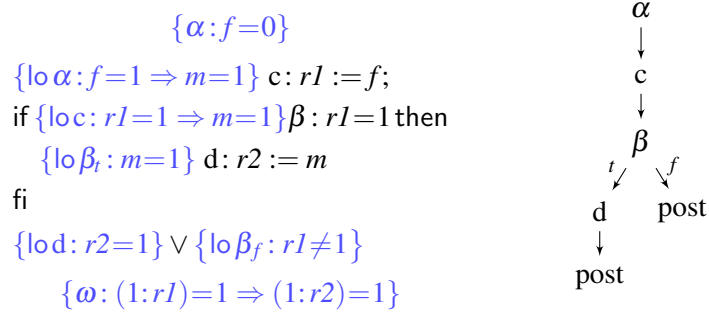}
%%	\begin{dot2tex}[file=conditionalsotree,dot,tikz,codeonly,styleonly,mathmode,options=-s]
%%		digraph G {
%%    		init[label="\<init>"]; /* this is a comment */
%%			c[label="\<labc>"];
%%    		init -> c;
%%    		beta[label="\<beta>"];
%%    		c -> beta;
%%			d[label="\<labd>"]
%%    		beta -> d[headlabel="_{t}",labelangle=25,labeldistance=3.0]; 
%%			beta -> f1[headlabel="_{f}"labelangle=-35,labeldistance=3.0];
%%    		f0[label="\<post>"]; f1[label="\<post>"];
%%    		d -> f0
%%		}
%% 	\end{dot2tex}
\end{tikzpicture} 
\end{minipage}
\caption{A conditional MP receiver and its \setorder{so} tree}
\figlabel{conditionalMPreceiver}
\end{figure}

Our example is a `conditional receiver' for MP which reads \<flag>, and then reads \<msg> only if \<flag> was 1. Unlabelled and unlaced we might write
\begin{equation}
\<r1>:=\<flag>;\; \<if> \<r1>=1 \<then> \<r2>:=\<msg> \<fi>
\eqnlabel{conditionalSCreceiver}
\end{equation}
Labelled, laced and embroidered we have the thread in \figref{conditionalMPreceiver}. In the same figure we show the thread's \<so> tree. We expand the conditional into a tree with two arms, and append \<post>, which stands for the thread postcondition, to each arm. We label the arrows exiting control expression \<beta> with $t$ and $f$ to indicate that one path is to be taken when \<beta> evaluates to \<true>, the other when it is \<false>. Note that the tree is derived entirely syntactically: it would have two arms even if \<beta> had a constant value,% -- and we shall see in \secref{tokenring} that this can be a problem.

Command \<labc> of \figref{conditionalMPreceiver} is laced after the initial assertion. We lace control expression \<beta> after \<labc> so that it can test what \<labc> read, and we lace \<labd> after $\<beta>_{t}$ so that it is only elaborated when \<labc> read 1. If execution follows the path including $\<beta>_{t}$ then, this proof claims, \<labd> will certainly read 1. If execution follows the other path then certainly $\<r1>!=1$. The thread postcondition is a disjunction of knots, one covering the `$\<beta>_{t}{->}\<labd>{->}\<post>$' path in the \<so> tree, the other the `$\<beta>_{f}{->}\<post>$' path.

\subsection{The constraint-coverage principle}

A thread program contains a single instance of each command and each control expression. The \<so> tree of the thread may contain multiple instances of a component; the program, in effect, conflates all those instances into one. In \figref{conditionalMPreceiver}, for example, the thread-postcondition component \<post> appears twice. Constraints apply between instances of their {\csource} and {\ctarget}; to conflate those instances and their constraints we must apply the disjunction of the instance-constraints to the program component. 

If a component were constrained on some paths but not all, the overall constraint would be \<true> -- i.e. no constraint. That would be useless as well as confusing.
\definition{Constraint-coverage principle}{constraintcoverage}{%!TEX root = ./Paper.tex
If a component is constrained at all, then for all paths in the \<so> tree to an instance of that component, there must be a disjunct in the component's knot which depends on one of the nodes of that path.} 
The program in \figref{conditionalMPreceiver} obeys the constraint-coverage principle. The {\csource}s of the constraints on \<labc> and \<beta> appear on every path to those components; the {\csource} of the constraint on \<labd> appears on every path to \<labd>; there's a disjunct in the thread-postcondition knot, which constrains termination, for each of the paths through the conditional. 

\begin{figure}
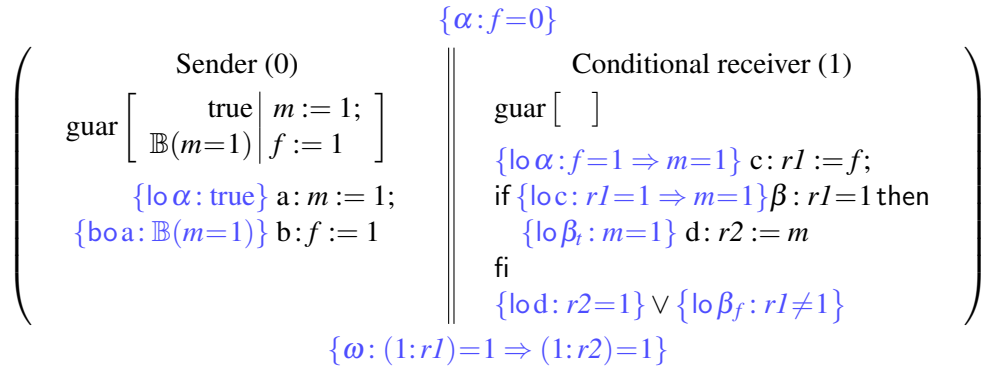

\centering
% proofs/MP_conditional.proof
$$\cols[c]
\assert{\<init>::\<flag>=0} \vspace{3pt} \\
\BRA[l||l]
\thr{Sender (0)}
  \guarantee{\interferenceg{\<true>}{\<msg>:=1}; \\ \interferenceg{\<Bfr>(\<msg>=1)}{\<flag>:=1}} \vspace{3pt} \\
  \!\!\cols[r@{\intfspace}l]
  \assertd{\<lo> \<init>:: \<true>} & \<laba>:: \<msg>:=1; \\
  \assertd{\<bo> \<laba>:: \<Bfr>(\<msg>=1)} & \<labb>:: \<flag>:=1
  \sloc
\rht
&
\thr{Conditional receiver (1)}
  \emptyguarantee \vspace{3pt} \\
  \assertd{\<lo> \<init>:: \<flag>=1=>\<msg>=1}\ \<labc>:: \<r1>:=\<flag>; \\
  \<if> \assertd{\<lo> \<labc>:: \<r1>=1=>\<msg>=1} \<beta>:: \<r1>=1 \<then> \\
  \pindent\assertd{\<lo> \<beta>_{t}:: \<msg>=1}\ \<labd>:: \<r2>:=\<msg> \\
  \<fi> \\
  \assertd{\<lo> \<labd>:: \<r2>=1} | \assertd{\<lo> \<beta>_{f}:: \<r1>!=1}
\rht
\KET \vspace{3pt} \\
\assert{\<final>::(1:::\<r1>)=1=>(1:::\<r2>)=1}
\sloc$$
\caption{Proof of conditional message-passing}

\figlabel{conditionalMPreceiverproof}
\end{figure}

\subsection{Proof of the conditional receiver}

The proof is in \figref{conditionalMPreceiverproof}. The sender is as in \figref{MPproof}. The precondition of command \<labc> in the receiver is implied by initial assertion \<init>, as before in \figref{MPproof}; it is stable against the sender's guarantee, as before; there is no entry in the guarantee, as before. The precondition $\<r1>=1=>\<msg>=1$ of control expression \<beta> is implied by the postcondition $(\<flag>=1=>\<msg>=1)@\<r1>=\<msg>$ of \<labc>, as with the precondition of \<labd> in \figref{MPproof}; it's stable, as before. 

\definition{Postcondition of a control expression}{controlexprpostcondition}{%!TEX root = ./Paper.tex
A control expression $\<gamma>::E$ with elaboration precondition $P$ has two postconditions, one for each of its outcomes. The $\<gamma>_{t}$ outcome has $P@E$; the $\<gamma>_{f}$ outcome has $P@!E$.} 
So for the $\<beta>_{t}{->}\<labd>$ constraint, the {\csource} postcondition is $(\<r1>=1=>\<msg>=1)@\<r1>=1$, which implies the $\<msg>=1$ embroidery. That embroidery is stable against the rely. The postcondition of \<labd> implies $\<r2>=1$, which is the embroidery on the \<labd>$->$\<post> constraint; once again it is stable. 

For the $\<beta>_{f}{->}$\<post> constraint the {\csource} postcondition is $(\<r1>=1=>\<msg>=1)@\<r1>!=1$, which implies $\<r1>!=1$, which is stable against the rely. The disjunction $\<r2>=1 | \<r1>!=1$ of the conjunctions of the assertions in each knot gives us the final postcondition that we require.

\subsection{Conditional lacing summary}

Lacing is already restricted so that it reinforces \<so>. The constraint-coverage principle further restricts it so that if a knot outside a conditional appeals to one of its arms, then it must be disjoined with a knot that appeals to the other arm. In lacing basic commands and control expressions our intention is to allow the programmer to express the minimal constraints that make the program work. Conditionals (and loops) don't exist at the machine level: there are only instructions and jumps. Focusing on commands and control expressions allows us to work in the same sort of way.

\subsection{\compilingmark{} Compiling constraints with a control-expression {\csource}}
\seclabel{nomoreisync}

As programmers we believe that barriers, which divide one part of an execution from another, are expensive, but that command orderings, which put one command before another, are cheap. Our \<lo> constraint corresponds to a command-ordering dependency, which makes it cheaper than \<bo>, which needs a barrier, which in turn is cheaper than \<uo>, which needs a very big barrier indeed.\footnote{Hardware architects encourage our belief that barriers are costly, and don't discourage our beliefs about their relative costs, but are tight-lipped about the magnitudes, so we're somewhat guessing.}

On two important architectures, Power and ARM, \<lo> constraints whose {\csource} is a control expression must be implemented with a branch plus a barrier (\<isync> on Power, \<isb> on ARM). It's the most lightweight of their barriers, but to a programmer it's annoying. Surely it shouldn't be necessary to use heavy machinery to constrain something as simple as a conditional read. We believe that the barrier slows execution considerably\footnote{And we've heard a hardware architect say so} so we'd like to avoid it if we can. In many cases, luckily, we can do so very easily.

\begin{figure}
\centering
$$\cols
\assertd{\<lo> \<init>} \<labc>::\<r1>:=\<flag>; \\
\<if> \assertd{\<lo> \<labc>} \<beta>::\<r1>=1 \<then> \\
\begin{progindent}
	\assertd{\<lo> \<labc>; \<lo> \<beta>_{t}} \<labd>::\<r2>:=\<msg>
\end{progindent} \\
\<fi> \\
\assertd{\<lo> \<labd>} | \assertd{\<lo> \<beta>_{f}}
\sloc$$
\caption{Alternative, more execution-efficient, lacing of the conditional MP receiver}
\figlabel{conditionalMPreceiverrelaced}
\end{figure}

In the MP receiver of \figref{MPproof}, $\<labd>::\<r2>:=\<msg>$ is \<lo>-constrained to be after $\<labc>::\<r1>:=\<flag>$. In \figref{conditionalMPreceiver} the same is true, but the constraint is indirect, through the control expression $\<beta>::\<r1>=1$. If we were to lace the thread slightly differently, making \<labd> depend on whatever \<beta> depends on, as in \figref{conditionalMPreceiverrelaced}, then we can argue that a barrier is unnecessary. It is irrelevant whether \<beta> is elaborated before, after or at the same time as \<labd>: if \<labd> is speculatively elaborated then, since it is on the same \<so> path as $\<beta>_{t}$ and \<lo>-follows the same command, successful speculation can only take place in a state in which \<beta> either has delivered or will deliver \<true>. The $\<beta>_{t}{->}\<labd>$ constraint is then operationally redundant and need not be implemented, though its embroidery may be safely used in the proof.

The \<isync>/\<isb> barrier is expensive, we believe, because it forces the hardware to order all \<so>-before-barrier reads before all \<so>-after-barrier reads. Given explicit lacing, we can see which reads need to be ordered, and we can use explicit \<lo> constraints between them. A clever compiler might be able to notice that implementation of \figref{conditionalMPreceiver} only needs an \<lo> constraint \<labc>$->$\<labd>. Expert programmers will know the trick, and will expect a compiler to implement it. But even if a compiler inserts a barrier when it doesn't really need to, we shall still get the result we require. 

Although it can easily be eliminated in simple examples, \<isync> is by no means redundant. In larger and more intricate examples it can be necessary to reserve a register for each multiply-constrained read in order to compile its \<lo> dependencies. At some point pressure on registers would become overwhelming, and then it would be best to use ctrl-\<isync>. 

One of our intentions in developing a program logic (at least for Parkinson and Bornat) was to explain weak memory to ourselves. The observation that you don't seem to need \<isync>/\<isb> very often seems to have escaped most observers, but is perfectly clear in logical terms.

\subsection{Loops}
\seclabel{loops}

Loops, like conditionals, need special lacing. Not only do we have to deal with constraint coverage, we also have to cope with \<lo>- and \<bo>-parallel interference of assignments from successive executions of the loop body.

%Loops in SC Hoare logic need \emph{loop invariants}, assertions which apply before, between and after executions of the loop body. A possible rule for a \<while> loop with invariant $P$ is
%\begin{equation}
%\inferR[\text{SC \<while>}]{\pre{P}\<while> E \<do> C \<od>\post{P@!E}}
%						   {\pre{P@E} C \post{P}}
%\eqnlabel{SCwhilerule}
%\end{equation}
%Our logic uses disjunction of knots including \emph{loopback}s, from {\csource}s later in a loop to a {\ctarget} earlier in the loop, to solve the problem of loop invariance.
%
%
%
%Loops, like conditionals, require special lacing. The precondition of a command in a loop often has to be a disjunction of a knot constrained by components outside the loop, and another knot constrained by components later in the loop.

We illustrate the problems, and their solutions, by considering an MP receiver which loops, waiting to see $\<flag>=1$, before it reads \<msg>. Unlabelled and unlaced we might write 
\begin{equation}
\<do> \<r1>:=\<flag> \<until> \<r1>=1;\; \<r2>:=\<msg>
\eqnlabel{MPloopSC}
\end{equation}
\Figref{MPloop} shows this receiver laced for weak memory, together with a prefix of its infinite \<so> tree. 

Consider first the tree. There is a path $\<init>{->}\<labc>{->}\<beta>$; from there via $\<beta>_{t}$ we reach \<labd>, but via $\<beta>_{f}$ we are back at \<labc>, and then on to \<beta> again. And so on, infinitely: each time round the loop there's a split into two paths. The infinite path which never takes the $\<beta>_{t}$ branch contains both \<beta> and \<labc> infinitely often.  

\begin{figure}
\centering
\begin{minipage}{200pt}
$$\cols[c]
\assert{\<init>::\<flag>=0} \vspace{3pt} \\
\cols 
\<do> \\
\begin{progindent}
	\assertd{\<lo> \<init>} |> \assertd{\<lo> \<labc>} \<labc>::\<r1>:=\<flag>
\end{progindent} \\
\<until> \assertd{\<lo> \<labc>} \<beta>::\<r1>=1; \\ 
\assertd{\<lo> \<beta>_{t}} \<labd>::\<r2>:=\<msg>
\sloc \vspace{5pt}\\
\assert{\<final>::(1:::\<r2>)=1} 
\sloc$$
\end{minipage}
\hstrut{20pt}
\begin{minipage}{100pt}
\begin{tikzpicture}[>=latex',scale=0.3]
	\input{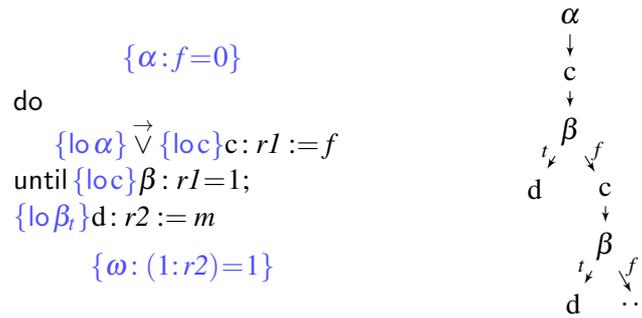}
%%	\begin{dot2tex}[file=loopmptree,dot,tikz,codeonly,styleonly,mathmode,options=-s]
%%		digraph G {
%%    		init[label="\<init>"]; /* this is a comment */
%%			c[label="\<labc>"];
%%    		init -> c;
%%    		beta[label="\<beta>"];
%%    		c -> beta;
%%			exit[label="\<labd>"];
%%    		beta -> exit[headlabel="_{t}",labelangle=25,labeldistance=3.0]; 
%%    		c1[label="\<labc>"]; 
%%			beta -> c1[headlabel="_{f}"labelangle=-35,labeldistance=3.0];
%%    		c1 -> beta1;
%%    		beta1[label="\<beta>"];
%%			exit1[label="\<labd>"];
%%    		beta1 -> exit1[headlabel="_{t}",labelangle=25,labeldistance=3.0]; 
%%    		c2[label="\dots"]; 
%%			beta1 -> c2[headlabel="_{f}"labelangle=-35,labeldistance=3.0];
%%		}
%% 	\end{dot2tex}
\end{tikzpicture} 
\end{minipage}
\caption{An MP receiver with a loop, and part of its \setorder{so} tree}
\figlabel{MPloop}
\end{figure}

A non-disjunctive constraint $\<init>{->}\<labc>$ in the program would have to stand for an infinite number of constraints in the tree, one to each instance of \<labc>, each carrying the same embroidery. That isn't impossible or even unreasonable, but it does produce a surprising result. A constraint which runs from \<init> to the second instance of \<labc>, for example, is \<lo> parallel with the first instance of \<labc>, which will interfere and may destabilise its embroidery. That's quite unlike the constraints which we've seen before: in the absence of loops, a constraint $\<laba>{->}\<labb>$ is never \<lo> parallel with either \<laba> or \<labb>.

What a component inside a loop requires is a stylised lacing which describes a constraint from the initial path to its first instance, and a constraint from that instance or something later in the loop to its next instance, and so on indefinitely. In the case of the looping receiver, the constraint on the first instance of \<labc> should come from \<init>, and subsequently each instance should take its constraint from the previous one. The effect is an infinity of constraints, all but one identical, and none of them \<lo> parallel with \<labc>. The overall constraint is the disjunction of the first constraint and one of the identical constraints.

All of that is indicated by the iterated constraint $\assertd{\<lo> \<init>} |> \assertd{\<lo> \<labc>}$ in \figref{MPloop}; the first disjunct covers paths from the beginning of execution and the second covers paths round the loop. The rest of the lacing is straightforward: \<beta> always follows \<labc>; \<labd> always follows $\<beta>_{f}$; thread postcondition \<post>, when we include one, always follows $\<labd>$. Note that each precondition has full coverage of the \<so>-tree paths to its component.

\definition{Iterated constraint}{iteratedconstraint}{%!TEX root = ./Paper.tex
A component \setlabelname{x} in a loop may have an iterated constraint $\setvar{knot1}|>\setvar{knot2}$. Constraints in \setvar{knot1} must cover all paths from the initial assertion to \setlabelname{x}; constraints in \setvar{knot2} must cover all paths from \setlabelname{x} to \setlabelname{x}.}

After all that preparation we discover that the proof in \figref{MPloopproof} doesn't need to use an iterated constraint. The embroidery of the constraint $\<init>{->}\<labc>$ doesn't mention \<r1>, so is immune to the interference of \<labc>, so it can constrain all its instances. (We don't apologise for introducing the iterated constraint through this example: several later examples will need it.)

\begin{figure}
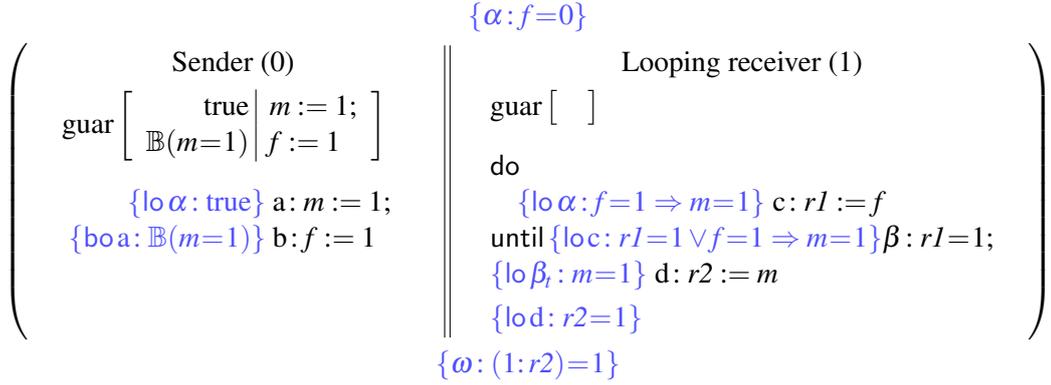

\centering{}
% proofs/MP_dountil.proof
$$\cols[c]
\assert{\<init>::\<flag>=0} \vspace{3pt} \\
\BRA[l||l]
\thr{Sender (0)}
  \guarantee{\interferenceg{\<true>}{\<msg>:=1}; \\ \interferenceg{\<Bfr>(\<msg>=1)}{\<flag>:=1}} \vspace{3pt} \vspace{3pt} \\
  \!\!\cols[r@{\intfspace}l]
  \assertd{\<lo> \<init>:: \<true>} & \<laba>:: \<msg>:=1; \\
  \assertd{\<bo> \<laba>:: \<Bfr>(\<msg>=1)} & \<labb>:: \<flag>:=1
  \sloc
\rht
&
\thr{Looping receiver (1)}
  \emptyguarantee \vspace{3pt} \vspace{3pt} \\
  \<do> \\
  \pindent\assertd{\<lo> \<init>:: \<flag>=1=>\<msg>=1}\ \<labc>:: \<r1>:=\<flag> \\
  \<until> \assertd{\<lo> \<labc>:: \<r1>=1|\<flag>=1=>\<msg>=1} \<beta>:: \<r1>=1; \\
  \assertd{\<lo> \<beta>_{t}:: \<msg>=1}\ \<labd>:: \<r2>:=\<msg> \vspace{3pt} \\
  \assertd{\<lo> \<labd>:: \<r2>=1}
\rht
\KET \vspace{3pt} \\
\assert{\<final>::(1:::\<r2>)=1}
\sloc$$
\caption{Proof of the looping receiver}
\figlabel{MPloopproof}
\end{figure}

In the proof the precondition $\<flag>=1=>\<msg>=1$ of \<labc> is implied by initial assertion \<init>, as before in \figref{MPproof}; it is stable against the rely, as before; it's stable against the interference of \<labc> itself (a new check); the constraint isn't \<lo>-parallel with \<labd>, as before; and there is nothing to write in the guarantee, as before. The precondition of \<beta> is implied by the postcondition of \<labc>, is stable against the rely, and the constraint isn't \<lo>-parallel with anything. The precondition of \<labd> is implied by the $\<beta>_{t}$ postcondition $(\<r1>=1=>\<msg>=1)@\<r1>=1$. The postcondition of \<labd> implies $\<r2>=1$, which is stable against anything, and the proof is complete. 

In passing we note that this example shows why our logic can't, in general, deal with termination. We can't prove that the receiver will ever read 1 from \<flag>: the rely only says that the sender \emph{might} sometimes write $\<flag>=1$. 

We also note that, using the technique of \secref{nomoreisync}, this example doesn't need a barrier to implement the $\<beta>_{t}{->}\<labd>$ constraint. Loops just got cheaper (sometimes).

\begin{figure}
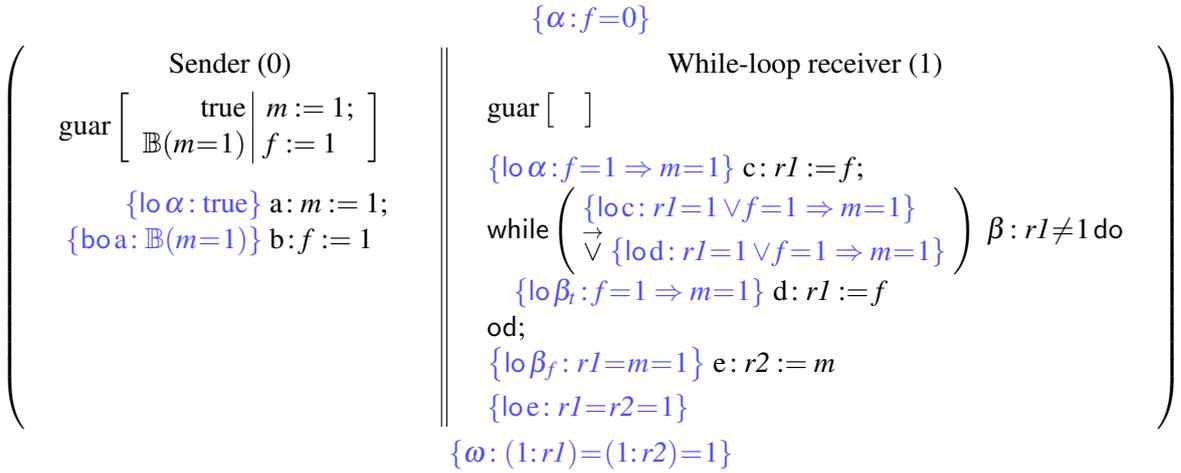

\centering
% proofs/MP_while.proof
$$\cols[c]
\assert{\<init>::\<flag>=0} \vspace{3pt} \\
\BRA[l||l]
\thr{Sender (0)}
  \guarantee{\interferenceg{\<true>}{\<msg>:=1}; \\ \interferenceg{\<Bfr>(\<msg>=1)}{\<flag>:=1}} \vspace{3pt} \vspace{3pt} \\
  \!\!\cols[r@{\intfspace}l]
  \assertd{\<lo> \<init>:: \<true>} & \<laba>:: \<msg>:=1; \\
  \assertd{\<bo> \<laba>:: \<Bfr>(\<msg>=1)} & \<labb>:: \<flag>:=1
  \sloc
\rht
&
\thr{While-loop receiver (1)}
  \emptyguarantee \vspace{3pt} \vspace{3pt} \\
  \assertd{\<lo> \<init>:: \<flag>=1=>\<msg>=1}\ \<labc>:: \<r1>:=\<flag>; \\
  \<while> \BRA \assertd{\<lo> \<labc>:: \<r1>=1|\<flag>=1=>\<msg>=1} \\
  				|> \;\assertd{\<lo> \<labd>:: \<r1>=1|\<flag>=1=>\<msg>=1}\KET \ 
				\<beta>:: \<r1>!=1 \<do> \\
  \pindent\assertd{\<lo> \<beta>_{t}:: \<flag>=1=>\<msg>=1}\ \<labd>:: \<r1>:=\<flag> \\
  \<od>; \\
  \assertd{\<lo> \<beta>_{f}:: \<r1>=\<msg>=1}\ \<labe>:: \<r2>:=\<msg> \vspace{3pt} \\
  \assertd{\<lo> \<labe>:: \<r1>=\<r2>=1}
\rht
\KET \vspace{3pt} \\
\assert{\<final>::(1:::\<r1>)=(1:::\<r2>)=1}
\sloc$$
\caption{MP receiver with a \setbracket{while} loop}
\figlabel{MPloopwhile}
\end{figure}

We used a \<do>-\<until> loop in our example because it avoids an invariant-intialising assignment and can illustrate both the possibility and the non-necessity of a loopback constraint. We can also handle a receiver with a \<while> loop: see \figref{MPloopwhile}. In this example we do need the loopback \<labd>$->$\<beta> in an iterated constraint, to ensure that components after the loop come after command \<labd> if it is elaborated. Without the loopback both the precondition of \<labe> and the thread postcondition are \<lo>-unstable against the interference of \<labd>.

\clearpage
\part{More precision and more orderings}
\partlabel{morestuff}
\thispagestyle{empty}
\clearpage
%!TEX root = ./Paper.tex

\section{Interference and stability}
\seclabel{interferenceNstability}
\seclabel{boloends}

In our examples so far, all variable occurrences in interference preconditions have been \<Bfr>-modal, and we have in effect used the SC stability \ruleref{SCstability} when dealing with external, \<lo>-parallel or \<bo>-parallel interference. But in general an interference precondition may contain amodal occurrences of variables, and then the SC rule won't work. It is time to confront the complexity of weak-memory stability checks and reveal the rules we actually use. 

In this section we also discuss `SC per location' aka `global coherence', with which some, perhaps most, architectures constrain propagation reordering, and `SC per register'. aka `register renaming'%, and the \<U> modality, which orders propagation more sternly than \<Bfr>
. Each of these mechanisms affects internal and external parallelism and therefore our treatment of stability.

We begin, once again, by considering the stability of assertion $P$ against interference \interferenced{Q}{x:=E}, taking into account \ruleref{quantifiedstability}, repeated here for convenience. 
\begin{tcolorbox}[before skip=3pt,top=3pt,bottom=3pt,lowerbox=ignored,colframe=red!50!black,colback=red!10!white, colbacktitle=red!20!white]%
%!TEX root = ./Paper.tex
$P$ is stable against \interferenced[\nvec{ns}]{Q}{x:=E} if it is stable against \interferenced{Q[\nvec{ns}\backslash\nvec{fs}]}{x:=E[\nvec{ns}\backslash\nvec{fs}]} where $\nvec{fs}$ are fresh names. Similarly for internal interference \interferenced[\nvec{ns}]{Q}{r:=E} and \interferenced[\nvec{ns}]{Q}{r:=x}.
\end{tcolorbox}

\subsection{Internal and external parallelism}

In SC execution the interfered-with $P$-asserting thread and the interfering $Q$-asserting thread see the same state of the store, so \ruleref{SCstability} considers the case when both $P$ and $Q$ hold. For \emph{internal} (\<lo>-parallel) stability, coming from an assigment in the same thread, both assertion and assignment operate within the same local `store' of buffers and caches so we can, in effect, use the SC rule -- that is, \ruleref{LOstability} of \secref{loparallelism}, repeated here for convenience. 
\begin{tcolorbox}[before skip=3pt,top=3pt,bottom=3pt,lowerbox=ignored,colframe=red!50!black,colback=red!10!white, colbacktitle=red!20!white]%
%!TEX root = ./Paper.tex
Constraint embroidery $P$ is LO stable against \<lo>-parallel assignment $A$ with precondition $Q$ if \vspace{2pt}\\
\hstrut{50pt} $\<sp>(P@Q,\;A)=>P$%
\end{tcolorbox}
Internal interference, just like external interference, is quotiented to hide registers (see \secref{SCregquotienting} for further discussion).

When interference comes from another thread we can't suppose that in general $Q$ must hold in the $P$-asserting thread at the instant when $x:=E$ interferes. Operationally this is because $x=E$ might arrive before all or some of the writes which underpin $Q$. In previous examples we have used \<Bfr>-modal $Q$ to eliminate propagation delay. Now we consider the writes which underpin amodal occurrences of variables in $Q$. Our treatment of stability must ensure that $P$ is stable in all the possible states corresponding to early or late propagation of those writes. We do that by \emph{hatting} amodal occurrences of variables to decouple them from the values of the same variables if they occur in $P$:
\ruledef{EXT stability}{EXTstability}{%!TEX root = ./Paper.tex
Constraint embroidery $P$ is EXT stable against \interferenced{Q}{x:=E} from another thread if \vspace{3pt} \\
\hstrut{50pt}$\<sp>(P@\widehat{Q},\;x:=E) => P$ 
}
$\widehat{Q}$ is a copy of $Q$ in which every free occurrence of a variable $v$ is renamed to $\<vhat>$, \emph{except} those inside a \<Bfr> or \<U> modality. 
\definition{Hatting}{hatdefinition}
{$$\cols[rcll] \widehat{\setvar{var}}				&=& \widehat{\setvar{var}} 						& \text{whether free or bound} \vspace{2pt} \\
\widehat{\setvar{reg}}				&=& \setvar{reg} \vspace{2pt} 					\\
\widehat{\setvar{const}}			&=& \setvar{const} \vspace{2pt} 				\\
\widehat{P\;\setbracket{binop}\;Q} 	&=& \widehat{P}\;\setbracket{binop}\;\widehat{Q}\vspace{2pt} \\
\widehat{\setbracket{unop}\;P} 		&=& \setbracket{unop}\;\widehat{P} 				\vspace{2pt} \\
\widehat{\setbracket{binder}\;v(P)}	&=& \setbracket{binder}\; v(\widehat{P}) 		\vspace{2pt} \\
\widehat{\<Bfr>(P)}					&=& \<Bfr>(P)@\widehat{P} 
 \sloc$$}
The intuition is that \<v> means `\<v> in the asserting thread, when the interference arrives' and \<vhat> means `\<v> in the interfering thread, when the interference was elaborated'. Note, then, that hatting operates even within binders: $\widehat{@*x(P(x))}$ is a remark about variables that held when the interference was constructed and is adequately described by $@*x(P(\widehat{x}))$ if we recognise that the binding quotients the identity of the variable, not its temporal or thread-specific occurrences. Expansion of the \<Bfr> modality allows the enclosed assertion to apply both when the interference arrives and when it was elaborated.

Observe that hatting separates the \<Bfr>-modal parts of $Q$, whose underpinning writes must be propagated before $x=E$, from the rest of $Q$, whose writes may or may not be propagated before $x=E$. Retaining the amodal writes in hatted form enables us to continue to use $E$, which may refer via registers to those hatted occurrences, to describe the value assigned by the interference. 

Hatting decouples amodal variable occurrences in $Q$ from similar occurrences in $P$. To impose some correspondence between hatted and unhatted versions -- say $y=\<yhat>$ -- would merely strengthen the left-hand side of the implication in \ruleref{EXTstability}. So if $P$ is stable when all amodal writes arrive late, then it is stable when some arrive on time.

Previous examples had no amodal variable occurrences in interference preconditions, so the interference precondition $Q$ was preserved by hatting, and the extra hatted formulas derived from \<Bfr> modalities were irrelevant in our examples, in which the assigned expression $E$ didn't depend on the values of variables. 

\subsection{In-flight interference}
\seclabel{bostability}

The EXT stability \ruleref{EXTstability} makes it clear that the important part of an interference precondition is its unhatted -- modal -- part. We might suppose, then, that the in-flight stability of \interferenced{P}{y:=F} against \<bo>-parallel \interferenced{Q}{x:=E} could be checked by
\begin{equation}
\<sp>(\widehat{P}@Q,\;x:=E) => \widehat{P}
\eqnlabel{bostabilitynotquite}
\end{equation}
But that would provide trivial stability in the case where the amodal parts of $Q$ contradict the modal parts of $P$. That can happen if $P$, modally, and $Q$, amodally, mention $x$, and/or if \<lo>-intervening assignments change other variables mentioned modally in $P$ and amodally $Q$. So we need to decouple the modal parts of $P$ from the amodal parts of $Q$. We can't simply hat $P$, because that would couple the amodal parts of $P$ and $Q$. Instead we hat $P$ with a different accent: $\dhatv{P}$ is just like $\widehat{P}$ in its operation.
\ruledef{In-flight (\setorder{bo}) stability}{BOstability}{%!TEX root = ./Paper.tex
Interference \interferenced{P}{y:=F} is BO stable against \<bo>-parallel \interferenced{Q}{x:=E} if \vspace{3pt}\\
\hstrut{50pt} $\<sp>(\widehat{P}@\dhatv{Q},\;x:=E)=>\widehat{P}$
}
In this rule there are unhatted variables, whose values are propagated before $y=F$ and $x=E$; hatted variables, whose values applied when $P$ was elaborated; and double-hatted variables, whose values applied when $Q$ was elaborated. 

This rule applies to the interference of assignments within a thread which are \<bo> parallel: that is, assignments to different variables which are \<so> but not \<bo> ordered. It also applies to interferences in the rely that come from different threads and assign to different variables: in that case we can think of $x$ as `$x$ in the interfered-with thread', $\widehat{x}$ as `$x$ in one interfering thread' and $\dhatv{x}$ as `$x$ in the other interfering thread'. 

BO parallelism asymmetrical within a thread, according to \defref{BOparallel}, repeated here for convenience. 
\begin{tcolorbox}[before skip=3pt,top=3pt,bottom=3pt,lowerbox=ignored,colframe=blue!50!black,colback=blue!10!white, colbacktitle=blue!20!white]%
%!TEX root = ./Paper.tex
If variable assignments to distinct variables are \<so>- but not \<bo>-ordered then the \<so>-later is \emph{bo parallel} with the \<so>-earlier and will interfere with its interference precondition.%
\end{tcolorbox}
The elaboration precondition of the later assignment will have, baked in, the effect of the first assignment.

\begin{figure}
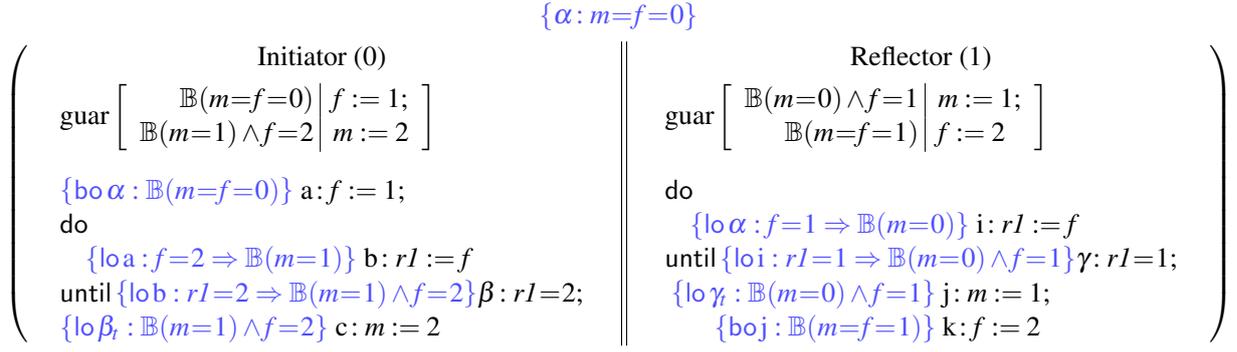

\centering
% proofs/MPdoubleparallel.proof
\scalebox{0.95}{$\cols[c]
\assert{\<init>::\<msg>=\<flag>=0} \vspace{3pt} \\
\BRA[l||l]
\thr{Initiator (0)}
  \guarantee{\interferenceg{\<Bfr>(\<msg>=\<flag>=0)}{\<flag>:=1}; \\ \interferenceg{\<Bfr>(\<msg>=1)@\<flag>=2}{\<msg>:=2}} \vspace{3pt} \vspace{3pt} \\
  \assertd{\<bo> \<init>: \<Bfr>(\<msg>=\<flag>=0)}\ \<laba>:: \<flag>:=1; \\
  \<do> \\
  \pindent\assertd{\<lo> \<laba>: \<flag>=2=>\<Bfr>(\<msg>=1)}\ \<labb>:: \<r1>:=\<flag> \\
  \<until> \assertd{\<lo> \<labb>: \<r1>=2=>\<Bfr>(\<msg>=1)@\<flag>=2} \<beta>:: \<r1>=2; \\
  \assertd{\<lo> \<beta>_{t}: \<Bfr>(\<msg>=1)@\<flag>=2}\ \<labc>:: \<msg>:=2
\rht
&
\thr{Reflector (1)}
  \guarantee{\interferenceg{\<Bfr>(\<msg>=0)@\<flag>=1}{\<msg>:=1}; \\ \interferenceg{\<Bfr>(\<msg>=\<flag>=1)}{\<flag>:=2}} \vspace{3pt} \vspace{3pt} \\
  \<do> \\
  \pindent\assertd{\<lo> \<init>: \<flag>=1=>\<Bfr>(\<msg>=0)}\ \<labi>:: \<r1>:=\<flag> \\
  \<until> \assertd{\<lo> \<labi>: \<r1>=1=>\<Bfr>(\<msg>=0)@\<flag>=1} \<gamma>:: \<r1>=1; \\
  \!\!\cols[r@{\intfspace}l]
  \assertd{\<lo> \<gamma>_{t}: \<Bfr>(\<msg>=0)@\<flag>=1} & \<labj>:: \<msg>:=1; \\
  \assertd{\<bo> \<labj>: \<Bfr>(\<msg>=\<flag>=1)} & \<labk>:: \<flag>:=2
  \sloc
\rht
\KET
\sloc$}
\caption{Spurious \setorder{bo} parallelism }
\figlabel{spuriousboparallelism}
\end{figure}

But why should we take any account at all of the \<Bfr>-modal parts of $Q$ in the BO stability check? The main reason is that the test for \<bo> parallelism in \defref{BOparallel} is too weak: it doesn't take account of what is called `detour' in~\citep{AlglaveetalHerdingcats2014}: causal chains between two assignments which may lead through more than one thread back to the first and include a \<bo> in one of the threads it passes through. In \figref{spuriousboparallelism}, for example, there is a chain \<laba>$->$\<labi>$->$\<gamma>$->$\<labj>$->$\<labk>$->$\<labb>$->$\<beta>$->$\<labc>. Execution must traverse that path: after \<laba> we have $\<flag>=1$ and the value of \<flag> won't change without the action of the reflector thread. The \<labj>$->$\<labk> constraint is \<bo>, and in the proof, even though \<laba> and \<labc> are \<bo> parallel according to \defref{BOparallel}, they have contradictory \<Bfr>-modal interference preconditions and so \<labc> won't destabilise the interference of \<laba>. %Indeed the only way that two apparently \<bo> parallel assignments can acquire contradictory \<Bfr>-modal interference preconditions is via a detour through another thread's \<bo> link.

%%Why does \ruleref{BOstability} quotient the amodal parts of $Q$ by hatting them? We have to use those amodal parts in order to give meaning to the assigned formula $F$ in $y:=F$. Quotienting them is the obvious answer. Hatting does the job, but it treats the two interference preconditions as if they were in different threads which, when we are considering \<bo>-parallel assignments in a single thread, is surprising.
%%
%%Finally, the asymmetry. In a single thread, the \<so>-later of two \<so> but not \<bo> ordered assignments is \<bo> parallel with the earlier, but not vice-versa. We observe that a set of assignments which are \<so> but not \<bo> ordered can be propagated in any order. In principle they could all be given similar interference precondition, but that would be difficult to arrange, and it wouldn't deal with the fact that, because of SCloc, assignments to the same variable can have different interference preconditions. The stable precondition is reached when an assignment is propagated last in the set, when all the others have overtaken it. The `undertaking' effect on an overtaking assignment need not be considered: in \figref{boinstabilitywithoutB}, for example, the sender's second interference can carry the stable precondition $\<msg>=1$ rather than the more accurate $\<msg>=0|\<msg>=1$. No harm will be done: the \<sat> test of \ruleref{EXTstability} won't have to deal with an assertion that stably insists on $\<msg>=0$ because $\<msg>=0$ isn't stable against the sender's other interference.

\subsection{A summary of stability checks}

Internal stability (within a thread): LO \ruleref{LOstability}.

External stability (inter-thread): EXT \ruleref{EXTstability}.

Stability of interference preconditions (within a thread; between threads in the rely): BO \ruleref{BOstability}.

There remain UEXT and UO stability, which deal with \<U>-modal and $\<sofar>$ assertions and are discussed in \secref{SBetc}.

%!TEX root = ./Paper.tex

\section{Implicit constraints}
\seclabel{implicitconstraints}

In our characterisation of weak memory, reordering of command executions within a thread applies to a path in the \<so> tree and is otherwise limited only by programmer-supplied constraints. In practice hardware architectures supply constraints of their own, but because our logic uses embroidery on explicit constraints, we can't easily take advantage of them. That isn't a problem: superfluous constraints compile to no code, and the absence of a constraint allows but does not enforce parallelism.

But there are two constraints which seem to be almost universally provided, and which therefore might be exploited: ordering of writes to a single location, and register renaming.

\subsection{SCloc (global coherence)}
\seclabel{SCloc}

%%So if $P$ can be interfered with by \interferenced{Q}{x:=E}, and both $P$ and $Q$ are stable as they will be in a valid proof, $P$ and $Q$ can't contradict each other. That implies that there must be a possible valuation of their variables in which both $P$ and $Q$ hold, which leads us to the EXT rule for external stability:
%%\ruledef{EXT stability}
%%{$$\cols
%%P \text{ is EXT-stable against } \interferenced{Q}{x:=E} \text{ if} \\
%%\quad \<sat>(P@Q)=>\<sp>(P@\widehat{Q},\;x:=E) => P 
%%\sloc$$
%%\rulelabel{EXTstability}}
%%Note that there can be no free registers in $Q$: see \ruleref{quantifiedstability}. Note also that we are not really introducing satisfaction into our logic: our interpretation of \ruleref{EXTstability} is that if $P@Q$ is unsatisfiable we have stability, but if we can't prove unsatisfiability we must prove the WM implication.

%We can strengthen our stability check beyond \ruleref{WMstability}. 
In many weak-memory architectures, including the major CPU architectures of x86, Power and ARM, there is a global constraint on the propagation of writes.
%\footnote{Although we know there are architectures where the constraint does not apply, we don't know whether that is by accident or design. We suspect that in any case it is difficult to implement.} 
If one thread `witnesses' two writes to a location in a particular order -- say $x=1$ before $x=2$ -- then no other thread will `witness' those writes in the opposite order. `Witnessing' covers both reading and writing, and the constraint doesn't require that other threads will witness those writes at all: only that if they are witnessed then they are witnessed in the same order. In detail it's more subtle than that, but in a logic in which nothing is in order unless it is \<lo> (or \<bo> or \<uo> or \<go>) constrained, it can be understood.

The constraint is called \emph{global coherence} in the Power and ARM architectures, and \emph{SC per location} in~\citep{AlglaveetalHerdingcats2014}; for brevity we call it SCloc. Under SCloc each thread witnesses, for each location, a subset of the partial order of all writes of all threads to that location. 

%%SCloc ensures that there cannot be simultaneous contradictory stable states. In particular at termination we can be sure that all threads will have received all of the writes that have been sent to them, so they will each see the same last write to each particular location. Without such a property we couldn't reason about the termination state of a program (see \secref{SClocalways}). But interference requires that the thread state corresponding to $P$ and the interfering thread state corresponding to $Q$ might also be simultaneous. Therefore their stable assertions about the values of the variables need not be identical, but cannot be contradictory. When $P$ is interfered with by \interferenced{Q}{x:=E} we know, therefore, that $P@Q$ -- without hatting -- can't be contradictory. Our check for stability can use that fact.
%%
%%We know also, for the same reasons, that the $P$-asserting thread and the $Q$-asserting thread can't disagree about the last write to $x$ before $x=E$. Unfortunately we can't require $x=\<xhat>$ -- that's too strong -- and we can't find a classical equivalent to the constructive $!(x!=\<xhat>)$, so we have to be silent about that. 
%%
%%Those considerations give us the EXT stability rule under SCloc.

%%\ruledef{EXT stability under SCloc}
%%{$$\cols
%%\text{Under SCloc, }P \text{ is EXT-stable against } \interferenced{Q}{x:=E} \text{ if} \vspace{3pt}\\
%%\quad \<sat>(P@Q) => \<sp>(P@\widehat{Q},\;x:=E) => P 
%%\sloc$$
%%\rulelabel{EXTstability}
%%}

It turns out (see \secref{SClocalways}) that termination reasoning in our logic depends on SCloc, and we believe that our use of the \<Bfr> and \<U> modalities would be unsound without it. %So \ruleref{EXTstability} is our EXT stability rule. We shall see below how the \<sat> chack helps with our proofs.
But all that is implicit: the visible effects of SCloc on our logic are only two:
\begin{enumerate}
\item Successive writes to the same location within a single thread will not be propagated out of \<lo> order. So the definition of \<bo> parallel in \defref{BOparallel} applies only to assignments to different variables.
\item We can reason about the overall coherence order of individual variables (see \secref{coherence}).
\end{enumerate}
SCloc is subtle. It's by no means true, for example, that on every architecture a write to a variable is always \<lo>-ordered with a \<so>-later read from the same variable. So we assume no implicit ordering, relying instead on explicit constraints.
 
%%\<Bo> stability also gets a slight boost. It is impossible, under SCloc, that writes to the same location from the same thread could overtake each other during propagation: they may not arrive anywhere other than in the order in which they were elaborated. So assignments to the same variable are not \<bo> parallel within a thread, and when building a rely from several guarantees we need only consider stability of interferences from different threads and different variables. We discuss \<bo> stability in more detail, and we explain its asymmetry, in\secref{bostabilityexplained}.
%%
%%SCloc entails implicit dependencies between a variable assignment and \<so>-later reads from or assignments to a variable and, vice-versa, between reads from and \<so>-later assignments to a variable. Different architectures apply these dependencies in different ways, some quite surprising. Reasoning in our logic requires explicit constraints which carry explicit embroidery, so we don't attempt to exploit implicit SCloc dependencies. Instead we include the constraints which make a proof work, whether or not they need special implementation.
%%
%%There is more to SCloc than a \<sat> check and implicit constraints, and in \secref{coherence} we show how to reason directly about the global coherence order.

\subsubsection{\problemmark{} RDW and CoRR litmus tests}

On Power and ARM the SCloc constraint demands that when two \<so>-ordered reads from the same location take their values from two different external writes, there is an implicit elaboration order between them -- i.e. an implicit \<lo> constraint. This principle is exploited in the CoRR0/1/2, RDW and RSW litmus tests of~\citep{PowerARMLitmusTests,MarangetSarkarSewellATutorialIntroductiontotheARMandPOWERRelaxedMemoryModels2013}. But it only applies when reading from \emph{external} writes, those generated in other threads. \citet{MarangetSarkarSewellATutorialIntroductiontotheARMandPOWERRelaxedMemoryModels2013} give a litmus test which shows that, on ARM at least, a read from an external write is not necessarily \<lo>-ordered with an \<so>-later internal write to the same variable (and we know of other more subtle examples which illustrate similar pecularities).

In a logic based on state assertions it would be difficult to make a distinction between internal and external writes, and we haven't tried to make one. RDW and RSW, which appeal to an implicit dependency, are therefore beyond us. But by including explicit \<lo> constraints in the observer threads of CoRR0/1/2, we can deal with the principle behind those tests. Our treatment is illustrated in several examples in \secref{coherence} and subsequently.

\subsubsection{A retreat}

In earlier versions of this work we made an argument from eventual consistency and SCloc that an assertion could not experience interference from an external assignment that originated from a contradictory stable state. That enabled us to strengthen the EXT stability check and permitted thread-local proofs of the LB litmus tests (\figref{LB}) and the token ring (\secref{tokenring}). We cannot maintain that argument; amongst other things, it made it impossible to have a simple treatment of BO stability. We have new proofs of those tests, but we have to use more powerful machinery: see \secref{LB} and \secref{tokenring}.

\begin{figure}
\centering
% proofs/SCreg.proof
$$\cols[c]
\thr{}
\guarantee{\interferenceg{\<true>}{y:=1}; \\ \interferenceg{\<true>}{x:=2}} \vspace{3pt} \\
\!\!\cols[r@{\intfspace}l]
 									& \<laba>:: \<r1>:=1; \\
\assertd{\<lo> \<laba>:: \<r1>=1} 	& \<labb>:: y:=\<r1>; \\
 									& \<labc>:: \<r1>:=2; \\
\assertd{\<lo> \<labc>:: \<r1>=2} 	& \<labd>:: x:=\<r1>
\sloc \vspace{3pt} \\
\assertd{\<lo> \<labd>:: x=2; \<lo> \<labb>:: y=1}
\rht
\sloc$$
\caption{Valid proof with SCreg, invalid without}
\figlabel{SCregproof}
\end{figure}

\subsection{Register renaming (SCreg)}
\seclabel{SCreg}

Our logic treats registers as local variables, invisible to other threads, and uses the LO stability \ruleref{LOstability} to deal with \<lo> parallelism of register assignments and assertions within a thread. Most modern architectures take a different approach, in which registers are more like single-assignment variables, and it makes a difference. The mechanism enables thread-local speculation of register assignments, and so it seems to be used in every architecture.

Consider, for example, the single-thread program in \figref{SCregproof}. According to our treatment so far, commands \<laba> and \<labb> are \<lo> parallel with the \<labc>$->$\<labd> constraint (\<so>-before but not \<lo>-before), and \<labc> and \<labd> are \<lo>-parallel with \<laba>$->$\<labb> (\<so>-after but not \<lo>-after). So by the rules of our logic \<laba> interferes with the $\<r1>=2$ embroidery on the \<labc>$->$\<labd> constraint, and \<labc> interferes with the $\<r1>=1$ embroidery of the \<laba>$->$\<labb> constraint. Each embroidery is obviously unstable in those circumstances. But in a machine which implements register assignment by taking a register from a pool and assigning it a value, and using that new register for \<so>-later commands until the next assignment to the `same' register, the \<r1> in commands \<laba> and \<labb> is not actually the same register as the \<r1> in commands \<labc> and \<labd>, so the \<lo> parallelism doesn't cause instability in either case.

Under \emph{register renaming} (SCreg) each reference to a register takes its value from the \<so>-latest assignment to that register and there can be no interference from other assignments using the same register name. That makes register interference very much ruled by \<so>. We can make our logic do the same thing if we said that \<lo> parallelism of a register assignment and a constraint requires the assignment to be \<so>-between the constraint's {\csource} and {\ctarget}. If we do that, our Arsenic proofchecker accepts the proof in \figref{SCregproof}.

In this paper we don't exploit SCreg, because our examples don't need it. Even if we did exploit it, we would still require explicit \<lo> constraints between a register assignment and subsequent uses of the register, because constraints carry embroidery, and embroidery is essential to proof. But since those constraints are free on SCreg machinery, we don't need to fear their cost.

\begin{figure}
\centering
% proofs/nothinair.proof
$$\cols[c]
\assert{\<init>::x=y=0} \vspace{3pt} \\
\BRA[l||l]
\thr{Thread 0}
  \<laba>:: \<r1>:=y; \\
  \<labb>:: x:=\<r1>
\rht
&
\thr{Thread 1}
  \<labc>:: \<r1>:=x; \\
  \<labd>:: y:=\<r1>
\rht
\KET \vspace{3pt} \\
\hcancel{\assert{\<final>::(0:::\<r1>)=(1:::\<r1>)=x=y=0}}
\sloc$$
\caption{42 from thin air?}
\figlabel{thinairq}
\end{figure}

\subsection{Quotienting internal interference}
\seclabel{SCregquotienting}
In checking internal interference -- LO, BO and UO -- we have to be careful with registers. Because of SCreg the same register name in different assertions may not refer to the same register. When dealing with inter-thread interference (EXT and UEXT) we quotient register names because threads don't share registers. In the absence of an analysis which could show whether two occurrences of the same register name refer to the same register, we quotient internal interference in just the same way that we do external interference. This causes considerable incompleteness (sob).

\subsection{An example that appeals to SCreg dependencies}

In the unconstrained program of \figref{thinairq}, we imagine that weak-memory hardware can choose the elaboration order of each thread. A machine might choose 42 as the initial value of \<r1> in thread 0 and, elaborating and propagating \<labb> first,  write $\<x>=42$. Then thread 2 might elaborate in \<so> order, reading $\<x>=42$ and writing $\<y>=42$. Finally command \<laba> of thread 0 could read $\<y>=42$. It hasn't followed \<so> order in thread 1, but it has used a reordering of the \<so> tree, so there is an execution which terminates with all of its registers and all of its variables containing 42. Given an unconstrained program it's quite possible and perfectly understandable: in weak memory we can't be sure to finish with $x=y=0$.

\begin{figure}
\centering
% proofs/nothinair.proof
$$\cols[c]
\assert{\<init>::x=y=0} \vspace{3pt} \\
\BRA[l||l]
\thr{Thread 0}
  \guarantee{\interferenceg{\<Bfr>(x=0)}{x:=0}} \vspace{3pt} \\
  \!\!\cols[r@{\intfspace}l]
  \assertd{\<lo> \<init>:: y=0} & \<laba>:: \<r1>:=y; \\
  \assertd{\<lo> \<laba>:: \<r1>=0;\<bo> \<init>:: \<Bfr>(x=0)} & \<labb>:: x:=\<r1>
  \sloc \\
  \assertd{\<lo> \<labb>:: \<r1>=x=0}
\rht
&
\thr{Thread 1}
  \guarantee{\interferenceg{\<Bfr>(y=0)}{y:=0}} \vspace{3pt} \\
  \!\!\cols[r@{\intfspace}l]
  \assertd{\<lo> \<init>:: x=0} & \<labc>:: \<r1>:=x; \\
  \assertd{\<lo> \<labc>:: \<r1>=0;\<bo> \<init>:: \<Bfr>(y=0)} & \<labd>:: y:=\<r1>
  \sloc \\
  \assertd{\<lo> \<labd>:: \<r1>=y=0}
\rht
\KET \vspace{3pt} \\
\assert{\<final>::(0:::\<r1>)=(1:::\<r1>)=x=y=0}
\sloc$$
\caption{Nothing from thin air with SCreg}
\figlabel{nothinair}
\end{figure}

But suppose we execute this machine on SCreg hardware: then there would be implicit \<lo> constraints $\<laba>{->}\<labb>$ and $\<labc>{->}\<labd>$. \Figref{nothinair} makes those constraints explicit, and shows that neither thread generates any effective interference, each writing 0 to a variable that is already 0.\footnote{This proof is a good example of circular rely/guarantee reasoning: thread 0 only writes 0 because thread 1 only writes 0 because thread 0 only writes 0 because \dots} There's no possibility, under SCreg, of a cycle in which both threads read and write a value plucked out of `thin air' (or rather, taken from an uninitialised register). 

The constrained program works whether or not there is SCreg, but on an SCreg machine the \<lo> constraints don't need implementation. On such machines there is no `thin air' execution of the apparently unconstrained \figref{thinairq}.

%!TEX root = ./Paper.tex

\section{Speculated propagation and the go ordering}
\seclabel{localspec}

\begin{figure}
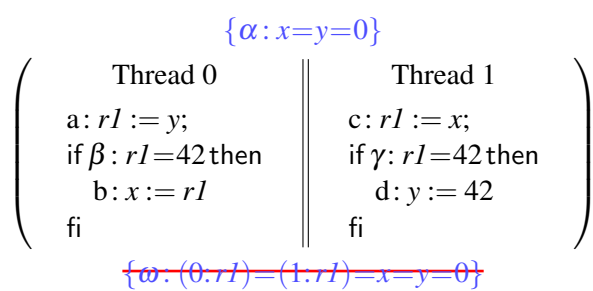

\centering
% proofs/C11_42.proof
$$\cols[c]
\assert{\<init>::x=y=0} \vspace{3pt} \\
\BRA[l||l]
\thr{Thread 0}
  \<laba>:: \<r1>:=y; \\
  \<if> \<beta>:: \<r1>=42 \<then> \\
  \pindent\<labb>:: x:=\<r1> \\
  \<fi>
\rht
&
\thr{Thread 1}
  \<labc>:: \<r1>:=x; \\
  \<if> \<gamma>:: \<r1>=42 \<then> \\
  \pindent\<labd>:: y:=42 \\
  \<fi>
\rht
\KET \vspace{3pt} \\
\hcancel{\assert{\<final>::(0:::\<r1>)=(1:::\<r1>)=x=y=0}}
\sloc$$
\caption{42 from slightly thicker air?}
\figlabel{42q}
\end{figure}

In \figref{42q}, adapted from the C11 standard~\citep{C2011}, there is a command which writes 42 so -- unlike \figref{thinairq} -- the hardware doesn't have to make it up. The standard doesn't prohibit an execution which finishes with $x=y=42$, but its authors express a hope that compiler writers will conspire to make it impossible.\footnote{The appeal is to compiler writers and not to hardware designers, and that's important. Our treatment is of a kind of high-level assembler. The problems of defining a programming language are different: see \secref{LibraryAbstractionandOptimisation}.} It seems a bit of a nightmare.

To see how the nightmare could be real we trace a possible execution of the unconstrained program. Command \<labd> in thread 1 is elaborated first and $y=42$ propagated; then thread 0 executes in \<so> order, reading $y=42$ , finding $\<r1>=42$ and writing $x=42$; then \<labc> of thread 1 reads 42 and control expression \<gamma> confirms that \<labd> indeed ought to be elaborated. That execution obeys the constraint on execution of threads that we have assumed throughout: each thread's execution follows a reordering of an \<so>-tree path. So perhaps the nightmare is real. If it is real it's important: \citet{BattyetalMathematizingC++concurrency2011} show that `satisfaction cycles', in which a command's execution is the cause of itself, prevent effective library abstraction in C11. In our nightmare the elaboration of $y:=42$ is eventually a cause, via thread 0, command \<labc> and control expression \<gamma>, of its own presence in the \<so> path used by thread 1.

But like most nightmares there's something odd about it. Suppose that \<labd> is speculatively elaborated and $y=42$ propagated from thread 1, but that thread 0 is quick off the mark, reading $\<y>=0$ from the initial state before $y=42$ reaches it. Then thread 0 won't write $\<x>=42$, and therefore \<labc> must read $\<x>=0$ from the initial state, \<gamma> will find $\<r1>!=42$ and execution must follow the $\<gamma>_{f}$ path, on which command \<labd> doesn't appear. So in those circumstances thread 1 has \emph{not} followed a reordering of an \<so>-tree path. It would either have to withdraw the propagation of $y=42$ and retreat to an execution in which it wrote nothing, or negotiate with thread 0 to re-execute reading 42 and then recapitulate its own execution of \<labc> and \<gamma>. 

This example shows that it can require inter-thread cooperation to allow withdrawal of speculated propagation. Such cooperation is difficult if one or more threads has read from and acted upon the propagation, as thread 0 did in our imagined execution. So far as we aware, and perhaps as a consequence of the difficulty, no architecture attempts speculated propagation: they speculate only elaboration, which is internal to a thread and which, because of register renaming, is relatively easily abandoned.

\begin{figure*}
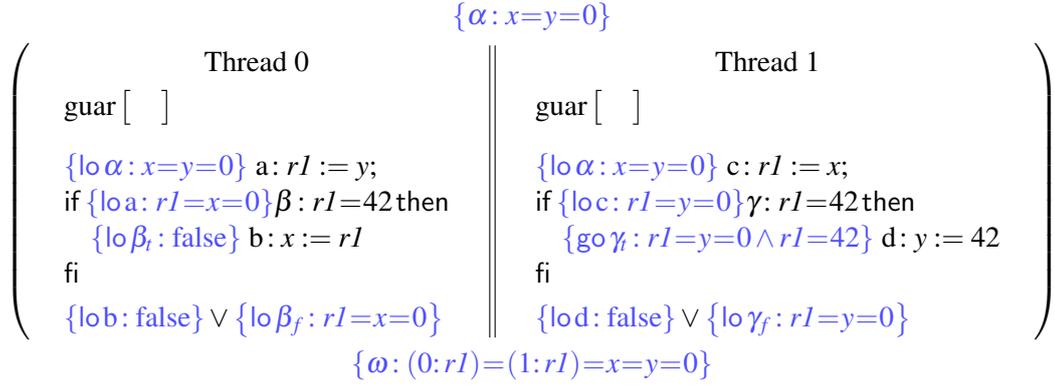

\centering
% proofs/C11_42.proof -- edited to eliminate LocalSpec=>
$$\cols[c]
\assert{\<init>::x=y=0} \vspace{3pt} \\
\BRA[l||l]
\thr{Thread 0}
  \emptyguarantee \vspace{3pt} \vspace{3pt} \\
  \assertd{\<lo> \<init>:: x=y=0}\ \<laba>:: \<r1>:=y; \\
  \<if> \assertd{\<lo> \<laba>:: \<r1>=x=0} \<beta>:: \<r1>=42 \<then> \\
  \pindent\assertd{\<lo> \<beta>_{t}:: \<false>}\ \<labb>:: x:=\<r1> \\
  \<fi> \vspace{3pt} \\
  \assertd{\<lo> \<labb>:: \<false>} | \assertd{\<lo> \<beta>_{f}:: \<r1>=x=0}
\rht
&
\thr{Thread 1}
  \emptyguarantee \vspace{3pt} \vspace{3pt} \\
  \assertd{\<lo> \<init>:: x=y=0}\ \<labc>:: \<r1>:=x; \\
  \<if> \assertd{\<lo> \<labc>:: \<r1>=y=0} \<gamma>:: \<r1>=42 \<then> \\
  \pindent\assertd{\<go> \<gamma>_{t}:: \<r1>=y=0@\<r1>=42}\ \<labd>:: y:=42 \\
  \<fi> \vspace{3pt} \\
  \assertd{\<lo> \<labd>:: \<false>} | \assertd{\<lo> \<gamma>_{f}:: \<r1>=y=0}
\rht
\KET \vspace{3pt} \\
\assert{\<final>::(0:::\<r1>)=(1:::\<r1>)=x=y=0}
\sloc$$
\caption{No thin air with SCreg dependencies and LocalSpec}
\figlabel{no42}
\end{figure*}

To deal with this problem, we add LocalSpec to the architecture parameters SCloc and SCreg. Under LocalSpec all speculation is local -- i.e. no speculated propagation of writes. A machine without LocalSpec can produce the execution we described. One with SCreg and LocalSpec cannot, as the proof in \figref{no42} shows. We have made explicit the SCreg-implicit \<lo> constraints $\<laba>{->}\<beta>_{t}{->}\<labb>$ and $\<labc>{->}\<gamma>_{t}$. We have also made explicit the LocalSpec-implicit \<go> constraint $\<gamma>_{t}{->}\<labd>$.

The target of a \<go> contraint must be a variable assignment, and the \emph{propagation} of the write generated by the target is constrained to occur after the \emph{elaboration} of the source. So \<go> doesn't fit in the \<lo>/\<bo>/\<uo> hierarchy, and in particular \<go> is not included in \<lo>. %\footnote{Note that \<go> constraints need not always be control-expression to variable assignment. In \tabref{orderingequivalences} we claim (so far without a soundness proof) that on ARM they are implicit between reads from and assignments to a single location.} 
In terms of proof a \<go> constraint's embroidery contributes primarily to the interference rather than the elaboration precondition of its target. The \<go> constraint in \figref{no42} provides an interference precondition of \<labd> which is \<false> -- i.e. it won't be propagated, ever. 

Execution follows a reordering of an \<so>-tree path. A path cannot contain an assignment that cannot be propagated, so the elaboration precondition of \<labd> must be \<false> as well. To provide this, the rules for forming a precondition are as follows:
\definition{Precondition from knot}{preconditionfromknot}
{\hstrut{5pt}\vspace{-3pt}
\begin{itemize*}
\item The overall precondition is the conjunction of a knot's embroidery or, if the knot is a disjunction of knots, the disjunction of their overall preconditions.
\item The elaboration precondition is the overall precondition, excluding the embroidery of \<go> constraints, conjoined with the satisfaction (\<sat>) of the overall precondition.
\item The interference precondition is either the overall precondition or, if the knot contains a square-bracketed \nonterminal{intfpre} assertion (see \tabref{lacedprograms}), the \nonterminal{intfpre} assertion, which must be implied by the overall precondition.
\end{itemize*}
}
If there are no \<go> constraints in a knot then the elaboration precondition is just the overall precondition. But in the case of command \<labd> in \figref{no42} we must exclude the embroidery of the \<go> constraint and conjoin the satisfaction of the overall precondition $\<sat>(\<r1>=y=0@\<r1>=42)$, which is \<false>. An assignment with a \<false> precondition needs no guarantee entry and causes no \<lo>-parallel or \<bo>-parallel instability; we have a proof.

Without the \<go> constraint the proof in \figref{no42} fails spectacularly. The preconditions of command \<labc> and control expression \<gamma> are \<lo>-parallel with command \<labd>, and therefore their embroidery is unstable.\footnote{They are \<lo>-parallel with the \<go> constraint in place, but the false elaboration precondition means no instability.} The guarantee should include \interferenced{\<true>}{y:=42}; and the embroidery of the \<labd>$->$thread-postcondition constraint can't be \<false>.

We can't prove that the nightmare execution can't happen on any weak-memory architecture; without LocalSpec our proof fails to show that it can't happen; and indeed we believe that it could happen in that case. But we can prove that it can't happen with commonly-applied architectural constraints -- with LocalSpec and with SCreg. 

\subsection{\problemmark{} PPOCA: a scary conditional}

\Figref{PPOCA} shows a version of MP with a receiver which conditionally writes and reads \<flag1> before reading \<msg>: note that the constraint $\<beta>_{t}{->}\<labd>$ is \<go> rather than \<lo>. It doesn't work, just because \<go> doesn't imply \<lo>. \Figref{litmus_PPOCA} shows an observed weak-memory execution with a chain of various implicit and explicit Power-specific orderings from command \<labc> to command \<labf> which don't compose to produce the equivalent of an \<lo> constraint. Experimentally the receiver can read 1 from \<flag> and then 0 from \<msg>. 

\begin{figure}
\centering
% proofs/PPOCAco.unproof
$$\cols[c]
\assert{\<init>::\<flag>=\<flag1>=0} \vspace{3pt} \\
\BRA[l||l]
\thr{Sender (0)}
  \!\!\cols[r@{\intfspace}l]
  \assertd{\<lo> \<init>} & \<laba>:: \<msg>:=1; \\
  \assertd{\<bo> \<laba>} & \<labb>:: \<flag>:=1
  \sloc
\rht
&
\thr{Faulty PPOCA receiver (1)}
  \assertd{\<lo> \<init>}\ \<labc>:: \<r1>:=\<flag>; \\
  \<if> \assertd{\<lo> \<labc>} \<beta>:: \<r1>=1 \<then> \\
  \pindent\!\!\cols[r@{\intfspace}l]
    \assertd{\<go> \<beta>_{t}} & \<labd>:: \<flag1>:=1; \\
    \assertd{\<lo> \<labd>} & \<labe>:: \<r2>:=\<flag1>; \\
    \assertd{\<lo> \<labe>} & \<labf>:: \<r3>:=\<msg>
    \sloc \\
  \<fi>
\rht
\KET \vspace{3pt} \\
\hcancel{\assert{\<final>::(1:::\<r1>)=1=>(1:::\<r3>)=1}}
\sloc$$
\caption{PPOCA: ineffective message-passing}
\figlabel{PPOCA}
\end{figure}

\begin{figure}
\centering
\includegraphics[scale=\picscale]{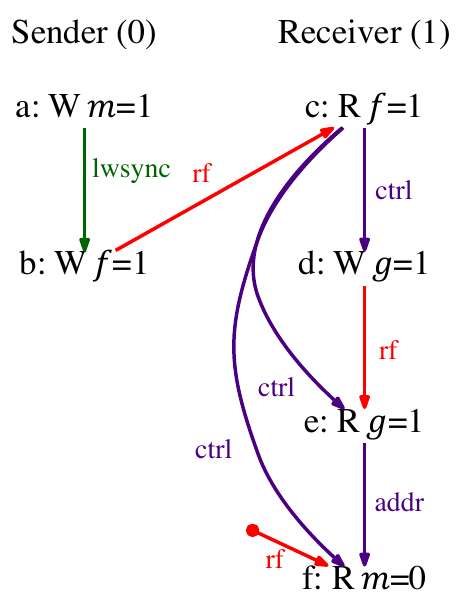}
\caption{PPOCA litmus test diagram}
\figlabel{litmus_PPOCA}
\end{figure}

\begin{figure}
\centering
% proofs/PPOCAco.unproof
$$\cols[c]
\assert{\<init>::\<flag>=\<flag1>=0} \vspace{3pt} \\
\BRA[l||l]
\thr{Sender (0)}
  \guarantee{\interferenceg{\<true>}{\<msg>:=1}; \\ \interferenceg{\<Bfr>(\<msg>=1)}{\<flag>:=1}} \vspace{3pt} \vspace{3pt} \\
  \!\!\cols[r@{\intfspace}l]
  \assertd{\<lo> \<init>:: \<true>} & \<laba>:: \<msg>:=1; \\
  \assertd{\<bo> \<laba>:: \<Bfr>(\<msg>=1)} & \<labb>:: \<flag>:=1
  \sloc
\rht
&
\thr{Faulty PPOCA receiver (1)}
  \guarantee{\interferenceg{\<msg>=1}{\<flag1>:=1}} \vspace{3pt} \vspace{3pt} \\
  \assertd{\<lo> \<init>:: \<flag>=1=>\<msg>=1}\ \<labc>:: \<r1>:=\<flag>; \\
  \<if> \assertd{\<lo> \<labc>:: \<r1>=1=>\<msg>=1} \<beta>:: \<r1>=1 \<then> \\
  \pindent\!\!\cols[r@{\intfspace}l]
    \assertd{\<go> \<beta>_{t}:: \<msg>=1} & \<labd>:: \<flag1>:=1; \\
    \assertd{\<lo> \<labd>:: \hcancel{\<msg>=1}} & \<labe>:: \<r2>:=\<flag1>; \\
    \assertd{\<lo> \<labe>:: \<msg>=1} & \<labf>:: \<r3>:=\<msg>
    \sloc \\
  \<fi> \vspace{3pt} \\
  \assertd{\<lo> \<labf>:: \<r3>=1} | \assertd{\<lo> \<beta>_{f}:: \<r1>!=1}
\rht
\KET \vspace{3pt} \\
\assert{\<final>::(1:::\<r1>)=1=>(1:::\<r3>)=1}
\sloc$$
\caption{PPOCA unproof}
\figlabel{PPOCAunproof}
\end{figure}

The issue is the difference between elaboration and propagation. In the unproof of \figref{PPOCAunproof} we have reproduced the orderings of the original: \<bo> \<laba>$->$\<labb> mimics Power lwsync; \<go> $\<beta>_{t}{->}\<labd>$ mimics the Power control dependency; \<lo> $\<labd>{->}\<labe>$ is implicit on Power; \<lo> $\<labe>{->}\<labf>$ mimics the address dependency of the original.\footnote{We haven't mimicked Power's control dependencies to register assignments \<labe> and \<labf>, because \<go> constraints can't target register assignments. But then control dependencies to register assignments don't have any force in the Power memory model either.} 

In \figref{PPOCAunproof} the elaboration precondition of \<labe>, according to \defref{preconditionfromknot}, is $\<sat>(\<msg>=1)$, which doesn't imply $\<msg>=1$: so the postcondition doesn't imply $\<msg>=1$; so we can't embroider $\<msg>=1$ on the \<lo> constraint $\<labd>{->}\<labe>$; so it's an unproof. But note that the interference precondition of \<labd> is $\<msg>=1$, as recorded in the guarantee.

\begin{figure}
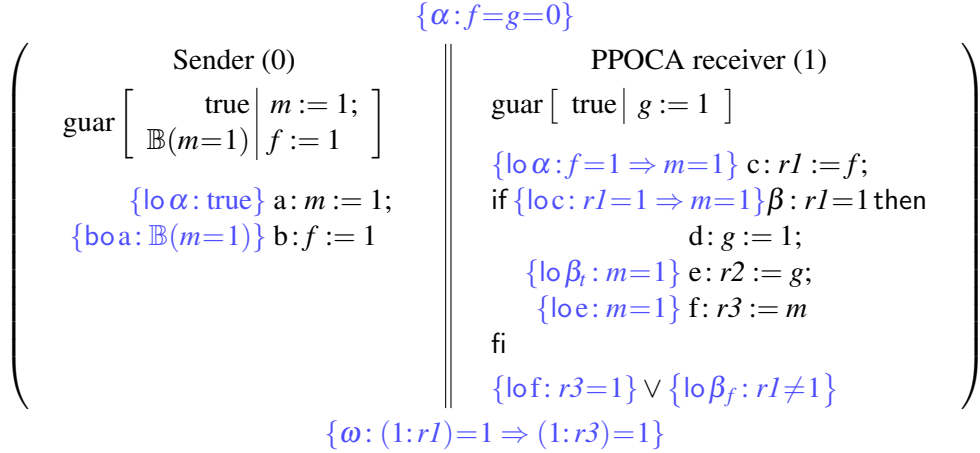

\centering
% proofs/PPOCA.proof
$$\cols[c]
\assert{\<init>::\<flag>=\<flag1>=0} \vspace{3pt} \\
\BRA[l||l]
\thr{Sender (0)}
  \guarantee{\interferenceg{\<true>}{\<msg>:=1}; \\ \interferenceg{\<Bfr>(\<msg>=1)}{\<flag>:=1}} \vspace{3pt} \vspace{3pt} \\
  \!\!\cols[r@{\intfspace}l]
  \assertd{\<lo> \<init>:: \<true>} & \<laba>:: \<msg>:=1; \\
  \assertd{\<bo> \<laba>:: \<Bfr>(\<msg>=1)} & \<labb>:: \<flag>:=1
  \sloc
\rht
&
\thr{PPOCA receiver (1)}
  \guarantee{\interferenceg{\<true>}{\<flag1>:=1}} \vspace{3pt} \vspace{3pt} \\
  \assertd{\<lo> \<init>:: \<flag>=1=>\<msg>=1}\ \<labc>:: \<r1>:=\<flag>; \\
  \<if> \assertd{\<lo> \<labc>:: \<r1>=1=>\<msg>=1} \<beta>:: \<r1>=1 \<then> \\
  \pindent\!\!\cols[r@{\intfspace}l]
     & \<labd>:: \<flag1>:=1; \\
    \assertd{\<lo> \<beta>_{t}:: \<msg>=1} & \<labe>:: \<r2>:=\<flag1>; \\
    \assertd{\<lo> \<labe>:: \<msg>=1} & \<labf>:: \<r3>:=\<msg>
    \sloc \\
  \<fi> \vspace{3pt} \\
  \assertd{\<lo> \<labf>:: \<r3>=1} | \assertd{\<lo> \<beta>_{f}:: \<r1>!=1}
\rht
\KET \vspace{3pt} \\
\assert{\<final>::(1:::\<r1>)=1=>(1:::\<r3>)=1}
\sloc$$
\caption{PPOCA effectively laced}
\figlabel{PPOCArelaced}
\end{figure}

Some litmus tests are subtle, but PPOCA frightens the horses, because it seems at first sight that conditionals don't do what they ought to. Once we realise that \<go> doesn't imply \<lo> the solution is obvious: use an \<lo> constraint either to \<labe>, as in the proof of a corrected PPOCA in \figref{PPOCArelaced}, or direct to \<labf>. We don't need to lace \<labd> because we don't care when it's elaborated. The horses may relax; the ordering they seek can easily be provided.   

\clearpage
\part{Temporal reasoning and coherence}
\partlabel{coherence}
\thispagestyle{empty}
\clearpage
%!TEX root = ./Paper.tex

Our treatment thus far has implicitly used temporal reasoning. Our \<Bfr> modality employs a specialised version of the temporal `\<since>' modality: that fact is implicit in the treatment of substitution and \<Bfr> (\defref{Bsubst}). But we need other modalities to be able to deal with many of the features of weak memory: x86's \<MFENCE>, Power's \<sync> and ARM's \<dsb> need a modality which applies across threads; `coherence' arguments need modalities which describe the global temporal order of writes; and several litmus tests are about events that occur during the period when a particular property holds.

All of our temporal reasoning is based on `\<since>': an assertion has held -- locally or globally -- since a barrier event or since execution started; or an assertion held at least once since execution started. 

Temporal modalities are very useful, but they introduce a difficulty: assertions which employ them can't simply be treated as descriptions of the current state of a thread. This undermines the justification of the \<Bfr> modality given in \asdref{whenceB}, and affects the treatment of termination. We deal with those difficulties in \partref{terminationlooseend}.

As usual we introduce our treatment with examples.

\section{A proof of SB using \setModality{U} and since}
\seclabel{SBetc}
\seclabel{since}

In introducing weak memory in \secref{weakmemoryexecutions}, our third litmus test in \figref{SBunlaced} was SB, abstracted from Dekker's algorithm for mutual exclusion. In weak memory, without constraints, both threads can read 0 as pictured in \figref{SBlitmus}. We can't make the program work by using \<lo> or \<bo> constraints: \<lo> has no effect on propagation, so both reads could happen before the writes propagate; \<bo> would do no more because the reads don't cause propagation. And in any case experiment shows that it takes a \<uo>-generating barrier in each thread to make the program work. 

%Our assertions and our reasoning are thread-local, but sometimes we can be sure that some part of the global state is under the control of a single thread. Then we can assert, in that thread, that all others see what it sees. 

\begin{figure}[b]
\centering
% proofs/SB2.proof
$$\cols[c]
\assert{\<init>::x=0@y=0} \vspace{3pt} \\
\BRA[l||l]
\thr{Thread 0}
  \!\!\cols[r@{\intfspace}l]
  \assertd{\<lo> \<init>} & \<laba>:: x:=1; \\
  \assertd{\<uo> \<laba>} & \<labb>:: \<r1>:=y
  \sloc
\rht
&
\thr{Thread 1}
  \!\!\cols[r@{\intfspace}l]
  \assertd{\<lo> \<init>} & \<labc>:: y:=1; \\
  \assertd{\<uo> \<labc>} & \<labd>:: \<r1>:=x
  \sloc
\rht
\KET \vspace{3pt} \\
\assert{\<final>::!((0:::\<r1>)=0@(1:::\<r1>)=0)}
\sloc$$
\caption{SB laced for weak memory with \setorder{uo} constraints}
\figlabel{SBlaced}
\end{figure}

\Figref{SBlaced} shows a laced version of SB, with \<uo> constraints \<laba>$->$\<labb> and \<labc>$->$\<labd>. \<Uo> provides \<lo> and \<bo> ordering, but does more than \<bo> in that the writes which underpin the {\csource} postcondition are propagated to all threads before the {\ctarget} elaborates. So every thread's cache contains $x=1$ before thread 0 reads $y$ from its own cache, and every thread's cache contains $y=1$ before thread 0 reads $x$. From the point of view of each thread we've established something like SC interleaving: the effect of \<laba> takes place everywhere before \<labb> is elaborated, and likewise for \<labc> before \<labd>.

Along with the \<uo> ordering goes a \<U> modality, and like the \<Bfr> modality on a \<bo> constraint, we wrap an assertion implied by the {\csource} postcondition in \<U> to get the constraint embroidery -- see \ruleref{inheritance}. 

$\<U>(P)$ asserts that $P$ holds in all threads since a barrier event; when we introduce it via a \<uo> constraint the barrier event happens after the elaboration of the constraint's source and before the elaboration of its target. In the proof of SB in \figref{SBproof}, we claim that $x=1$ holds in all threads when command \<labb> is elaborated. If \<labb> then reads 0 from $y$, there was therefore an instant in thread 0 in which simultaneously $x=1$ held in every thread and $y=0$ held in thread 0. $\<U>(x=1) \<since> y=0$ is a temporal modality which claims that the coincidence occurred and that $\<U>(x=1)$ has held continuously ever since. Similarly in thread 1 we claim $\<U>(y=1) \<since> x=0$. The thread-history diagram in \figref{SBstateshistory} illustrates why those claims can't both be true: if the $\<U>(x=1)$ epoch starts late enough for thread 1 to see $x=0$ when $y=1$ already holds everywhere, then thread 0 can't later see $y=0$ -- and a similar diagram would show that if thread 0 sees $y=0$ then thread 1 can't see $x=0$. Mutual exclusion, as required, but so far only an informal proof. 

\begin{figure}
\centering
% proofs/SB2.proof
$$\cols[c]
\assert{\<init>::x=0@y=0} \vspace{3pt} \\
\BRA[l||l]
\thr{Thread 0}
  \guarantee{\interferenceg{\<true>}{x:=1}} \vspace{3pt} \\
  \!\!\cols[r@{\intfspace}l]
  \assertd{\<lo> \<init>:: \<true>} & \<laba>:: x:=1; \\
  \assertd{\<uo> \<laba>:: \<U>(x=1)} & \<labb>:: \<r1>:=y
  \sloc \\
  \assertd{\<lo> \<labb>:: \<r1>=0=>(\<U>(x=1) \<since> y=0)}
\rht
&
\thr{Thread 1}
  \guarantee{\interferenceg{\<true>}{y:=1}} \vspace{3pt} \\
  \!\!\cols[r@{\intfspace}l]
  \assertd{\<lo> \<init>:: \<true>} & \<labc>:: y:=1; \\
  \assertd{\<uo> \<labc>:: \<U>(y=1)} & \<labd>:: \<r1>:=x
  \sloc \\
  \assertd{\<lo> \<labd>:: \<r1>=0=>(\<U>(y=1) \<since> x=0)}
\rht
\KET \vspace{3pt} \\
\assert{\<final>::!((0:::\<r1>)=0@(1:::\<r1>)=0)}
\sloc$$
\caption{A weak-memory proof of SB}
\figlabel{SBproof}
\end{figure}

\begin{figure}
\centering
\includegraphics[scale=\picscale]{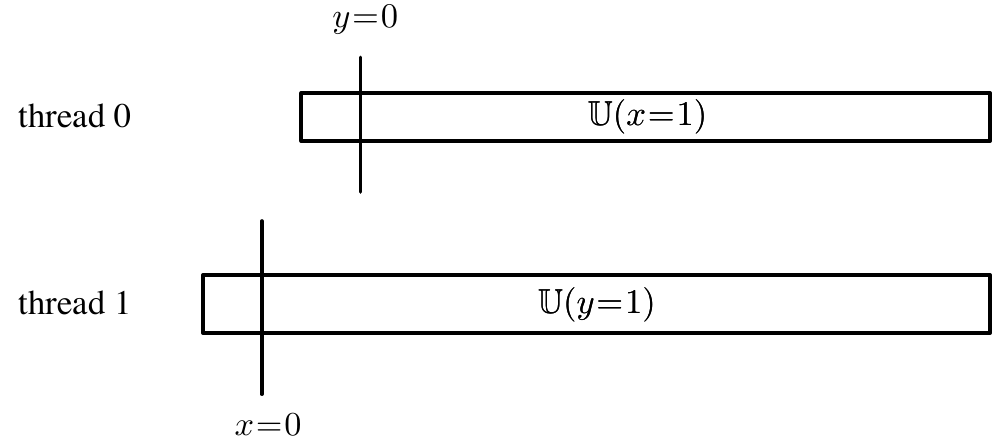}
\caption{An impossible SB thread history}
\figlabel{SBstateshistory}
\end{figure}

Since the two threads of SB are similar, swapping the r\^oles of \<x> and \<y>, we check only thread 0 of the proof in \figref{SBproof}. The precondition \<true> of command \<laba> is stable against any interference, and \<laba> gives us the guarantee of thread 0. The command's postcondition implies $x=1$, which is stable because thread 1 writes only to $y$; we decorate that postcondition with \<U> in a \<uo>-ordering to \<labb>. 

To deal with the pre and post conditions of \<labb> we have to consider the formal properties of \<U>. It shares all the properties in \defref{Bproperties}, reading \<U> for \<Bfr>, plus the following:
\definition{Additional properties of the \setModality{U} modality}{Uproperties}
{$$\cols[rcl]
\<U>(P)			&=>& \<Bfr>(P)	\\
\<U>(\<Bfr>(P))	&==& \<U>(P)	\\
\<Bfr>(\<U>(P))	&==& \<U>(P)	

\sloc$$}

Substitution and hatting with \<U> is just as with \<Bfr>.
\definition{Substitution and hatting of the \setModality{U} modality}{Usubsthat}
{$$\cols[rcll]
\<U>(P)[x\backslash \<x'>] 		&==& \hook{\<U>(P)}@P[x\backslash \<x'>]\\
\hook{\<U>(P)}@P				&=>& \<U>(P) \\
\<U>(P)[r\backslash \<r'>] 		&==& \<U>(P[r\backslash \<r'>])
 \vspace{5pt} \\
\widehat{\<U>(P)}				&==& \<U>(P)@\widehat{P}
\sloc$$}
So the precondition $\<U>(x=1)$ of \<labb> is stable against the interference of thread 1 because thread 1 doesn't interfere with $x$. The instantaneous postcondition of \<labb> is 
\begin{equation}
\<U>(x=1)@\<r1>=y
\eqnlabel{SBunstablepost}
\end{equation}
which we may weaken to 
\begin{equation}
\<r1>=0=>(\<U>(x=1)@y=0)
\eqnlabel{SBunstablepost2}
\end{equation}
This says that when $\<r1>=0$ thread 0 has witnessed the case we are interested in, in which it wrote 1 and read 0. But this assertion is unstable because thread 1 can write $y=1$. We further weaken to
\begin{equation}
\<r1>=0=>(\<U>(x=1) \<since> y=0)
\eqnlabel{SBstablepost}
\end{equation}
which is stable against thread 1's interference despite the fact that it mentions $y=0$. That's the magic of `\<since>'. To understand it we need to look at the formal properties of `\<since>'.

$P \<since> Q$ holds if $P@Q$ holds, or if $P@Q$ held in the past and $P$ has held ever since. It's established by a temporal coincidence; it implies $P$; and it implies that $Q$ happened, which we capture with the $\<ouat>$ -- `once upon a time' -- modality, to be discussed in \secref{ouat}.
\definition{Properties of \setbracket{since}}{Sinceproperties}
{$$\cols[rcl] P@Q				&=>&	P \<since> Q \\
P \<since> Q	&=>&	P @ \<ouat>(Q)
 \sloc$$}
In substitution it behaves like \<Bfr> and \<U> (no surprise since they are each a version of \<since>). It is stable if its first component is stable. It's a local assertion, applying in a single thread, so it can be hatted.
\definition{Substitution and hatting of \setbracket{since}}{Sincesubsthat}
{$$\cols[rcll]
(P \<since> Q)[x\backslash\<x'>] 	&==& 	\hook{(P \<since> Q)}@P[x\backslash\<x'>]  			\\
\hook{(P \<since> Q)}@P 			&=>&	(P \<since> Q)										\\
(P \<since> Q)[r\backslash\<r'>] 	&==& 	P[r\backslash\<r'>] \<since> Q[r\backslash\<r'>] \vspace{5pt} \\
%!TEX root = ./Paper.tex
\widehat{P \<since> Q}				&==&	\widehat{P \<since> Q}

\sloc$$}
By the first two lines of this definition, $\<r1>=0=>(\<U>(x=1) \<since> y=0)$ is stable if $\<U>(x=1)$ is stable, which it is.

Formal reasoning allows us to justify thread postconditions which involve \<U> and \<since>. But we haven't yet described the machinery which allows us to conclude formally that it is impossible for both threads in \figref{SBproof} to read 0: i.e. that it is impossible at termination that $\<U>(x=1) \<since> y=0$ could hold in thread 0 and $\<U>(y=1) \<since> x=0$ in thread 1. The formal machinery is discussed in \partref{terminationlooseend} and in our description of the SMT embedding of our logic in \appxref{embedding}. For the moment we rely on the informal argument we gave above, referring to the thread history diagram of \figref{SBstateshistory}. 

\section{The \setModality{S}ofar modality}

As well as \<U>, it is convenient to introduce \<sofar>. $\<Sofar>(P)$ means that $P$ has held since execution started. It has all the properties of \defref{Bproperties}, reading \<sofar> for \<Bfr>, and of \defref{Uproperties}, reading \<sofar> for \<U>. In addition we can say that if $P$ has held since the start of execution, then it must certainly have held when interference was propagated or elaborated. 
\definition{Additional properties of the \setModality{S}ofar modality}{Sofarproperties}
{$$\cols[rcl]
\<sofar>(P)			&=>& \<U>(P) \\
\<sofar>(\<U>(P))	&=& \<sofar>(P)	\\
\<U>(\<sofar>(P))	&=& \<sofar>(P)	\\
\<sofar>(P) 		&=>& \widehat{P} \\ 
\<sofar>(P) 		&=>& \dhatv{P}
%%\<sofar>(P) 		&=>& \widetilde{P} \\
%%\<sofar>(P) 		&=>& \dtildev{P}
\sloc
$$}
%$\widetilde{P}$ and $\dtildev{P}$ are hattings associated with the UEXT and UO instabilities, introduced below.
In substitution \<sofar> behaves like \<since>, \<U> and \<Bfr>; it's unaffected by hatting because (unlike \<U>) $\<sofar>(P)=>\widehat{P}$.
\definition{Substitution and hatting of the \setModality{S}ofar modality}{Sofarsubsthat}
{$$\cols[rcll]
\<sofar>(P)[x\backslash \<x'>] 		&==& \hook{\<sofar>(P)}@P[x\backslash \<x'>]\\
\hook{\<sofar>(P)}@P				&=>& \<sofar>(P) \\
\<sofar>(P)[r\backslash \<r'>] 		&==& \<sofar>(P[r\backslash \<r'>])
 \vspace{5pt} \\
%!TEX root = ./Paper.tex
\widehat{\<sofar>(P)}					&==& \<sofar>(P)

\sloc$$}

\begin{figure}[b]
\centering
% proofs/UEXT.unproof
% proofs/UEXT.unproof
$$\cols[c]
\assert{\<init>::\<msg>=\<flag>=0} \vspace{3pt} \\
\BRA[l||l]
\thr{Lo sender (0)}
  \guarantee{\interferenceg{\<true>}{\<msg>:=1}; \\ \interferenceg{\<msg>=1}{\<flag>:=1}} \vspace{3pt} \vspace{3pt} \\
  \!\!\cols[r@{\intfspace}l]
  \assertd{\<lo> \<init>: \<true>} & \<laba>:: \<msg>:=1; \\
  \assertd{\<lo> \<laba>: \<msg>=1} & \<labb>:: \<flag>:=1
  \sloc
\rht
&
\thr{UEXT-unstable receiver (1)}
  \emptyguarantee \vspace{3pt} \vspace{3pt} \\
  \!\!\cols[r@{\intfspace}l]
  \assertd{\<uo> \<init>: \hcancel{\<sofar>(\<msg>=0)@\<flag>=0|\<msg>=1}} & \<labc>:: \<r1>:=\<flag>; \\
  \assertd{\<lo> \<labc>: \<r1>=1=>\<msg>=1} & \<labd>:: \<r2>:=\<msg>
  \sloc \vspace{3pt} \\
  \assertd{\<lo> \<labd>: \<r1>=1=>\<r2>=1}
\rht
\KET \vspace{3pt} \\
\assert{\<final>::(1:::\<r1>)=1=>(1:::\<r2>)=1}
\sloc$$
\caption{An unproof of message-passing, displaying UEXT instability}
\figlabel{mattsexample}
\end{figure}

\section{Interference without propagation -- UEXT instability}

The \<U> and \<sofar> modalities introduce a new kind of instability. Suppose we claim, in thread A, that $\<U>(x=1)$. For that claim to be stable, for it to hold no matter how long we wait for thread A to make a move, it must be impossible for any other thread B to write $x=0$. We don't have to wait for that write to percolate from B to A: if at any instant $x=0$ in thread B, then at that same instant $!\<U>(x=1)$ in thread A. 

Consider the unproof in \figref{mattsexample}. The sender is unproblematic, and its guarantee is straightforward -- but note that because it uses an \<lo> constraint \<laba>$->$\<labb>, rather than \<bo> as in MP, the second line of its guarantee has an amodal interference precondition. 

In the receiver the precondition of command \<labc> is implied by the initial state: we suppose that our examples are set up so that when each thread starts the initial state has already been propagated to all threads -- i.e. $\<sofar>(\<msg>=\<flag>=0)$ holds. Then because of the modality properties in \defref{Bproperties}, reading \<sofar> for \<Bfr>, $\<sofar>(\<msg>=\<flag>=0)$ implies $\<sofar>(\<msg>=0)@\<flag>=0$, which we can weaken to $(\<sofar>(\<msg>=0)@\<flag>=0)|\<msg>=1$, the precondition of \<labc>.

That precondition is straightforwardly EXT stable against the first line of the sender's guarantee:
\begin{equation}
\squeezecols{2.5pt}
\cols
	& \<sp>(((\<sofar>(\<msg>=0)@\<flag>=0)|\<msg>=1)@\widehat{\<true>},\;\<msg>:=1) \vspace{2pt}\\
=	& ((\hook{\<sofar>(\<msg>=0)}@\<msg'>=0@\<flag>=0)|\<msg'>=1)@\<true>@\<msg>=1 \\
=>	& \<msg>=1 \\
=>	& (\<sofar>(\<msg>=0)@\<flag>=0)|\<msg>=1
\sloc
\eqnlabel{mattsexamplestab1}
\end{equation}
It's more subtly stable against the second line:
\begin{equation}
\cols
	& \<sp>(((\<sofar>(\<msg>=0)@\<flag>=0)|\<msg>=1)@\widehat{\<msg>=1},\;\<flag>:=1) \vspace{2pt}\\
=	& ((\<sofar>(\<msg>=0)@\<flag'>=0)|\<msg>=1)@\<msghat>=1@\<flag>=1 \vspace{2pt}\\
=> 	& ((\<sofar>(\<msg>=0)@\<msghat>=0@\<flag'>=0)|\<msg>=1)@\<msghat>=1@\<flag>=1 \\
=>	& \<msg>=1 \\
=>	& (\<sofar>(\<msg>=0)@\<flag>=0)|\<msg>=1
\sloc
\eqnlabel{mattsexamplestab2}
\end{equation}
-- observe that the first implication step uses one of the properties of \<sofar>. If $m=0$ has held since the beginning of execution then it certainly held when the interference was elaborated -- i.e. $\<sofar>(\<msg>=0)=>\<msghat>=0$.

The precondition of \<labd> is implied by the postcondition of \<labc>:
\begin{equation}
\cols
	& \<sp>((\<sofar>(\<msg>=0)@\<flag>=0)|\<msg>=1,\;\<r1>:=\<flag>) \\
=	& ((\<sofar>(\<msg>=0)@\<flag>=0)|\<msg>=1)@\<r1>=\<flag> \\
=>	& \<flag>!=0=>\<msg>=1)@\<r1>=\<flag> \\
=> 	& \<r1>=1=>\<msg>=1
\sloc
\eqnlabel{mattsexampleimp1}
\end{equation}
and is stable because $\<msg>=1$ is stable. The rest of the proof is straightforward.

\Figref{mattsexample} has to be an unproof: it claims that we can achieve message-passing with an \<lo>-constrained sender. Once again we've overlooked a cause of instability. We can imagine that $\<flag>=0$ and $\<msg>=0$ in the receiver, yet $\<msg>=1$ in the sender, after the first of its interferences has been elaborated but not propagated. At that instant $\<sofar>(\<msg>=0)$ is already untrue in the receiver since it claims that $\<msg>=0$ holds in every thread. But \<msg> is not yet 1 in the receiver, and so at that instant the precondition of \<labc> is false. The EXT stability check supposes that the receiver's precondition holds until the interference is received; our operational intuition says that \<U>-modal and \<sofar>-modal assertions may not hold for that long.

This is then an example in which UEXT stability -- stability against interference without propagation -- really matters. Even though the semantics of \<U> and \<sofar> have to be global, we can still check UEXT stability locally.
\ruledef{UEXT stability}{UEXTstability}{%!TEX root = ./Paper.tex
Constraint embroidery $P$ is UEXT stable against \interferenced{Q}{x:=E} from another thread if \vspace{3pt}\\
\hstrut{50pt}$\<sp>(\widetilde{P}@Q,\;x:=E) => \widetilde{P}$ 
}
where $\widetilde{P}$ replaces both amodal and \<Bfr>-modal occurrences of \<v> with \<vtilde>, leaving only \<U>-modal and \<sofar>-modal occurrences untouched. 

Observe that the `twiddling' in UEXT stability \ruleref{UEXTstability} is on the opposite side of the conjunction to the `hatting' in EXT stability \ruleref{EXTstability}: there we hatted $Q$ because we were checking in the interfered-with thread; here we twiddle $P$ because we are checking in the interfering thread.

The only difference between hatting and twiddling is the treatment of \<Bfr>:
\definition{\setModality{B} and \ensuremath{(\;{\tilde{}}\;)}}{Btilde}
{$$\cols[rcll]%!TEX root = ./Paper.tex
\widetilde{\<Bfr>(P)}				&==& \widetilde{\<Bfr>(P)}
\sloc$$}
-- $\<Bfr>(P)$ is local, $\<U>(P)$ and \<sofar>(P) are not.

The precondition of \<labc> in \figref{mattsexample} is UEXT unstable against the first line of the sender's guarantee:
\begin{equation}
\cols
	& \<sp>((\<sofar>(\<msg>=0)@\<flag>=0)|\<msg>=1)\sptilde@\<true>,\;\<msg>:=1) \\
=	& \<sp>((\<sofar>(\<msg>=0)@\<flagtilde>=0)|\<msgtilde>=1,\;\<msg>:=1) \vspace{2pt}\\
=	& ((\hook{\<sofar>(\<msg>=0)}@\<msg'>=0@\<flagtilde>=0)|\<msgtilde>=1)@\<msg>=1 \\
\not=>	& (\<sofar>(\<msg>=0)@\<flagtilde>=0)|\<msgtilde>=1 \\
= 	& ({\<sofar>(\<msg>=0)@\<flag>=0)|\<msg>=1})\sptilde
\sloc
%\eqnlabel{label}
\end{equation}
So when we take UEXT stability into account, the precondition of \<labc> in the receiver of \figref{mattsexample} is unstable and the proof fails, as experiment tells us it should.

\section{In-flight UO instability}

%\begin{figure}
%\centering
%$$\cols[c]
%\assert{\<init>::\<msg>=0@\<flag>=0} \vspace{3pt}\\
%\BRA[l||l]
%\thr{uoloSender (0)} 
%	\guarantee{\interferenceg{\<U>(\<flag>=0)}{\<msg>:=1} \\ \interferenceg{\<msg>=1}{\<flag>:=1}}\\ 
%	\assertd{\<uo> \<init>::\<U>(\<flag>=0)} \\ \<laba>::\<msg>:=1 \\ 
%	\assertd{\<lo> \<laba>::\<msg>=1} \\ \<labb>::\<flag>:=1 
%\rht
%&
%\thr{multiReceiver (1)} 
%	\emptyguarantee \\
%	\assertd{\<lo>\<init>::\<true>} \\ \<labc>::\<r1>:=\<flag> \\ 
%	\assertd{\<lo> \<labc>::\<r1>=1=>\<flag>=1}\\ \<labd>::\<r2>:=\<msg> \\
%	\assertd{\<lo> \<labd>::\<r1>=1@\<r2>=0=>\<flag>=1\<msg>=0}\\ \<labe>::\<r3>:=\<msg> \\
%	\assertd{\<lo> \<labe>::\<r1>=1@\<r2>=0=>\<r3>=0}
%\rht
%\KET \vspace{3pt}\\ 
%\assert{\<final>:: (1:::\<r1>)=1@(1:::\<r2>)=0 => (1:::\<r3>)=0} 
%\sloc$$
%\caption{An unproof which neglects \setorder{uo} parallelism between two interferences}
%\figlabel{uoinstability2}
%\end{figure}

\<Uo> parallelism is a second kind of in-flight parallelism. 
\definition{\setorder{Uo} parallelism}{UOparallel}{If variable assignments are \<so>- but not \<uo>-ordered then the \<so>-later is \emph{uo parallel} with the \<so>-earlier and will interfere with the \<so>-earlier's interference precondition. A variable assignment is also \<uo> parallel with itself.}
Note the similarity to \<bo> parallelism (\defref{BOparallel}). One difference is that an assignment is \<uo> parallel with itself: we explore this in the example below. Another is that \<uo> parallelism applies between assignments to the same variable.
\ruledef{UO stability}{UOstability}{%!TEX root = ./Paper.tex
Interference \interferenced{P}{x:=E} is UO stable against \<uo>-parallel \interferenced{Q}{y:=F} if \vspace{3pt}\\
\hstrut{50pt}$\<sp>(\widetilde{P}@\dtildev{Q},\;y:=F) => \widetilde{P}$
}
$\dtildev{Q}$ is like $\widetilde{Q}$ except that $v$ is replaced by $\dtildev{v}$ (and, as with a single twiddle, $\<sofar>(P)=>\dtildev{P}$). Note the similarity to BO stability \ruleref{BOstability}.

Consider the sender of \figref{uoinstability1}: it's exactly the sender of MP except that we have given command \<laba> a \<U>-modal precondition, which is validly inherited from the initial state and is stable because the receiver makes no interference. The sender's guarantee is constructed as usual from the elaboration preconditions. The program works because it has more than the constraints of MP -- \<uo> rather than \<bo>, and \<bo> includes \<uo> -- but the proof is misleading.

\begin{figure}
\centering
% proofs/uo-unstable-interference.unproof
$$\cols[c]
\assert{\<init>::\<msg>=\<flag>=0} \vspace{3pt} \\
\BRA[l||l]
\thr{Uo-unstable Sender (0)}
  \guarantee{\interferenceg{\hcancel{\<U>(\<msg>=0)}}{\<msg>:=1}; \\ \interferenceg{\<Bfr>(\<msg>=1)}{\<flag>:=1}} \vspace{3pt} \vspace{3pt} \\
  \!\!\cols[r@{\intfspace}l]
  \assertd{\<uo> \<init>:: \<U>(\<msg>=0)} & \<laba>:: \<msg>:=1; \\
  \assertd{\<bo> \<laba>:: \<Bfr>(\<msg>=1)} & \<labb>:: \<flag>:=1
  \sloc
\rht
&
\thr{Receiver (1)}
  \emptyguarantee \vspace{3pt} \vspace{3pt} \\
  \!\!\cols[r@{\intfspace}l]
  \assertd{\<lo> \<init>:: \<flag>=1=>\<msg>=1} & \<labc>:: \<r1>:=\<flag>; \\
  \assertd{\<lo> \<labc>:: \<r1>=1=>\<msg>=1} & \<labd>:: \<r2>:=\<msg>
  \sloc \vspace{3pt} \\
  \assertd{\<lo> \<labd>:: \<r1>=1=>\<r2>=1}
\rht
\KET \vspace{3pt} \\
\assert{\<final>::(1:::\<r1>)=1=>(1:::\<r2>)=1}
\sloc$$
\caption{Unproof which neglects \setorder{uo} self-parallelism of interference}
\figlabel{uoinstability1}
\end{figure}

The problem is with the first line of the sender's guarantee. \interferenced{\<U>(\<msg>=0)}{\<msg>:=1} is a nonsense, because as soon as $\<msg>:=1$ is elaborated, its interference precondition is false, so we can't say that it will hold when $\<msg>=1$ is propagated to the receiver. The assignment overtakes \emph{its own interference precondition}. 

\Ruleref{UOstability} says that an assignment is \<uo> parallel with itself. In the case of command \<laba> the interference precondition isn't UO stable:
\begin{equation}
\cols
	& \<sp>(\widetilde{\<U>(\<msg>=0)}@\dtilde{\<U>(\<msg>=0)}, \; \<msg>:=1) \\
==	& \<sp>(\<U>(\<msg>=0)@\widetilde{m}=0@\<U>(\<msg>=0)@\dtildev{m}=0, \; \<msg>:=1) \\
==	& \<sp>(\<U>(\<msg>=0)@\widetilde{m}=0@\dtildev{m}=0, \; \<msg>:=1) \\
==	& \hook{\<U>(\<msg>=0)}@\<msg'>=0@\widetilde{m}=0@\dtildev{m}=0@\<msg>=1 \\
\not=>	& \<U>(\<msg>=0)@\widetilde{m}=0 \\
==	& \widetilde{\<U>(\<msg>=0)}
\sloc\eqnlabel{uoinstabilitycheck}
\end{equation}
The first line of the guarantee is wrong, and its precondition should be weakened.

It's possible to construct a receiver which shows that relying on the sender's interference as written in \figref{uoinstability1} is unsound, but it would be tedious to show it. 

\section{A proof of LB using \setModality{S}ofar}
\seclabel{LB}

The first of our three weak-memory litmus tests was LB (\figref{LB}). It was the easiest to explain in operational terms, but it's the hardest of the three to prove, needing heavy temporal machinery despite being apparently so simple. The laced version of LB is in \figref{LBlaced}. It uses no more than \<lo> ordering in each thread.

\begin{figure}
\centering
% proofs/LB.proof
$$\cols[c]
\assert{\<init>::x=y=0} \vspace{3pt} \\
\BRA[l||l]
\thr{Thread 0}
  \!\!\cols[r@{\intfspace}l]
  \assertd{\<lo> \<init>} & \<laba>:: \<r1>:=y; \\
  \assertd{\<lo> \<laba>} & \<labb>:: x:=1
  \sloc
\rht
&
\thr{Thread 1}
  \!\!\cols[r@{\intfspace}l]
  \assertd{\<lo> \<init>} & \<labc>:: \<r1>:=x; \\
  \assertd{\<lo> \<labc>} & \<labd>:: y:=1
  \sloc
\rht
\KET \vspace{3pt} \\
\assert{\<final>::!((0:::\<r1>)=1@(1:::\<r1>)=1)}
\sloc$$
\caption{LB, laced}
\figlabel{LBlaced}
\end{figure}

\begin{figure}
\centering
% proofs/LB.proof
$$\cols[c]
\assert{\<init>::x=y=(0,0)} \vspace{3pt} \\
\BRA[l]
\thr{Thread 0}
  \guarantee{\interferenceg{\<Bfr>(x=(0,0))@y=(1,0)}{x:=1,1}; \\ \interferenceg{\<Bfr>(x=(0,0))}{x:=1,0}} \vspace{3pt} \vspace{3pt} \\
  \!\!\cols[r@{\intfspace}l]
  \assertd{\<lo> \<init>: \<sofar>(x=(0,0)@(y=(0,0)|y=(1,0)))} & \<laba>:: \<r1>,\_:=y; \\
  \assertd{\<lo> \<laba>: \BRA \<sofar>(x=(0,0)@(y=(0,0)|y=(1,0))) \\
  						       @\;(\<r1>=0|\<r1>=1@y=(1,0))
						  \KET} & \\ \ipre{\<Bfr>(x=(0,0))@(\<r1>=0|\<r1>=1@y=(1,0))} & \<labb>:: x:=1,\<r1>
  \sloc \vspace{3pt} \\
  \assertd{\<lo> \<labb>: \<r1>=1=>y=(1,0)@x=(1,1)} \vspace{3pt} \\
  \rely{\interferenceg{\<Bfr>(y=(0,0))@x=(1,0)}{y:=1,1}; \\ \interferenceg{\<Bfr>(y=(0,0))}{y:=1,0}} \vspace{3pt} \vspace{3pt} \\
\rht
\KET \vspace{3pt} \\
\assert{\<final>::!((0:::\<r1>)=1@(1:::\<r1>)=1)}
\sloc$$
\caption{A proof of LB thread 0}
\figlabel{LBproof}
\end{figure}

The proof of thread 0 is in \figref{LBproof}, with a rely from thread 1. It uses auxiliary extension of $x$ and $y$, as in the proof of WRC (\figref{WRCaux}), to allow us to distinguish write-1-after-read-0 from write-1-after-read-1. It uses \<sofar>, which allows us to reason about the state in which interference is elaborated and propagated. Because \<sofar> is susceptible to self-UO instability, command \<labb> has an explicit interference precondition.

Our proof is essentially circular. Thread 0 can't see $y=(1,1)$ unless it first writes $x=(1,0)$ or $x=(1,1)$, because reading from one of those writes is the only way that thread 1 can make $\<r1>=1$. So before thread 0 writes to $x$ we can be sure that $y!=(1,1)$ -- i.e. $y=(0,0)|y=(1,0)$. And then only the second of those possibilities will allow it to write $x=(1,1)$. That's captured in its guarantee. And then we know by symmetry that thread 1 must have a similar guarantee, which is thread 1's rely. And that gives us our circular proof: they can't both read after the other writes. 

Formally, the precondition of \<laba> is EXT stable against the first line of the sender's guarantee because $\<sofar>(x=(0,0))=>\<xhat>=(0,0)$:
\begin{equation}
\cols
	& \<sp>(\<sofar>(x=(0,0)@(y=(0,0)|y=(1,0)))@(\<Bfr>(y=(0,0))@x=(1,0))\sphat,\;y:=1,1) \\
==	& \<sp>(\<sofar>(x=(0,0)@(y=(0,0)|y=(1,0)))@\<Bfr>(y=(0,0))@\<yhat>=(0,0)@\<xhat>=(1,0),\;y:=1,1) \\
=> 	& \<sp>(\<sofar>(x=(0,0)@(y=(0,0)|y=(1,0)))@\<xhat>=(0,0)@\<Bfr>(y=(0,0))@\<yhat>=(0,0)@\<xhat>=(1,0),\;y:=1,1) \\
=>	& \<xhat>=(0,0)@\<xhat>=(1,0) \\
==	& \<false> \\
=>	& \<sofar>(x=(0,0)@(y=(0,0)|y=(1,0)))
\sloc\eqnlabel{LBstab1}
\end{equation}
It's UEXT stable against the same interference because $\<sofar>(x=(0,0))=>\dtildev{x}=(0,0)$. It's EXT and UEXT stable against the second line of the rely because $y=(1,0)$ is stable; it's a register assigment so we don't have to consider BO or UO instability, and it makes no entry in the guarantee.

The postcondition of command \<laba> is $\<sofar>(x=(0,0)@(y=(0,0)|y=(1,0)))@(\<r1>,\_)=y$ which implies the precondition of \<labb>, which is EXT and UEXT stable against the guarantee for the same reasons as before. But as an interference it would be self-UO unstable so we weaken $\<sofar>(x=(0,0)@(y=(0,0)|y=(1,0)))$ to $\<Bfr>(x=(0,0))$ and we have a self-UO-stable interference precondition. Considering the two cases $\<r1>=1$ and $\<r1>=0$ give us the first and second lines of the guarantee. 

The thread postcondition says that if $\<r1>=1$ we must have $y=(1,0)$; the postcondition of thread 1 would insist that if $\<r1>=1$ in thread 1 we must have $y=(1,1)$; so we can't have $\<r1>=1$ in both threads and we have a proof.

This is the simplest proof we have been able to construct. It requires temporal cross-thread reasoning, and perhaps that is just the way it has to be. The use of \<sofar> might seem to be restrictive, working only with short and simple litmus tests. But in truth any write can be made unique with just enough auxiliary padding, so though awkward it isn't an incompleteness.

\section{The \setModality{O}uat modality}
\seclabel{ouat}

$\<U>(P)$ says that $P$ has held everywhere since a barrier; $\<Sofar>(P)$ that $P$ has held everywhere since execution started. $\<Bfr>(P)$ says that $P$ has held locally since a barrier. There's clearly a missing local modality.

Rather than make a local analogue of \<sofar>, we prefer to make a negated version. $\<ouat>(P)$ says that $P$ has held locally sometime since execution started -- an `existence' rather than a `universal' modality. %\footnote{It's equivalent to `not((not $P$) has held locally since execution started)'. It's that double negative that puts us off.}
It's extremely useful when reasoning about the witnessed order of writes to a single location.

Its most spectacular property is that as a positive assertion it's immune to variable interference -- if $P$ happened, it happened, and variable interference can't change that. %\footnote{Immunity to interference requires special treatment in the SMT embedding of our proofchecker: see \secref{embedding}.} 
A negative assertion is less stable. Being thread local, it can be hatted and twiddled.
\definition{Substitution and hatting of \setModality{O}uat}{Ouatsubsthat}
{$$\cols[rcll]
\<ouat>(P)[x\backslash{}\<x'>]		&==&	\hook{\<ouat>(P)} \\
\hook{\<ouat>(P)}|P					&<=>&	\<ouat>(P) \\
!\hook{\<ouat>(P)}@!P				&<=>&	!\<ouat>(P) \\
\<ouat>(P)[r\backslash{}\<r'>]	 	&==&	\<ouat>(P[r\backslash{}\<r'>]) \vspace{5pt} \\
%!TEX root = ./Paper.tex
\widehat{\<ouat>(P)}					&==&	\widehat{\<ouat>(P)}

\sloc$$}

Because it is an existence variety of \<since> rather than a universal, \<ouat>'s properties are de Morgan opposites of the properties of the other modalities. 
\definition{Properties of \setModality{O}uat}{Ouatproperties}{%!TEX root = ./Paper.tex
$$\cols[rcll]
\<ouat>(P)			&<<=& P  \\
\<ouat>(P|Q)		&==& \<ouat>(P)|\<ouat>(Q)  \\
\<ouat>(\<ouat>(P))	&==& \<ouat>(P) \\
\<ouat>(P@Q) 		&=>& \<ouat>(P)@\<ouat>(Q)	\\
\<ouat>(P) 			&=>& P						& \text{(if } P \text{ doesn't mention variables)} \\
\<ouat>(P@Q) 		&<<=& \<ouat>(P)@\<ouat>(Q)	& \text{(if } P \text{ or } Q \text{ don't mention variables)} \\
(P=>Q)				&=>& \<ouat>(P)=>\<ouat>(Q)	& \text{(if } P=>Q \text{ is a tautology)} \vspace{5pt} \\
\<ouat>(P)			&=>& !\<sofar>(!P)
\sloc$$}

%!TEX root = ./Paper.tex

\section{Reasoning about coherence order}
\seclabel{coherence}

Several litmus tests and, we believe, many real-world algorithms\footnote{One of the origins of this work was Bornat's attempt to verify an idempotent work-stealing algorithm of~\citet{MichaeletalIdempotentworkstealing2009}. That attempt foundered because he couldn't see a way to reason about coherence. Now, four years later, he can't yet reason about the heap. Ho hum, investigations continue.} exploit the `global coherence' or `SC per location' constraint SCloc. In those tests and algorithms certain outcomes are impossible because they would require two threads to witness writes to memory in different orders.

Coherence is so important that we have decided to address it explicitly in our logic. Since the constraint applies to writes rather than to values written, we have to ensure that, for variables about which we assert an ordering of writes, each write has a unique value. That involves temporal reasoning and, in many cases, auxiliary information. 

\begin{figure}
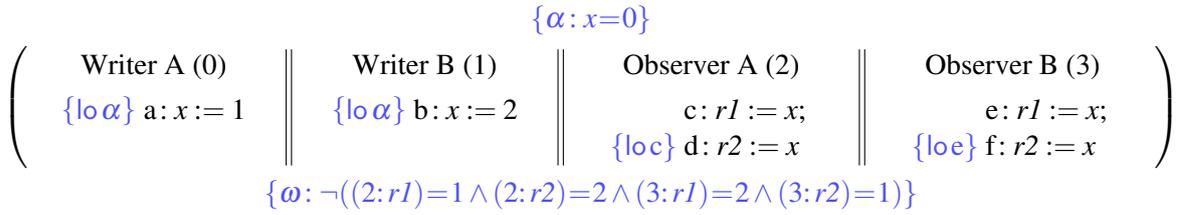

\centering
% proofs/CoRR2.proof
$$\cols[c]
\assert{\<init>::x=0} \vspace{3pt} \\
\BRA[l||l||l||l]
\thr{Writer A (0)}
  \assertd{\<lo> \<init>}\ \<laba>:: x:=1
\rht
&
\thr{Writer B (1)}
  \assertd{\<lo> \<init>}\ \<labb>:: x:=2
\rht
&
\thr{Observer A (2)}
  \!\!\cols[r@{\intfspace}l]
   & \<labc>:: \<r1>:=x; \\
  \assertd{\<lo> \<labc>} & \<labd>:: \<r2>:=x
  \sloc
\rht
&
\thr{Observer B (3)}
  \!\!\cols[r@{\intfspace}l]
   & \<labe>:: \<r1>:=x; \\
  \assertd{\<lo> \<labe>} & \<labf>:: \<r2>:=x
  \sloc
\rht
\KET \vspace{3pt} \\
\assert{\<final>::!((2:::\<r1>)=1@(2:::\<r2>)=2@(3:::\<r1>)=2@(3:::\<r2>)=1)}
\sloc$$
\caption{CoRR2 laced}
\figlabel{CoRR2laced}
\end{figure}

\subsection{A first coherence example}

Consider \figref{CoRR2laced}. Writer threads 0 and 1 write different values to $x$; observer threads 2 and 3 read $x$ twice, each potentially reading from two different writes. Under SCloc each observer thread sees a fraction of the partial order of writes to $x$, and it is impossible that they could see the same writes in different orders.\footnote{In the original litmus test this outcome, which involves successive reads from different external writes in each observer thread, would provide implicit dependencies between commands \<labc> and \<labd> and between \<labe> and \<labf>. For reasons explained in \secref{SCloc} we must use an explicit \<lo> constraint between the commands in each observer thread, but we believe that nevertheless our treatment captures the essence of the test.}

\begin{figure}
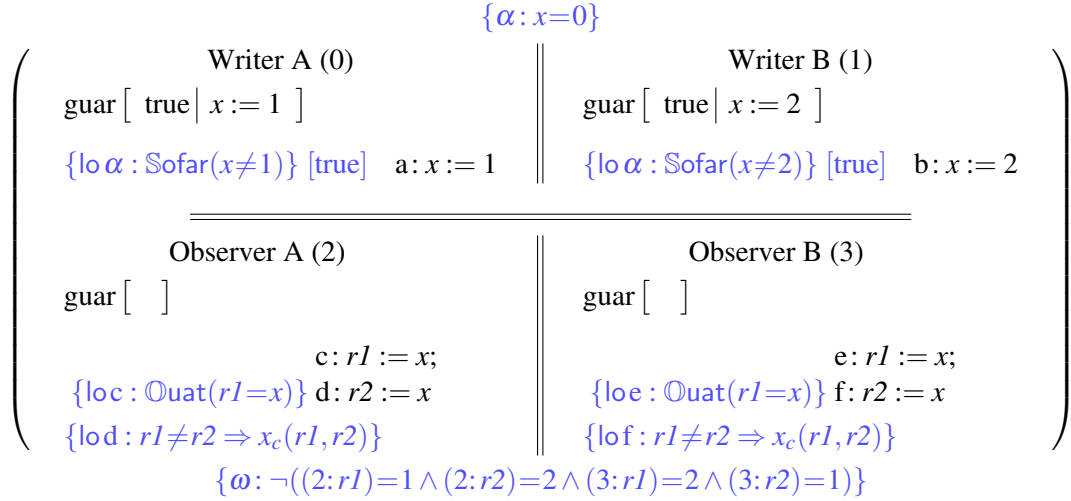

\centering
% proofs/CoRR2.proof
$$\cols[c]
\assert{\<init>::x=0} \vspace{3pt} \\
\BRA[l||l]
\thr{Writer A (0)}
  \guarantee{\interferenceg{\<true>}{x:=1}} \vspace{3pt} \vspace{3pt} \\
  \assertd{\<lo> \<init>: \<sofar>(x!=1)}\ \ipre{\<true>} & \<laba>:: x:=1
\rht
&
\thr{Writer B (1)}
  \guarantee{\interferenceg{\<true>}{x:=2}} \vspace{3pt} \vspace{3pt} \\
  \assertd{\<lo> \<init>: \<sofar>(x!=2)}\ \ipre{\<true>} & \<labb>:: x:=2
\rht \vspace{3pt} \\
\multicolumn{2}{c}{\begin{minipage}{260pt} $\cols \hstrut{260pt} \\[-8pt] \hline \hline \sloc$ \end{minipage}} \vspace{3pt} \\ 
\thr{Observer A (2)}
  \emptyguarantee \vspace{3pt} \vspace{3pt} \\
  \!\!\cols[r@{\intfspace}l]
   & \<labc>:: \<r1>:=x; \\
  \assertd{\<lo> \<labc>: \<ouat>(\<r1>=x)} & \<labd>:: \<r2>:=x
  \sloc \vspace{3pt} \\
  \assertd{\<lo> \<labd>: \<r1>!=\<r2>=>x_{c}(\<r1>,\<r2>)}
\rht
&
\thr{Observer B (3)}
  \emptyguarantee \vspace{3pt} \vspace{3pt} \\
  \!\!\cols[r@{\intfspace}l]
   & \<labe>:: \<r1>:=x; \\
  \assertd{\<lo> \<labe>: \<ouat>(\<r1>=x)} & \<labf>:: \<r2>:=x
  \sloc \vspace{3pt} \\
  \assertd{\<lo> \<labf>: \<r1>!=\<r2>=>x_{c}(\<r1>,\<r2>)}
\rht
\KET \vspace{3pt} \\
\assert{\<final>::!((2:::\<r1>)=1@(2:::\<r2>)=2@(3:::\<r1>)=2@(3:::\<r2>)=1)}
\sloc$$
\caption{CoRR2 proof}
\figlabel{CoRR2proof}
\end{figure}

The two writers clearly write different values, each different from the initial-state value, and there are no other assignments to $x$ in the program, so we can see that this is an example in which we can identify writes according to the values that they contain. %In general, however, we would have to add auxiliary extension to make writes distinct -- messy, but necessary.

The proof in \figref{CoRR2proof} contains, in the postcondition of each observer, a coherence assertion $x_{c}(\<r1>,\<r2>)$, which claims that there are values $\setvar{W1}$ and $\setvar{W2}$ such that $\<r1>=\setvar{W1}@\<r2>=\setvar{W2}$, and the write $x=\setvar{W1}$ is globally ordered before $x=\setvar{W2}$. 

For proofchecking purposes we require that every assignment $x:=E$ to a variable for which there is a coherence assertion anywhere in the program must have a precondition which implies $\<sofar>(x!=E)$, which guarantees that writes to that variable are unique. This necessarily causes self-UO instability, and to avoid that we shall always need an explicit interference precondition.
\begin{aside}
There is an argument that $!\<ouat>(x=E)$, which would not provoke self-UO instability, would be sufficient to guarantee uniqueness of writes. Although it's not the case that $!\<ouat>(x=E)=>\<sofar>(x!=E)$, we can argue that if there is any assignment $x:=E$ that could be executed in any other thread, or earlier in the same thread, then we could never prove $!\<ouat>(x=E)$. In the absence of a soundness proof, this counts as a bonzer wheeze, all of which we've decided to avoid in version 2 of Lace logic. $\<sofar>(x!=E)$ (see especially the embedding description in \secref{embedding}) is far more obviously adequate.
\end{aside} 
Thus the elaboration precondition of command \<laba> in the writer asserts $\<sofar>(x!=1)$, the elaboration precondition of \<labb> asserts $\<sofar>(x!=2)$, and each command has an interference precondition \<true>. Both elaboration preconditions are implied by the initial state, which establishes $\<sofar>(x=0)$. The elaboration precondition of \<laba> is stable against the interference of Writer B:
\begin{equation}
\cols
	& \<sp>(\<sofar>(x!=1)@\widehat{\<true>},\;x:=2) \\
==	& \hook{\<sofar>(x!=1)}@\<x'>!=1@\<true>@x=2 \\
=>	& \hook{\<sofar>(x!=1)}@x!=1 \\
=>	& \<sofar>(x!=1) 
\sloc
\eqnlabel{CoRR2stab1}
\end{equation}
and its interference precondition \<true> is trivially implied by $\<sofar>(x!=1)$. The preconditions of \<labb> in writer B are valid for similar reasons.

In our treatment coherence is not a modality, it's a relation between values. The meaning of coherence assertions is defined by axioms. 
\definition{Coherence axioms}{coherenceaxioms}{\begin{tabular}{rl}
membership		& $x_c(A,B)=>\<cv>(x)$ \\
irreflexive 	& $x_c(A,B) => A!=B$ \\
transitive 		& $x_c(A,B)@x_c(B,C)=>x_c(A,C)$ \\
antisymmetric 	& $x_c(A,B)=>!x_c(B,A)$ \\         
observed 		& $\<ouat>(\<ouat>(x=A)@x=B)@A!=B@\<cv>(x)=>x_c(A,B)$ \\ 
%%history 		& $x_c(A,B)=>!(\<ouat>(x=A@\<ouat>(x=B)))$
\end{tabular}
}
We require that the unique-write test is applied to variables $v$ for which $v_{c}(\_,\_)$ appears in any assertion, and implicitly $\<cv>(v)$ holds only for such variables. The `membership' axiom then ensures that we can't deduce $x_c(A,B)$ for any other variable.

Coherence is irreflexive, transitive and antisymmetric; $x_c(A,B)$ is observed if in some thread's history there was a state in which $x=B$ and previously $x=A$; by antisymmetry $x_c(A,B)$ implies that no thread has ever seen $x=B$ before $x=A$. Because coherence is a global relation, we do not allow deduction of local observations from coherence. In hooking and hatting, only $A$ and $B$ in $x_c(A,B)$ are affected.
\definition{Coherence hooking and hatting}{coherencehookinghatting}{
$$\cols[rcll]
%!TEX root = ./Paper.tex
(v_c(A,B))[x\backslash \<x'>]		&==& 	v_c(A[x\backslash \<x'>],B[x\backslash \<x'>]) \\
(v_c(A,B))[r\backslash \<r'>]		&==& 	v_c(A[r\backslash \<x'>],B[x\backslash \<r'>]) \vspace{5pt} \\

\widehat{v_c(A,B)}					&==&	v_c(\widehat{A},\widehat{B}) 
\sloc$$}

In observer A of \figref{CoRR2proof} the instantaneous postcondition of \<labc> is $\<r1>=x$, which is unstable, but which implies the stable $\<ouat>(x=\<r1>)$. Then the postcondition of \<labd> is $\<ouat>(x=\<r1>)@\<r2>=x$ which, provided \<r1> and \<r2> have different values, implies $x_{c}(\<r1>,\<r2>)$ by the `observed' axiom of \defref{coherenceaxioms}. Observer B is similar. It is impossible, by the `antisymmetric' axiom, that we could have $x_c(1,2)$ in one observer and $x_c(2,1)$ in the other, which gives us the program postcondition.

\begin{figure}
\centering
% proofs/S.proof
$$\cols[c]
\assert{\<init>::\<msg>=\<flag>=0} \vspace{3pt} \\
\BRA[l||l]
\thr{Thread 0}
  \!\!\cols[r@{\intfspace}l]
  \assertd{\<lo> \<init>} & \<laba>:: \<msg>:=1; \\
  \assertd{\<bo> \<laba>} & \<labb>:: \<flag>:=1
  \sloc
\rht
&
\thr{Thread 1}
  \!\!\cols[r@{\intfspace}l]
  \assertd{\<lo> \<init>} & \<labc>:: \<r1>:=\<flag>; \\
  \assertd{\<lo> \<labc>;\<lo> \<init>} & \<labd>:: \<msg>:=2
  \sloc
\rht
\KET \vspace{3pt} \\
\assert{\<final>::!((1:::\<r1>)=1@\<msg>=1)}
\sloc$$
\caption{Litmus test S: \setModality{B} transmits variable history}
\figlabel{Slaced}
\end{figure}

\subsection{Coherence and \setModality{B}}

Consider \figref{Slaced}, a version of the litmus test S.\footnote{The original uses $x$ and $y$ rather than \<msg> and \<flag>, and swaps 2 and 1. We think our version shows its inheritance from MP more clearly. Note that the initial state needs $\<msg>=0$ to make sure that writes to \<msg> are unique.} It is almost MP, except that thread 2 doesn't read \<msg>, it writes it. But then $\<msg>=2$ is not stable in thread 0, and we have to reason about coherence (or use auxiliary information) to derive the conclusion.

\begin{figure}
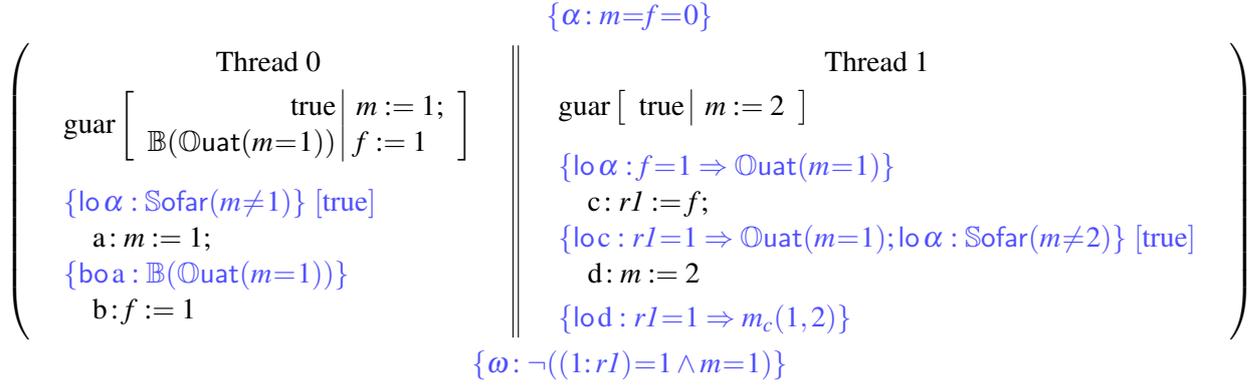

\centering
%% proofs/S.proof
$$\cols[c]
\assert{\<init>::\<msg>=\<flag>=0} \vspace{3pt} \\
\BRA[l||l]
\thr{Thread 0}
  \guarantee{\interferenceg{\<true>}{\<msg>:=1}; \\ \interferenceg{\<Bfr>(\<ouat>(\<msg>=1))}{\<flag>:=1}} \vspace{3pt} \vspace{3pt} \\
  \assertd{\<lo> \<init>: \<sofar>(\<msg>!=1)}\ \ipre{\<true>} \\ \quad \<laba>:: \<msg>:=1; \\
  \assertd{\<bo> \<laba>: \<Bfr>(\<ouat>(\<msg>=1))} \\ \quad \<labb>:: \<flag>:=1
\rht
&
\thr{Thread 1}
  \guarantee{\interferenceg{\<true>}{\<msg>:=2}} \vspace{3pt} \vspace{3pt} \\
  \assertd{\<lo> \<init>: \<flag>=1=>\<ouat>(\<msg>=1)} \\ \quad \<labc>:: \<r1>:=\<flag>; \\
  \assertd{\<lo> \<labc>: \<r1>=1=>\<ouat>(\<msg>=1);\<lo> \<init>: \<sofar>(\<msg>!=2)}\ \ipre{\<true>} \\ \quad \<labd>:: \<msg>:=2 \vspace{3pt} \\
  \assertd{\<lo> \<labd>: \<r1>=1=>\<msg>_{c}(1,2)}
\rht
\KET \vspace{3pt} \\
\assert{\<final>::!((1:::\<r1>)=1@\<msg>=1)}
\sloc$$
\caption{Proof of S}
\figlabel{Sproof}
\end{figure}

The proof of S in \figref{Sproof} is straightforward. It contains two coherence assertions for $\<msg>$, so we need to prove $\<sofar>(\<msg>!=1)$ in the elaboration precondition of \<laba> and $\<sofar>(\<msg>!=2)$ in the precondition of \<labd>. $\<sofar>(\<msg>!=1)$ in thread 0 is stable against \interferenced{\<true>}{\<msg>:=2} by an argument very similar to \eqnref{CoRR2stab1}, 
%%\begin{equation}
%%\cols
%%	& \<sp>(\<sofar>(\<msg>!=1)@\widehat{\<true>},\;\<msg>:=2) \\
%%=	& \hook{\<sofar>(\<msg>!=1)}@\<msg'>!=1@\<msg>=2 \\
%%=>	& \hook{\<sofar>(\<msg>!=1)}@\<msg>!=1 \\
%%=>	& \<sofar>(\<msg>!=1)
%%\sloc
%%\eqnlabel{Sstab1}
%%\end{equation}
and similarly $\<sofar>(\<msg>!=2)$ in thread 1 is stable against \interferenced{\<true>}{\<msg>:=1}.

The postcondition of \<laba> implies $\<msg>=1$, which implies $\<ouat>(\<msg>=1)$, which is stable. That's decorated with \<Bfr> to make the embroidery of the \<laba>$->$\<labb> constraint. Each assignment, with its elaboration precondition, is copied to the guarantee.

In thread 1 the preconditions of \<labc> and \<labd> are stable because $\<ouat>(\<msg>=1)$ is stable and $\<sofar>(\<msg>!=2)$ is stable both against \interferenced{\<true>}{\<msg>:=1}, as noted above, and against \interferenced{\dots}{\<flag>:=1}. The postcondition of \<labd> implies $(\<r1>=1=>\<ouat>(\<msg>=1))@\<msg>=2$, which, after some rearrangement and then by the `observed' axiom, implies the thread postcondition. The program postcondition follows by an argument about eventual consistency which we shall justify in \secref{PMS}: once the dust has settled, and all the writes have been propagated, $\<msg>_{c}(1,2)=>\<msg>!=1$.

%%Note that in this proof $\<sofar>(x!=2)$, an assertion about the history of thread 0, is transmitted, using a \<Bfr>-modal interference precondition, to thread 1. That's sound, we believe: when thread 1 reads $y$ it is causally connected to the history of thread 0 before the elaboration of $y:=1$. We shall see in \secref{PMS} that without that causal connection threads do not necessarily see each others' histories.

%%\begin{figure}
%%\centering
%%% proofs/2+2W.proof
%%$$\cols[c]
%%\assert{\<init>::x=0@y=0} \vspace{3pt} \\
%%\BRA[l||l]
%%\thr{Writer A (0)}
%%  \!\!\cols[r@{\intfspace}l]
%%  \assertd{\<lo> \<init>} & \<laba>:: x:=1; \\
%%  \assertd{\<bo> \<laba>} & \<labb>:: y:=2
%%  \sloc
%%\rht
%%&
%%\thr{Writer B (1)}
%%  \!\!\cols[r@{\intfspace}l]
%%  \assertd{\<lo> \<init>} & \<labc>:: y:=1; \\
%%  \assertd{\<bo> \<labc>} & \<labd>:: x:=2
%%  \sloc
%%\rht
%%\KET \vspace{3pt} \\
%%\assert{\<final>::!(x=y=1)}
%%\sloc$$
%%\caption{Litmus test 2+2W}
%%\figlabel{2+2W}
%%\end{figure}

\begin{figure}
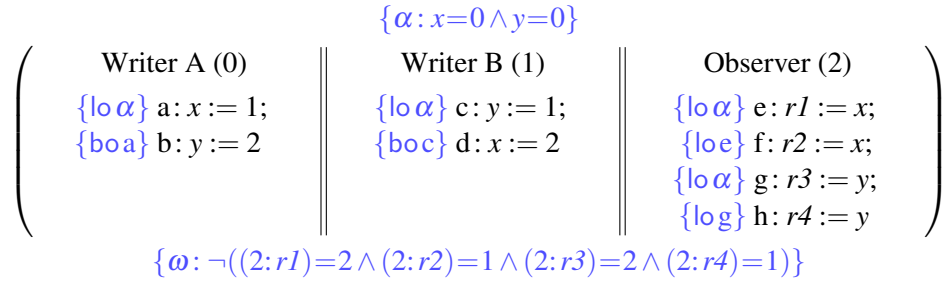

\centering
% proofs/2+2W.proof
$$\cols[c]
\assert{\<init>::x=0@y=0} \vspace{3pt} \\
\BRA[l||l||l]
\thr{Writer A (0)}
  \!\!\cols[r@{\intfspace}l]
  \assertd{\<lo> \<init>} & \<laba>:: x:=1; \\
  \assertd{\<bo> \<laba>} & \<labb>:: y:=2
  \sloc
\rht
&
\thr{Writer B (1)}
  \!\!\cols[r@{\intfspace}l]
  \assertd{\<lo> \<init>} & \<labc>:: y:=1; \\
  \assertd{\<bo> \<labc>} & \<labd>:: x:=2
  \sloc
\rht
&
\thr{Observer (2)}
  \!\!\cols[r@{\intfspace}l]
  \assertd{\<lo> \<init>} & \<labe>:: \<r1>:=x; \\
  \assertd{\<lo> \<labe>} & \<labf>:: \<r2>:=x; \\
  \assertd{\<lo> \<init>} & \<labg>:: \<r3>:=y; \\
  \assertd{\<lo> \<labg>} & \<labh>:: \<r4>:=y
  \sloc
\rht
\KET \vspace{3pt} \\
\assert{\<final>::!((2:::\<r1>)=2@(2:::\<r2>)=1@(2:::\<r3>)=2@(2:::\<r4>)=1)}
\sloc$$
\caption{Litmus test 2+2W with observer thread, laced}
\figlabel{2+2Wlaced}
\end{figure}

\subsection{2+2W}

\Figref{2+2Wlaced} is a subtle problem. The online litmus test \citep{2+2W+lwsyncs} has two writer threads with \<bo> ordering in each (but no observer thread) and requires on termination $!(x=y=1)$. The litmus test diagram in \citep{PowerARMLitmusTests} requires that with \<bo> ordering we cannot have both $x_{c}(2,1)$ and $y_{c}(2,1)$ which, as we shall see in \secref{PMS}, implies on termination $!(x!=2@y!=2)$. Obviously neither writer can terminate with $x=0|y=0$. So all of this boils down to $(x=1@y=2)|(x=2@y=1)$. Operational reasoning explains why either requirement must be satisfied: if $x=2$ reaches the store before $x=1$, for example, then $y=1$, which \<bo>-precedes $x=2$, must already have arrived and therefore will be before $y=2$, which \<bo>-succeeds $x=1$ -- so on termination $x=1$ is the last write to $x$ and $y=2$ is the last write to $y$. And vice-versa if $y=1$ is the last write to $y$, $x=2$ will be the last write to $x$.

\begin{figure}
\centering
% proofs/2+2W.proof
$$\cols[c]
\assert{\<init>::x=0@y=0} \vspace{3pt} \\
\BRA[c]
\cols[l||l]
\thr{Writer A (0)}
  \guarantee{\interferenceg{\<Bfr>(\<sofar>(y!=2))}{x:=1}; \\ \interferenceg{\<Bfr>(\<ouat>(x=1))}{y:=2}} \vspace{3pt} \vspace{3pt} \\
  \!\!\cols[r@{\intfspace}l]
  \assertd{\<lo> \<init>:: \<Bfr>(\<sofar>(y!=2))} & \<laba>:: x:=1; \\
  \assertd{\<bo> \<laba>:: \<Bfr>(\<ouat>(x=1))} & \<labb>:: y:=2
  \sloc
\rht
&
\thr{Writer B (1)}
  \guarantee{\interferenceg{\<Bfr>(\<sofar>(x!=2))}{y:=1}; \\ \interferenceg{\<Bfr>(\<ouat>(y=1))}{x:=2}} \vspace{3pt} \vspace{3pt} \\
  \!\!\cols[r@{\intfspace}l]
  \assertd{\<lo> \<init>:: \<Bfr>(\<sofar>(x!=2))} & \<labc>:: y:=1; \\
  \assertd{\<bo> \<labc>:: \<Bfr>(\<ouat>(y=1))} & \<labd>:: x:=2
  \sloc
\rht
\sloc \vspace{3pt} \\
\begin{minipage}{340pt} $\cols \hstrut{340pt} \\[-8pt] \hline \hline \sloc$ \end{minipage} \vspace{3pt} \\ 
\thr{Observer (2)}
  \emptyguarantee \vspace{3pt} \vspace{3pt} \\
  \!\!\cols[r@{\intfspace}l]
  \assertd{\<lo> \<init>:: x=2=>\<ouat>(y=1)} & \<labe>:: \<r1>:=x; \vspace{2pt} \\
  \assertd{\<lo> \<labe>:: \<r1>=2=>\BRA \<ouat>(y=1)@\<ouat>(x=2)\\@\;(x=1=>!\<ouat>(y=1@\<ouat>(y=2)))\KET} & \<labf>:: \<r2>:=x; \vspace{2pt} \\
  \assertd{\<lo> \<init>:: y=2=>\<ouat>(x=1)} & \<labg>:: \<r3>:=y; \vspace{2pt} \\
  \assertd{\<lo> \<labg>:: \<r3>=2=>\BRA \<ouat>(x=1)@\<ouat>(y=2)\;@\\(y=1=>!\<ouat>(x=1@\<ouat>(x=2)))\KET} & \<labh>:: \<r4>:=y
  \sloc \vspace{3pt} \\
  \Assertd{\<lo> \<labf>:: \<r1>=2@\<r2>=1=>\<ouat>(x=1@\<ouat>(x=2))@!\<ouat>(y=1@\<ouat>(y=2)); \\
  		   \<lo> \<labh>:: \<r3>=2@\<r4>=1=>\<ouat>(y=1@\<ouat>(y=2))@!\<ouat>(x=1@\<ouat>(x=2))
		  }
\rht
\KET \vspace{3pt} \\
\assert{\<final>::!((2:::\<r1>)=2@(2:::\<r2>)=1@(2:::\<r3>)=2@(2:::\<r4>)=1)}
\sloc$$
\caption{2+2W proof}
\figlabel{2+2Wproof}
\end{figure}

The proof in \figref{2+2Wproof} imitates the operational reasoning because we haven't yet found another way. The observer reads $x$ twice in sequence and, internally in parallel, reads $y$ twice. There's a lot of \<lo>-parallel interference but the registers are distinct so there is no \<lo>-parallel instability. 

Since the commands which read $x$ and the commands which read $y$, and their embroidery, are symmetrical, we check the $x$-reading \<labe> and \<labf>. The precondition $x=2=>\<ouat>(y=1)$ of \<labe> is stable because $\<ouat>(y=1)$ is stable, and it's non-trivially established by the interference \interferenced{\<Bfr>(\<ouat>(y=1))}{x:=2} of writer B. Then the postcondition $(x=2=>\<ouat>(y=1))@\<r1>=x$ implies $\<r1>=2=>\<ouat>(x=2)$. Then, we have stability of $x=1=>!\<ouat>(y=1@\<ouat>(y=2))$ against the interference of writer A:
\begin{equation}
\cols
	& \<sp>((x=1=>!\<ouat>(y=1@\<ouat>(y=2)))@\widehat{\<Bfr>(\<sofar>(y!=2))}@x=\<xhat>,\;x:=1) \\
=	& (\<x'>=1=>!\<ouat>(y=1@\<ouat>(y=2)))@\<Bfr>(\<sofar>(y!=2))@\<x'>=\<xhat>@x=1 \\
=>	& \<sofar>(y!=2)@x=1 \\
=	& !\<ouat>(y=2)@x=1 \\
=	& !\<ouat>(\<ouat>(y=2))@x=1 \\
=>	& !\<ouat>(y=1@\<ouat>(y=2))@x=1 \\
=>	& x=1=>!\<ouat>(y=1@\<ouat>(y=2))
\sloc
\eqnlabel{2+2Wstab.1}
\end{equation}
\begin{equation}
\cols
	& \<sp>((x=1=>!\<ouat>(y=1@\<ouat>(y=2)))@\widehat{\<Bfr>(\<ouat>(x=1))}@y=\<yhat>,\;y:=2) \\
=	& \<sp>((x=1=>\<sofar>(y!=1|\<sofar>(y!=2)))@\<Bfr>(\<ouat>(x=1))@y=\<yhat>,\;y:=2) \\
=>	& (x=1=>\hook{\<sofar>(y!=1|\<sofar>(y!=2))}@(\<y'>!=1|\hook{\<sofar>(y!=2)}@\<y'>!=2))@y=2 \\
=>	& (x=1=>\hook{\<sofar>(y!=1|\<sofar>(y!=2))})@y!=1 \\
=>	& (x=1=>\hook{\<sofar>(y!=1|\<sofar>(y!=2))})@(y!=1|\<sofar>(y!=2)) \\
=>	& x=1=>\<sofar>(y!=1|\<sofar>(y!=2)) \\
=	& x=1=>!\<ouat>(y=1@\<ouat>(y=2))
\sloc
\eqnlabel{2+2Wstab.2}
\end{equation}

Stability against interference from writer B is trivial, in the first case because we can appeal to $\<ouat>(x=2)$, and in the other because $x=2=>x!=1$
\begin{align}
&{\cols
	& \<sp>\BRA \BRA \<ouat>(x=2)@(x=1=>!\<ouat>(y=1@\<ouat>(y=2)))\\
				@\; \widehat{\<Bfr>(\<sofar>(x!=2))}@y=\<yhat>\KET,\; y:=1
		   \KET \\
=>	& \<ouat>(x=2)@\<Bfr>(\<sofar>(x!=2)) \\
=>	& \<ouat>(x=2)@\<sofar>(x!=2) \\
=	& \<false> \\
=>	& x=1=>!\<ouat>(y=1@\<ouat>(y=2))
\sloc} \eqnlabel{2+2Wstab.3} \\
&{\cols
	& \<sp>((x=1=>!\<ouat>(y=1@\<ouat>(y=2)))@\widehat{\<Bfr>(\<ouat>(y=1))}@y=\<yhat>,\;x:=2) \\
=	& (\<x'>=1=>!\<ouat>(y=1@\<ouat>(y=2)))@\<Bfr>(\<ouat>(y=1)@y=\<yhat>@x=2 \\
=>	& x!=1 \\
=> 	& x=1=>!\<ouat>(y=1@\<ouat>(y=2))
\sloc} \eqnlabel{2+2Wstab.4}
\end{align}

Then the postcondition of \<labe> can say that if we read $x=2$ and then $x=1$ we have the makings of $x_{c}(2,1)$ but we cannot have seen the makings of $y_{c}(2,1)$ -- but note that we can't deduce $!y_{c}(2,1)$ from the coherence axioms of \defref{coherenceaxioms}. The postcondition of \<labh> says that if we read $y=2$ followed by $y=1$ we have the opposite; and we have a contradiction of sorts. It's not the contradiction we wanted, in terms of the values of $x$ and $y$, but it's sort of convincing: there can't be a thread which reads 2 then 1 from $x$ and also 2 then 1 from $y$, which is what the litmus test diagram seems to say.

The proof would not have worked with two observer threads. It is entirely possible that one thread can see $x_{c}(2,1)$ but fail to see $y_{c}(2,1)$. That doesn't mean that no other thread can see $y_{c}(2,1)$. Putting the two observer threads into one makes it possible to make a contradiction.

%!TEX root = ./Paper.tex

\section{SCloc is not optional}
\seclabel{SClocalways}

Consider the program in \figref{nonSCloctermination}: two MPs muddled together. There's a possible execution in which receiver A reads sender B's flag but sender A's message, while receiver B reads A's flag and B's message. The program postcondition says this shouldn't happen. If it did happen we would have, stably, $m=1$ in receiver A and $m=2$ in receiver B. A proof is shown in \figref{nonSClocterminationproof}. The internal deductions are impeccable; this seems to be an impossible outcome. %These two states are `simultaneous' in the sense that if we wait long enough after termination they will both hold at the same time.

Under SCloc this outcome really is impossible. When receiver A reads $\<flag>=2$ we can deduce $\<ouat>(\<msg>=2)$, so when it reads $\<msg>=1$ we know $\<msg>_{c}(2,1)$. In receiver B, similarly, if we read $\<flag>=1$ and $\<msg>=2$ we know $\<msg>_{c}(1,2)$. The global coherence constraint prohibits that outcome, and it would be easy to make a proof to show that it's formally impossible.

\begin{figure}
\squeezecols{3pt}
\centering
% proofs/nonSCloctermination.proof
$$\cols[c]
\assert{\<init>::m=0@f=0} \vspace{3pt} \\
\BRA[l||l||l||l]
\thr{Sender A (0)}
  \!\!\cols[r@{\intfspace}l]
  \assertd{\<lo> \<init>} & \<laba>:: m:=1; \\
  \assertd{\<bo> \<laba>} & \<labb>:: f:=1
  \sloc
\rht
&
\thr{Sender B (1)}
  \!\!\cols[r@{\intfspace}l]
  \assertd{\<lo> \<init>} & \<laba>:: m:=2; \\
  \assertd{\<bo> \<laba>} & \<labb>:: f:=2
  \sloc
\rht
&
\thr{Receiver A (2)}
  \!\!\cols[r@{\intfspace}l]
   & \<labe>:: \<r1>:=f; \\
  \assertd{\<lo> \<labe>} & \<labf>:: \<r2>:=m
  \sloc
\rht
&
\thr{Receiver B (3)}
  \!\!\cols[r@{\intfspace}l]
   & \<labe>:: \<r1>:=f; \\
  \assertd{\<lo> \<labe>} & \<labf>:: \<r2>:=m
  \sloc
\rht
\KET \vspace{3pt} \\
\assert{\<final>::!((2:::\<r1>)=2@(2:::\<r2>)=1@(3:::\<r1>)=1@(3:::\<r2>)=2)}
\sloc$$
\caption{A termination puzzle}
\figlabel{nonSCloctermination}
\end{figure}

\begin{figure}
\squeezecols{3pt}
\centering
% proofs/nonSCloctermination.proof
$$\cols[c]
\assert{\<init>::\<msg>=0@\<flag>=0} \vspace{3pt} \\
\BRA[l||l]
\thr{Sender A (0)}
  \guarantee{\interferenceg{\<Bfr>(\<sofar>(\<flag>!=1))}{\<msg>:=1}; \\ \interferenceg{\<Bfr>(\<ouat>(\<msg>=1))}{\<flag>:=1}} \vspace{3pt} \vspace{3pt} \\
  \!\!\cols[r@{\intfspace}l]
  \assertd{\<lo> \<init>:: \<Bfr>(\<sofar>(\<flag>!=1))} & \<laba>:: \<msg>:=1; \\
  \assertd{\<bo> \<laba>:: \<Bfr>(\<ouat>(\<msg>=1))} & \<labb>:: \<flag>:=1
  \sloc
\rht
&
\thr{Sender B (1)}
  \guarantee{\interferenceg{\<Bfr>(\<sofar>(\<flag>!=2))}{\<msg>:=2}; \\ \interferenceg{\<Bfr>(\<ouat>(\<msg>=2))}{\<flag>:=2}} \vspace{3pt} \vspace{3pt} \\
  \!\!\cols[r@{\intfspace}l]
  \assertd{\<lo> \<init>:: \<Bfr>(\<sofar>(\<flag>!=2))} & \<laba>:: \<msg>:=2; \\
  \assertd{\<bo> \<laba>:: \<Bfr>(\<ouat>(\<msg>=2))} & \<labb>:: \<flag>:=2
  \sloc
\rht \vspace{3pt} \\
\multicolumn{2}{c}{\begin{minipage}{360pt} $\cols \hstrut{360pt} \\[-8pt] \hline \hline \sloc$ \end{minipage}} \vspace{3pt} \\
\thr{Receiver A (2)}
  \emptyguarantee \vspace{3pt} \vspace{3pt} \\
  \!\!\cols[r@{\intfspace}l]
   & \<labe>:: \<r1>:=\<flag>; \\
  \assertd{\<lo> \<labe>:: \<r1>=2=>\<ouat>(\<flag>=2)} & \<labf>:: \<r2>:=\<msg>
  \sloc \vspace{3pt} \\
  \assertd{\<lo> \<labf>:: \<r1>=2@\<r2>=1=>\<ouat>(\<flag>=2)@\<msg>=1}
\rht
&
\thr{Receiver B (3)}
  \emptyguarantee \vspace{3pt} \vspace{3pt} \\
  \!\!\cols[r@{\intfspace}l]
   & \<labe>:: \<r1>:=\<flag>; \\
  \assertd{\<lo> \<labe>:: \<r1>=1=>\<ouat>(\<flag>=1)} & \<labf>:: \<r2>:=\<msg>
  \sloc \vspace{3pt} \\
  \assertd{\<lo> \<labf>:: \<r1>=1@\<r2>=2=>\<ouat>(\<flag>=1)@\<msg>=2}
\rht
\KET \vspace{3pt} \\
\assert{\<final>::!((2:::\<r1>)=2@(2:::\<r2>)=1@(3:::\<r1>)=1@(3:::\<r2>)=2)}
\sloc$$
\caption{Apparently-contradictory stable termination states}
\figlabel{nonSClocterminationproof}
\end{figure}

But suppose that we don't have SCloc, that there is no global constraint on propagation. Then it would appear that there is an execution of this program which has contradictory stable terminal states of the threads. The informal treatment of termination which we've used up to this point, as well as each of the formal treatments of termination which we present in \secref{PMS}, would say that the execution is impossible, which would be unsound. But even if we could patch that problem, there's a deeper one.

Our treatment of interference uses assertions to stand for the writes that underpin them. If different threads could receive writes in different orders it is clear that it would not be correct to assume that if $P$ holds stably before a \<uo> barrier then after the barrier $P$ holds in every thread. The writes underpinning $P$ might be propagated everywhere, but some other writes, superseded in $P$, might be propagated to some of those threads during that process. With SCloc that can't happen; without it we have no valid treatment of \<U>. Similar remarks apply to the use of assertions with \<Bfr>. SCloc is the only mechanism that we are aware of which allows our treatment. 

It would seem that we must, reluctantly and in the absence of a better solution, assume SCloc. There is a loss: it means that our logic does not apply to distributed systems, which are very like weak memory but don't have SCloc, or to the non-SCloc GPU architectures which we believe may currently exist. 

\var{aux1,aux2}
\begin{figure}
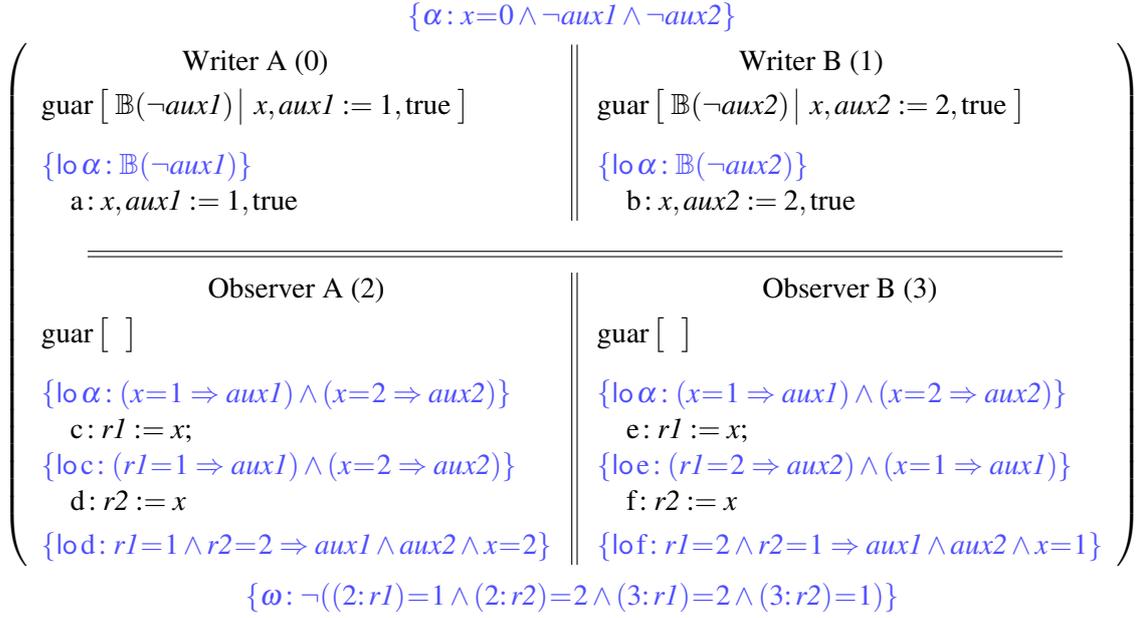

\squeezecols{2.5pt}
\centering
% proofs/CoRR2_aux.proof
$$\cols[c]
\assert{\<init>::x=0@!\<aux1>@!\<aux2>} \vspace{3pt} \\
\BRA[l||l]
\thr{Writer A (0)}
  \guarantee{\interferenceg{\<Bfr>(!\<aux1>)}{x,\<aux1>:=1,\<true>}} \vspace{3pt} \vspace{3pt} \\
  \assertd{\<lo> \<init>:: \<Bfr>(!\<aux1>)} \\ \quad \<laba>:: x,\<aux1>:=1,\<true>
\rht
&
\thr{Writer B (1)}
  \guarantee{\interferenceg{\<Bfr>(!\<aux2>)}{x,\<aux2>:=2,\<true>}} \vspace{3pt} \vspace{3pt} \\
  \assertd{\<lo> \<init>:: \<Bfr>(!\<aux2>)} \\ \quad \<labb>:: x,\<aux2>:=2,\<true>
\rht \vspace{3pt} \\
\multicolumn{2}{c}{\begin{minipage}{360pt} $\cols \hstrut{360pt} \\[-8pt] \hline \hline \sloc$ \end{minipage}} \vspace{3pt} \\
\thr{Observer A (2)}
  \emptyguarantee \vspace{3pt} \vspace{3pt} \\
  \assertd{\<lo> \<init>:: (x=1=>\<aux1>)@(x=2=>\<aux2>)} \\ \quad \<labc>:: \<r1>:=x; \\
  \assertd{\<lo> \<labc>:: (\<r1>=1=>\<aux1>)@(x=2=>\<aux2>)} \\ \quad \<labd>:: \<r2>:=x \vspace{3pt} \\
  \assertd{\<lo> \<labd>:: \<r1>=1@\<r2>=2=>\<aux1>@\<aux2>@x=2}
\rht
&
\thr{Observer B (3)}
  \emptyguarantee \vspace{3pt} \vspace{3pt} \\
  \assertd{\<lo> \<init>:: (x=1=>\<aux1>)@(x=2=>\<aux2>)} \\ \quad \<labe>:: \<r1>:=x; \\
  \assertd{\<lo> \<labe>:: (\<r1>=2=>\<aux2>)@(x=1=>\<aux1>)} \\ \quad \<labf>:: \<r2>:=x \vspace{3pt} \\
  \assertd{\<lo> \<labf>:: \<r1>=2@\<r2>=1=>\<aux1>@\<aux2>@x=1}
\rht
\KET \vspace{5pt} \\
\assert{\<final>::!((2:::\<r1>)=1@(2:::\<r2>)=2@(3:::\<r1>)=2@(3:::\<r2>)=1)}
\sloc$$
\caption{CoRR2 with auxiliaries}
\figlabel{CoRR2auxproof}
\end{figure}

\subsection{Coherence from unrepeatable writes}

The proof in \figref{nonSClocterminationproof} relies on unrepeatable writes: once an observer has seen $f=1$ from writer A, for example, we can be sure both that $m=1$ has been propagated to it, and that it won't be propagated again. That's achieved by the the \<bo> ordering in the writer, and enforced by the $\<Bfr>(\<sofar>(\<flag>!=1))$ interference precondition in its guarantee.

We can achieve a similar effect in a different example, once again potentially permitting contradictory stable terminal states, using auxiliary information rather than temporal reasoning. We use \emph{composite writes} in which an `actual write' and an `auxiliary write' are elaborated simultaneously and propagated simultaneously (see \secref{auxiliaries}). \Figref{CoRR2auxproof} is a version of CoRR2; in the final assertion $(2:::\<r1>)=1@(2:::\<r2>)=2$ implies that stably $x=2$ in observer A and $(3:::\<r1>)=2@(3:::\<r2>)=1$ implies that stably $x=1$ in observer B. Under SCloc, of course, we have $x_{c}(1,2)$ in observer A and $x_{c}(2,1)$ in observer B, so it can't happen. Without SCloc it is another example of the possibility of simultaneous contradictory stable states, which we cannot soundly deal with.

\clearpage
\part{Termination and a loose end}
\partlabel{terminationlooseend}
\thispagestyle{empty}
\clearpage
%!TEX root = ./Paper.tex

We have based our treatment of interference on the \<Bfr> modality. A local assertion $P$ is transmuted by a \<bo>-inducing barrier into $\<Bfr>(P)$, an assertion which guarantees propagation of the writes underlying $P$ before writes generated by later assignments. We were silent about what the `underlying' or `underpinning' writes might be for a particular $P$. The treatment was plausible for atemporal $P$ at least. We used a similar argument when introducing the \<U> modality: the writes underlying $P$ are transmitted to all threads, and $\<U>(P)$ claims that $P$ therefore holds in all threads.

Now that we've introduced temporal assertions that argument is no longer sound. Temporal assertions can describe historical coincidences -- $\<ouat>(x=y)$, for example -- that can't be said to be described by `underlying writes', and won't be reliably propagated to other threads. This is a considerable difficulty, which affects our treatment of \<Bfr>, \<U> and termination. 

We resolve the difficulty partly by prohibiting problematic temporal assertions in \<Bfr> and \<U>, but to treat termination we must grasp the nettle and define what would be propagated by $\<U>(P)$ if $P$ contains temporal coincidences.

\section{\ensuremath{\downarrow\,}\setvar{P}, the propagatable version of \setvar{P}}
\seclabel{downP}

Some assertions plainly describe the writes which make up the current state and can obviously be propagated to other threads: $x=0$ is an example. Some do not: $\<ouat>(x=0)$ is an example. But we've used $\<Bfr>(\<ouat>(\<msg>=1))$ in \figref{Sproof} in the proof of S: were we right to do so? We believe so, because SCloc aka coherence is a feature of our logic. The write(s) that are described by $\<ouat>(\<msg>=1)$ will either include $\<msg>=1$ or a coherence-successor of it, and that's a useful description. But multivariate historical assertions are a problem: $\<Bfr>(\<ouat>(\<msg>=1@\<flag>=0))$ wouldn't do what it purports to do. By propagating writes from the current state -- $\<msg>=1$ or a coherence-successor, $\<flag>=0$ or a coherence-successor -- it's impossible to guarantee that in the receiving threads' history it was ever the case that $\<msg>=1@\<flag>=0$.

We can, however, define $|/P$, an assertion implied by $P$ that can be sure to be propagated by $\<Bfr>(P)$ or $\<U>(P)$. Our definition is incomplete, but that's ok: it just restricts the proofs that we can make. It's complicated by the need to distinguish negative and positive occurrences. In problematic cases we replace the assertion with a fresh boolean variable. $||/P$ -- the negative-position case -- is just like $|/P$ except for the modality cases shown.
\definition{\ensuremath{\downarrow}\setvar{P} -- positive occurrences -- and $||/P$ -- negative occurrences}{downP}
{%!TEX root = ./Paper.tex
$$\cols[rcll]
|/P								&==&	P									& \text{if }P\text{ does not have multiple free variables} \\
|/(!P)							&==&	! ||/P 								\\
%%|/(P=>Q)						&==&	|/(!P|Q) 							\\
%%|/(P<=>Q)						&==&	|/((P=>Q)@(Q=>P)) 					\\
|/(\setbracket{unop}\;P)		&==&	\setbracket{unop}\;|/P 				\\
|/(P\;\setbracket{binop}\;Q)	&==& 	|/P\;\setbracket{binop}\;|/Q 		\\
|/(@*v(P))						&==& 	@*v(|/P)							& \text{noting that }v\text{ is a bound variable} \\
|/(|*v(P))						&==& 	|*v(|/P)							& \text{noting that }v\text{ is a bound variable} \vspace{5pt}\\
|/\<Bfr>(P)						&==&	\<Bfr>(|/P)							\\
|/\<U>(P)						&==&	\<U>(|/P)							\\
|/\<Ouat>(P@Q)					&==&	|/\<Ouat>(P)@|/\<Ouat>(Q)			\\
|/\<Ouat>(P|Q)					&==&	|/\<Ouat>(P)| |/\<Ouat>(Q)			\\
|/\<Ouat>(P)					&==&	\setbracket{fresh}\;\setvar{boolv}	\\
|/(P\<since>Q)					&==&	|/P@|/\<Ouat>(Q)					\\
|/\<sofar>(P@Q)					&==&	|/\<sofar>(P)@|/\<sofar>(Q)			\\
|/\<sofar>(P)					&==&	P									\vspace{5pt} \\

||/\<Bfr>(P)					&==&	\setbracket{fresh}\;\setvar{boolv}	\\
||/\<U>(P)						&==&	\setbracket{fresh}\;\setvar{boolv}	\\
||/\<ouat>(P)					&==&	\setbracket{fresh}\;\setvar{boolv}	\\
||/(P\<since>Q)					&==&	\setbracket{fresh}\;\setvar{boolv}	\\
||/\<sofar>(P)					&==&	\setbracket{fresh}\;\setvar{boolv}	
\sloc$$}
In computing $|/P$ and $||/P$ we exploit equivalences, treating $P=>Q$ as $!P|Q$, $P<=>Q$ as $(P=>Q)@(Q=>P)$, $!(P@Q)$ as $!P|!Q$, $!(P|Q)$ as $!P@!Q$.

The most important point to note is that in propagation we lose multivariate historical coincidences like $\<ouat>(x=y)$ or $\<ouat>(x=1@\<ouat>(y=2))$. But single-variable coincidences like $\<ouat>(x=1@\<ouat>(x=2))$ are preserved: they are the province of SCloc.

\section{A proper treatment of \setModality{B} and \setModality{U}}

We have a choice in dealing with \<Bfr>- and \<U>-modal temporal assertions. We could say that $\<Bfr>(P)=>|/P$ and $\<U>(P)=>|/P$, and modify our treatment of hooking and hatting to suit, but that would complicate all our definitions. Or we could note that multivariate temporal coincidences in \<Bfr>- or \<U>-modal positions don't mean what they seem to say, and prohibit them. Which is what we have done.
\definition{Restricted \setModality{B}, \setModality{U} and initial assertion}{BUrestriction}{%!TEX root = ./Paper.tex
A propagatable assertion $P$ is one for which $P$ is equivalent to $|/P$. An initial-state assertion must be propagatable. In $\<Bfr>(P)$ and $\<U>(P)$, $P$ must be propagatable.}
Justification for restriction of the initial assertion, and non-restriction of \<sofar>, is given in the next section. 

\section{A proper treatment of termination}
\seclabel{PMS}

The same question of propagation of assertions arises in our treatment of termination. In early examples we pretended that all that was necessary was to take the conjunction of the thread postconditions. In the presence of temporal assertions, that's now obviously inadequate.

We have supposed throughout that our examples are executed within a test harness which ensures that all threads see the initial state before they start, and which somehow brings their final states together when they have all finished.

Initialisation is straightforward: the test harness sets up the variables to satisfy the initial assertion, executes a \<uo>-inducing barrier like \<MFENCE>, \<sync> or \<dsb>, and then sets a flag variable which each thread waits to see. In effect it sets up $\<sofar>(\<init>)$ for the initial assertion \<init>. But since we are relying on it being propagated from the initialising thread to each test-program thread, it must be a propagatable assertion -- i.e. no temporal coincidences. But then, the initial state being established, it's possible to note coincidences: so from an initial assertion $x=y=0$ it's reasonable to deduce $\<sofar>(\<ouat>(x=y))$ -- i.e. \<sofar> assertions don't need to be propagatable. 

We can imagine three mechanisms which would deal with termination. They all require that each thread stores the values of its registers into unique shared variables -- the $(n:::\setvar{reg})$ variables of our final assertions -- and signals that it's finished by writing to a thread-specific flag variable. The test harness loops until it sees that all the termination flags have been set and then somehow waits long enough for all the writes of all the threads to have been propagated to it before it inspects the final state. There seem to be three ways to `wait long enough':
\begin{enumerate*}
\item The controlling thread might wait a millisecond or so (these machines are very fast); \label{PMS.wait}
\item Each test thread might place a \<bo>-inducing barrier before writing its flag; \label{PMS.bo}
\item Each test thread might place a \<uo>-inducing barrier before writing its flag. \label{PMS.uo}
\end{enumerate*}
Because we had no way of deciding which mechanism was best we called the problem the \emph{purple mystic sync} or PMS. Our problem was to devise a logical treatment which fitted any and all of the three mechanisms.

%%\begin{figure}
%%\centering
%%% proofs/R.proof
%%$$\cols[c]
%%\assert{\<init>::x=y=0} \vspace{3pt} \\
%%\BRA[l||l]
%%\thr{Thread 0}
%%  \<laba>:: x:=1; \\
%%  \<labb>:: y:=1
%%\rht
%%&
%%\thr{Thread 1}
%%  \<labc>:: y:=2; \\
%%  \<labd>:: \<r1>:=x
%%\rht
%%\KET \vspace{3pt} \\
%%\assert{\<final>::!((1:::\<r1>)=0@y=2)}
%%\sloc$$
%%\caption{Litmus test R, unlaced}
%%\figlabel{R}
%%\end{figure}

\begin{figure}
\centering
% proofs/R.proof
% proofs/R.proof
$$\cols[c]
\assert{\<init>::x=y=0} \vspace{3pt} \\
\BRA[l||l]
\thr{Thread 0}
  \!\!\cols[r@{\intfspace}l]
   & \<laba>:: x:=1; \\
  \assertd{\<uo> \<laba>;\<lo> \<init>} & \<labb>:: y:=1
  \sloc
\rht
&
\thr{Thread 1}
  \!\!\cols[r@{\intfspace}l]
  \assertd{\<lo> \<init>} & \<labc>:: y:=2; \\
  \assertd{\<uo> \<labc>} & \<labd>:: \<r1>:=x
  \sloc
\rht
\KET \vspace{3pt} \\
\assert{\<final>::!((1:::\<r1>)=0@y=2)}
\sloc$$
\caption{Litmus test R: \setorder{uo} ordering essential}
\figlabel{Rlaced}
\end{figure}

\subsection{R: a problematic litmus test}
Our difficulties in doing so have to do with temporal assertions. \Figref{Rlaced} shows a litmus test which illustrates the challenge. By design on Power and ARM, and experimentally, it is impossible for thread 1 to read 0 from $x$ if globally there is $y_{c}(1,2)$ coherence and there is a \<uo>-inducing barrier between the commands of each thread. Crucially, the test fails with any weaker constraints -- it fails with combinations of a \<bo>-inducing barrier and an \<lo>-inducing address dependency as in \figref{R+uo+lo.unproof}, for example.

\begin{figure}
\centering
% proofs/R.proof
$$\cols[c]
\assert{\<init>::x=y=0} \vspace{3pt} \\
\BRA[l||l]
\thr{Thread 0}
  \guarantee{\interferenceg{\<true>}{x:=1}; \\ \interferenceg{\<true>}{y:=1}} \vspace{3pt} \vspace{3pt} \\
   \\ \quad \<laba>:: x:=1; \\
  \assertd{\<uo> \<laba>: \<U>(x=1);\<lo> \<init>: \<sofar>(y!=1)}\ \ipre{\<true>} \\ \quad \<labb>:: y:=1 \vspace{3pt} \\
  \assertd{\<lo> \<labb>: \BRA (\<U>(x=1) \<since> y=1)\\@\;(y=2=>y_{c}(1,2))\KET}
\rht
&
\thr{Thread 1}
  \guarantee{\interferenceg{\<true>}{y:=2}} \vspace{3pt} \vspace{3pt} \\
  \assertd{\<lo> \<init>: \<sofar>(y!=2)}\ \ipre{\<true>} \\ \quad \<labc>:: y:=2; \\
  \assertd{\<uo> \<labc>: \<U>((y=2|y=1)@\<ouat>(y=2))} \\ \quad \<labd>:: \<r1>:=x \vspace{3pt} \\
  \assertd{\<lo> \<labd>: \<r1>=0=>(\<U>(y=2|y=1@y_{c}(2,1)) \<since> x=0)}
\rht
\KET \vspace{3pt} \\
\assert{\<final>::!((1:::\<r1>)=0@y=2)}
\sloc$$
\caption{Proof of R with \setorder{uo} constraints in each thread}
\figlabel{Rproof}
\end{figure}

\begin{figure}
\centering
\includegraphics[scale=\picscale]{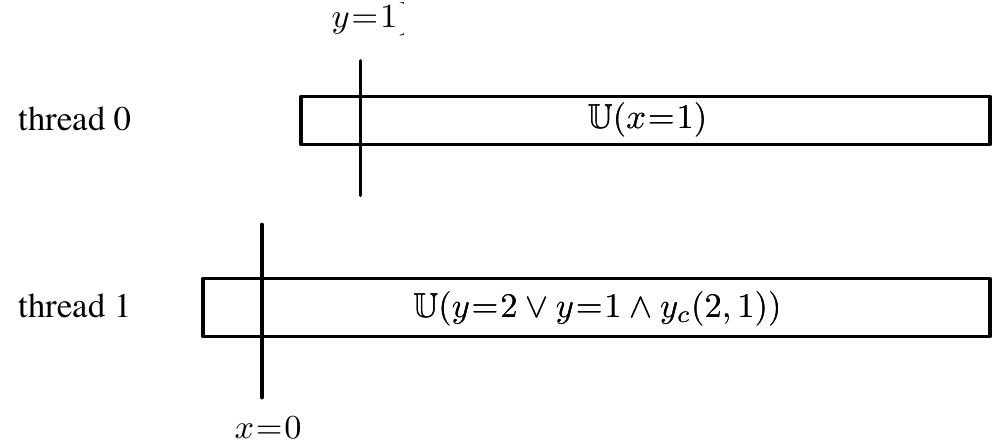}
\caption{The thread histories of \figref{Rproof}}
\figlabel{Rthreadhistories}
\end{figure}

\Figref{Rproof} is a successful proof. The crucial point is that the strongest postcondition of \<labc> implies $y=2$, which implies $y=2|y=1@y_{c}(2,1)$. That is stable against the interference of thread 0: \interferenced{\<true>}{x:=1} is irrelevant because it doesn't write to $y$, but \interferenced{\<Bfr>(x=1)}{y:=1} needs checking.
\begin{equation}
\cols
	& \<sp>((y=2|y=1@y_{c}(2,1))@\widehat{\<Bfr>(x=1)},\;y:=1) \\
==	& (\<y'>=2|\<y'>=1@y_{c}(2,1))@\<Bfr>(x=1)@\<xhat>=1@y=1 \\
=>	& (\<y'>=2|\<y'>=1@y_{c}(2,1))@y=1 \\
==	& \<y'>=2@y=1|\<y'>=1@y=1@y_{c}(2,1) \\
=>	& \<ouat>(y=2)@y=1|\<y'>=1@y=1@y_{c}(2,1) \\
=>	& y_{c}(2,1)@y=1 \\
=> 	& y=2|y=1@y_{c}(2,1)
\sloc
\eqnlabel{Rstab.1}
\end{equation}
The universalised version is EXT and UEXT stable, following similar reasoning.

The thread histories are pictured in \figref{Rthreadhistories}. Plainly $x=0$ in thread 1 must precede the $\<U>(x=1)$ period; plainly $\<U>(y=2|y=1@y_{c}(2,1))$ holds when $y=1$ in thread 0, and therefore we have $y_{c}(2,1)$ from that instant. Plainly? Hardly, and anyway only informally. We need a formal treatment.

\begin{figure}
\centering
% proofs/R+uo+lo.unproof
$$\cols[c]
\assert{\<init>::x=y=0} \vspace{3pt} \\
\BRA[l||l]
\thr{Thread 0}
  \guarantee{\interferenceg{\<true>}{x:=1}; \\ \interferenceg{\<Bfr>(x=1)}{y:=1}} \vspace{3pt} \vspace{3pt} \\
  \assertd{\<lo> \<init>: \<true>} \\ \quad \<laba>:: x:=1; \\
  \assertd{\<lo> \<init>: !\<ouat>(y=1);\<uo> \<laba>: \<U>(x=1)} \\ \quad \<labb>:: y:=1 \vspace{3pt} \\
  \assertd{\<lo> \<labb>: \<U>(x=1) \<since> y=1}
\rht
&
\thr{Thread 1}
  \guarantee{\interferenceg{\<true>}{y:=2}} \vspace{3pt} \vspace{3pt} \\
  \assertd{\<lo> \<init>: !\<ouat>(y=2)} \\ \quad \<labc>:: y:=2; \\
  \assertd{\<lo> \<labc>: y=2|y=1@x=1} \\ \quad \<labd>:: \<r1>:=x \vspace{3pt} \\
  \assertd{\<lo> \<labd>: \<r1>=0=>\<ouat>(x=0@y=2)}
\rht
\KET \vspace{3pt} \\
\hcancel{\assert{\<final>::!((1:::\<r1>)=0@y=2)}}
\sloc$$
\caption{An unproof of R with \setorder{uo} in thread 0, \setorder{lo} in thread 1}
\figlabel{R+uo+lo.unproof}
\end{figure}

\begin{figure}
\centering
\includegraphics[scale=\picscale]{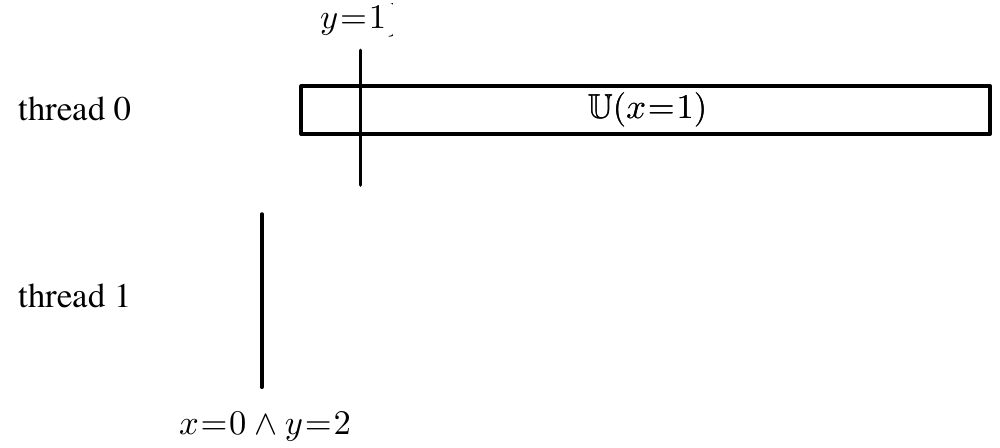}
\caption{Thread histories of \figref{R+uo+lo.unproof}}
\figlabel{R+uo+lo.histories}
\end{figure}

To begin to unpick the problem, we present in \figref{R+uo+lo.unproof} an unproof of the same test with \<lo> ordering in thread 1. This time the postcondition of \<labc> is $y=2|y=1@x=1$, which is clearly stable. The threads' postconditions describe the thread histories pictured in \figref{R+uo+lo.histories}. Plainly $x=0@y=2$ in thread 1 occurs globally before the $\<U>(x=1)$ epoch, which contains $y=1$. So $y=2$ occurs in thread 1 before $y=1$ in thread 0, which seems to suggest that we might have $y_{c}(2,1)$. Yet experiment shows that we would be wrong to think so, because this program \emph{can} terminate with $(1:::\<r1>)=0@y=2$. Plainly we are beyond `plainly'. Formality is essential.

\subsection{The termination problem solved}

The conclusion of thread 1 in \figref{R+uo+lo.unproof} involves a multivariate temporal coincidence: $\<ouat>(x=0@y=2)$. None of our PMS mechanisms would propagate this to a controlling thread. Instead it would receive $|/\<ouat>(x=0@y=2)$, which is $\<ouat>(x=0)@\<ouat>(y=2)$, and it couldn't rely on reasoning like \figref{R+uo+lo.histories}: the $x=0$ event has to come before the $\<U>(x=1)$ epoch, but $y=2$ doesn't.

Our treatment of PMS is as follows, where $Q@@@@n$ means `$Q$ in thread $n$'. 
\definition{The PMS assertion}{PMSassertion}{%!TEX root = ./Paper.tex
If the final assertions of the threads are $P_{0}\dots P_{n-1}$, in which register occurrences are translated into $(\setvar{threadnum}:::\setvar{reg})$, and the program-final assertion is \<final>, then the  assertion \vspace{5pt}  \\
$
\hstrut{20pt} P_{0}@@@@0 @ P_{1}@@@@1 @ \dots @ P_{n-1}@@@@(n-1) @ (|/P_{0} \;@ |/P_{1}@ \dots @ |/P_{n-1})@@@@n => \<final>@@@@n
$}
In checking this assertion we allow an additional \emph{pms coherence} axiom.
\definition{PMS coherence axiom}{pmscoherenceaxiom}{\begin{tabular}{rl}
pms coherence		& $x_c(A,B)=>\<cv>(x)@x!=A$
\end{tabular}
}
The PMS assertion allows contradictions between thread-local claims (this is how the termination of SB is handled, for example), but doesn't allow thread-local coincidences to be observed in the controlling thread. The extra coherence axiom allows us to exploit the fact that, when all the program activity is over and writes have been propagated everywhere, all threads see the same last write to each variable.

Thus we check in the proof of \figref{Rproof}
\begin{equation}
\BRA ((\<U>(x=1) \<since> y=1)@(y=2=>y_c(1,2)))@@@@0 \\
	 @ \; ((1:\<r1>)=0=>(\<U>(y=2|y=1@y_c(2,1)) \<since> x=0))@@@@1 \\
	 @ \; (\<U>(x=1)@\<ouat>(y=1)@(y=2=>y_c(1,2)))@@@@2 \\
	 @ \; (((1:\<r1>)=0=>\<U>(y=2|y=1@y_c(2,1))@\<ouat>(x=0)))@@@@2
\KET
=>(!((1:\<r1>)=0@y=2))@@@@2
\eqnlabel{PMSassertion_R}
\end{equation}
In this example there's a contradiction between the \<since> assertions in threads 1 and 2 in just the case that $(1:\<r1>)=0@y=2$.

In the case of \figref{R+uo+lo.unproof} the PMS assertion is
\begin{equation}
\BRA (\<U>(x=1) \<since> y=1)@@@@0 \\
	 @ \; ((1:::\<r1>)=0=>\<ouat>(x=0@y=2))@@@@1 \\
	 @ \; (\<U>(x=1)@\<ouat>(y=1))@@@@2 \\
	 @ \; (((1:::\<r1>)=0=>\<ouat>(x=0)@\<ouat>(y=2)))@@@@2
\KET =>(!((1:::\<r1>)=0@y=2))@@@@2
\eqnlabel{PMSassertion_R+uo+lo.unproof}
\end{equation}
There's nothing in that assertion which constrains the order of arrival of $y=1$, $x=0$ and $y=2$ in the controlling thread 2. So the assertion doesn't hold, as required.

\clearpage
\part{A worked example}
\thispagestyle{empty}
\clearpage
%!TEX root = ./Paper.tex

\section{The token ring}
\seclabel{tokenring}

\begin{figure}
\centering
\scalebox{0.9}{\raisebox{-5.7em}{\begin{minipage}[b][19em]{150pt}
\begin{lstlisting}^^J
volatile _Bool latch [WORKERS];^^J
volatile _Bool flag [WORKERS];^^J
^^J
void worker(int i) \{^^J
\ \ while (!latch[i]);^^J
\ \ for (;;) \{^^J
\ \ \ \ latch[i] = 0;^^J
\ \ \ \ if (flag[i]) \{^^J
\ \ \ \ \ \ flag[i] = 0;^^J
\ \ \ \ \ \ flag[(i+1)\%WORKERS] = 1;^^J
\ \ \ \ \ \ latch[(i+1)\%WORKERS] = 1;^^J
\ \ \ \ \}^^J
\ \ \ \ while (!latch [i]);^^J
\ \ \}^^J
\}
\end{lstlisting}
\end{minipage}}}
\caption{Token ring (from~\protect\citep{AlglaveetalSoftwareVerificationforWeakMemoryviaProgramTransformation2013})}%\citep{AlglaveetalSoftwareVerificationforWeakMemoryviaProgramTransformation2013})}
\figlabel{Jade'stokenring}
\end{figure}

The token ring problem was posted on the web as a bug report~\citep{waitlatchisvulnerable2007}, found by Michael Tautschnig, and a model-checked solution was reported in~\citep{AlglaveetalSoftwareVerificationforWeakMemoryviaProgramTransformation2013}. A token (in an element of an array confusingly called `flag') is passed around a ring of numbered worker threads. Each thread 
\begin{enumerate}
\item waits until it sees its own element in the array of latches has been set;
\item zeroes that latch element;
\item reads from its own element in the array of tokens;
\item if the token value it sees is non-zero: 
\begin{enumerate}
\item zeroes its own token element;
\item and sets the next thread's latch and token.
\end{enumerate}
\end{enumerate}
 If a thread sees a zero token then it doesn't pass anything on and circulation stops. A simplification of the original is shown in \figref{Jade'stokenring}. The bug report observed that the special procedure calls being used to read and set latch and token elements, each of which invoked some barriers, didn't provide all the orderings necessary to avoid a hang-up on a Power machine with eight hardware threads. 

In essence each worker is an MP reader, reading latch[i] and flag[i] rather than \<flag> and \<msg>, and then an MP writer, writing flag[i+1] and latch[i+1] rather than $\<msg>$ and $\<flag>$. There's a need to reset latch[i] and flag[i] once they've been read and before they have been written again by another thread, and the time to do that is before writing latch[i+1]. Alglave et al. found bugs by model checking and proposed that those bugs would go away by imposing \<lo> order between the reads of latch[i] and flag[i], via a Power address dependency, and \<bo> between the writes of flag[i+1] and latch[i+1], via a Power \<lwsync>. With those modifications the algorithm model-checked for small sizes of ring, and tested on the hardware that the bug-reporters were using. It seems plausible, given our analysis of MP in the proofs so far, that it should work for a ring of any size running for as long as you like. But weak memory is scary, and it might still bite us. We need a proof that it can't.

Even though it's such a simple program, the token ring is a challenging example. We have solved it in two abstracted guises.

\begin{figure}
\centering
$$\cols
\assert { \<latch>=\<token>=0 } \vspace{3pt}\\
\guarantee{\interferenceg[A]{A\%W=0@\<token>=\<latch>=A}{\<token>:=A+1};\\
	       \interferenceg[A]{A\%W=0@\<token>=A+1@\<latch>=A}{\<latch>:=A+1}
	      }\\
\<while> \<true> \<do> \\
\begin{progindent}
	\assertd{\<latch>\%W=0=>\<latch>=\<token>} \\
	\<do> \<r1> := \<latch> \<until> \<r1>\%W=0; \\
    \assertd{\<r1>\%W=0@\<r1>=\<latch>=\<token>} \\
    \<if> \<r1>=\<r2> \<then> \\
    \quad \<r2> := \<token>; \<token> := \<r2>+1; \<latch> := \<r1>+1 \\
    \<fi> 
%	\assertd{\<latch>\%W=0=>\<latch>=\<token>} 
\end{progindent} \\
\<od> \vspace{3pt} \\
\rely{\interferenceg[B]{B\%W!=0@\<token>=\<latch>=B}{\<token>:=B+1};\\
	  \interferenceg[B]{B\%W!=0@\<token>=B+1@\<latch>=B}{\<latch>:=B+1}
	 }
\sloc$$
\caption{Worker 0 of a single-latch, single-token $W$-ring ($W>=2$), SC with assertions}
\figlabel{SCsingletokenring}
\end{figure}

\var{rauxN}
\begin{figure}
\centering
% proofs/tokenringsingleifthenaux.proof
$$\cols[c]
\assert{\<init>::\<latch>=\<token>=0} \vspace{3pt} \\
\thr{Thread 0}
  \<labx>:: \<rauxN>:=0; \\
  \<while>  \<beta>:: \<true> \<do> \\
  \pindent\cols
    \<do> \\
    \pindent\assertd{\<bo> \<init>;\<lo> \<labx>} |> (\assertd{\<lo> \<labe>} | \assertd{\<lo> \<delta>_{f}})\ \<labb>:: \<r1>:=\<latch> \\
    \<until> \assertd{\<lo> \<labb>} \<gamma>:: \<r1>\%W=0; \\
    \assertd{\<lo> \<gamma>_{t}}\ \<labc>:: \<r2>:=\<token>; \\
    \<if> \assertd{\<lo> \<labc>} \<delta>:: \<r1>=\<r2> \<then> \\
    \pindent\!\!\cols[r@{\intfspace}l]
      \assertd{\<lo> \<delta>_{t}} & \<labd>:: \<token>:=\<r2>+1; \\
      \assertd{\<lo> \<labd>} & \<laby>:: \<rauxN>:=\<r1>+W; \\
      \assertd{\<lo> \<laby>;\<bo> \<labd>} & \<labe>:: \<latch>:=\<r1>+1
      \sloc \\
    \<fi>
    \sloc \\
  \<od>
\rht
\sloc$$
\caption{Worker 0 of a single-latch, single-token $W$-ring ($W>=2$), laced, with auxiliary assignments}
\figlabel{lacedsingletokenring}
\end{figure}

\var{loopinv,lhseen}
\begin{figure*}
\centering
$$\cols[c]
\assert{\<init>::\<latch>=\<token>=0} \vspace{3pt} \\
\BRA
\thr{Thread 0}
  \guarantee{\interferenceg[A]{A\%W=0@\<latch>=A@\<Bfr>(\<token>=A)}{\<token>:=A+1}; \\ \interferenceg[A]{A\%W=0@\<Bfr>(\<latch>=A=\<token>-1)}{\<latch>:=A+1}} \vspace{3pt} \vspace{3pt} \\
  \macro{\<loopinv>}{\BRA \<rauxN>>=0@\<rauxN>\%W=0 \\
  						  @ \;|* L (\<rauxN>-W<L=\<latch><=\<rauxN>@\<sofar>(\<latch><=\<rauxN>)@\<Bfr>(L<=\<token><=\<rauxN>))\KET} \vspace{3pt} \\
  \macro{\<lhseen>}{\<r1>\%W=0@0<=\<latch>=\<r1>=\<rauxN>@\<sofar>(\<latch><=\<r1>)@\<Bfr>(\<token>=\<r1>)} \vspace{3pt} \\
  \<labx>:: \<rauxN>:=0; \\
  \<while>  \<beta>:: \<true> \<do> \\
  \pindent\cols
    \<do> \\
    \pindent\BRA \assertd{\<bo> \<init>: \<sofar>(\<latch>=\<token>=0);\<lo> \<labx>: \<rauxN>=0} \\
    		|> \;(\assertd{\<lo> \<labe>: \<loopinv>} | \assertd{\<lo> \<delta>_{f}: \<false>})\KET\ \<labb>:: \<r1>:=\<latch> \\
    \<until> \assertd{\<lo> \<labb>: \<loopinv>@(\<r1>\%W=0=>\<r1>=\<latch>)} \<gamma>:: \<r1>\%W=0; \\
    \assertd{\<lo> \<gamma>_{t}: \<lhseen>}\ \<labc>:: \<r2>:=\<token>; \\
    \<if> \assertd{\<lo> \<labc>: \<lhseen>@\<r2>=\<token>} \<delta>:: \<r1>=\<r2> \<then> \\
    \pindent\!\!\cols[r@{\intfspace}l]
      \assertd{\<lo> \<delta>_{t}: \<lhseen>@\<r1>=\<r2>=\<token>} & \\ \ipre{\<r1>\%W=0@\<latch>=\<r1>=\<r2>@\<Bfr>(\<token>=\<r1>)} & \<labd>:: \<token>:=\<r2>+1; \\
      \assertd{\<lo> \<labd>: \<r1>=\<rauxN>} & \<laby>:: \<rauxN>:=\<r1>+W; \\
      \Assertd{\<lo> \<laby>: \<rauxN>=\<r1>+W; \\
      		   \<bo> \<labd>: \<Bfr>(\<r1>\%W=0@0<=\<latch>=\<r1>=\<token>-1)@\<sofar>(\<latch><=\<r1>)} & \\ \ipre{\<Bfr>(\<r1>\%W=0@0<=\<latch>=\<r1>=\<token>-1)} & \<labe>:: \<latch>:=\<r1>+1
      \sloc \\
    \<fi>
    \sloc \\
  \<od> \vspace{3pt} \\
  \rely{\interferenceg[C]{C\%W!=0@\<latch>=C@\<Bfr>(\<token>=C)}{\<token>:=C+1}; \\ \interferenceg[C]{C\%W!=0@\<Bfr>(\<latch>=C<\<token>)}{\<latch>:=C+1}} \vspace{3pt}
\rht \\
\begin{minipage}{340pt} $\cols \hstrut{400pt} \\[-8pt] \hline \hline \sloc$ \end{minipage} \vspace{3pt} \\ 
\thr{Thread 1}
  \guarantee{\interferenceg[D]{D\%W=1@\<latch>=D@\<Bfr>(\<token>=D)}{\<token>:=D+1}; \\ \interferenceg[D]{D\%W=1@\<Bfr>(\<latch>=D=\<token>-1)}{\<latch>:=D+1}} \vspace{3pt}
\rht\\
\begin{minipage}{340pt} $\cols \hstrut{400pt} \\[-8pt] \hline \hline \sloc$ \end{minipage} \vspace{3pt} \\ 
\thr{Thread 2}
  \guarantee{\interferenceg[E]{E\%W=2@\<latch>=E@\<Bfr>(\<token>=E)}{\<token>:=E+1}; \\ \interferenceg[E]{E\%W=2@\<Bfr>(\<latch>=E=\<token>-1)}{\<latch>:=E+1}} \vspace{3pt}
\rht
\KET
\sloc$$
\caption{Proof of worker 0 in a single-latch, single-token $W$-ring ($W=3$)}
\figlabel{singletokenring3proof}
\end{figure*}

\subsection{A single-latch, single-token ring}

One workder from our first abstraction, unlaced as if for SC execution, is in \figref{SCsingletokenring}; $W$ is the size of the ring. As in C, the `$\%$' operator is \setbracket{mod}. In place of arrays of latches and tokens there is a single latch variable \<latch> and a single token variable \<token>. Latch and token each carry an ever-increasing integer -- in effect, auxiliary information. The assertions are intended to make it clear that this is an MP receiver followed by an MP sender. Since the value of \<token> is never decreased, we don't stop at a zero-value token. Instead our criterion for success in the control expression of the conditional is $\<r1>=\<r2>$ -- i.e. that \<latch> and \<token> are in step. The rely is the union of the guarantees of workers with $B\%W=1$, $B\%W=2$, $\dots$, $B\%W=W-1$.

The problems of the weak-memory proof are to do with in-flight interference, because the \<token> writes of one worker can overtake the \<latch> writes of another, and we have to be careful about \<bo> parallelism between one iteration of the loop and the next. %For a more convicing proof we would need to add an auxiliary register to make it clear that circulating values do increase.

\Figref{lacedsingletokenring} shows the single-latch single-token algorithm laced for weak memory. We have added an auxiliary register \<rauxN> which records the next \<latch>/\<token> value that the worker should deal with. Command \<labb>'s knot is an iterated constraint, first knot from the initial assertion \<init> and the initialisation of \<rauxN>, second knot from the loop-final command \<labe> or failure of the control expression \<delta>. Otherwise the lacing is straightforward: \<lo> throughout apart from (as in MP) \<bo> from \<labe> to \<labf>. 

The proof of one worker in a three-worker ring is in \figref{singletokenring3proof}.\footnote{Alone amongst the proofs in this paper, this one has not been completely machine-checked. The SMT solver Z3, which underpins the Arsenic proof checker, can't deal with the modal arithmetic if $W$ is constrained only to be greater than 1, but it is happy if we set $W$ equal to a positive integer of any size we like. That is, we can machine-check the proof for very large values of $W$, and we are confident anyway that it is a proof, but we don't have the stamp of mechanical approval.} We include macros \<loopinv> and \<lhseen> to shorten the constraint embroidery. Confidence that circulation never stops because latch and token get out of step comes from the fact that $\<delta>_{f}$ has postcondition \<false>, i.e. the test $\<r1>=\<r2>$ always succeeds. Note that, for the first time, this example includes purely auxiliary assignments in $\<labx>::\<rauxN>:=0$ and $\<laby>:: \<rauxN>:=\<r1>+W$; the constraints \<labx>$->$\<labb> and \<laby>$->$\<labe> are therefore restricted to not mention regular variables in their embroidery, so that when the auxiliary assignment is deleted, no assertions about regular variable values, which could affect \<bo> or \<uo> propagation, are deleted as well.

The first interesting point is the rely. In SC proofs the rely of one thread is simply the union of the guarantees of the other threads. In weak memory that's not enough, because of \<bo> instability between \<token> and \<latch> interference. The \<token> interference of worker 2 \<bo>-interferes with the \<latch> interference of worker 1:
\begin{equation}
\setlength{\arraycolsep}{2.5pt}
\cols
	&  \<sp>\BRA (D\%W=1@\<Bfr>(\<latch>=D=\<token>-1))\sphat@(E\%W=2@\<latch>=E@\<Bfr>(\<token>=E))\,\dhat{}\,, \\ 
				         \quad \<token>:=E+1
			\KET \vspace{3pt} \\
	&  \<sp>\BRA D\%W=1@\<Bfr>(\<latch>=D=\<token>-1)@\widehat{\<latch>}=D=\widehat{\<token>}-1
				 @E\%W=2@\dhatv{\<latch>}=E@\<Bfr>(\<token>=E)@\dhatv{\<token>}=E, \\ 
				         \quad \<token>:=E+1
			\KET \vspace{3pt} \\
=	&  \BRA 
        D\%W=1@\hook{\<Bfr>(\<latch>=D=\<token>-1)}@\<latch>=D=\<token'>-1@\widehat{\<latch>}=D=\widehat{\<token>}-1 \\
        @\; E\%W=2@\dhatv{\<latch>}=E@\hook{\<Bfr>(\<token>=E)}@\<token'>=E@\dhatv{\<token>}=E \\
		@\; \<token>=E+1
	   \KET  \vspace{3pt} \\
=>	& D\%W=1@E=D+1@\hook{\<Bfr>(\<latch>=D=\<token>-1)}@\<latch>=D@\<token>=E+1 \\
\not=> & D\%W=1@\<latch>=D@\<Bfr>(\<latch>=D=\<token>-1)
\sloc
\eqnlabel{singletokenringovertaking} 
\end{equation}
The same problem would arise in larger rings, and $W=3$ is the smallest to show it. We fix the problem by weakening the \<latch> interference in the rely to $\<Bfr>(\<latch>=C<\<token>)$.

The initial assertion plus initialisation of \<rauxN> establish the loop invariant \<loopinv>, which is stable because the rely can only increase \<latch> and \<token> until, eventually, $\<latch>\%W=0$; the bound on \<token> in \<loopinv> means that when \<latch> hits that limit, so has \<token>. The precondition of commands \<labc> and \<labd> are stable against the rely because $\<r1>\%W=0$. Command \<labe> is stable against the first line of the rely by the property of \<sofar>: $\<sofar>(\<latch><=\<r1>)=>\widehat{\<latch>}<=\<r1>$. The instantaneous postcondition of \<labe> implies \<loopinv> -- trivially, because at that instant $\<latch>\%W!=0@\<rauxN>\%W=0@\<rauxN>=\<latch>+W-1=\<token>+W-2$.

There is a little bit of \<bo> parallelism: command \<labe>, on one iteration of the loop, is \<bo>-parallel with command \<labd>, on the next iteration. But because one demands $\<Bfr>(\<latch>=\<token>)$ and the other $\<Bfr>(\<latch>=\<token>-1)$, and both require $\<latch>\%W=0$, \<bo> stability is straightforward.

Alglave et al. inserted a \<bo>-inducing barrier and an \<lo>-inducing address dependency into the original algorithm. The \<bo> $\<labd>{->}\<labe>$ constraint corresponds to the barrier; the \<lo> $\<labb>{->}{\<labc>}$ constraint (via $\<gamma>_{t}$) corresponds to the address dependency. There are a lot of other constraints, but they are all implicitly present in Alglave et al.'s presentation, or can be ignored
\begin{itemize*}
\item \<bo> $\<init>{->}\<labb>$ is provided by our supposed test harness; 
\item \<lo> $\<labe>{->}\<labf>$ is implicit on Power because of SCloc;
\item \<lo> $\<delta>_{f}{->}\<labb>$ is embroidered with \<false>;
\item \<lo> $\<labb>{->}\<gamma>$, \<lo> ${\<labc>{->}\<delta>}$ (via $\<gamma>_{t}$) are implicit on Power because of SCreg;
\item constraints to, from and through auxiliary assignments are themselves auxiliary.
\end{itemize*}
So we have added nothing that Alglave et al. did not. It would appear that our constraints are not excessive.

Nevertheless, the single-token single-latch ring is an imperfect abstraction of the original. Although it shows latch and token circulating as they should, it doesn't directly address the zero-token hangup which was the cause of the original bug report, branching on what is in effect auxiliary information in token and latch. And it doesn't show that you can use arrays of latches and tokens to get the right result.

\labelname{labghbar[ghbar]}
\begin{figure}
\centering
% proofs/tokenring4ifthen.proof
$$\cols[c]
\assert{\<init>::\<latch0>=1@\<latch1>=\<latch2>=\<latch3>=0@\<token0>=1@\<token1>=\<token2>=\<token3>=0} \vspace{3pt} \\
\thr{Thread 0}
  \<while>  \<beta>:: \<true> \<do> \\
  \pindent\cols
    \<do> \\
    \pindent\assertd{\<lo> \<init>} |> (\assertd{\<lo> \<labh>} | \assertd{\<lo> \<delta>_{f};\<lo> \<labc>})\ \<labb>:: \<r1>:=\<latch0> \\
    \<until> \assertd{\<lo> \<labb>} \<gamma>:: \<r1>=1; \\
    \!\!\cols[r@{\intfspace}l]
    \assertd{\<lo> \<gamma>_{t}} & \<labc>:: \<latch0>:=0; \\
    \assertd{\<lo> \<gamma>_{t}} & \<labd>:: \<r2>:=\<token0>;
    \sloc \\
    \<if> \assertd{\<lo> \<labd>} \<delta>:: \<r2>!=0 \<then> \\
    \pindent\!\!\cols[r@{\intfspace}l]
      \assertd{\<lo> \<delta>_{t}} & \<labf>:: \<token0>:=0; \\
      \assertd{\<lo> \<delta>_{t}} & \<labg>:: \<token1>:=\<r2>+1; \\
      \assertd{\<bo> \<labf>;\<bo> \<labg>;\<bo> \<labc>} & \<labh>:: \<latch1>:=1
      \sloc \\
    \<fi>
    \sloc \\
  \<od>
\rht
\sloc$$
\caption{Worker 0 of a multi-latch multi-token four-ring, laced}
\figlabel{lacedtokenring}
\end{figure}

\subsection{A multi-latch, multi-token-variable ring}
\var{rauxN[rxN],auxP[xP],auxQ[xQ]}
\Figref{lacedtokenring} shows worker 0 from an abstraction with four workers, each with their own latch and token variables: it's a transparent abstraction from \figref{SCsingletokenring}, if you allow four variables to stand for a four-element array. We chose four workers to show that the proof is considerably independent of the size of the ring. The lacing is mostly straightforward, if a little too tight (from \<labh> to \<labb>) and in one respect a little peculiar (from \<labc>, through the `\<false>' branch of \<delta> to \<labb>): we discuss those oddities later.

The proof obligations, on top of those shared with the single-variable abstraction, are to show that it's impossible for spurious repeated interference `from the past' -- from workers which have already contributed to the current state -- to reverse the change of $\<latch0>$ at command \<labc> and \<token0> at command \<labf>; and to show that interference `from the future' can't prematurely undermine the assignment to \<token1> in command \<labg>. 

Our proof is quite large. We present it in several parts. It depends on auxiliary variables \setvar{auxP} and \setvar{auxQ} -- abbreviated to \<auxP> and \<auxQ> to fit the proof on the page -- as well as \setvar{rauxN} inherited from the single-variable version -- abbreviated to \<rauxN>.
\var{notyet,loopinv,lhseen,tkseen}

\begin{figure}
\centering
$$\cols
  \macro[(N)]{\<notyet>}{\<sofar>(\<auxP><=N@\<auxQ><=N)} \vspace{3pt} \\
  \macro{\<loopinv>}{\BRA \<rauxN>\%W=0@\<rauxN>>=0@\<rauxN>-W<=\<auxQ><=\<auxP><=\<rauxN>@\<notyet>(\<rauxN>) \\@\;
        				  (\<latch0>=0|\<latch0>=1@\<Bfr>(\<token0>=\<rauxN>+1@\<auxP>=\<rauxN>)@\<auxQ>=\<rauxN>)
				     \KET} \vspace{3pt} \\
  \macro{\<lhseen>}{\<rauxN>=\<auxQ>@\<rauxN>\%W=0@\<rauxN>>=0@\<Bfr>(\<auxP>=\<rauxN>)@\<notyet>(\<rauxN>)} \vspace{3pt} \\
  \macro{\<tkseen>}{\<r2>=\<auxQ>+1>0@\<lhseen>} 
\sloc$$
\caption{token four-ring macros}
\figlabel{tokenring4macros}
\end{figure}

\begin{figure}
\centering
% proofs/tokenring4ifthen.proof
$$\cols[c]
\assert{\<init>::\<latch0>=1@\<latch1>=\<latch2>=\<latch3>=0@\<token0>=1@\<token1>=\<token2>=\<token3>=0@\<auxP>=\<auxQ>=0} \vspace{3pt} \\
\thr{Thread 0}
  \guarantee{\interferenceg[A]{A\%W=0@A>=0@\<Bfr>(\<auxP>=A)}{\<latch0>:=0}; \\ \interferenceg[A]{A\%W=0@A>=0@\<Bfr>(\<auxP>=A@\<token0>=A+1)@\<auxQ>=A}{\<token0>:=0}; \\ \interferenceg[A]{A\%W=0@A>=0@\<Bfr>(\<auxP>=A)@\<auxQ>=A}{\<token1>:=A+2}; \\ \interferenceg[A]{A\%W=0@A>=0@\<Bfr>(\<auxP>=A@\<auxQ>=A)}{\<auxP>:=A+1}; \\ \interferenceg[A]{A\%W=0@A>=0@\<Bfr>(\<auxP>=A+1@\<auxQ>=A@\<token1>=A+2)}{\<latch1>,\<auxQ>:=1,A+1}} \vspace{3pt} \vspace{3pt} \\
  \<labx>:: \<rauxN>:=0; \\
  \<while>  \<beta>:: \<true> \<do> \\
  \pindent\cols
    \<do> \\
    \pindent\BRA \assertd{\<lo> \<init>: \<sofar>(\<latch0>=1@\<token0>=1@\<auxP>=\<auxQ>=0);\<lo> \<labx>: \<rauxN>=0} \\
    			 |>\; (\assertd{\<lo> \<labh>: \<loopinv>} | \assertd{\<lo> \<delta>_{f}: \<false>;\<lo> \<labc>: \<true>})
			\KET\ \<labb>:: \<r1>:=\<latch0> \vspace{3pt}\\
    \<until> \assertd{\<lo> \<labb>: \<loopinv>@(\<r1>=1=>\<lhseen>@\<latch0>=1)} \<gamma>:: \<r1>=1; \vspace{3pt}\\
    \!\!\cols[r@{\intfspace}l]
    \assertd{\<lo> \<gamma>_{t}: \<lhseen>@\<latch0>=1} \ \ipre{\<rauxN>\%W=0@\<rauxN>>=0@\<Bfr>(\<auxP>=\<rauxN>)} & \<labc>:: \<latch0>:=0; \vspace{3pt}\\
    \assertd{\<lo> \<gamma>_{t}: \<lhseen>@\<Bfr>(\<token0>=\<auxP>+1)} & \<labd>:: \<r2>:=\<token0>; \vspace{3pt}
    \sloc \\
    \<if> \assertd{\<lo> \<labd>: \<r2>=\<rauxN>+1>0@\<lhseen>@\<Bfr>(\<token0>=\<r2>)} \<delta>:: \<r2>!=0 \\
    \<then> \vspace{3pt}\\
    \pindent\!\!\cols[r@{\intfspace}l]
      \assertd{\<lo> \<delta>_{t}: \<r2>=\<rauxN>+1>0@\<lhseen>@\<Bfr>(\<token0>=\<r2>)} & \vspace{1.3pt}\\ 
      \ipre{\<rauxN>\%W=0@\<rauxN>>=0@\<Bfr>(\<auxP>=\<token0>-1=\<rauxN>)@\<auxQ>=\<rauxN>} & \<labf>:: \<token0>:=0; \vspace{3pt}\\
      \assertd{\<lo> \<delta>_{t}: \<r2>=\<rauxN>+1>0@\<lhseen>} & \vspace{1.3pt}\\ 
      \ipre{\<r2>=\<rauxN>+1@\<rauxN>\%W=0@\<rauxN>>=0@\<Bfr>(\<auxP>=\<rauxN>)@\<auxQ>=\<rauxN>} & \<labg>:: \<token1>:=\<r2>+1; \vspace{3pt}\\
      \assertd{\<lo> \<labc>: \<latch0>=0;\<lo> \<labf>: \<r2>=\<auxP>+1>0@\<token0>=0@\<auxP>\%W=0;\<lo> \<labg>: \<lhseen>} & \<laby>:: \<rauxN>:=\<rauxN>+W; \vspace{3pt}\\
      \Assertd{\<lo> \<labf>: \<token0>=0@\<Bfr>(\<auxP>\%W=0); \\
      		   \<lo> \<labg>: |* N (N\%W=0@N>=0@\<Bfr>(\<auxP>=N)@\<token1>=\<r2>+1=N+2@\<notyet>(N)); \\
		       \<lo> \<labc>: \<latch0>=0@\<auxQ>=\<auxP>@\<auxP>\%W=0;\\
		       \<lo> \<laby>: \<rauxN>=\<auxP>+W@\<auxP>\%W=0} & \<labghbar>:: \<skip>;\vspace{3pt} \\
      \assertd{\<bo> \<labghbar>: \BRA |* N (N\%W=0@N>=0@\<latch0>=0@\<rauxN>=N+W \\@\; \<r2>=N+1@\<Bfr>(\<auxP>=N@\<auxQ>=N)@\<notyet>(N))\KET} & \vspace{1.3pt}\\ 
      \ipre{|* N (N\%W=0@N>=0@\<r2>=N+1@\<rauxN>=N+W@\<Bfr>(\<auxP>=N@\<auxQ>=N))} & \<labz>:: \<auxP>:=\<r2>;\vspace{3pt} \\
      \Assertd{\<bo> \<labz>: \BRA \<rauxN>\%W=0@\<rauxN>>=W@\<Bfr>(\<auxP>=\<rauxN>-W+1\\@\;\<auxQ>=\<rauxN>-W)@\<sofar>(\<auxQ><=\<rauxN>-W)\KET;\\
               \<bo> \<labghbar>: 
               \BRA \<latch0>=0@\<Bfr>(\<token1>=\<r2>+1@\<auxQ>=\<r2>-1=\<rauxN>-W) \\@\; \<sofar>(\<auxP><=\<rauxN>-W+1@\<auxQ><=\<rauxN>-W)@\<rauxN>\%W=0\KET} & \vspace{1.3pt}\\ 
               \ipre{\<r2>=\<rauxN>-W+1@\<rauxN>\%W=0@\<rauxN>>=W@\<Bfr>(\<auxP>=\<auxQ>+1=\<token1>-1=\<r2>)} & \<labh>:: \<latch1>,\<auxQ>:=1,\<r2>
      \sloc \\
    \<fi>
    \sloc \\
  \<od>
\rht
\sloc$$
\caption{Proof of worker 0 of a token three-ring (macros in \figref{tokenring4macros}, rely in \figref{tokenring4rely})}
\figlabel{tokenring4proof}
\end{figure}

\begin{figure}
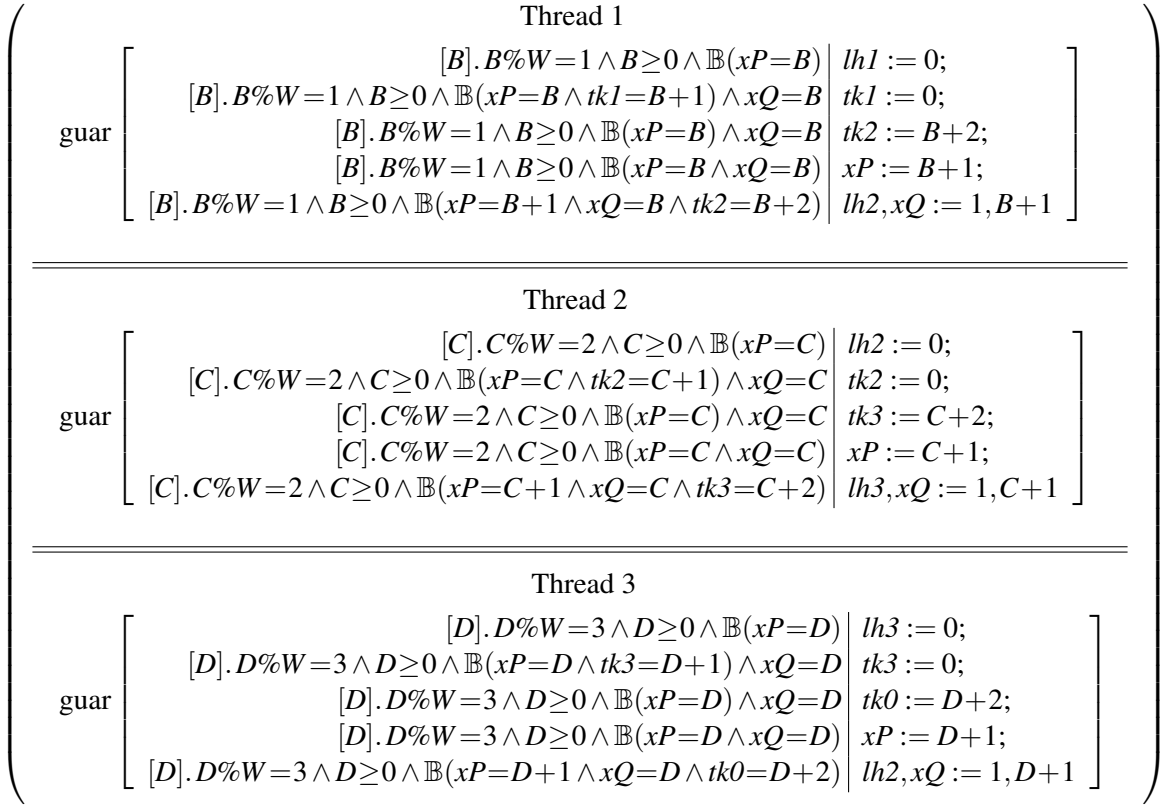

\centering
$$  
\BRA
\thr{Thread 1}
  \guarantee{\interferenceg[B]{B\%W=1@B>=0@\<Bfr>(\<auxP>=B)}{\<latch1>:=0}; \\ 
             \interferenceg[B]{B\%W=1@B>=0@\<Bfr>(\<auxP>=B@\<token1>=B+1)@\<auxQ>=B}{\<token1>:=0}; \\ 
             \interferenceg[B]{B\%W=1@B>=0@\<Bfr>(\<auxP>=B)@\<auxQ>=B}{\<token2>:=B+2}; \\ 
             \interferenceg[B]{B\%W=1@B>=0@\<Bfr>(\<auxP>=B@\<auxQ>=B)}{\<auxP>:=B+1}; \\ 
             \interferenceg[B]{B\%W=1@B>=0@\<Bfr>(\<auxP>=B+1@\<auxQ>=B@\<token2>=B+2)}{\<latch2>,\<auxQ>:=1,B+1}} \vspace{3pt}
\rht \\
\begin{minipage}{340pt} $\cols \hstrut{400pt} \\[-8pt] \hline \hline \sloc$ \end{minipage} \vspace{3pt} \\ 
\thr{Thread 2}
  \guarantee{\interferenceg[C]{C\%W=2@C>=0@\<Bfr>(\<auxP>=C)}{\<latch2>:=0}; \\ 
             \interferenceg[C]{C\%W=2@C>=0@\<Bfr>(\<auxP>=C@\<token2>=C+1)@\<auxQ>=C}{\<token2>:=0}; \\ 
             \interferenceg[C]{C\%W=2@C>=0@\<Bfr>(\<auxP>=C)@\<auxQ>=C}{\<token3>:=C+2}; \\ 
             \interferenceg[C]{C\%W=2@C>=0@\<Bfr>(\<auxP>=C@\<auxQ>=C)}{\<auxP>:=C+1}; \\ 
             \interferenceg[C]{C\%W=2@C>=0@\<Bfr>(\<auxP>=C+1@\<auxQ>=C@\<token3>=C+2)}{\<latch3>,\<auxQ>:=1,C+1}} \vspace{3pt}
\rht \\
\begin{minipage}{340pt} $\cols \hstrut{400pt} \\[-8pt] \hline \hline \sloc$ \end{minipage} \vspace{3pt} \\ 
\thr{Thread 3}
  \guarantee{\interferenceg[D]{D\%W=3@D>=0@\<Bfr>(\<auxP>=D)}{\<latch3>:=0}; \\ 
             \interferenceg[D]{D\%W=3@D>=0@\<Bfr>(\<auxP>=D@\<token3>=D+1)@\<auxQ>=D}{\<token3>:=0}; \\ 
             \interferenceg[D]{D\%W=3@D>=0@\<Bfr>(\<auxP>=D)@\<auxQ>=D}{\<token0>:=D+2}; \\ 
             \interferenceg[D]{D\%W=3@D>=0@\<Bfr>(\<auxP>=D@\<auxQ>=D)}{\<auxP>:=D+1}; \\ 
             \interferenceg[D]{D\%W=3@D>=0@\<Bfr>(\<auxP>=D+1@\<auxQ>=D@\<token0>=D+2)}{\<latch2>,\<auxQ>:=1,D+1}} \vspace{3pt}
\rht
\KET
$$
\caption{Rely of worker 0 of a token four-ring}
\figlabel{tokenring4rely}
\end{figure}

\Figref{tokenring4proof} shows the proof. It's intricate, and there are some new tricks to reveal. 

First, the rely is computed from the guarantees of the other three threads, shown in \figref{tokenring4rely}. There is some \<bo> instability in this collection: the second line of thread 2's guarantee, which assigns \<token2>:=0, interferes with the precondition of the last line of thread 1's guarantee. But in considering thread 0 we can ignore the effect of assignments to \<token2>, because thread 0 takes no notice of \<token2>. Similarly the second line of thread 3's guarantee interferes with the last line of thread 2's guarantee, and can similarly be ignored. 

Second, there is a \<bo> constraint from command \<labz>, which is an auxiliary assignment, to command \<labh>, which is a regular assignment (albeit in parallel with an auxiliary assignment). Auxiliary-to-regular constraints mustn't add any new orderings to the program: that's ok because everything that \<labz> depends on, so does \<labh>. The embroidery on an auxiliary-to-regular constraint can't affect the order of propagation: that's ok because the \<labz>$->$\<labh> embroidery only mentions auxiliary variables. But note that the barrier between \<labz> and \<labh> is an auxiliary device just like an auxiliary assignment: in execution terms it isn't really there.

We've introduced a \<skip> command labelled \<labghbar>, which represents the (possibly necessary) barrier between \<labg> and \<labh> which implements the \<bo> constraint. It's purely an explanatory device.

The proof itself relies mostly on the devices already used in the simpler case of the single-variable token ring. Knowing that \<auxP> and \<auxQ> are between $\<rauxN>-W$ and \<rauxN> stops interference from the past; saying that $\<sofar>(\<auxP><=\<rauxN>@\<auxQ><=\<rauxN>)$ stops interference from the future. We also use $\<Bfr>(\<auxP>=\<rauxN>)$ in the precondition of the commands up to \<labghbar>: the auxiliary assignment to \<auxP> has to be \<bo>-ordered with those commands -- hence after \<labghbar> -- but also \<bo>-ordered before \<labh> -- hence the auxiliary \<bo> constraint.

We can see that if we ignore assignments to variables which aren't used in the proof of thread 0, all that survives of the interference of thread 2 in the rely of thread 0 are its writes to \<auxP> and \<auxQ>, each guarded by a precondition which ensures, by modulo $W$ arithmetic, that they can never effectively interfere with anything in thread 0. That makes it clear that we could make a proof for five, six, \dots or any number of threads. Not a formal argument, of course, but perhaps good enough.

There's one deficiency in this proof. There's no reason why \<labh> should be constrained to elaborate before \<labb> (on the next iteration). It's clear that \<labb> can't read $\<latch0>=1$ until the effect of \<labh>'s assignment $\<latch1>:=1$ has been propagated to thread 2 and the consequences have filtered right round the ring -- so the extra \<lo> ordering used in our proof won't introduce any inefficiency, at least. It's clear also that \<labh> has to elaborate before its own next elaboration, so it would be reasonable to \<lo>-constrain it to occur before \<labghbar> rather than \<labb>. But to do so would introduce all sorts of \<lo> instability into the proof, requiring still more auxiliary gymnastics. Although we believe such a proof would be possible, we haven't yet attempted to make it.

\subsection{Impossible \setorder{so} paths}
\seclabel{impossiblepaths}

The single-latch single-token algorithm of \figref{lacedsingletokenring} and the multi-variable algorithm of \figref{lacedtokenring} both use a conditional \<delta> which is provably certain to deliver \<true>. In the proof of \figref{singletokenring3proof} there is a constraint $\<delta>_{f}{->}\<labb>$ which carries embroidery \<false>. There is a path in the \<so> tree which includes $\<delta>_{f}$ and then \<labb>, even though it is provably impossible that no execution could take that path, so the lacing is required.

In the proof of the single-latch single-token ring that spurious constraint has no effect: its embroidery vanishes in the disjunctive precondition of command \<labb>. In the multi-latch multi-token proof of \figref{tokenring4proof} things are more complicated, because it is conjoined with an \<lo> constraint $\<labc>{->}\<labb>$. Without that extra constraint the proof is invalid, because there is an \<so>-tree path $\<labc>{->}\<labd>{->}\<delta>_{f}{->}\<beta>_{t}{->}\<laba>{->}\<labb>$ and, without the \<lo> constraint, the rules of \<lo> parallelism say that \<labc>, which zeroes the latch, interferes with and destabilises the precondition of control expression \<gamma> on the next iteration of the loop.

We have to include a spurious \<lo> constraint $\<labc>{->}\<labb>$, to counter \<lo>-parallel instability on an \<so>-tree path that can never be followed. We could argue that no damage has been done to the implementation: SCloc is an assumption of our logic, and so the constraint is perhaps implicit and doesn't require implementation; or we might elide it before we present the program to a compiler, since it is in a knot whose overall assertion is \<false>. But we're not sure we can assume the implicit ordering (ARM is going down some frightening roads), and eliding lacing derived syntactically from a verified proof feels wrong.

We don't believe that we have a solution to the problem of impossible \<so> tree paths and their spurious lacing. At present we acknowledge an incompleteness in our logic in that spurious lacing is sometimes required.

\clearpage
\part{And finally}
\thispagestyle{empty}
\clearpage
%!TEX root = ./Paper.tex

\section{Auxiliary variables and registers}
\seclabel{auxiliaries}

Owicki-Gries-style proofs of concurrent algorithms have always needed to use auxiliary state. Our logic is no different. Auxiliary variable names start with \<aux>; auxiliary register names start with \<raux>. The conditions for sound use of auxiliaries are
\begin{enumerate*}
\item an expression assigned to an regular register or variable may not mention an auxiliary register;
\item a value from an auxiliary variable may not be assigned to an regular register;
\item a control expression may not mention an auxiliary register;
\item dependencies to or from auxiliary assignments may not induce any additional orderings between regular commands;
\item dependencies to or from auxiliary assignments may not induce any additional propagation orderings of regular variables.
\end{enumerate*}
In our notation an expression cannot mention a variable, whether auxiliary or regular, and we cannot assign from a variable to a variable. The first three conditions, then, are versions of the conditions on SC Owicki-Gries proofs, but the fourth and fifth are new for weak memory. Together all five conditions ensure that when auxiliaries are deleted from a program the values assigned to regular registers and variables, the lacing of regular commands and control expressions, and the propagation of regular variable writes will be unchanged.

In our Arsenic proof-checker we check the fourth condition by ensuring that for each auxiliary-to-regular constraint, for each of the immediate regular predecessors of the auxiliary command and for each constrained path between each of those predecessors and the regular target of the constraint, there is a regular path to match each auxiliary path. 

We require in our logic that embroidery on all auxiliary-to-regular constraints does not mention regular variables: that ensures the fifth condition and, because of the fourth condition, doesn't introduce any incompleteness. 

In weak memory the problem of auxiliaries is propagation. If we assume that assignments to auxiliaries are instantaneously propagated to other threads we can produce unsound proofs. If we put special orderings in our programs to make them propagate before or after other variables, we violate our fourth condition. Our solution is to allow auxiliary writes to `piggy-back' on regular writes. Auxiliary writes can be made coincidental in elaboration and propagation with a particular regular write and the two can be propagated together as a composite.  

We have two logical devices for auxiliary writes. 
\emph{Auxiliary extension} pretends that the value of an regular variable is a tuple of an regular part and some auxiliary parts where, by convention, the regular part comes first. We allow an extended write and an extended read: 
\begin{align}
x&:=E,\setvar{Eaux},\setvar{Eaux'},\dots \\
r,\setvar{raux},\setvar{raux'},\dots&:=x
\eqnlabel{extendedwrite}
\end{align}
The pretence is that the extended value is propagated as a unit to other threads. This allows simultaneous atomic assignments, but only when involving at most one regular (non-auxiliary) write.

\emph{Composite writes} allow an regular variable to be assigned at the same time as one or more auxiliaries. 
\begin{align}
x,\setvar{aux},\setvar{aux'},\dots:=E,\setvar{Eaux},\setvar{Eaux'},\dots 
\eqnlabel{compositewrite}
\end{align}
The writes generated by the assignment are propagated together as a unit. Variables assigned in a composite write may also be assigned in the normal way, and we don't provide a composite read.

%!TEX root = ./Paper.tex

\section{Inclusion of interference in guarantee (or rely)}

Most of the time an assignment corresponds to a single entry in the guarantee, using the assignment precondition as precondition in the entry. But sometimes an assignment corresponds to more than one entry, and sometimes the contribution of several assignments can be combined into a number of entries. We also have to check inclusion of the interferences of other threads when a thread has an explicit rely.

\word{effect[intf]}
\var{v1,vn}
\definition{Effect of a single interference}{intfeffect}{%!TEX root = ./Paper.tex
$$\cols
\<effect>(\interferenced{Q}{x:=E}) =@= Q @ x'=E @ \<v1>=\<v1>'\;@\;\dots\;@\;\<vn>=\<vn>' \vspace{3pt}\\
\quad \text{where }\<v1>,\dots,\<vn>=\nvec{vs}\backslash{}x \vspace{3pt}\\
\qquad \text{where }\nvec{vs} \text{ are the free variables of the guarantee } \bigcup \text{ free variables of } Q \text{ } \bigcup \text{ } \{x\} \text{.}
\sloc$$}
Inclusion is tested against the disjunction of the effects of the guarantee, including the base case that nothing changes.
\var{g1,gn}
\ruledef{Inclusion of interference}{intfinclusion}{%!TEX root = ./Paper.tex
Interference \interferenced{Q}{x:=E} is included in the guarantee \setword{guar}[\<g1>;\dots;\<gn>] if \vspace{3pt} \\
\hstrut{20pt}$\<effect>(\interferenced{Q}{x:=E})=>\<effect>(\<g1>)|\dots|\<effect>(\<gn>)|(\<v1>=\<v1>'\;@\;\dots\;@\;\<vn>=\<vn>')$\vspace{3pt}\\
\hstrut{30pt} where \<v1>,\dots,\<vn>=\nvec{vs}, the free variables of the guarantee $\bigcup$ free variables of $Q$ $\bigcup$ $\{x\}$.
}
The same rule applies to an explicit rely and the individual entries in other threads' guarantees: the effect of each entry must be included in the effect of the rely.

%!TEX root = ./Paper.tex

\section{Library Abstraction and Optimisation}
\seclabel{LibraryAbstractionandOptimisation}

Our logic, if it explains anything, explains weak-memory hardware to programmers. It doesn't solve the problems that face those defining a language such as C11~\citep{C2011}, and it doesn't solve the problems with the C11 definition uncovered by \citet{battyetalLibraryabstractionC++} because it doesn't deal with library abstraction.

The problem of library abstraction in this context is to define the effects of library code, which operates on shared memory, in such a way that library calls can be embedded in a program that otherwise operates on local memory only, and to be able to reason individually about the effects of such library calls. Our treatment so far doesn't point a way towards library abstraction because the interaction between shared-memory accesses in separate library calls can't be eliminated without heavy \<uo>-inducing barriers. Even though the program outside the library won't cause a problem, the context of other library calls is destabilising, absent those barriers.

Our use of constraints also doesn't sit well with compiler optimisation. We don't know, yet, how optimisations should deal with constraints. The intricacy of our proofs makes it clear that optimisers might have to leave constraints alone unless they have very good meta-level proofs of the properties of their optimisations. That is an argument for library abstraction, of course: if we could isolate library code and constrain optimisations within it, that might be a practicable solution.

The truth is that we have very little to say at this stage about library abstraction and optimisation, and still less to say about how to define programming languages which use one and are subject to the other.

%!TEX root = ./Paper.tex

\section{Deficiencies}
\seclabel{deficiencies}

New Lace logic is work in progress. It doesn't yet have a proof of soundness. %The treatment of stability, in particular, requires proof.

Our logic deals only with variables. We hope to move on to heap using separation.

We cannot deal with library abstraction or compiler optimisation.

Our logic can't handle the RDW litmus test. 

We have never had more than a sketch of a treatment of sychronising assignment (\<lwarx> / \<stwcx> on Power, \<ldrex> / \<strex> on ARM, \<CAS> on x86). We haven't yet attempted to handle synchronising assignment in new Lace. 

Except for the token ring, we haven't yet verified any non-litmus examples.
%!TEX root = ./Paper.tex

\section*{Acknowledgements}
\seclabel{acknowledgements}

Thanks to Peter Sewell, Susmit Sarkar, Mike Dodds, Mark Batty, Scott Owens and John Wickerson for extensive discussions when we first envisaged a program logic for Power. Thanks to Derek Williams and Luc Maranget (and most of those above) for putting up with Bornat's ignorant questions about Power.

Thanks to Peter O'Hearn for helping us to begin to make a program logic. 
Thanks to Noam Riznetsky for telling us about \<since>.
Thanks to Nikolaj Bj\o{}rner, Josh Berdine and Cristoph Wintersteiger for Z3 help. 

Thanks for ever to Crary and Sullivan for pointing the way out of the machine-specificity swamp.

\clearpage
\phantomsection

%!TEX root = ./Paper.tex

\clearpage
\appendix
%!TEX root = ./Paper.tex

\section{Summary of the logic}
\appxlabel{summary}

The notation is described in \tabref{lacedprograms} and \tabref{proglang}. In text input for the Arsenic proof checker, \nonterminal{knot} brackets are ``\texttt{\{*}'' and ``\texttt{*\}}'', and \nonterminal{intfpre} brackets are ``\texttt{[*}'' and ``\texttt{*]}''.

Auxiliary extension of variables, restrictions on auxiliary assignment and auxiliary-to-regular lacing are described in \secref{auxiliaries} and are not summarised here.

\begin{keeptogether}
Constraints run between labelled primitive components -- assignments and control expressions. Constraint lacing may only reinforce the dependencies of the \<so> tree. Where a component occurs more than once in the \<so> tree, it may need a disjunctive constraint constructed according to \defref{constraintcoverage}. Components in loops may require iterated constraints.
\begin{tcolorbox}[before skip=3pt,top=3pt,bottom=3pt,lowerbox=ignored,colframe=blue!50!black,colback=blue!10!white, colbacktitle=blue!20!white]%
%!TEX root = ./Paper.tex
If a component is constrained at all, then for all paths in the \<so> tree to an instance of that component, there must be a disjunct in the component's knot which depends on one of the nodes of that path.%
\end{tcolorbox}
\begin{tcolorbox}[before skip=3pt,top=3pt,bottom=3pt,lowerbox=ignored,colframe=blue!50!black,colback=blue!10!white, colbacktitle=blue!20!white]%
%!TEX root = ./Paper.tex
A component \setlabelname{x} in a loop may have an iterated constraint $\setvar{knot1}|>\setvar{knot2}$. Constraints in \setvar{knot1} must cover all paths from the initial assertion to \setlabelname{x}; constraints in \setvar{knot2} must cover all paths from \setlabelname{x} to \setlabelname{x}.%
\end{tcolorbox}
\end{keeptogether}

\begin{keeptogether}
Interference and elaboration preconditions are derived from knots according to \defref{preconditionfromknot}.
\begin{tcolorbox}[before skip=3pt,top=3pt,bottom=3pt,lowerbox=ignored,colframe=blue!50!black,colback=blue!10!white, colbacktitle=blue!20!white]%
\begin{itemize*}
\item The overall precondition is the conjunction of a knot's embroidery or, if the knot is a disjunction of knots, the disjunction of their overall preconditions.
\item The elaboration precondition is the overall precondition, excluding the embroidery of \<go> constraints, conjoined with the satisfaction (\<sat>) of the overall precondition.
\item The interference precondition is either the overall precondition or, if the knot contains a square-bracketed \nonterminal{intfpre} assertion (see \tabref{lacedprograms}), the \nonterminal{intfpre} assertion, which must be implied by the overall precondition.
\end{itemize*}%
\end{tcolorbox}
\end{keeptogether}

\begin{keeptogether}
The embroidery of a constraint is derived from the postcondition of its source component. Component postconditions are derived according to definitions \ref{def:postconditions} and \ref{def:controlexprpostcondition}. Strongest postconditions are defined in \defref{sp}. 
\begin{tcolorbox}[before skip=3pt,top=3pt,bottom=3pt,lowerbox=ignored,colframe=blue!50!black,colback=blue!10!white, colbacktitle=blue!20!white]%
The postcondition of the initial-state assertion is the assertion itself. The postcondition of \<skip> is its precondition. The postcondition of  $\<assert> P$ is the conjunction of its precondition and $P$. For assignments we use \emph{strongest postconditions}~\citep{DijkstraDisciplineofprogramming1976}.%
\end{tcolorbox}
\begin{tcolorbox}[before skip=3pt,top=3pt,bottom=3pt,lowerbox=ignored,colframe=blue!50!black,colback=blue!10!white, colbacktitle=blue!20!white]%
%!TEX root = ./Paper.tex
A control expression $\<gamma>::E$ with elaboration precondition $P$ has two postconditions, one for each of its outcomes. The $\<gamma>_{t}$ outcome has $P@E$; the $\<gamma>_{f}$ outcome has $P@!E$.%
\end{tcolorbox}
\begin{tcolorbox}[before skip=3pt,top=3pt,bottom=3pt,lowerbox=ignored,colframe=blue!50!black,colback=blue!10!white, colbacktitle=blue!20!white]%
$$\cols[lcl]
\<sp>(P,\;x:=E) &=@=& P[x\backslash \<x'>]@x=E \\
\<sp>(P,\;r:=E) &=@=& P[r\backslash \<r'>]@r=E[r\backslash \<r'>] \\
\<sp>(P,\;r:=x) &=@=& P[r\backslash \<r'>]@r=x
\sloc$$ %
\end{tcolorbox}
\end{keeptogether}

\begin{keeptogether}
Assertions embroidered on stitches must be inherited as described in \ruleref{inheritance}. 
\begin{tcolorbox}[before skip=3pt,top=3pt,bottom=3pt,lowerbox=ignored,colframe=red!50!black,colback=red!10!white, colbacktitle=red!20!white]%
%!TEX root = ./Paper.tex
If a constraint's source postcondition is $P$ and the constraint's embroidery is $Q$ then we require \vspace{-5pt}
\begin{itemize*}
\item on an \<lo> or \<go> constraint, $P=>Q$;
\item on a \<bo> constraint, for some $R$, $P=>R$ and $\<Bfr>(R)=>Q$;
\item on a \<uo> constraint, for some $R$, $P=>R$ and $\<U>(R)=>Q$.
\end{itemize*}%
\end{tcolorbox}
\end{keeptogether}

\begin{keeptogether}
\<Lo> parallelism is defined in \defref{LOparallelism}. Embroidery must be stable against \<lo> parallel (internal) interference by \ruleref{LOstability}. 
\begin{tcolorbox}[before skip=3pt,top=3pt,bottom=3pt,lowerbox=ignored,colframe=blue!50!black,colback=blue!10!white, colbacktitle=blue!20!white]%
An assignment \<laba> is \emph{lo parallel} with a constraint \<labb>$->$\<labc> if tbere is an \<so> path connecting \<laba>, \<labb> and \<labc> in which \<laba> is not constrained to come before \<labb> or after \<labc>.%
\end{tcolorbox}
\begin{tcolorbox}[before skip=3pt,top=3pt,bottom=3pt,lowerbox=ignored,colframe=red!50!black,colback=red!10!white, colbacktitle=red!20!white]%
%!TEX root = ./Paper.tex
Constraint embroidery $P$ is LO stable against \<lo>-parallel assignment $A$ with precondition $Q$ if \vspace{2pt}\\
\hstrut{50pt} $\<sp>(P@Q,\;A)=>P$%
\end{tcolorbox}
\end{keeptogether}

\begin{keeptogether}
Embroidery must be stable against external interference by rules \ref{rule:EXTstability} and \ref{rule:UEXTstability}.
\begin{tcolorbox}[before skip=3pt,top=3pt,bottom=3pt,lowerbox=ignored,colframe=red!50!black,colback=red!10!white, colbacktitle=red!20!white]%
%!TEX root = ./Paper.tex
Constraint embroidery $P$ is EXT stable against \interferenced{Q}{x:=E} from another thread if \vspace{3pt} \\
\hstrut{50pt}$\<sp>(P@\widehat{Q},\;x:=E) => P$ 
\end{tcolorbox}
\begin{tcolorbox}[before skip=3pt,top=3pt,bottom=3pt,lowerbox=ignored,colframe=red!50!black,colback=red!10!white, colbacktitle=red!20!white]%
%!TEX root = ./Paper.tex
Constraint embroidery $P$ is UEXT stable against \interferenced{Q}{x:=E} from another thread if \vspace{3pt}\\
\hstrut{50pt}$\<sp>(\widetilde{P}@Q,\;x:=E) => \widetilde{P}$ 
\end{tcolorbox} 
\end{keeptogether}

\begin{keeptogether}
Hatting (and double-hatting) is defined in definitions \ref{def:hatdefinition}, \ref{def:Usubsthat}, \ref{def:Sincesubsthat}, \ref{def:Sofarsubsthat}, \ref{def:Ouatsubsthat} and \ref{def:coherencehookinghatting}.
\begin{colsdefn}
\widehat{\setvar{var}}				&=& \widehat{\setvar{var}} 						& \text{whether free or bound} \vspace{2pt} \\
\widehat{\setvar{reg}}				&=& \setvar{reg} \vspace{2pt} 					\\
\widehat{\setvar{const}}			&=& \setvar{const} \vspace{2pt} 				\\
\widehat{P\;\setbracket{binop}\;Q} 	&=& \widehat{P}\;\setbracket{binop}\;\widehat{Q}\vspace{2pt} \\
\widehat{\setbracket{unop}\;P} 		&=& \setbracket{unop}\;\widehat{P} 				\vspace{2pt} \\
\widehat{\setbracket{binder}\;v(P)}	&=& \setbracket{binder}\; v(\widehat{P}) 		\vspace{2pt} \\
\widehat{\<Bfr>(P)}					&=& \<Bfr>(P)@\widehat{P} 
 \vspace{5pt} \\
\widehat{\<U>(P)}				&==& \<U>(P)@\widehat{P} \vspace{5pt} \\
%!TEX root = ./Paper.tex
\widehat{P \<since> Q}				&==&	\widehat{P \<since> Q}
 \vspace{5pt} \\
%!TEX root = ./Paper.tex
\widehat{\<sofar>(P)}					&==& \<sofar>(P)
 \vspace{5pt} \\
%!TEX root = ./Paper.tex
\widehat{\<ouat>(P)}					&==&	\widehat{\<ouat>(P)}
 \vspace{5pt} \\
\widehat{v_c(A,B)}					&==&	v_c(\widehat{A},\widehat{B})
\end{colsdefn}
\end{keeptogether}

\begin{keeptogether}
Twiddling (and double-twiddling) is the same as hatting except for \<Bfr> (\defref{Btilde}).
\begin{colsdefn}
%!TEX root = ./Paper.tex
\widetilde{\<Bfr>(P)}				&==& \widetilde{\<Bfr>(P)}

\end{colsdefn}
\end{keeptogether}

\begin{keeptogether}
Stability against quotiented interference is tested by \ruleref{quantifiedstability}.
\begin{tcolorbox}[before skip=3pt,top=3pt,bottom=3pt,lowerbox=ignored,colframe=red!50!black,colback=red!10!white, colbacktitle=red!20!white]%
%!TEX root = ./Paper.tex
$P$ is stable against \interferenced[\nvec{ns}]{Q}{x:=E} if it is stable against \interferenced{Q[\nvec{ns}\backslash\nvec{fs}]}{x:=E[\nvec{ns}\backslash\nvec{fs}]} where $\nvec{fs}$ are fresh names. Similarly for internal interference \interferenced[\nvec{ns}]{Q}{r:=E} and \interferenced[\nvec{ns}]{Q}{r:=x}.
\end{tcolorbox}
\end{keeptogether}

\begin{keeptogether}
The effect of strongest-post substitutions on modalities is defined in definitions \ref{def:Bsubst}, \ref{def:Usubsthat}, \ref{def:Sincesubsthat}, \ref{def:Sofarsubsthat}, and \ref{def:Ouatsubsthat} -- all the same except for \<ouat>. Its effect on coherence assertions is described in \defref{coherencehookinghatting}.
\begin{colsdefn}
\<modality>(P)[x\backslash \<x'>] 	&==& \hook{\<modality>(P)}@P[x\backslash \<x'>] \vspace{2pt}\\
\hook{\<modality>(P)}@P 				&=>& \<modality>(P) \vspace{2pt}  \\
\<modality>(P)[r\backslash \<r'>] 	&==& \<modality>(P[r\backslash \<r'>]) 		\vspace{5pt} \\
\<ouat>(P)[x\backslash{}\<x'>]		&==&	\hook{\<ouat>(P)} \\
\hook{\<ouat>(P)}|P					&<=>&	\<ouat>(P) \\
!\hook{\<ouat>(P)}@!P				&<=>&	!\<ouat>(P) \\
\<ouat>(P)[r\backslash{}\<r'>]	 	&==&	\<ouat>(P[r\backslash{}\<r'>])\vspace{5pt} \\
%!TEX root = ./Paper.tex
(v_c(A,B))[x\backslash \<x'>]		&==& 	v_c(A[x\backslash \<x'>],B[x\backslash \<x'>]) \\
(v_c(A,B))[r\backslash \<r'>]		&==& 	v_c(A[r\backslash \<x'>],B[x\backslash \<r'>])
\end{colsdefn}
\end{keeptogether}

The rely of a thread is the union of the guarantees of all other threads. In forming the rely we must check BO stability with \ruleref{BOstability} between all interferences not from the same thread and not affecting the same variable, and UO stability between all interferences not from the same thread.

\begin{keeptogether}
\<Bo> parallelism is defined in \defref{BOparallel} and BO stability in \ruleref{BOstability}.
\begin{tcolorbox}[before skip=3pt,top=3pt,bottom=3pt,lowerbox=ignored,colframe=blue!50!black,colback=blue!10!white, colbacktitle=blue!20!white]%
%!TEX root = ./Paper.tex
If variable assignments to distinct variables are \<so>- but not \<bo>-ordered then the \<so>-later is \emph{bo parallel} with the \<so>-earlier and will interfere with its interference precondition.%
\end{tcolorbox}
\begin{tcolorbox}[before skip=3pt,top=3pt,bottom=3pt,lowerbox=ignored,colframe=red!50!black,colback=red!10!white, colbacktitle=red!20!white]%
%!TEX root = ./Paper.tex
Interference \interferenced{P}{y:=F} is BO stable against \<bo>-parallel \interferenced{Q}{x:=E} if \vspace{3pt}\\
\hstrut{50pt} $\<sp>(\widehat{P}@\dhatv{Q},\;x:=E)=>\widehat{P}$
\end{tcolorbox}
\end{keeptogether}
\begin{keeptogether}
\<Uo> parallelism is defined in \defref{UOparallel} and UO stability in \ruleref{UOstability}.
\begin{tcolorbox}[before skip=3pt,top=3pt,bottom=3pt,lowerbox=ignored,colframe=blue!50!black,colback=blue!10!white, colbacktitle=blue!20!white]%
If variable assignments are \<so>- but not \<uo>-ordered then the \<so>-later is \emph{uo parallel} with the \<so>-earlier and will interfere with the \<so>-earlier's interference precondition. A variable assignment is also \<uo> parallel with itself.%
\end{tcolorbox}
\begin{tcolorbox}[before skip=3pt,top=3pt,bottom=3pt,lowerbox=ignored,colframe=red!50!black,colback=red!10!white, colbacktitle=red!20!white]%
%!TEX root = ./Paper.tex
Interference \interferenced{P}{x:=E} is UO stable against \<uo>-parallel \interferenced{Q}{y:=F} if \vspace{3pt}\\
\hstrut{50pt}$\<sp>(\widetilde{P}@\dtildev{Q},\;y:=F) => \widetilde{P}$
\end{tcolorbox}
\end{keeptogether}
\begin{keeptogether}
Inclusion in a guarantee is tested by \ruleref{intfinclusion} with the support of \defref{intfeffect}.
\begin{tcolorbox}[before skip=3pt,top=3pt,bottom=3pt,lowerbox=ignored,colframe=red!50!black,colback=red!10!white, colbacktitle=red!20!white]%
%!TEX root = ./Paper.tex
Interference \interferenced{Q}{x:=E} is included in the guarantee \setword{guar}[\<g1>;\dots;\<gn>] if \vspace{3pt} \\
\hstrut{20pt}$\<effect>(\interferenced{Q}{x:=E})=>\<effect>(\<g1>)|\dots|\<effect>(\<gn>)|(\<v1>=\<v1>'\;@\;\dots\;@\;\<vn>=\<vn>')$\vspace{3pt}\\
\hstrut{30pt} where \<v1>,\dots,\<vn>=\nvec{vs}, the free variables of the guarantee $\bigcup$ free variables of $Q$ $\bigcup$ $\{x\}$.
\end{tcolorbox}
\begin{tcolorbox}[before skip=3pt,top=3pt,bottom=3pt,lowerbox=ignored,colframe=blue!50!black,colback=blue!10!white, colbacktitle=blue!20!white]%
%!TEX root = ./Paper.tex
$$\cols
\<effect>(\interferenced{Q}{x:=E}) =@= Q @ x'=E @ \<v1>=\<v1>'\;@\;\dots\;@\;\<vn>=\<vn>' \vspace{3pt}\\
\quad \text{where }\<v1>,\dots,\<vn>=\nvec{vs}\backslash{}x \vspace{3pt}\\
\qquad \text{where }\nvec{vs} \text{ are the free variables of the guarantee } \bigcup \text{ free variables of } Q \text{ } \bigcup \text{ } \{x\} \text{.}
\sloc$$%
\end{tcolorbox}
\end{keeptogether}

\begin{keeptogether}
\<Bfr> and \<U> modalities are restricted in use by \defref{BUrestriction}.
\begin{tcolorbox}[before skip=3pt,top=3pt,bottom=3pt,lowerbox=ignored,colframe=blue!50!black,colback=blue!10!white, colbacktitle=blue!20!white]%
%!TEX root = ./Paper.tex
A propagatable assertion $P$ is one for which $P$ is equivalent to $|/P$. An initial-state assertion must be propagatable. In $\<Bfr>(P)$ and $\<U>(P)$, $P$ must be propagatable.%
\end{tcolorbox}
\end{keeptogether}

\begin{keeptogether}
$|/P$ (positive occurrences) and $||/P$ (negative occurrernces) are defined in \defref{downP}. 
\begin{tcolorbox}[before skip=3pt,top=3pt,bottom=3pt,lowerbox=ignored,colframe=blue!50!black,colback=blue!10!white, colbacktitle=blue!20!white]%
%!TEX root = ./Paper.tex
$$\cols[rcll]
|/P								&==&	P									& \text{if }P\text{ does not have multiple free variables} \\
|/(!P)							&==&	! ||/P 								\\
%%|/(P=>Q)						&==&	|/(!P|Q) 							\\
%%|/(P<=>Q)						&==&	|/((P=>Q)@(Q=>P)) 					\\
|/(\setbracket{unop}\;P)		&==&	\setbracket{unop}\;|/P 				\\
|/(P\;\setbracket{binop}\;Q)	&==& 	|/P\;\setbracket{binop}\;|/Q 		\\
|/(@*v(P))						&==& 	@*v(|/P)							& \text{noting that }v\text{ is a bound variable} \\
|/(|*v(P))						&==& 	|*v(|/P)							& \text{noting that }v\text{ is a bound variable} \vspace{5pt}\\
|/\<Bfr>(P)						&==&	\<Bfr>(|/P)							\\
|/\<U>(P)						&==&	\<U>(|/P)							\\
|/\<Ouat>(P@Q)					&==&	|/\<Ouat>(P)@|/\<Ouat>(Q)			\\
|/\<Ouat>(P|Q)					&==&	|/\<Ouat>(P)| |/\<Ouat>(Q)			\\
|/\<Ouat>(P)					&==&	\setbracket{fresh}\;\setvar{boolv}	\\
|/(P\<since>Q)					&==&	|/P@|/\<Ouat>(Q)					\\
|/\<sofar>(P@Q)					&==&	|/\<sofar>(P)@|/\<sofar>(Q)			\\
|/\<sofar>(P)					&==&	P									\vspace{5pt} \\

||/\<Bfr>(P)					&==&	\setbracket{fresh}\;\setvar{boolv}	\\
||/\<U>(P)						&==&	\setbracket{fresh}\;\setvar{boolv}	\\
||/\<ouat>(P)					&==&	\setbracket{fresh}\;\setvar{boolv}	\\
||/(P\<since>Q)					&==&	\setbracket{fresh}\;\setvar{boolv}	\\
||/\<sofar>(P)					&==&	\setbracket{fresh}\;\setvar{boolv}	
\sloc$$%
\end{tcolorbox}
\end{keeptogether}

\begin{keeptogether}
Coherence is defined in \defref{coherenceaxioms}.
\begin{tcolorbox}[before skip=3pt,top=3pt,bottom=3pt,lowerbox=ignored,colframe=blue!50!black,colback=blue!10!white, colbacktitle=blue!20!white]%
\begin{tabular}{rl}
membership		& $x_c(A,B)=>\<cv>(x)$ \\
irreflexive 	& $x_c(A,B) => A!=B$ \\
transitive 		& $x_c(A,B)@x_c(B,C)=>x_c(A,C)$ \\
antisymmetric 	& $x_c(A,B)=>!x_c(B,A)$ \\         
observed 		& $\<ouat>(\<ouat>(x=A)@x=B)@A!=B@\<cv>(x)=>x_c(A,B)$ \\ 
%%history 		& $x_c(A,B)=>!(\<ouat>(x=A@\<ouat>(x=B)))$
\end{tabular}
\end{tcolorbox}
\end{keeptogether}

\begin{keeptogether}
The program-final assertion is checked according to \defref{PMSassertion}, with the assistance of the additional coherence axiom in \defref{pmscoherenceaxiom}.
\begin{tcolorbox}[before skip=3pt,top=3pt,bottom=3pt,lowerbox=ignored,colframe=blue!50!black,colback=blue!10!white, colbacktitle=blue!20!white]%
%!TEX root = ./Paper.tex
If the final assertions of the threads are $P_{0}\dots P_{n-1}$, in which register occurrences are translated into $(\setvar{threadnum}:::\setvar{reg})$, and the program-final assertion is \<final>, then the  assertion \vspace{5pt}  \\
$
\hstrut{20pt} P_{0}@@@@0 @ P_{1}@@@@1 @ \dots @ P_{n-1}@@@@(n-1) @ (|/P_{0} \;@ |/P_{1}@ \dots @ |/P_{n-1})@@@@n => \<final>@@@@n
$%
\end{tcolorbox}
\begin{tcolorbox}[before skip=3pt,top=3pt,bottom=3pt,lowerbox=ignored,colframe=blue!50!black,colback=blue!10!white, colbacktitle=blue!20!white]%
\begin{tabular}{rl}
pms coherence		& $x_c(A,B)=>\<cv>(x)@x!=A$
\end{tabular}
\end{tcolorbox}
\end{keeptogether}

\begin{keeptogether}
The properties of modalities are set out in definitions \ref{def:Bproperties}, \ref{def:Uproperties}, \ref{def:Sinceproperties}, \ref{def:Sofarproperties} and \ref{def:Ouatproperties}. \<Bfr>, \<U>, \<since> and \<Sofar> share some properties
\begin{colsdefn}
\<modality>(P)			&=>& P  \\
\<modality>(P@Q)			&==& \<modality>(P)@\<modality>(Q)  \\
\<modality>(\<modality>(P))	&==& \<modality>(P) \\
\<modality>(P|Q) 		&<<=& \<modality>(P)|\<modality>(Q)	\\
\<modality>(P) 			&<<=& P						& \text{(if } P \text{ doesn't mention variables)} \\
\<modality>(P|Q) 		&=>& \<modality>(P)|\<modality>(Q)	& \text{(if } P \text{ or } Q \text{ don't mention variables)} \\
(P=>Q)				&=>& \<modality>(P)=>\<modality>(Q)	& \text{(if } P=>Q \text{ is a tautology)}
\end{colsdefn}
to which they add some of their own:
\begin{colsdefn}
\<U>(P)			&=>& \<Bfr>(P)	\\
\<U>(\<Bfr>(P))	&==& \<U>(P)	\\
\<Bfr>(\<U>(P))	&==& \<U>(P)	
\vspace{5pt} \\
P@Q				&=>&	P \<since> Q \\
P \<since> Q	&=>&	P @ \<ouat>(Q)
\vspace{5pt} \\
\<sofar>(P)			&=>& \<U>(P) \\
\<sofar>(\<U>(P))	&=& \<sofar>(P)	\\
\<U>(\<sofar>(P))	&=& \<sofar>(P)	\\
\<sofar>(P) 		&=>& \widehat{P} \\ 
\<sofar>(P) 		&=>& \dhatv{P}
%%\<sofar>(P) 		&=>& \widetilde{P} \\
%%\<sofar>(P) 		&=>& \dtildev{P}

\end{colsdefn}
($\<sofar>(P)$ also implies $\widetilde{P}$ and $\dtilde{P}$).
\end{keeptogether}
\begin{keeptogether}
\<Ouat> is different:
\begin{tcolorbox}[before skip=3pt,top=3pt,bottom=3pt,lowerbox=ignored,colframe=blue!50!black,colback=blue!10!white, colbacktitle=blue!20!white]%
%!TEX root = ./Paper.tex
$$\cols[rcll]
\<ouat>(P)			&<<=& P  \\
\<ouat>(P|Q)		&==& \<ouat>(P)|\<ouat>(Q)  \\
\<ouat>(\<ouat>(P))	&==& \<ouat>(P) \\
\<ouat>(P@Q) 		&=>& \<ouat>(P)@\<ouat>(Q)	\\
\<ouat>(P) 			&=>& P						& \text{(if } P \text{ doesn't mention variables)} \\
\<ouat>(P@Q) 		&<<=& \<ouat>(P)@\<ouat>(Q)	& \text{(if } P \text{ or } Q \text{ don't mention variables)} \\
(P=>Q)				&=>& \<ouat>(P)=>\<ouat>(Q)	& \text{(if } P=>Q \text{ is a tautology)} \vspace{5pt} \\
\<ouat>(P)			&=>& !\<sofar>(!P)
\sloc$$%
\end{tcolorbox}
\end{keeptogether}

\clearpage
%!TEX root = ./Paper.tex

\var{hi,bev,name,boundv,v0,vn,tn,himin,hatI,dhatI}
\word{val,co,fv}
\section{\semanticsmark{} An SMT embedding}
\appxlabel{embedding}

The Arsenic proof checker translates proof obligations into a spatio-temporal logic of threads and events. We use Microsoft Z3~\citep{MouraBjornerZ32008} as the proof-checking oracle.

\begin{figure}
\centering
\includegraphics[scale=0.7]{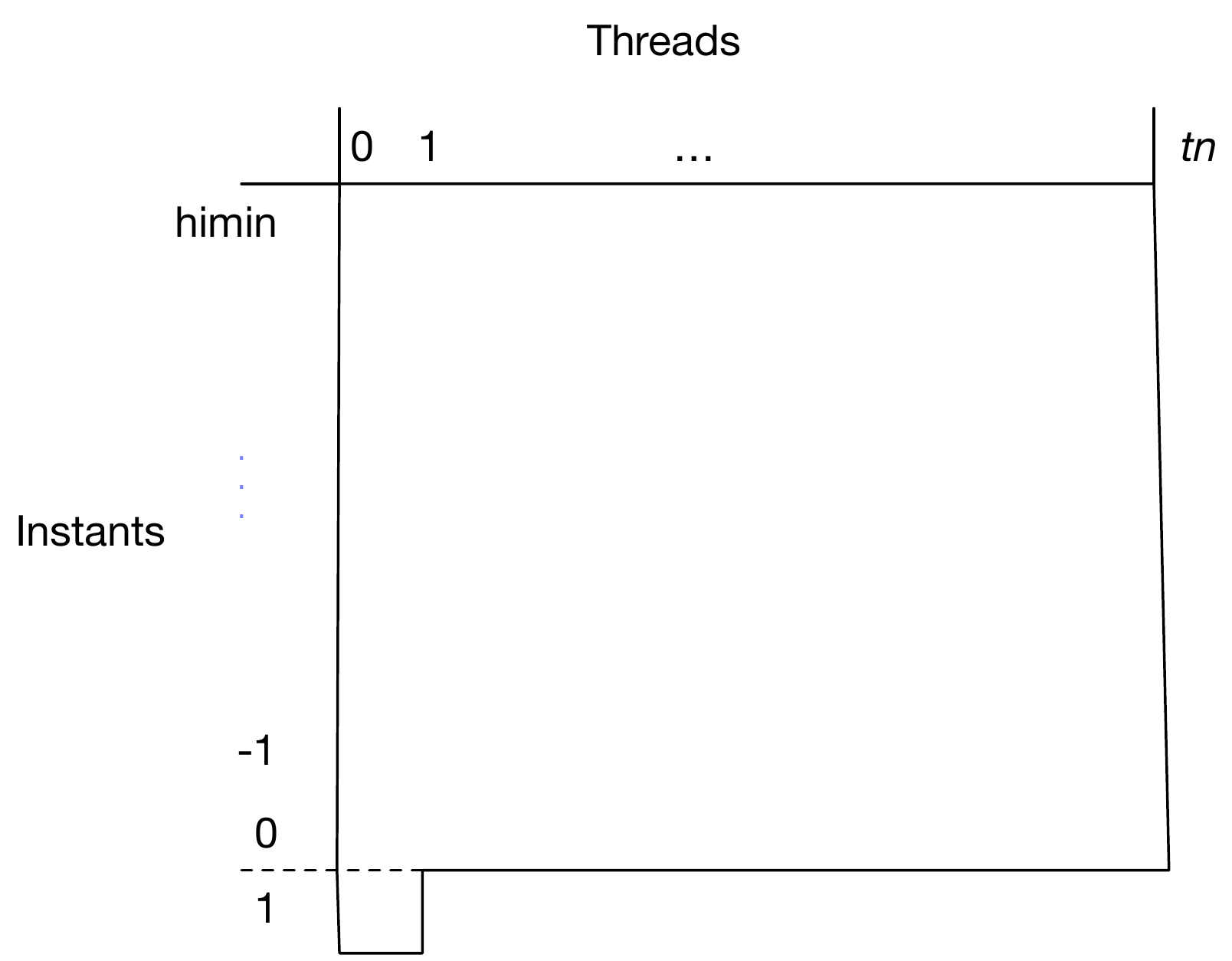}
\caption{threads \ensuremath{\times} instants}
\figlabel{threadsXinstants}
\end{figure}

The semantic domain of the embedding is an array of threads \ensuremath{\times} instants, pictured in \figref{threadsXinstants}. Thread numbers $0\dots\<tn>-1$ don't correspond to thread numbers in the source program, except in the special case of checking the PMS assertion. The instantaneous effect of an assignment, notionally in thread 0, is to create the special $(0,1)$ element: $(0,0)$ is therefore the `hooked' state and $(0,1)$ the plain (unhatted, unhooked) state.

\begin{itemize}
\item If a query contains hooking and/or hatting and/or twiddling, then \<tn> is 2 (threads 0 and 1); with double hatting and/or double twiddling, it is 3; in either case `now' is state 1 and the initial instant is \<himin>.
\item Otherwise, if a query contains thread-numbered formulae $P@@@@n$ -- which can happen in the PMS assertion -- then \<tn> is the greatest thread number plus one, and `now' is 0. 
\item Otherwise, if a query contains \<sofar> or \<U> then \<tn> is 2 and `now' is 0.
\item Otherwise, \<tn> is 1 and `now' is 0.
\end{itemize}
 
In \figref{embedding} $|[P|]^{T}_{I}$ is the value of $P$ in thread $T$ at instant $I$. \<Fandw> -- `far and wide' -- is a modality invented to aid the description of the embedding. 

The basis of the embedding is the treatment of variables. $|[x|]^{T}_{I}$ is $\<val>_{\tau(x)}(x,T,I)$, where $\<val>_{\tau(x)}$ is a function whose result type is the type of $x$ in the source program, and $x$ in the embedded formula is an uninterpreted name. A hatted variable $\<xhat>$ is embedded as $\<val>_{\tau(x)}(x,1,\<hatI>)$, where $\setvar{\<hatI>}$ is constrained to be negative; similarly $\dhatv{x}$ is $\<val>_{\tau(x)}(x,2,\setvar{\<dhatI>})$. Since hatting and twiddling never occur together in the same proof obligation, we can use the same embedding for twiddling as for hatting. Hooked and hatted modalities are shifted from $(1,0)$ to $(0,0)$, $(1,\<hatI>)$ and $(1,\<dhatI>)$ just like variables. Register hooking is renaming.

Translation of `\<since>' is the basis of our embedding of modalities. \<Bfr> and \<U> are versions of `\<since>' which use a special `boundary event' variable \<bev>. To provide $\<sofar>(P)=>\<Bfr>(P)$, $\<sofar>(P)=>\<U>(P)$ and $\<Bfr>(\<true>)$ = $\<U>(\<true>)$ = $\<true>$ the embedding ensures that \<bev> holds at \<himin>, a state constrained to be before any other state.

Coherence assertions are translated very like variables; \<cv> assertions are unchanged.

Because state $(0,1)$, and state $(0,0)$, from which it derives, use different variable embeddings, \<sp> requires careful treatment. The definition of \<sp> and variable assignment that the proof-checker uses makes it clear that only the assigned variable changes value.
\definition{Strongest post \setword{sp} in the presence of temporal assertions}{sptemporal}
{$$\cols
\quad \<sp>(P,\;x:=E) =@= P[x\backslash \<x'>]@x=E@\<y'>=y@\<z'>=z\;@\;\dots \vspace{2pt }\\
\text{where } y,z,\dots \text{ are the variables, other than } x \text{, free in } P \text{.}
\sloc$$}

\begin{figure}
\centering
$$\cols[rcll]
|[x|]^{T}_{I} 							&=& \<val>_{\tau(x)}(x,T,I) \vspace{2pt} \\
|[r|]^{T}_{I} 							&=& r \vspace{2pt} \\
|[\text{constant}|]^{T}_{I} 			&=& \text{constant}  \vspace{8pt} \\

|[\hook{r}|]^{0}_{0} 					&=& r' \vspace{2pt} \\
|[\hook{P}|]^{0}_{1}  					&=& |[P|]^{0}_{0} 			& \text{variables and modalities} \vspace{8pt} \\

|[\widehat{P}|]^{0}_{1} 				&=& |[{P}|]^{1}_{\<hatI>} 	& \text{variables and modalities} \vspace{2pt} \\
|[\dhat{P}|]^{0}_{1} 					&=& |[P|]^{2}_{\<dhatI>} 	& \text{variables and modalities} \vspace{8pt} \\

|[\tilde{P}|]^{0}_{1} 					&=& |[{P}|]^{1}_{\<hatI>} 	& \text{variables and modalities} \vspace{2pt} \\
|[\dtilde{P}|]^{0}_{1} 					&=& |[P|]^{2}_{\<dhatI>} 	& \text{variables and modalities} \vspace{8pt} \\

|[P \<since> Q|]^{T}_{I} 				&=& |*j\BRA \<himin><=j<=I@|[Q|]^{T}_{j} @ @*j'\BRA j<=j'<=I=>|[P|]^{T}_{j'}\KET\KET 														\vspace{8pt} \\

|[\<Bfr>(P)|]^{T}_{I} 					&=& |[P \<since> \<bev>|]^{T}_{I} \vspace{2pt} \\
|[\<U>(P)|]^{T}_{I} 					&=& |[\<Fandw>(P) \<since> \<bev>|]^{T}_{I} \vspace{8pt} \\

|[\<sofar>(P)|]^{T}_{I} 				&=& @*j(\<himin><=j<=I=>|[\<Fandw>(P)|]^{T}_{j}) \vspace{2pt} \\
|[\<ouat>(P)|]^{T}_{I} 					&=& |*j(\<himin><=j<=I@|[P|]^{T}_{j}) \vspace{8pt} \\

|[\<Fandw>(P)|]^{0}_{1} 				&=& |[P|]^{0}_{1} \vspace{2pt} \\
|[\<Fandw>(P)|]^{T}_{I} 				&=& @*t(0<=t<\<tn>=>|[P|]^{t}_{I}) \vspace{8pt} \\

|[\setbracket{binder}\;x(P)|]^{T}_{I} 	&=& \setbracket{binder}\;x(|[P|]^{T}_{I}) \vspace{2pt} \\
|[P\;\setbracket{binop}\;Q|]^{T}_{I} 	&=& |[P|]^{T}_{I}\;\setbracket{binop}\;|[Q|]^{T}_{I} \vspace{2pt} \\
|[\setbracket{unop}\;P|]^{T}_{I} 		&=& \setbracket{unop}\;|[P|]^{T}_{I} \vspace{8pt} \\

|[P@@@@{n}|]^{T}_{0}					&=& |[P|]^{n}_{0} \vspace{8pt} \\

|[x_{c}(A,B)|]^{T}_{I}					&=& \<co>_{\tau(x)}(x,|[A|]^{T}_{I},|[B|]^{T}_{I})\vspace{2pt} \\
|[\<cv>(x)|]^{T}_{I}					&=& \<cv>(x)
\sloc$$
\caption{Rules of embedding}
\figlabel{embedding}
\end{figure}

\begin{figure}
\centering
$$\cols[rcll]
|[(P\<since>Q)\<since>R|]^{T}_{I}	&=& |[P\<since>\;((P\<since>Q)@R)|]^{T}_{I} \vspace{2pt} \\
|[\<sofar>(P)\<since>Q|]^{T}_{I} 	&=& |[\<sofar>(P)@\<ouat>(Q)|]^{T}_{I} \vspace{2pt} \\
|[\<U>(\<U>(P))|]^{T}_{I} 			&=& |[\<U>(P)|]^{T}_{I} \vspace{2pt} \\
|[\<U>(\<Bfr>(P))|]^{T}_{I} 		&=& |[\<U>(P)|]^{T}_{I} \vspace{2pt} \\
|[\<U>(\<sofar>(P))|]^{T}_{I} 		&=& |[\<sofar>(P)|]^{T}_{I} \vspace{2pt} \\
|[\<U>(P\<since>Q)|]^{T}_{I} 		&=& |[\<Fandw>(P)\<since>(\<Fandw>(P)@Q)|]^{T}_{I} \vspace{2pt} \\
|[\<sofar>(\<U>(P))|]^{T}_{I}		&=& |[\<sofar>(P)|]^{T}_{I} \vspace{2pt} \\
|[\<sofar>(\<Bfr>(P))|]^{T}_{I}		&=& |[\<sofar>(P)|]^{T}_{I} \vspace{2pt} \\
|[\<sofar>(\<sofar>(P))|]^{T}_{I}	&=& |[\<sofar>(P)|]^{T}_{I} 
%%|[\<U>(\<sofar>(P))|]^{N,T}_{I}		&= |[\<sofar>(P)|]^{N,T}_{I} \vspace{2pt} \\
%%|[\<sofar>(\<U>(P))|]^{N,T}_{I}		&=& |[\<sofar>(\<Fandw>(P))|]^{N,N}_{I} \vspace{2pt} \\
%%|[\<U>(!\<sofar>(P))|]^{N,T}_{I}		&= |[!\<sofar>(P)|]^{N,T}_{I} \vspace{2pt} \\
\sloc$$
\caption{Embedding equivalences}
\figlabel{embeddingequivalences}
\end{figure}
To reduce the stress on Z3 caused by alternation of $|*$ and $@*$ quantifiers we employ some equivalences, shown in \figref{embeddingequivalences}, many of which derive from definitions \ref{def:Uproperties}, \ref{def:Sinceproperties}, \ref{def:Sofarproperties} and \ref{def:Ouatproperties}.

%%\subsection{`Pure' and temporally-constant assertions}
%%
%%From the embedding, plus the assumption $@*t(|*i(|[\<bev>|]^{t}_{i}))$, it follows that a modality applied to an assertion which is temporally constant is equivalent to the assertion. `Pure' assertions -- those which which don't mention variables -- are temporally constant for sure. Hence some of the rules in \secref{embroideryandB}.

\end{document}